\documentclass[a4paper,12pt]{article}
\usepackage[utf8]{inputenc}
\usepackage{mathtools}
\usepackage{amsmath}
\usepackage[affil-it]{authblk}
\usepackage{float}
\usepackage{subcaption}
\usepackage{verbatim}
\usepackage[english]{babel}

\usepackage{geometry}

\usepackage{graphicx}
\graphicspath{{figures/}}

\usepackage[draft]{hyperref}
\hypersetup{colorlinks=true,urlcolor=blue,citecolor=blue,linkcolor=blue}  

\usepackage[bottom]{footmisc} 

\usepackage{natbib}
\newcommand{\killpunct}[1]{} 

\usepackage[center]{titlesec}
\titleformat{\section}{\normalfont\fontsize{12}{15}\bfseries}{\thesection}{1em}{\centering}
\titleformat{\subsection}{\normalfont\fontsize{12}{15}\bfseries}{\thesubsection}{1em}{\centering}
\makeatletter
\g@addto@macro\normalsize{%
	\setlength\abovedisplayskip{5pt}
	\setlength\belowdisplayskip{11pt}
	\setlength\abovedisplayshortskip{5pt}
	\setlength\belowdisplayshortskip{11pt}
}
\makeatother

\title{Funding liquidity, credit risk and unconventional monetary policy in the Euro area: A GVAR approach}
\author{Graziano Moramarco\thanks{Contact: Department of Economics, University of Bologna - Piazza Scaravilli 2, 40126 Bologna, Italy; e-mail: graziano.moramarco@unibo.it.  The author gratefully acknowledges financial support from the Italian Ministry of Education, University and Research (PRIN 2017, Grant 2017TA7TYC).}} 
\affil{Department of Economics, University of Bologna \\ and Johns Hopkins University SAIS Europe}
\date{}

\begin{document}

	\maketitle
	
	\begin{abstract}
		
		This paper investigates the transmission of funding liquidity shocks, credit risk shocks and unconventional monetary policy within the Euro area. To this aim, we estimate a financial GVAR model for Germany, France, Italy and Spain on monthly data over the period 2006-2017. The interactions between repo markets, sovereign bonds and banks' CDS spreads are analyzed, explicitly accounting for the country-specific effects of the ECB's asset purchase programmes. Impulse response analysis signals marginally significant core-periphery heterogeneity, flight-to-quality effects and spillovers between liquidity conditions and credit risk. Simulated reductions in ECB programmes tend to result in higher government bond yields and bank CDS spreads, especially for Italy and Spain, as well as in falling repo trade volumes and rising repo rates across the Euro area. However, only a few responses to shocks achieve statistical significance.
		
	\end{abstract}
	

	\pagebreak
	
	\section{Introduction}
	
	Large increases in government bond yield spreads and banks' CDS spreads were among the main symptoms of the Euro area debt crisis. Both heightened credit risk and liquidity concerns contributed to fuel the sovereign-banking vicious circle. In response to market distress, the European Central Bank (ECB) implemented unprecedented policy interventions. Restoring adequate liquidity conditions was a stated goal of these interventions. For instance, the ECB launched the Securities Markets Programme (SMP) to ensure ``depth and liquidity" to ``dysfunctional" markets, and extended its long-term refinancing operations (LTRO) to ``support bank lending and liquidity".\footnote{See \url{https://www.ecb.europa.eu/press/pr/date/2010/html/pr100510.en.html}.}
	
	In this paper, we address the empirical relationships between funding liquidity, sov\-\-ereign bond yields and credit risk in the banking industry within the Euro area, as well as the effects of the ECB's asset purchase programmes. To this aim, we estimate a financial global vector autoregressive (GVAR) model (\citealt{pesaranetal04}) for Germany, France, Italy and Spain on monthly data from 2006 to 2017.
	The endogenous variables of the model are interest rates and trading volumes in the repo markets, sovereign bond yields and CDS spreads for large banks.
	Trading volume in the repo market is taken as a proxy for funding liquidity, while CDS spreads isolate credit risk components in the banking sector, which are closely related to sovereign risk in the Euro area. Also, the presence of government yields in the model allows to characterize the empirical behavior of the market value of collateral. Collateralized borrowing, especially through the repo market, is an important source of funding for traders and financial institutions, and is a key determinant of market liquidity.  

	To account for the effects of the ECB policy interventions, we include both the policy interest rate and the amounts of assets purchased by the ECB as weakly exogenous variables in all four country-specific models composing the GVAR.
	We distinctly consider the different asset purchase programmes carried out by the ECB.
	First, the long-term refinancing operations (LTRO), which during the global financial crisis were extended to 6-month and 12-month maturities. 
	Second, the Securities Markets Programme (SMP), introduced in May 2010.
	In September 2012 the ECB also launched the Outright Monetary Transactions (OMT) programme, promising unlimited government bond purchases for distressed countries that complied with the rules of the European Stability Mechanism (ESM). However, no OMTs have been actually implemented, so that there are no purchased volumes associated with this programme. Nonetheless, since the mere announcement of the programme is thought to have been considerably effective in reducing yield spreads, we include a dummy variable which takes on value 1 in all periods following the announcement that the ECB would undertake outright transactions in secondary sovereign bond markets (i.e., from August 2012) and value 0 in previous periods. 
	Finally, in late 2014 the ECB launched its asset purchase programme (APP), which was then expanded in early 2015. The APP consists of four sub-programmes: the third covered bond purchase programme (CBPP3, Oct. 2014), the asset-backed securities purchase programme (ABSPP, Nov. 2014), the public sector purchase programme (PSPP, Mar. 2015) and the corporate sector purchase programme (CSPP, Jun. 2016). We include in the GVAR model the total APP volumes, summing up the four sub-programmes.
	
	We perform impulse response analysis to assess the dynamic cross-country effects of shocks to repo markets, sovereign bond yields and bank CDS spreads. Moreover, we simulate reductions in the ECB programmes in order to quantify the effects of policy interventions on the endogenous variables. 
	The results suggest marginally significant core-periphery heterogeneity and flight-to-quality phenomena, as well as spillover effects between funding liquidity and credit risk, and bank-sovereign feedback effects in Italy and Spain. 
	Also, reductions in the ECB programmes tend to be associated with increases in government yields and bank CDS spreads, especially in Italy and Spain, and with declines in funding liquidity and increases in repo rates in all countries. 
	However, the estimates are subject to substantial uncertainty: less than one fifth of the estimated responses to shocks are statistically significant at the 10\% level in at least one period, and very few achieve significance at the 5\% level.
	
	The paper relates to the literature on liquidity in times of crisis (e.g., \citealt{brunnermeierpedersen08};  \citealt{adrianetal16}), on the effects of central banks' asset purchases on yields (e.g., \citealt{krishnamurthyetal17}; \citealt{depooteretal16}) and on financial GVAR models (\citealt{giesetuxen07}; \citealt{chudickfratzscher11,chudickfratzscher12}; \citealt{grosskok13}; \citealt{grayetal13}). 
	
	The remainder of the paper is organized as follows: Section 2 introduces the data, Section 3 illustrates the GVAR model, Section 4 presents the results, and Section 5 concludes.

	\section{Data}
	
	The country-specific financial variables considered in our analysis are repo interest rates, repo trading volumes, 10-year government bond yields and 1-year bank CDS indices. 
	
	Time series of repo interest rates and trading volumes have been provided by the MTS/BrokerTec RepoFunds Rate initiative, using data from repo transactions executed on either the BrokerTec or the MTS electronic platforms.\footnote{\url{https://www.mtsmarkets.com/repofunds-rate.}} Repo rates are volume-weighted average rates and represent the effective cost of funding for the majority of repo trades in each of the relevant sovereign bond markets. Data for France, Germany and Italy start in 2006. Data for Spain are only available from 2012 onwards, so that including them would dramatically reduce the sample length. For this reason, in the sub-model for Spain we only include the 10-year government bond yield and the 1-year bank CDS index as domestic endogenous variables. 
	
	Data on 10-year sovereign bond yields are from the OECD.
	
	Country-specific 1-year bank CDS indices are calculated as simple averages of individual CDS spreads across large domestic banks. For Germany, we consider Deutsche Bank and Commerzbank. For France: Cr\'edit Agricole, BNP Paribas and Soci\'et\'e G\'en\'erale. For Italy: Mediobanca, Unicredit, Intesa Sanpaolo and Monte dei Paschi di Siena. For Spain: Santander, Banco Popular Espa\~{n}ol and Banco Sabadell. All data on bank-specific CDS are provided by Datastream.
	
	In each country model, Euro-area-wide monetary policy variables are included as weakly exogenous. These reflect both conventional and unconventional monetary policy operations. First of all, we include the policy rate, i.e., the interest rate on the ECB's main refinancing operations. Cumulative volumes of assets purchased by the ECB under its purchase programmes are also considered. 
	
	In every country-specific model (VARX*) composing the GVAR, domestic macroeconomic variables are related to their foreign counterparts. These foreign variables are built as cross-country weighted averages. The weights we use are based on data on gross bilateral financial positions. In particular, we use total portfolio liability data from the IMF's Coordinated Portfolio Investment Survey (CPIS) of December 2015. As an example, when building foreign variables for Italy, the weight assigned to Germany is equal to the amount of Italian liabilities towards Germany, as a share of the total liabilities of Italy towards France, Germany and Spain (hence, the weights sum to 1).

	Data on repos, sovereign bonds, CDS spreads and the ECB policy rate are available at a daily frequency. Data on ECB programmes are available at a weekly frequency.
	To smooth out the high volatility exhibited by daily/weekly data, we conduct our analysis on monthly data, obtained by averaging the original higher-frequency data within the month. The estimation sample spans the period from January 2006 to March 2017. 
	As regards variable transformations, repo trading volumes are included in logarithms, while the ECB's purchased assets are transformed as follows: given the level $x_t$, the variable $\ln(1+x_t)$ is used in the model (simple log cannot be used, since the level can assume value 0). The other variables are included in levels.

	\section{The GVAR model}
	
	\subsection{The model}
	
	First introduced by \citet{pesaranetal04}, the GVAR model results from the aggregation of country-specific VARX* models.
	Consider a generic country $i$, with $i=1,...,N$, where $N$ is the total number of countries in the GVAR.
	Denote with $\mathbf{x}_{it}$ the  $k_i \times 1$ vector of domestic macroeconomic variables of country $i$ at time $t$, with $\mathbf{x}_{it}^\ast$ the $k_{i}^* \times 1$ vector of foreign variables and with $\mathbf{d}_t$ the vector of common exogenous variables. In this paper,  $\mathbf{x}_{it}$ comprises domestic repo rates and trading volumes (except for Spain), the 10-year government bond yield and the 1-year bank CDS index, while $\mathbf{d}_t$ comprises the ECB policy rate, the amounts of assets purchased under the LTRO, SMP and APP and a dummy variable for the OMT programme. 
	The VARX* model for country $i$ can be written as:
	\begin{equation}\label{eq:VARX} 
		\mathbf{x}_{it} = \mathbf{a_0}_{i} + \mathbf{a_1}_{i} {\it t} + \mathbf{\Phi}_{i} \mathbf{x}_{i,t-j} + \mathbf{\Lambda}_{0i} \mathbf{x}_{i,t}^\ast + \mathbf{\Lambda}_{1i} \mathbf{x}_{i,t-1}^\ast + \boldsymbol{\Psi}_{0i} \mathbf{d}_{t} + \boldsymbol{\Psi}_{1i} \mathbf{d}_{t-1} + \boldsymbol{\nu}_{it}
	\end{equation} 
	where $\mathbf{a_0}_{i}$ and $\mathbf{a_1}_{i}$ are vectors of deterministic constants and trends, respectively, $\mathbf{\Phi}_{i}$, $\mathbf{\Lambda}_{li}$ and $\boldsymbol{\Psi}_{li}$, with $l=0,1$, are matrices of parameters and $ \boldsymbol{\nu}_{it} $ is the vector of i.i.d. errors.
	Let's call $ \textstyle k=\sum_{i}^{N} k _{i}$ the total number of national variables in the global economy. Domestic and foreign variables can be expressed in terms of the $k \times 1$ global vector of stacked endogenous variables $\mathbf{x}_t$:
	\begin{equation}
		\begin{pmatrix}
			\mathbf{x}_{it} \\
			\mathbf{x}_{it}^\ast
		\end{pmatrix}
		= 
		\mathbf{W}_i 
		\begin{bmatrix}
			\mathbf{x}_{1t} \\
			\bf \vdots \\
			\mathbf{x}_{Nt} 
		\end{bmatrix} 
		= \mathbf{W}_i \mathbf{x}_t
	\end{equation}
	where $\mathbf{W}_i$ is the $ (k_i + k_i^*)  \times k $ matrix of weights based on bilateral financial positions.

	Stacking all country-specific VARX* models delivers a vector autoregressive representation of the global economy:
	\begin{equation}
		\mathbf{G} \mathbf{x}_t = \mathbf{a}_0 + \mathbf{a}_1 {\it t} + \mathbf{H} \mathbf{x}_{t-1} + \boldsymbol{\Psi}_0 \mathbf{d}_t  + \boldsymbol{\Psi}_1 \mathbf{d}_{t-1} + \boldsymbol{\nu}_{t}
	\end{equation} 
	where $\mathbf{a_0} = \left( \mathbf{a_0}'_{1}, \mathbf{a_0}'_{2}, \cdots, \mathbf{a_0}'_{N} \right)' $ and $\mathbf{a_1}=\left( \mathbf{a_1}'_{1}, \mathbf{a_1}'_{2}, \cdots, \mathbf{a_1}'_{N} \right)'$ are the $k \times 1$ stacked vectors of global constants and trends, respectively, $ \boldsymbol{\nu}_{t} = \left( \boldsymbol{\nu}'_{1t}, \boldsymbol{\nu}'_{2t}, \cdots, \boldsymbol{\nu}'_{Nt} \right)' $ is the $k \times 1$ stacked vector of errors and
	\begin{gather}
		\hspace{-10pt}  
		\mathbf{G} = \begin{pmatrix} \left(\mathbf{I}_1, -\boldsymbol{\Lambda}_{01} \right) \mathbf{W}_1 \\  \left(\mathbf{I}_2, -\boldsymbol{\Lambda}_{02} \right) \mathbf{W}_2 \\  \vdots \\  \left(\mathbf{I}_N, -\boldsymbol{\Lambda}_{0N} \right) \mathbf{W}_N \end{pmatrix},
		\hspace{10pt}
		\mathbf{H} = \begin{pmatrix} \left(\boldsymbol{\Phi}_{11}, \boldsymbol{\Lambda}_{11} \right) \mathbf{W}_1 \\  \left(\boldsymbol{\Phi}_{12}, \boldsymbol{\Lambda}_{12} \right) W_2 \\ \vdots \\  \left(\boldsymbol{\Phi}_{1N}, \boldsymbol{\Lambda}_{1N} \right) \mathbf{W}_N \end{pmatrix},
		\hspace{10pt}
		\boldsymbol{\Psi}_0 = \begin{pmatrix} \boldsymbol{\Psi}_{01} \\  \boldsymbol{\Psi}_{02} \\ \vdots \\  \boldsymbol{\Psi}_{0N} \end{pmatrix},
		\hspace{10pt}
		\boldsymbol{\Psi}_1 = \begin{pmatrix} \boldsymbol{\Psi}_{11} \\  \boldsymbol{\Psi}_{12} \\ \vdots \\  \boldsymbol{\Psi}_{1N} \end{pmatrix}
	\end{gather}    
	\noindent The reduced form of the global model is obtained by inverting the $\mathbf{G}$ matrix:
	\begin{equation}\label{eq:GVAR}
		\mathbf{x}_t = \mathbf{c_0} + \mathbf{c_1} {\it t} + \mathbf{F} \mathbf{x}_{t-1} + \boldsymbol{\Gamma}_0 \mathbf{d}_t + \boldsymbol{\Gamma}_1 \mathbf{d}_{t-1} + \boldsymbol{\epsilon}_{t}
	\end{equation}    
	\noindent where:
	\begin{gather}
		\begin{matrix}
			\bf c_0 = G^{-1} a_0, & \quad \bf c_1 = G^{-1} a_1, &  \quad \bf F = G^{-1} H,
		\end{matrix}\\[0.0cm]
		\begin{matrix}
			\boldsymbol{\Gamma}_0 = \mathbf{G}^{-1} \boldsymbol{\Psi}_0,  \quad \boldsymbol{\Gamma}_1 = \mathbf{G}^{-1} \boldsymbol{\Psi}_1
		\end{matrix}		
	\end{gather}
	and $  \boldsymbol{\epsilon}_{t}= \mathbf{G^{-1}} \boldsymbol{\nu}_{t}$ are the reduced-form global shocks. Finally, let $\boldsymbol{\Sigma}$ denote the covariance matrix of $\boldsymbol{\nu}_{t}$.

	\subsection{Impulse response analysis}
	The tool typically used to analyze the dynamic properties of a GVAR model consists in the generalized impulse response functions (GIRF), which trace out the response of the model at different future periods to a shock in one variable of one country at time $t$. This shock is assumed to equal one estimated standard error of the variable. The estimated error covariance matrix is then used to assign consistent shocks to all other variables in the model at time $t$. Finally, the dynamic properties of the GVAR determine how the variables' responses to the initial shock evolve through time.

	Formally, the global vector of $h$-step-ahead GIRF with respect to a shock in the endogenous variable $q$ in country $i$ at time $t$ is defined as:
	\begin{equation}
		\mathbf{GIRF}_{iq}(h) = E(\mathbf{x}_{t+h} | \nu_{iqt} = \sigma_{iq}, \mathcal{I}_{t-1} ) -  E(\mathbf{x}_{t+h} | \mathcal{I}_{t-1} ) 
	\end{equation}
	where $\nu_{iqt}$ denotes the element of $\boldsymbol{\nu}_{t}$ associated with variable $q$ in country $i$, $\sigma_{iq}$ is the standard deviation of $\nu_{iqt}$, and $\mathcal{I}_{t-1} = (\mathbf{x}_{t-1},\mathbf{x}_{t-2},...) $ is the information set at time $t-1$.
	Assuming that variable $q$ in country $i$ is the $j$-th element of the global endogenous vector $\mathbf{x}_t$, the GIRF vector can be expressed as:  
	\begin{equation}
		\mathbf{GIRF}_{j}(h) = \frac{1}{\sigma_{j}} \mathbf{F}^h \mathbf{G}^{-1} \boldsymbol{\Sigma} \mathbf{s}_j
	\end{equation}
	where $\sigma_{j} \equiv \sigma_{iq}$ and $\mathbf{s}_j$ is a $k \times 1$ selection vector, i.e., a vector in which the $j$-th element is equal to 1 and all other elements are zeros.
	
	Analogously, generalized impulse response functions can be derived with respect to shocks to the exogenous variables. However, this would require specifying a dynamic process for the exogenous variables. As will be explained in the following section, we adopt an alternative approach to assess the impact of shocks to the ECB asset purchase programmes.

	\section{Results}	
	This section presents the results in terms of dynamic responses to shocks in the variables of the GVAR model.\footnote{The GVAR is estimated in error correction form (see \citealt{pesaranetal04}; \citealt{deesetal07a}). Details on cointegration properties are available upon request.}
	When interpreting the results, it is important to stress that the response functions should not be given any causal or structural interpretation. They just capture empirical dynamic correlations between the variables.
	
	The effects of shocks to the endogenous variables are assessed by means of GIRF functions computed over a horizon of 24 months after the time of the shock, which is denoted as time 0.
	To assess the responses of the endogenous variables to shocks in the exogenous variables, we follow a different approach. We simulate the model over the period 2015M1-2017M3 under alternative ``shock scenarios" for the exogenous variables. In each scenario, a single ECB programme is shocked. Shocking the LTRO, SMP and APP consists in assuming that the amount of assets held by the ECB under the relevant programme was 50\% lower than the amount actually held. The shock scenario for the OMT programme is defined by setting the OMT dummy variable to 0 over the 2015M1-2017M3 period, thereby assuming the termination of the programme. Finally, the responses of the endogenous variables are computed as the differences between the time profiles of the variables in the shock simulation and the baseline profiles, i.e., the dynamic forecasts of the model over the 2015M1-2017M3 period obtained by considering the actual values for the ECB variables. 
	
	For all impulse response functions, we report point estimates as well as 90\% confidence intervals obtained using the bootstrap procedure for GVAR models developed by \citet{deesetal07a, deesetal07b}, with 1000 replications.
	Confidence intervals are generally very large, and only a small share of responses are statistically significant at the 10\% level. At the 5\% level, the only significant effects include the responses of Euro area banks' CDS spreads to Italian repo rate shocks, the response of the Italian repo rate to a LTRO reduction, and the responses of CDS spreads to a simulated termination of the OMT programme, all with positive sign.
	
	\subsection{Shocks to the endogenous variables}

\begin{figure}[H]
		\centering		
		\caption{GIRF of repo trade volume, shock to GER government yield}
		\begin{minipage}{.5\textwidth}
			\hspace{-15pt}
			\includegraphics[scale=0.15]{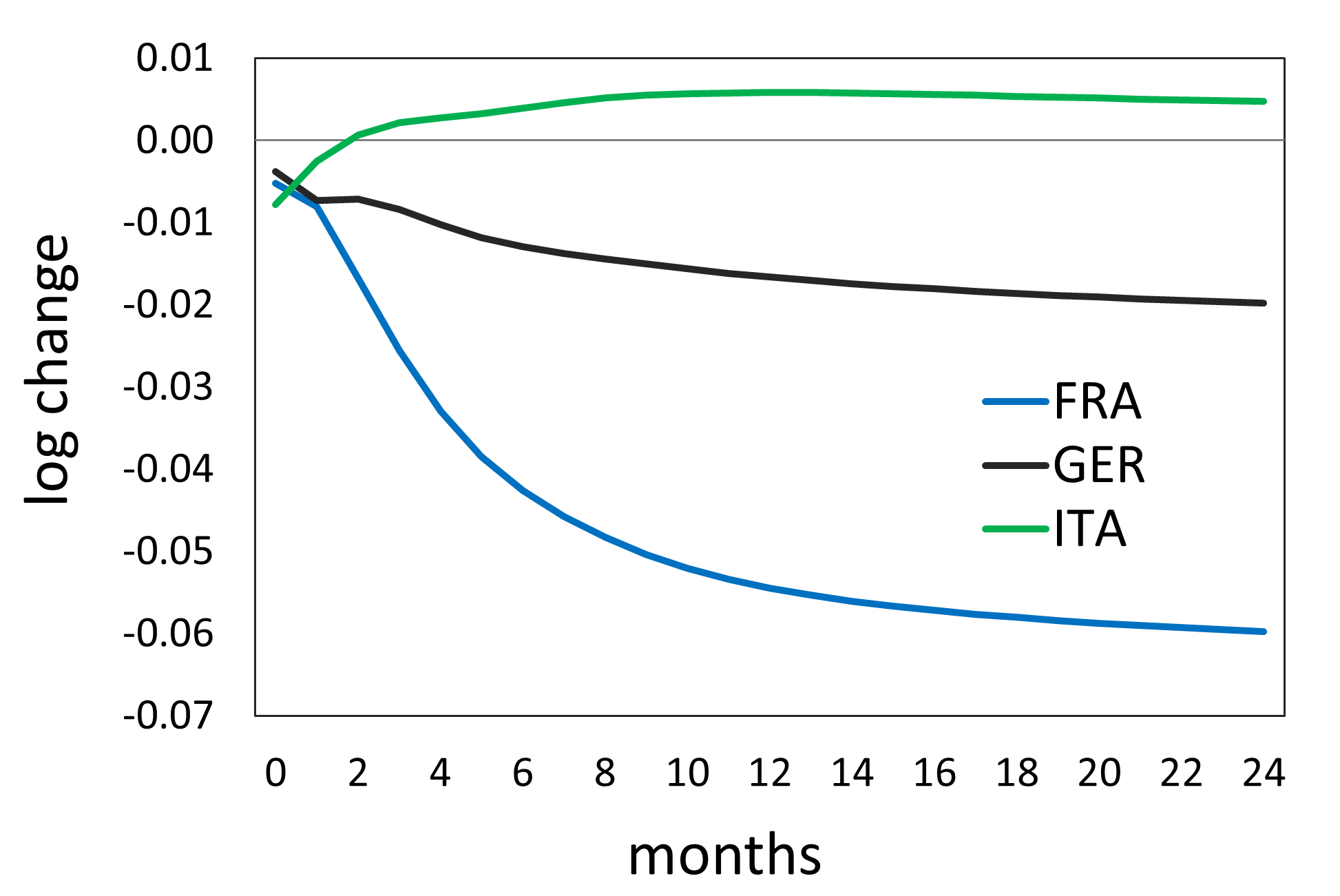}
		\end{minipage}%
		\begin{minipage}{.60\textwidth}
			\includegraphics[scale=0.11]{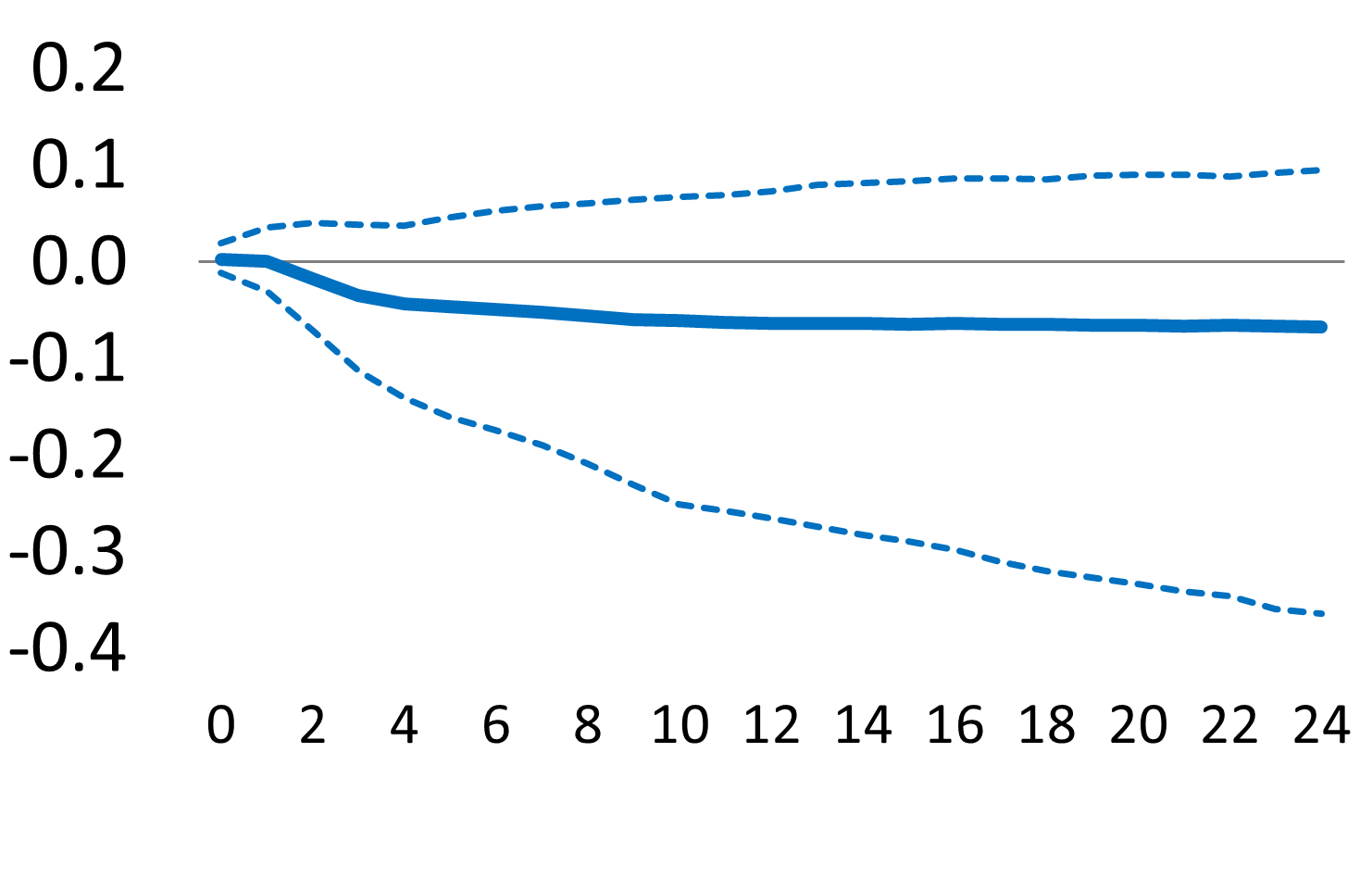}
			\includegraphics[scale=0.11]{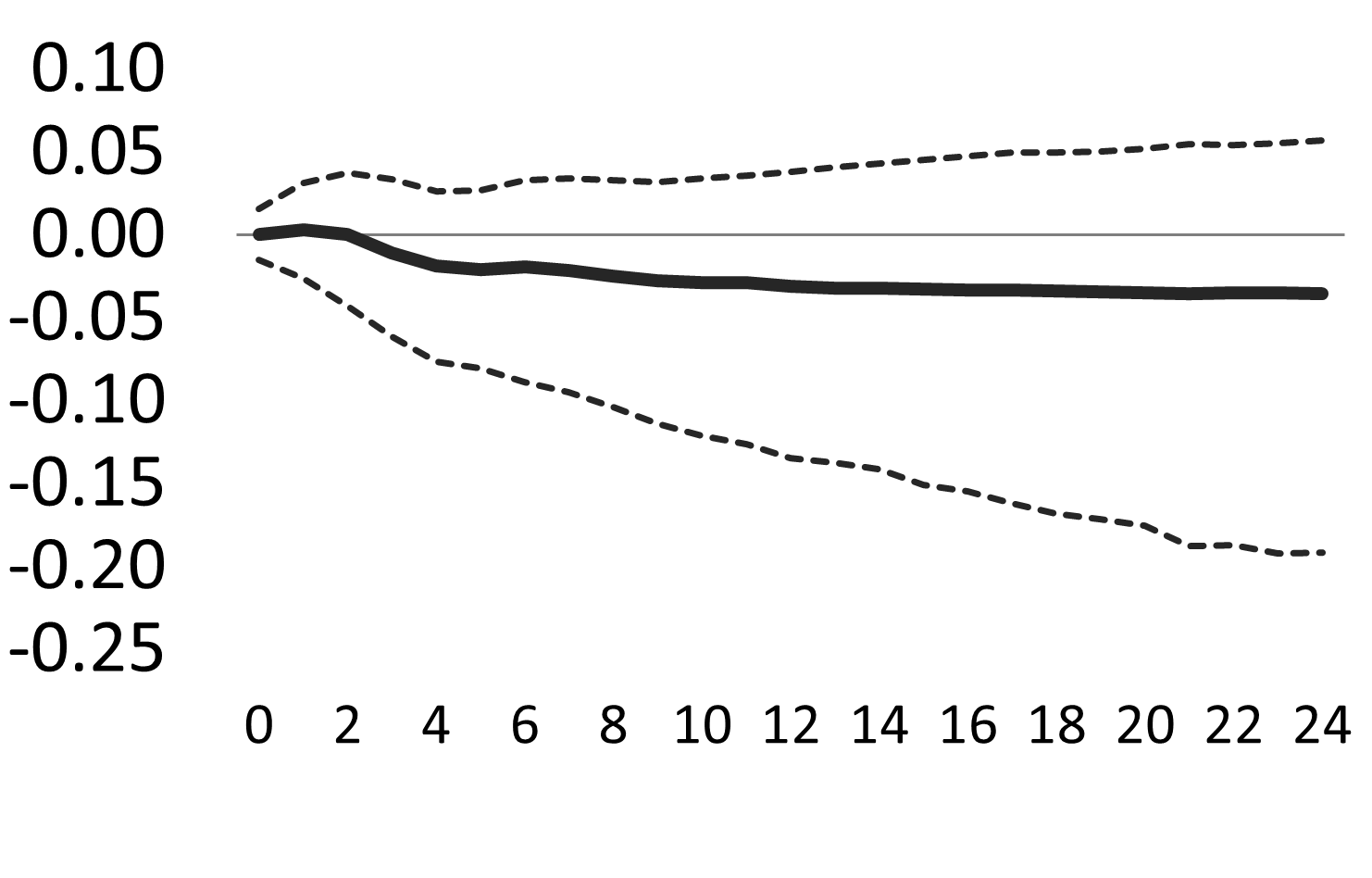}\\
			\includegraphics[scale=0.11]{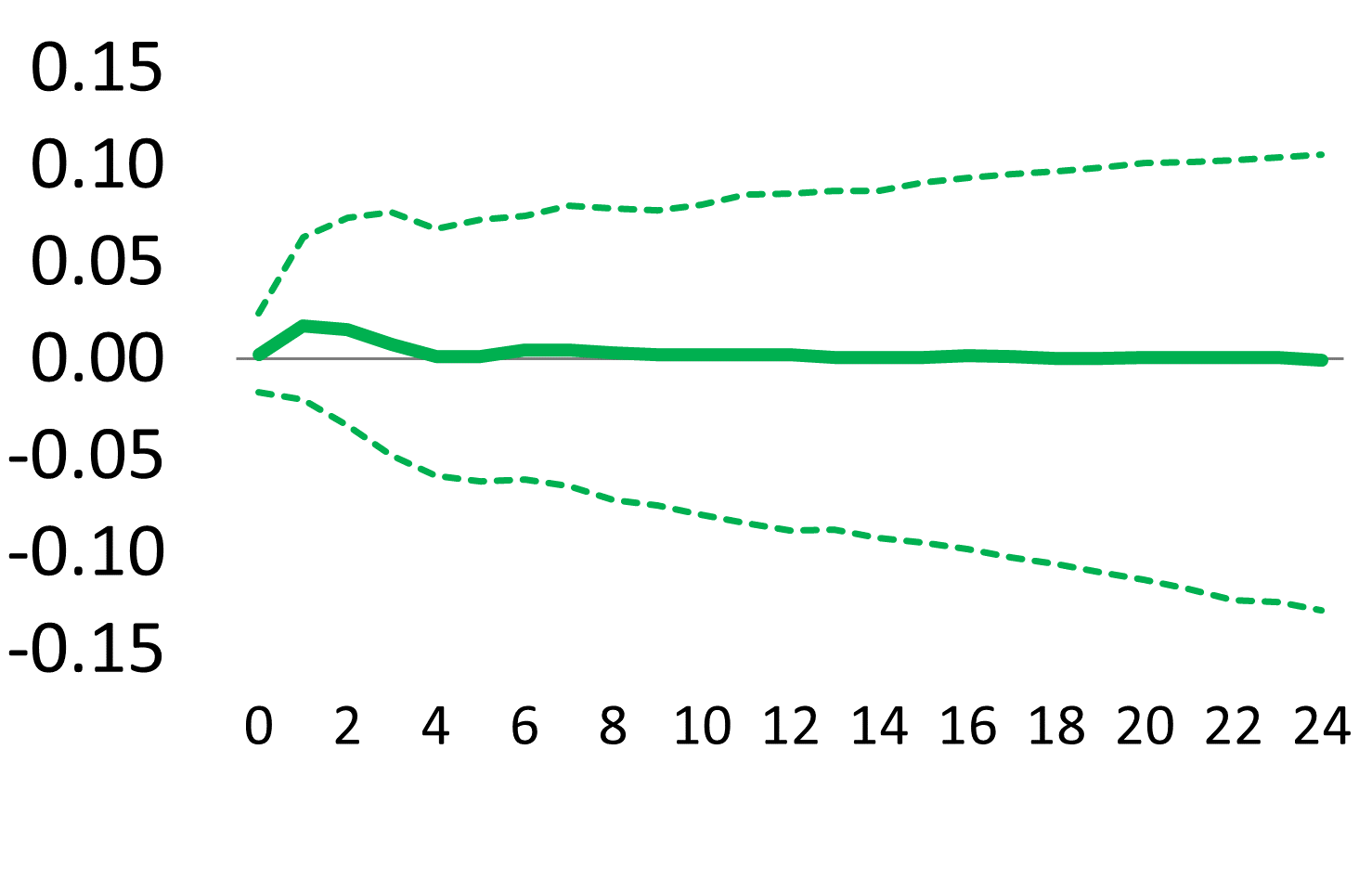}
		\end{minipage}%
		\subcaption*{\textit{Notes}: The figure shows the estimated generalized impulse responses to a positive 1 s.e. shock (left panel) and bootstrap 90\% confidence intervals and medians (right panels).}
		\label{fig:girf_gegov_vol}
	\end{figure}

\begin{figure}[H]
		\centering		
		\caption{GIRF of government yields, shock to domestic repo trade volume}
		\begin{minipage}{.5\textwidth}
			\hspace{-15pt}
			\includegraphics[scale=0.15]{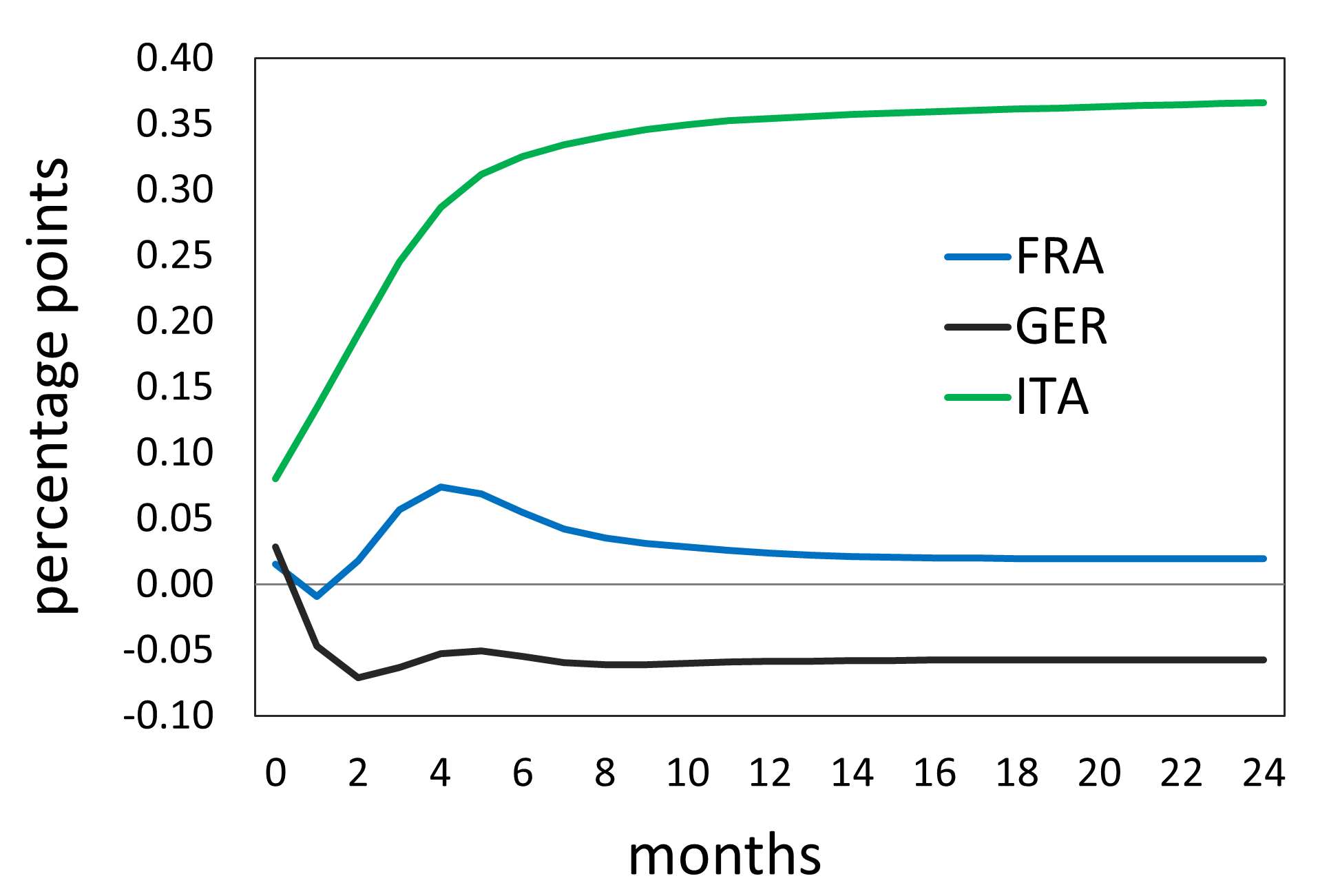}
		\end{minipage}%
		\begin{minipage}{.60\textwidth}
			\includegraphics[scale=0.11]{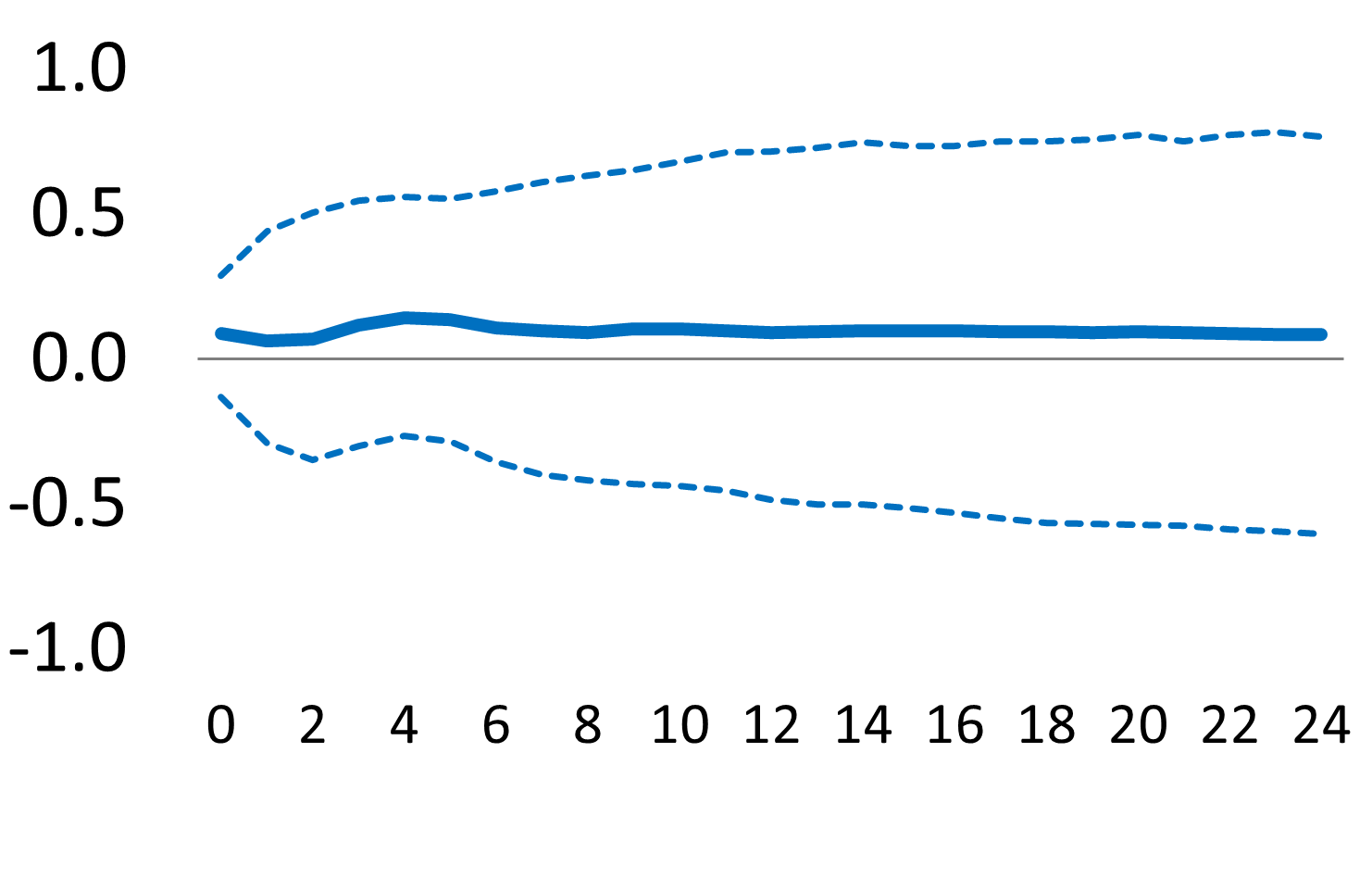}
			\includegraphics[scale=0.11]{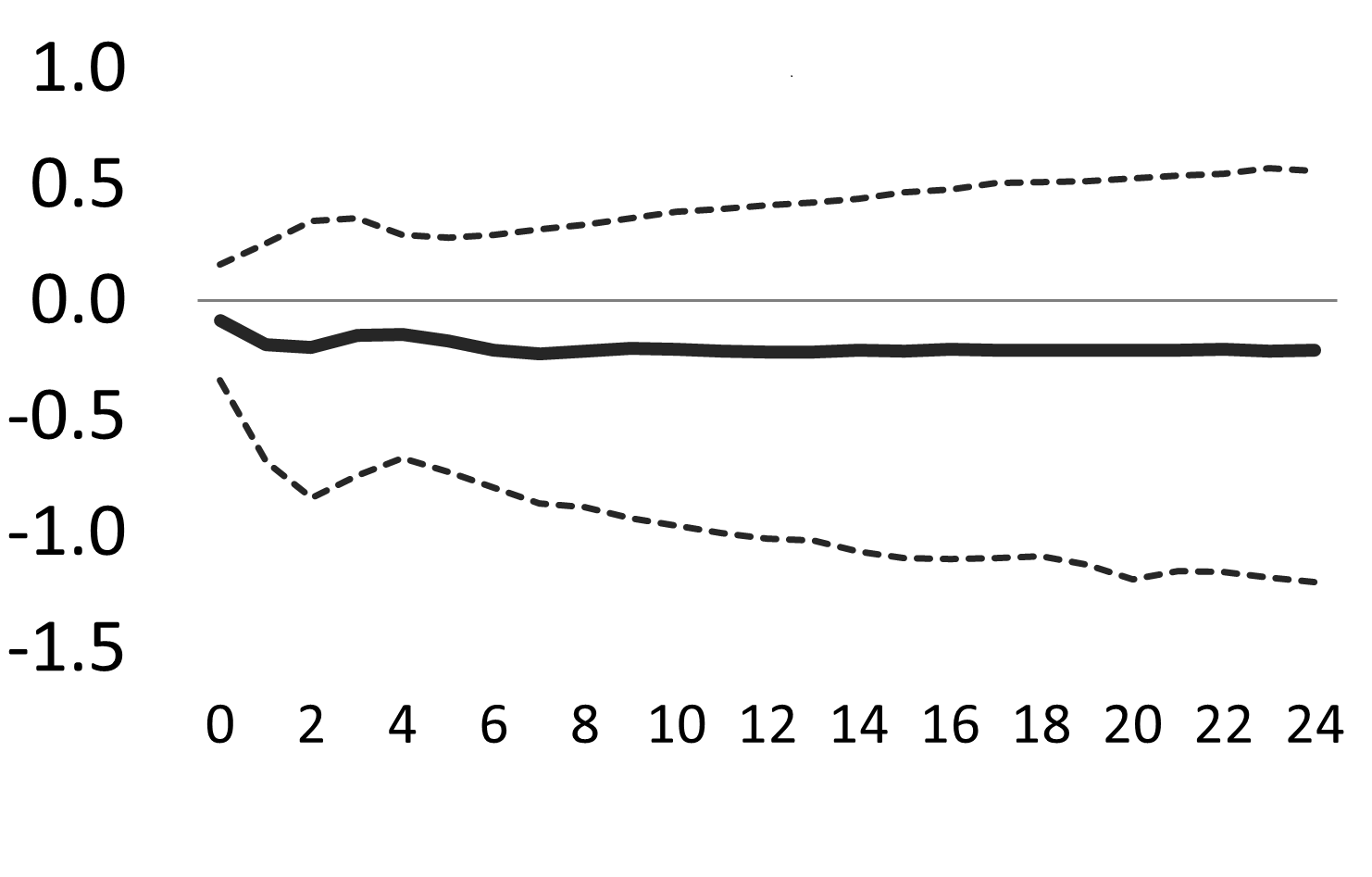}\\
			\includegraphics[scale=0.11]{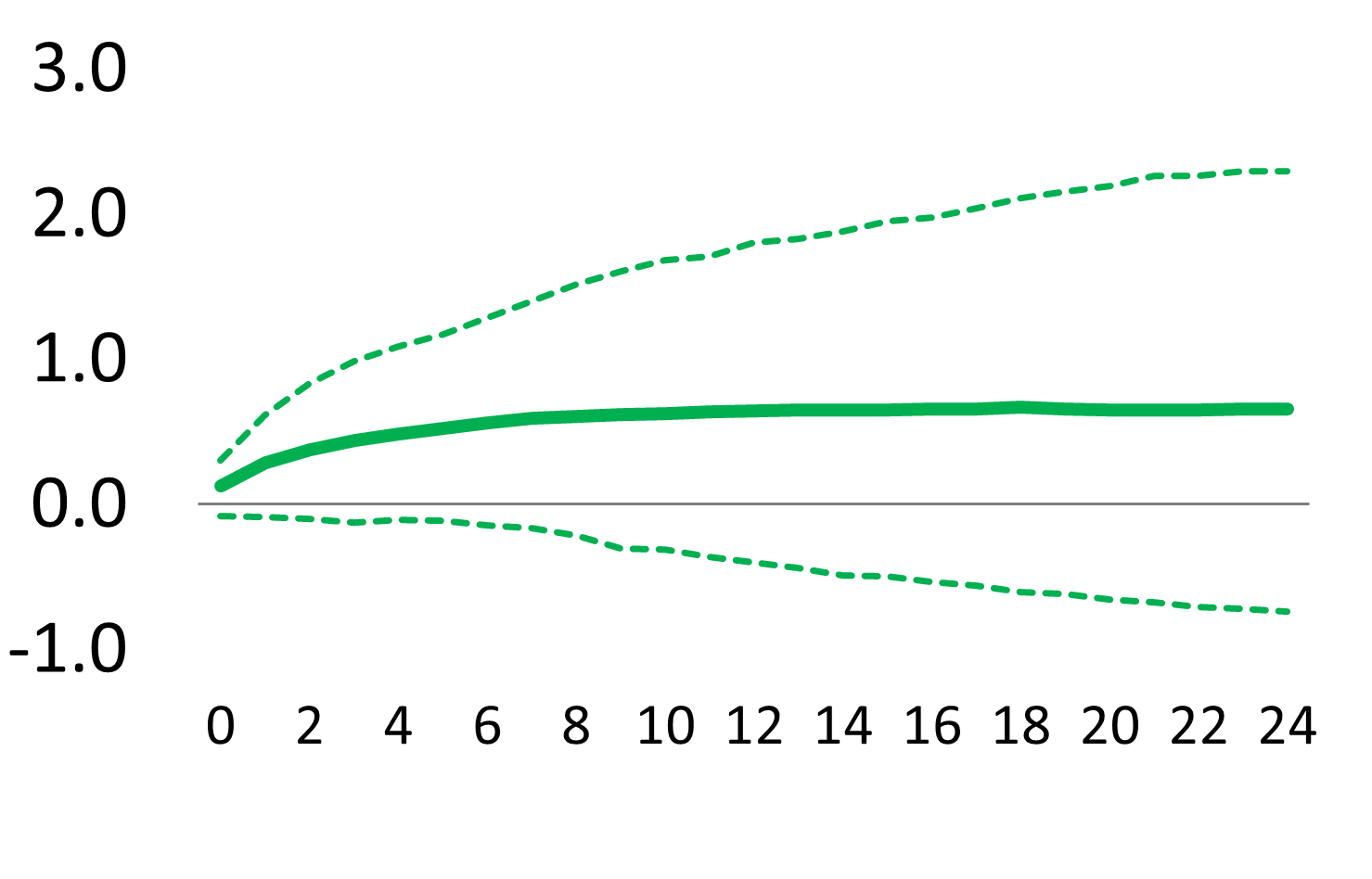}
		\end{minipage}%
		\subcaption*{\textit{Notes}: The figure shows the estimated generalized impulse responses to a positive 1 s.e. shock (left panel) and bootstrap 90\% confidence intervals and medians (right panels).}
		\label{fig:girf_vol_gov}
	\end{figure}

\begin{figure}[H]
		\centering		
		\caption{GIRF of bank CDS spreads, shock to domestic repo trade volume}
		\begin{minipage}{.5\textwidth}
			\hspace{-15pt}
			\includegraphics[scale=0.15]{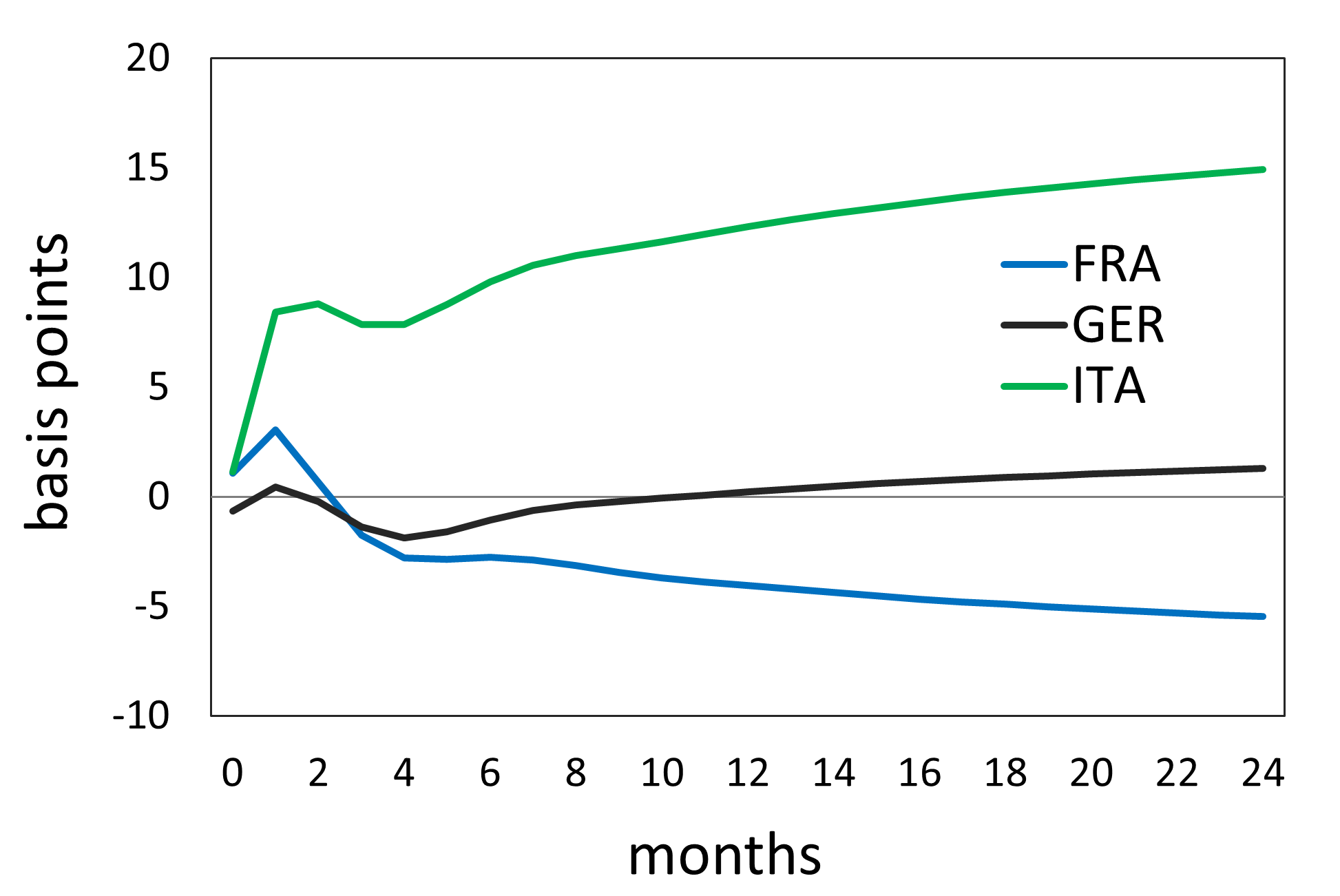}
		\end{minipage}%
		\begin{minipage}{.60\textwidth}
			\includegraphics[scale=0.11]{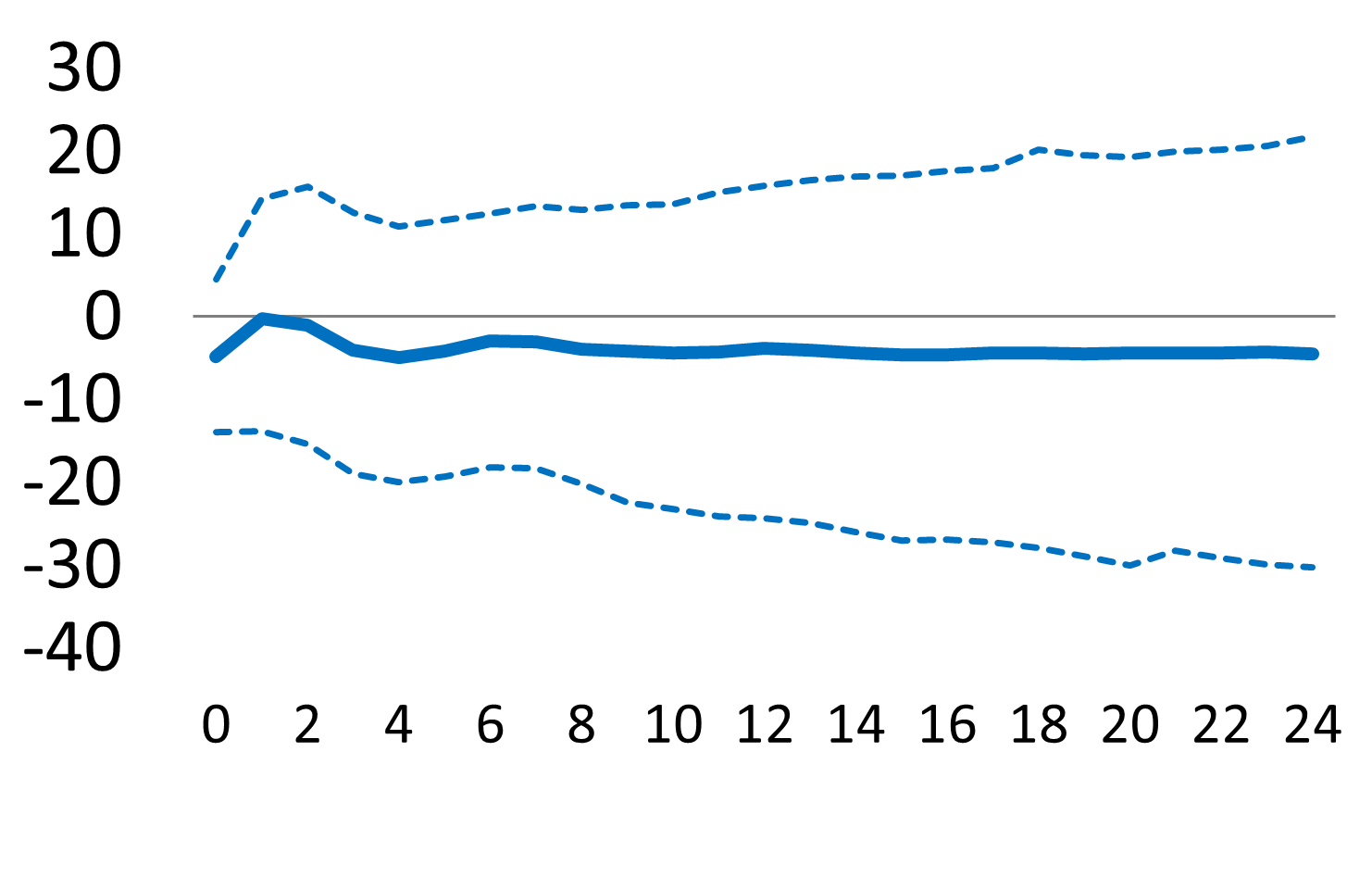}
			\includegraphics[scale=0.11]{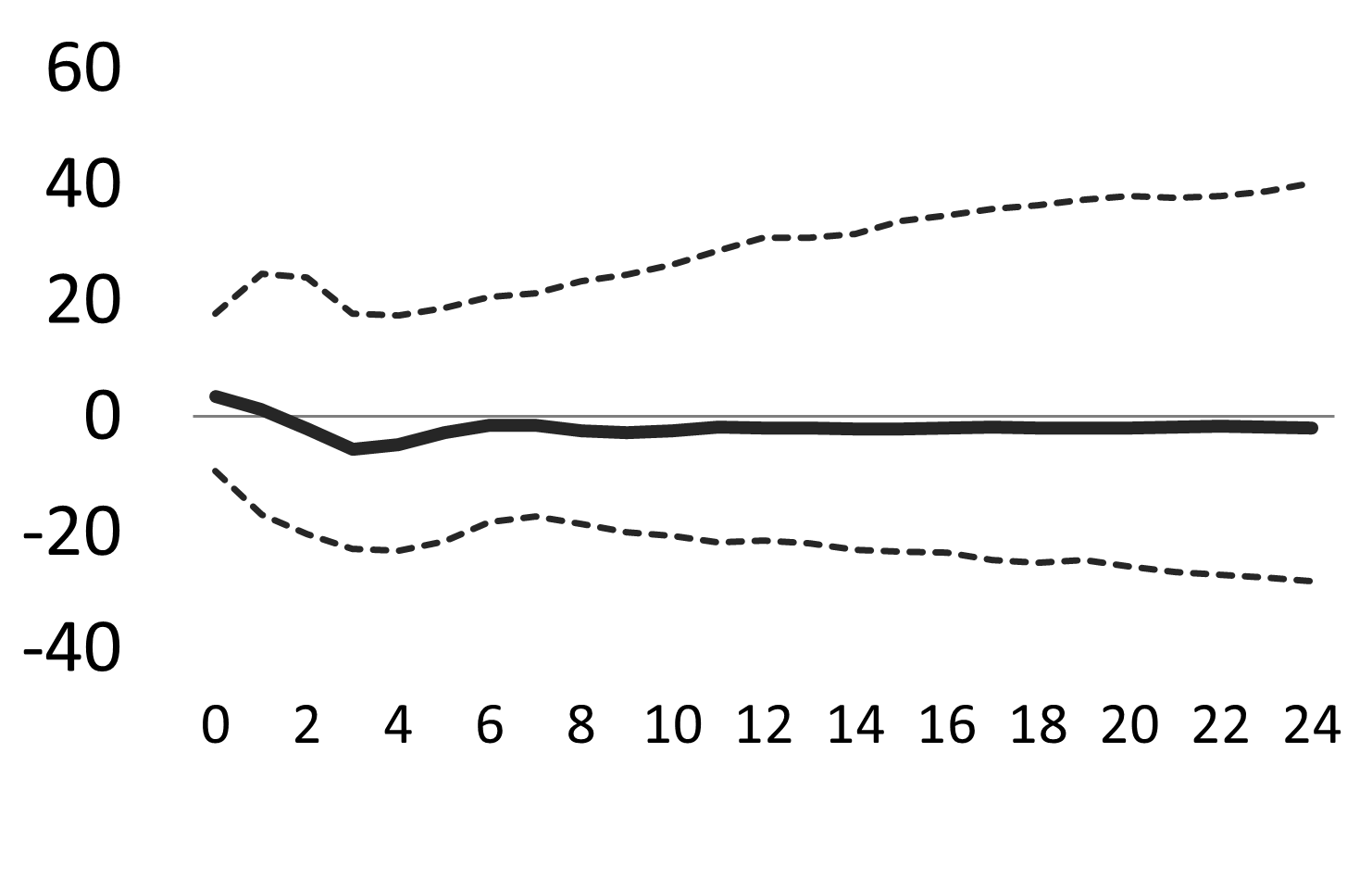}\\
			\includegraphics[scale=0.11]{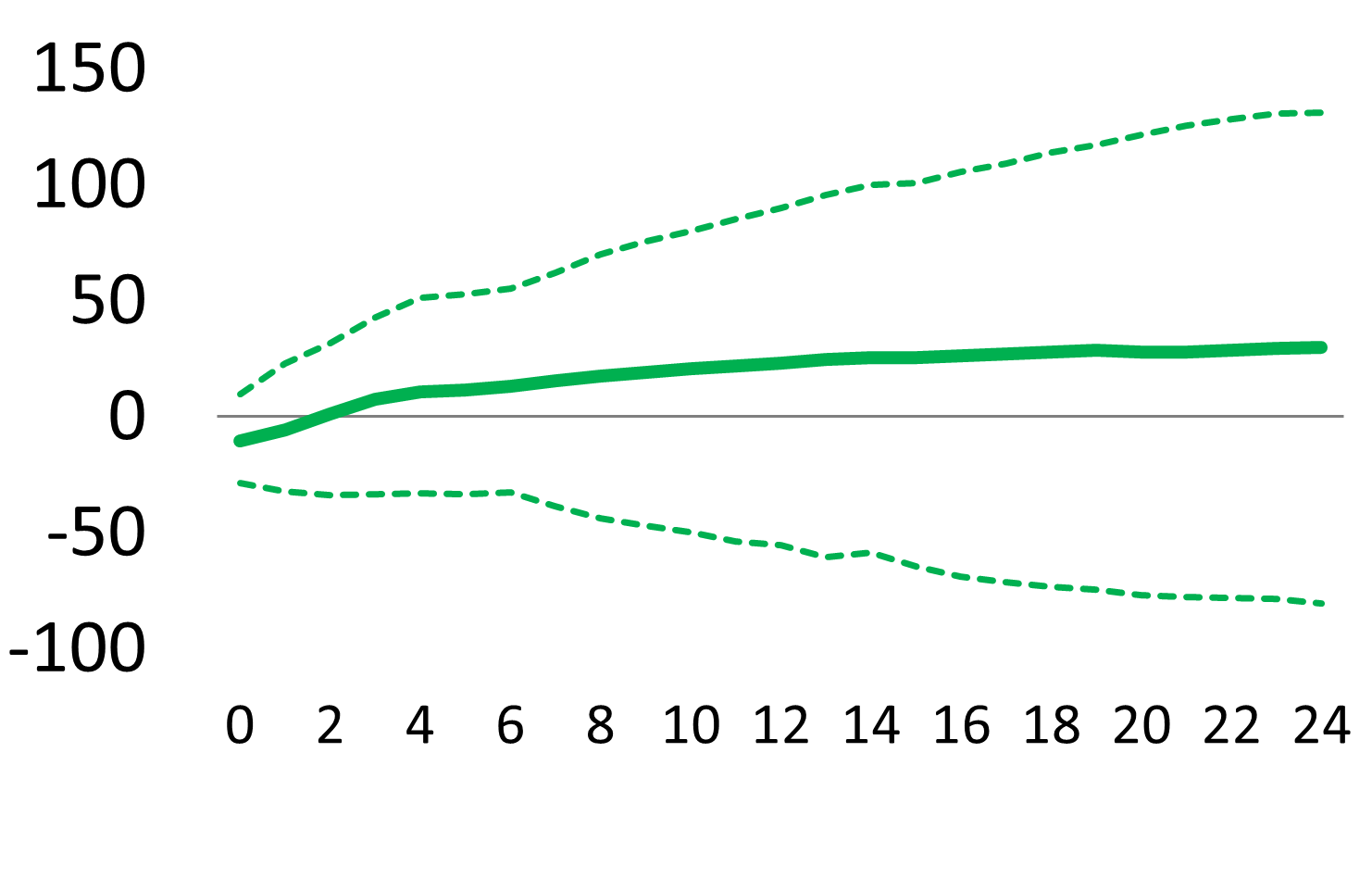}
		\end{minipage}%
		\subcaption*{\textit{Notes}: The figure shows the estimated generalized impulse responses to a positive 1 s.e. shock (left panel) and bootstrap 90\% confidence intervals and medians (right panels).}
		\label{fig:girf_vol_cds}
	\end{figure}

\begin{figure}[H]
		\centering		
		\caption{GIRF of bank CDS spreads, shock to ITA repo rate}
		\begin{minipage}{.5\textwidth}
			\hspace{-15pt}
			\includegraphics[scale=0.15]{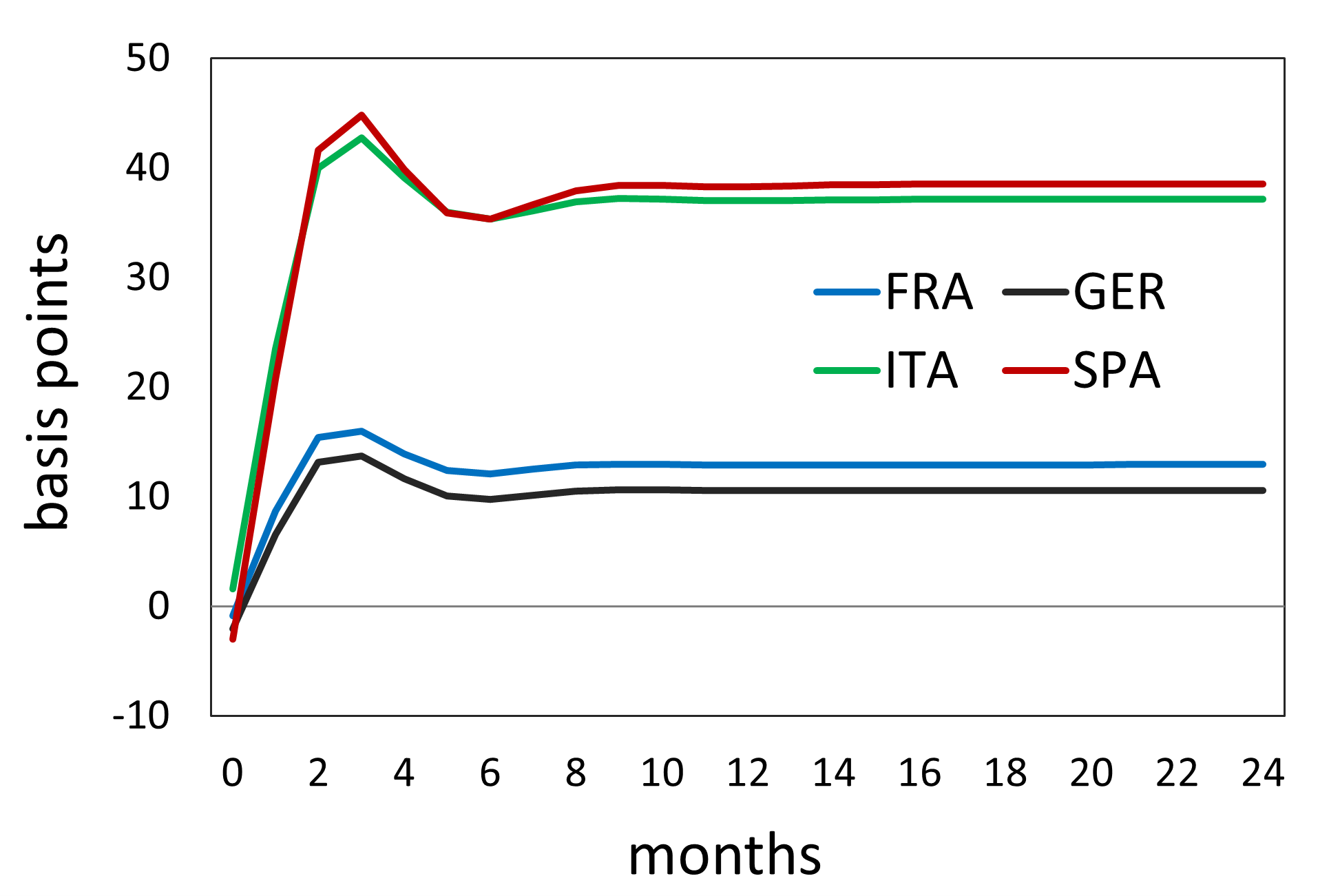}
		\end{minipage}%
		\begin{minipage}{.60\textwidth}
			\includegraphics[scale=0.11]{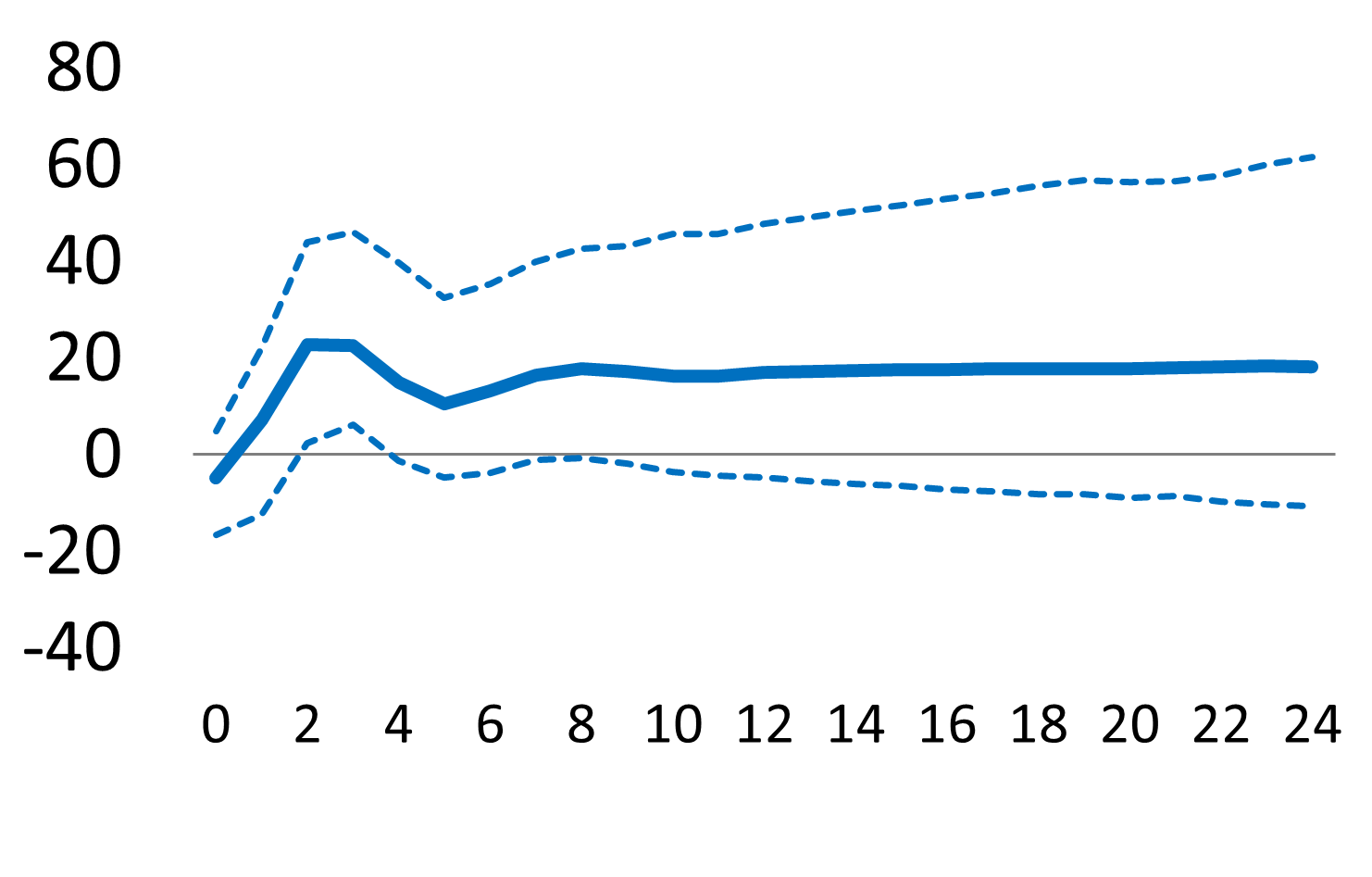}
			\includegraphics[scale=0.11]{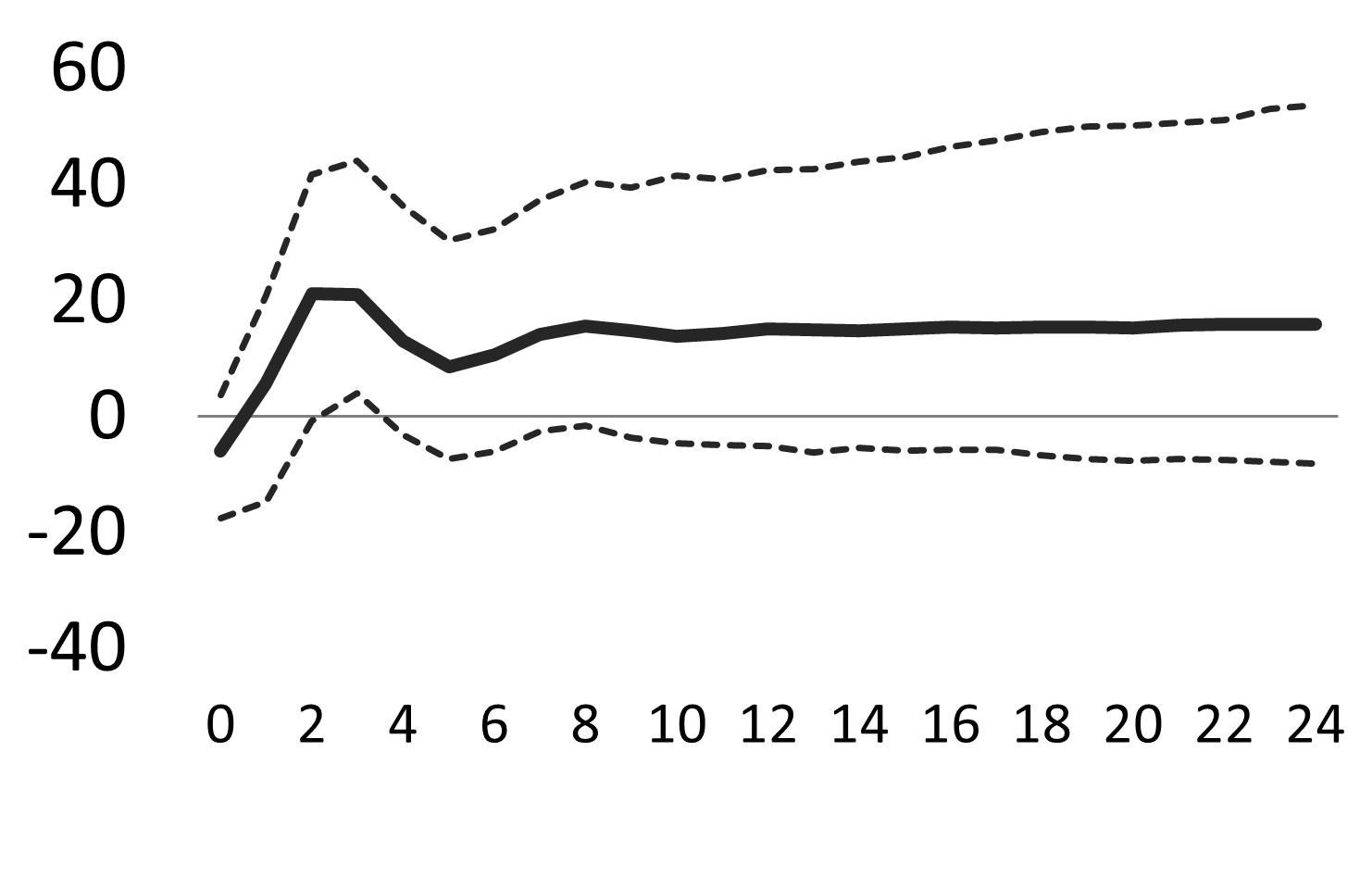}\\
			\includegraphics[scale=0.11]{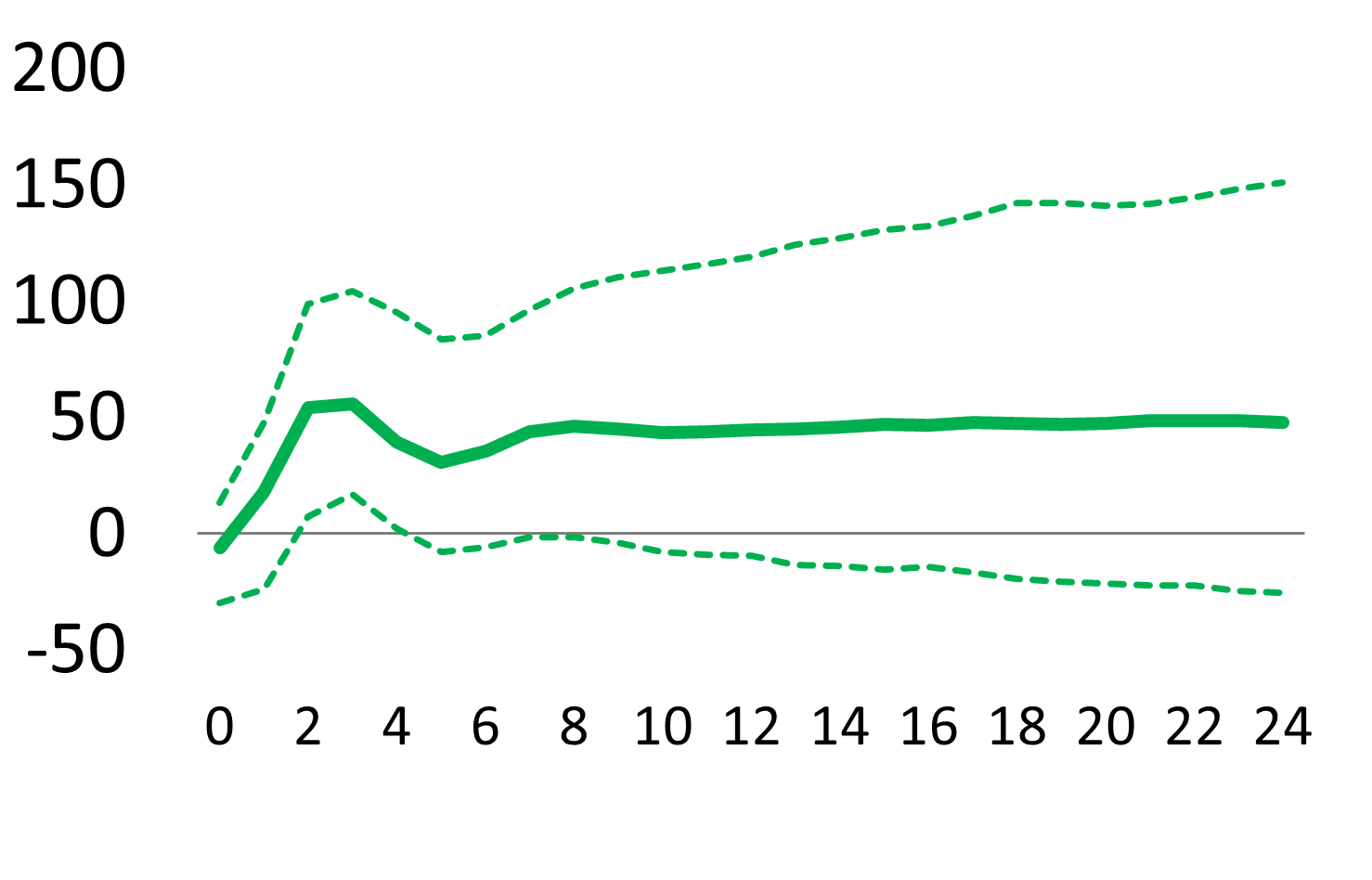}
			\includegraphics[scale=0.11]{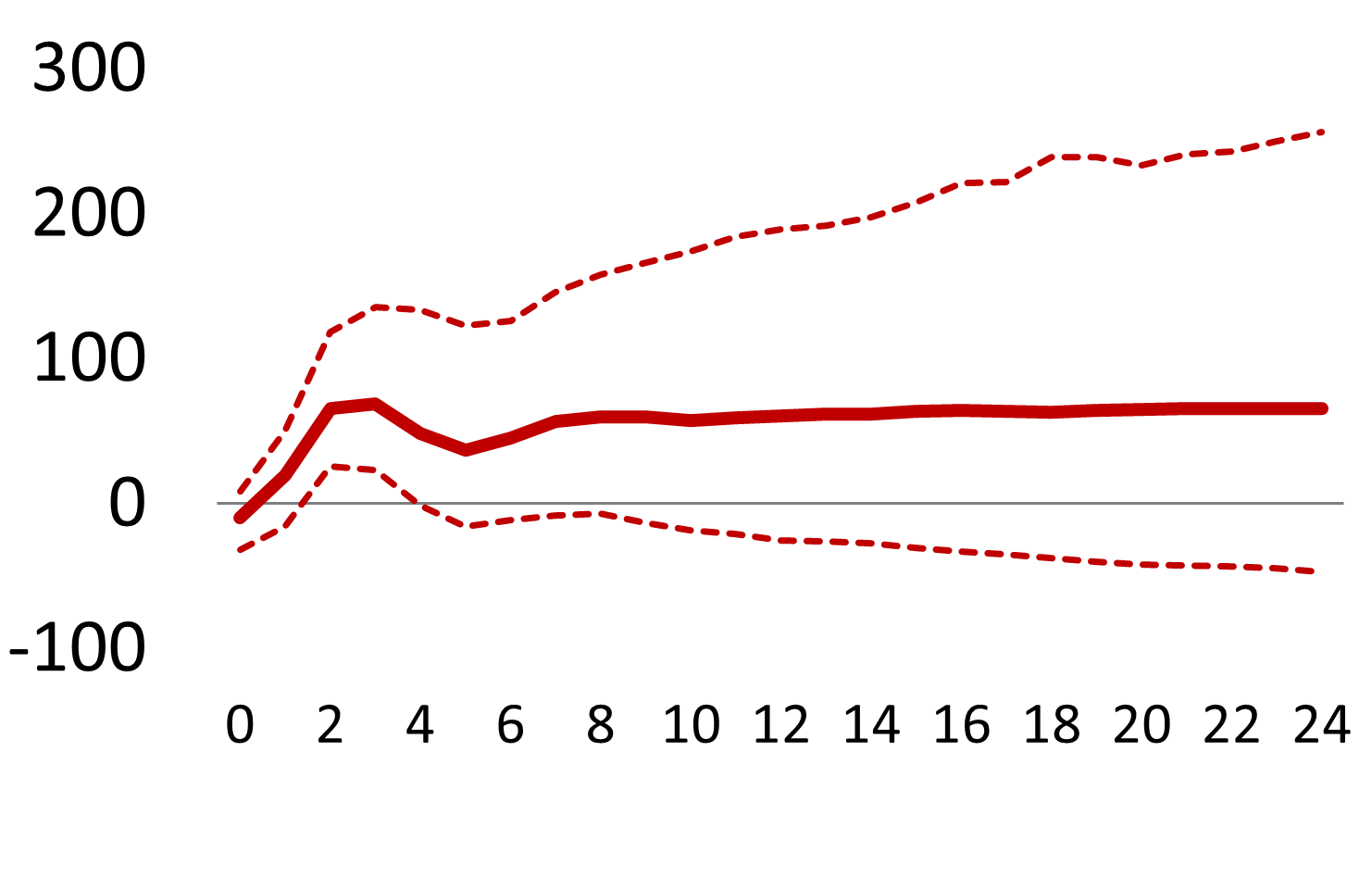}
		\end{minipage}%
		\subcaption*{\textit{Notes}: The figure shows the estimated generalized impulse responses to a positive 1 s.e. shock (left panel) and bootstrap 90\% confidence intervals and medians (right panels).}
		\label{fig:girf_itrep_cds}
	\end{figure}

	Figures \ref{fig:girf_gegov_vol}-\ref{fig:girf_itrep_cds} summarize the empirical relationships between repo market conditions and government yields/bank CDS spreads. 
	Figure \ref{fig:girf_gegov_vol} plots the responses of repo trading volumes to a positive one-standard-error shock in the yield of the European safe asset, i.e., the German Bund (point estimates are displayed in the left panel, bootstrap confidence intervals and medians in the right panels). The estimated response is negative for Germany (-2\%, approximately, after 2 years) and France (-6\%), but not for Italy.
	Figures \ref{fig:girf_vol_gov}-\ref{fig:girf_vol_cds} show that a positive shock in domestic repo trading volume tends to be followed by increasing government yields and bank CDS spreads in Italy and decreasing yields and CDS spreads in Germany. 
	These results appear to capture flight-to-quality effects.
	Figure \ref{fig:girf_itrep_cds} shows the responses of bank CDS spreads to a shock in the Italian repo market rate. The figure highlights a distinction between Italy and Spain, on one side, and Germany and France, on the other: after a couple of months, Italian and Spanish CDS spreads exhibit significant increases of around 40 basis points (bps), while German and French CDS spreads increase by only 10-15 bps.   

\begin{figure}[H]
		\centering		
		\caption{GIRF of bank CDS spreads, shock to domestic government yield}
		\begin{minipage}{.5\textwidth}
			\hspace{-15pt}
			\includegraphics[scale=0.15]{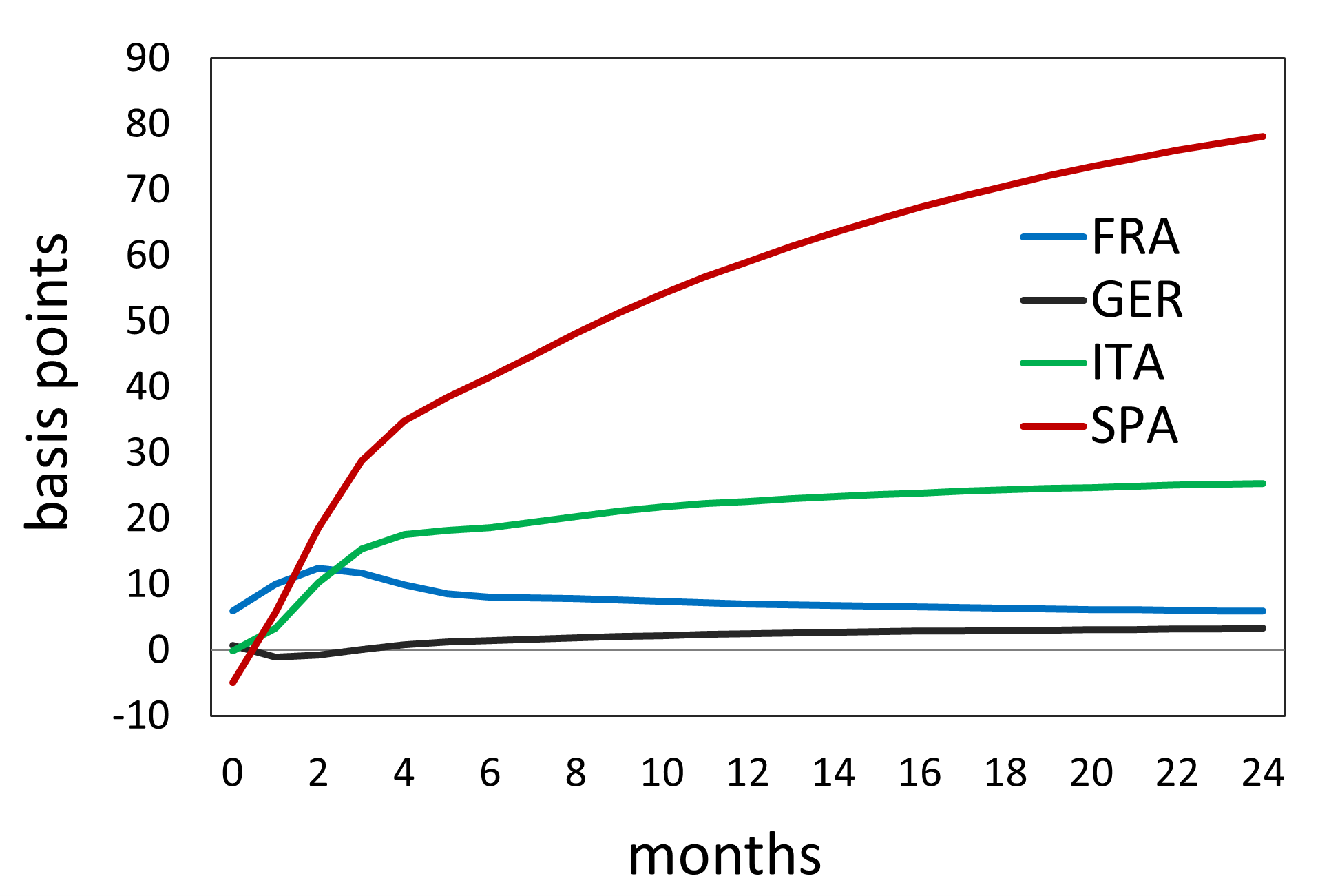}
		\end{minipage}%
		\begin{minipage}{.60\textwidth}
			\includegraphics[scale=0.11]{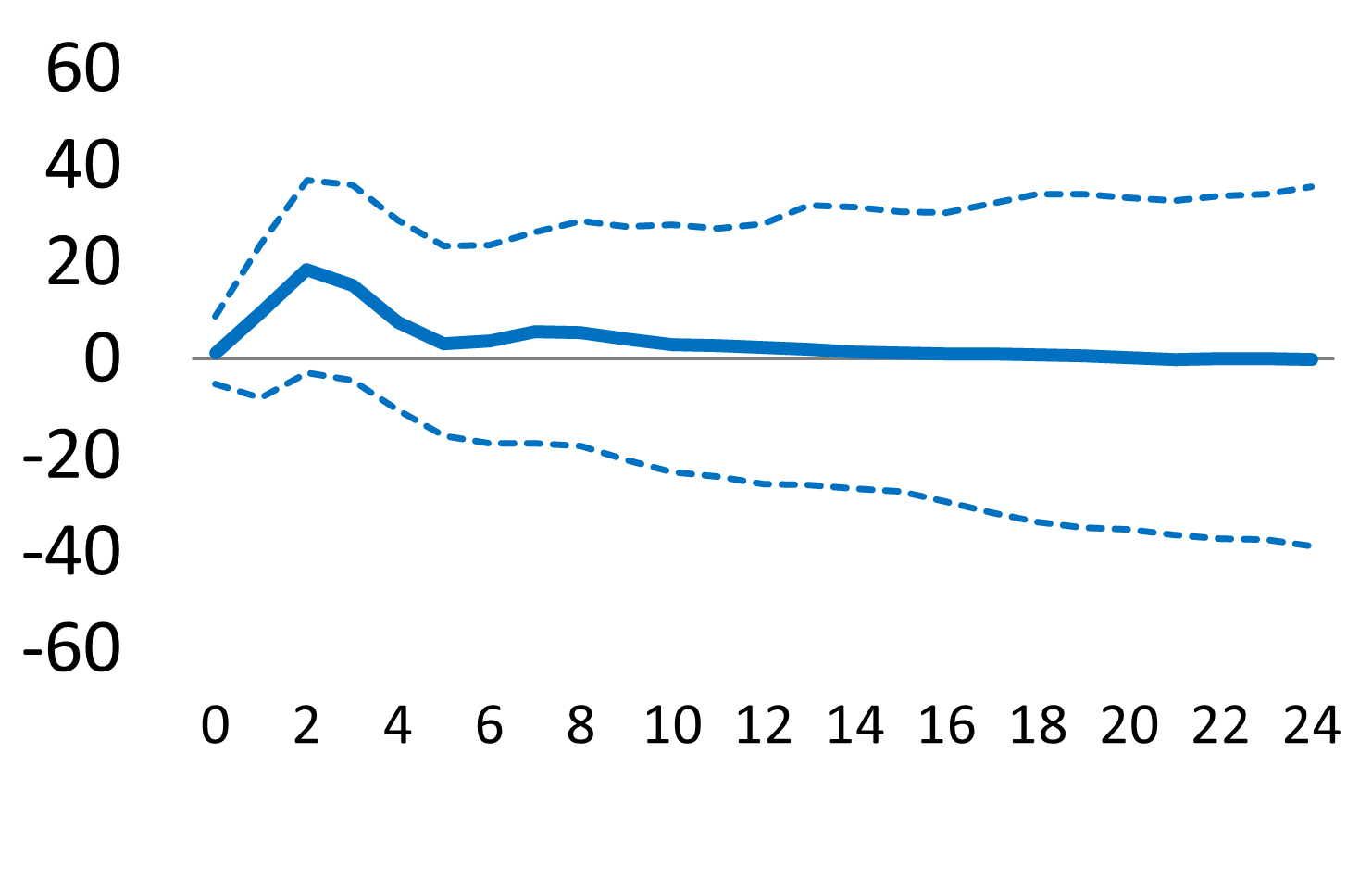}
			\includegraphics[scale=0.11]{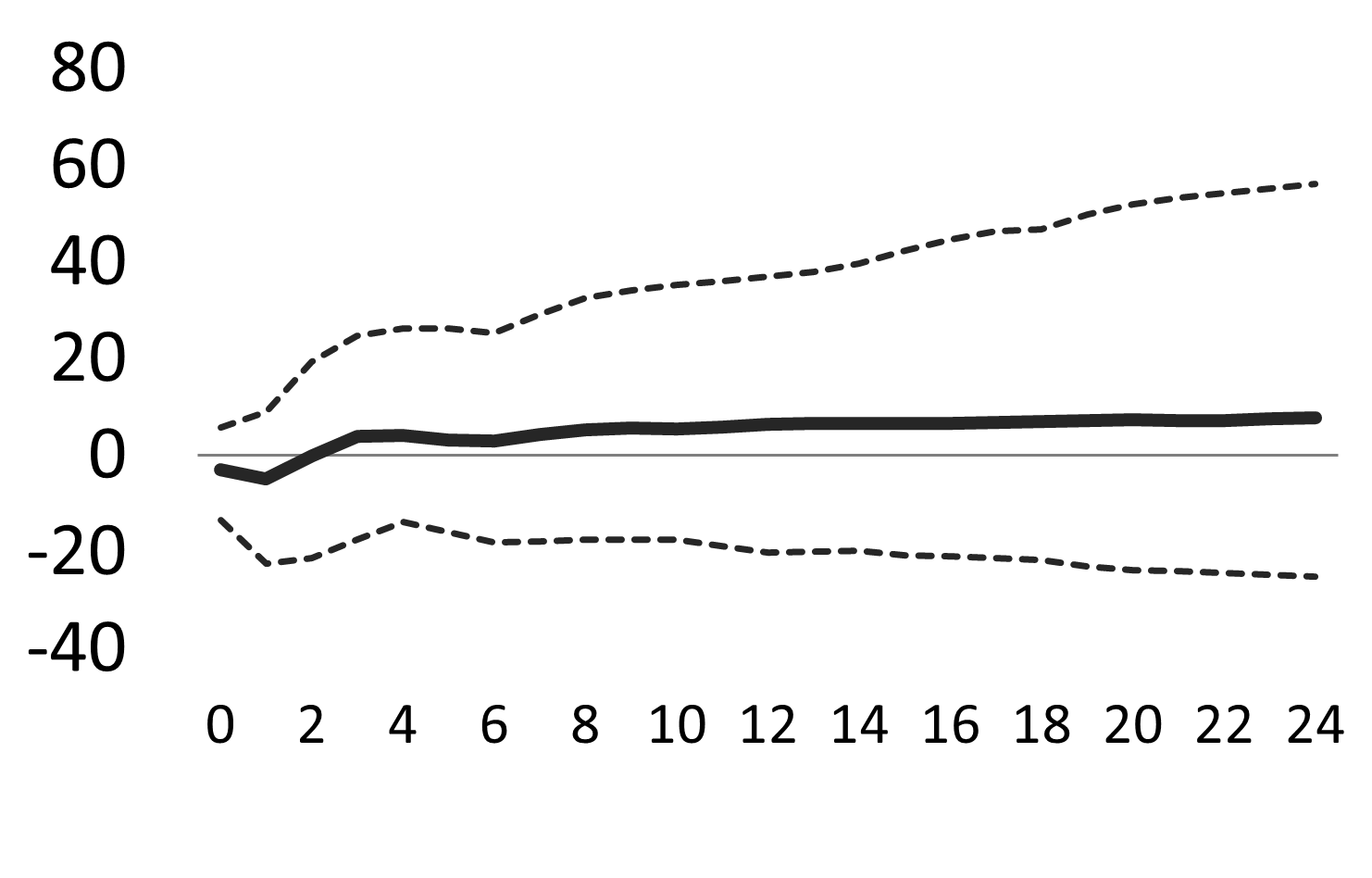}\\
			\includegraphics[scale=0.11]{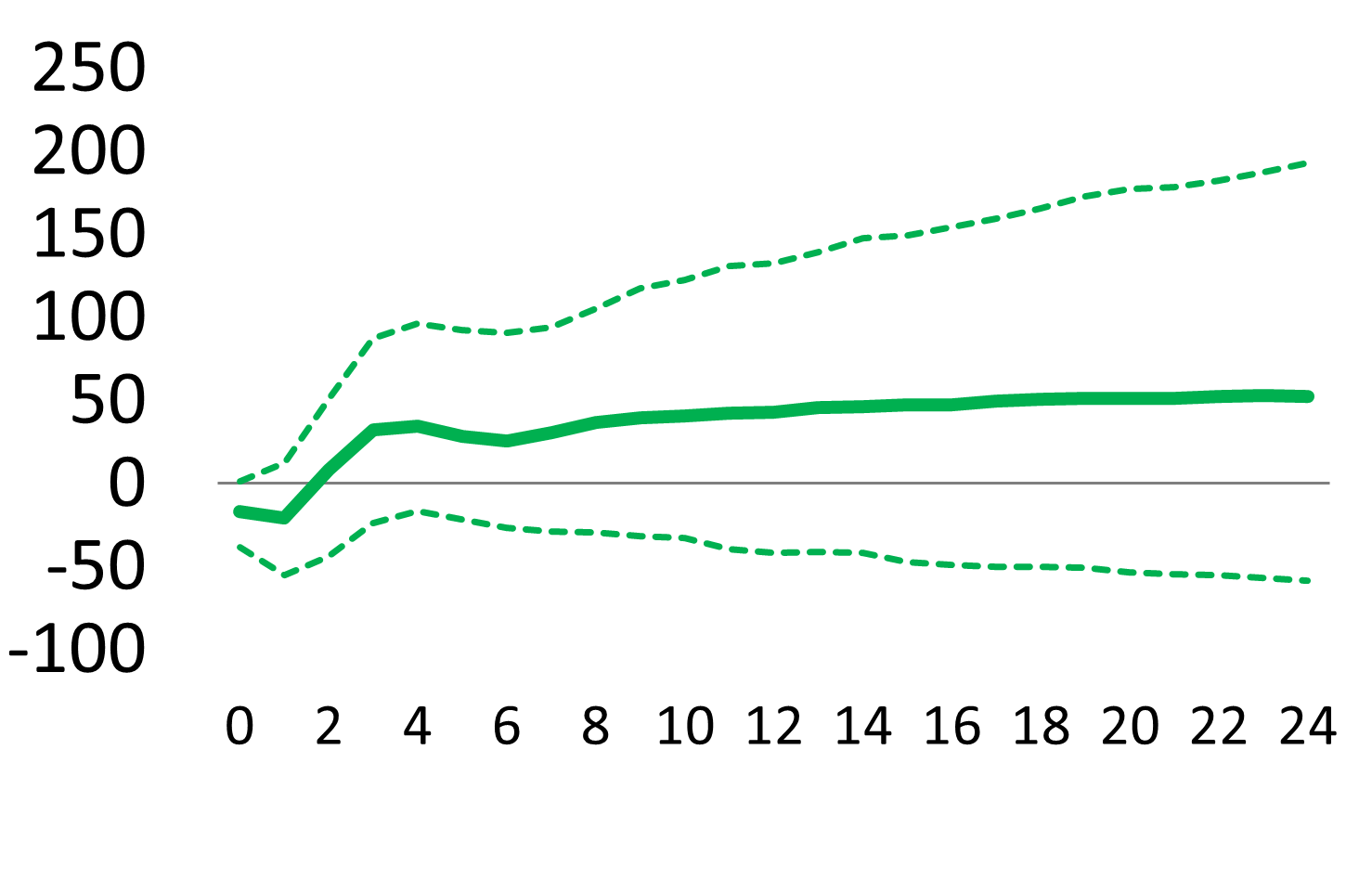}
			\includegraphics[scale=0.11]{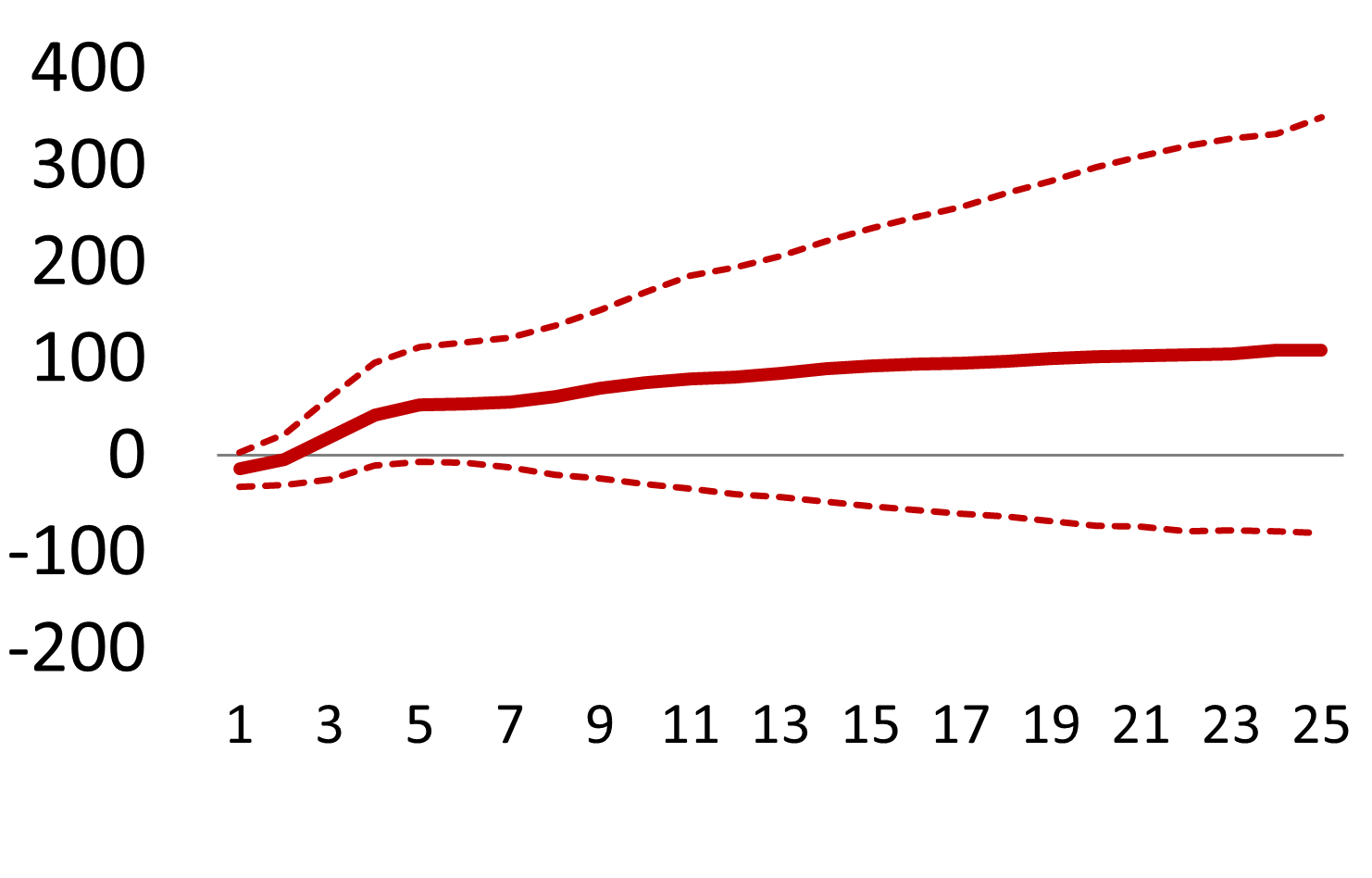}
		\end{minipage}%
		\subcaption*{\textit{Notes}: The figure shows the estimated generalized impulse responses to a positive 1 s.e. shock (left panel) and bootstrap 90\% confidence intervals and medians (right panels).}
		\label{fig:girf_gov_cds}
	\end{figure}

\begin{figure}[H]
		\centering		
		\caption{GIRF of bank CDS spreads, shock to GER government yield}
		\begin{minipage}{.5\textwidth}
			\hspace{-15pt}
			\includegraphics[scale=0.15]{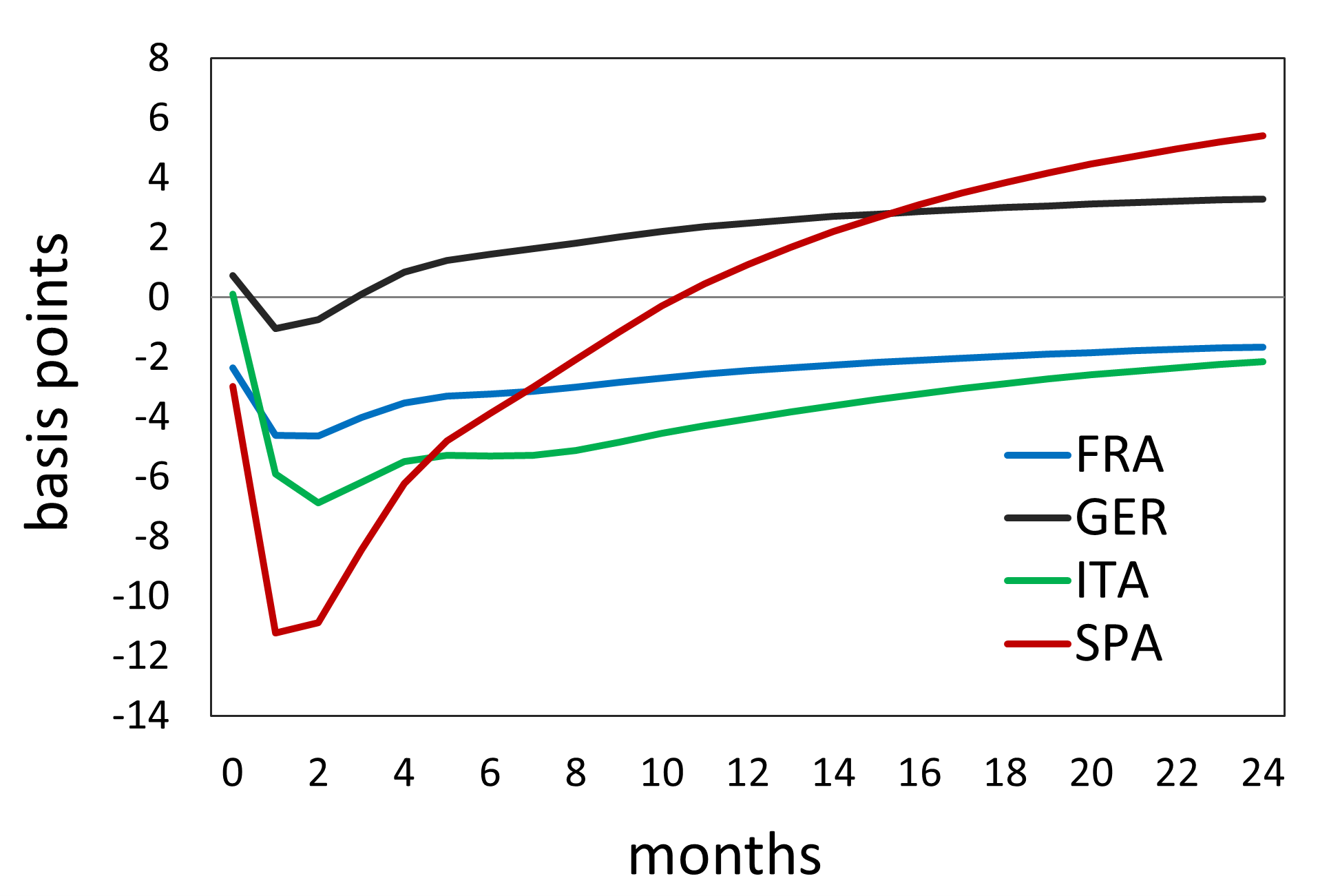}
		\end{minipage}%
		\begin{minipage}{.60\textwidth}
			\includegraphics[scale=0.11]{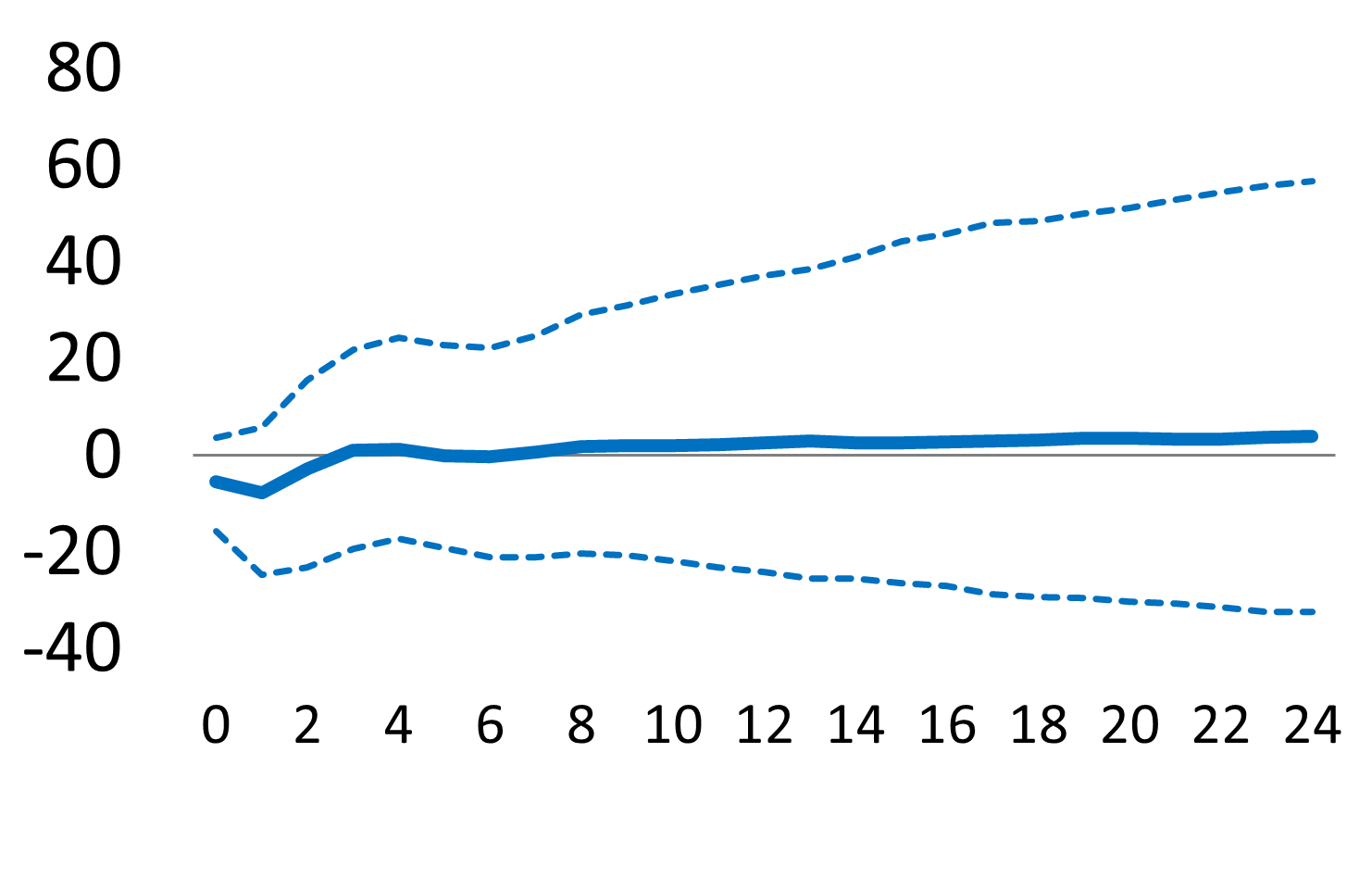}
			\includegraphics[scale=0.11]{girf_gegov_gecds_boot.png}\\
			\includegraphics[scale=0.11]{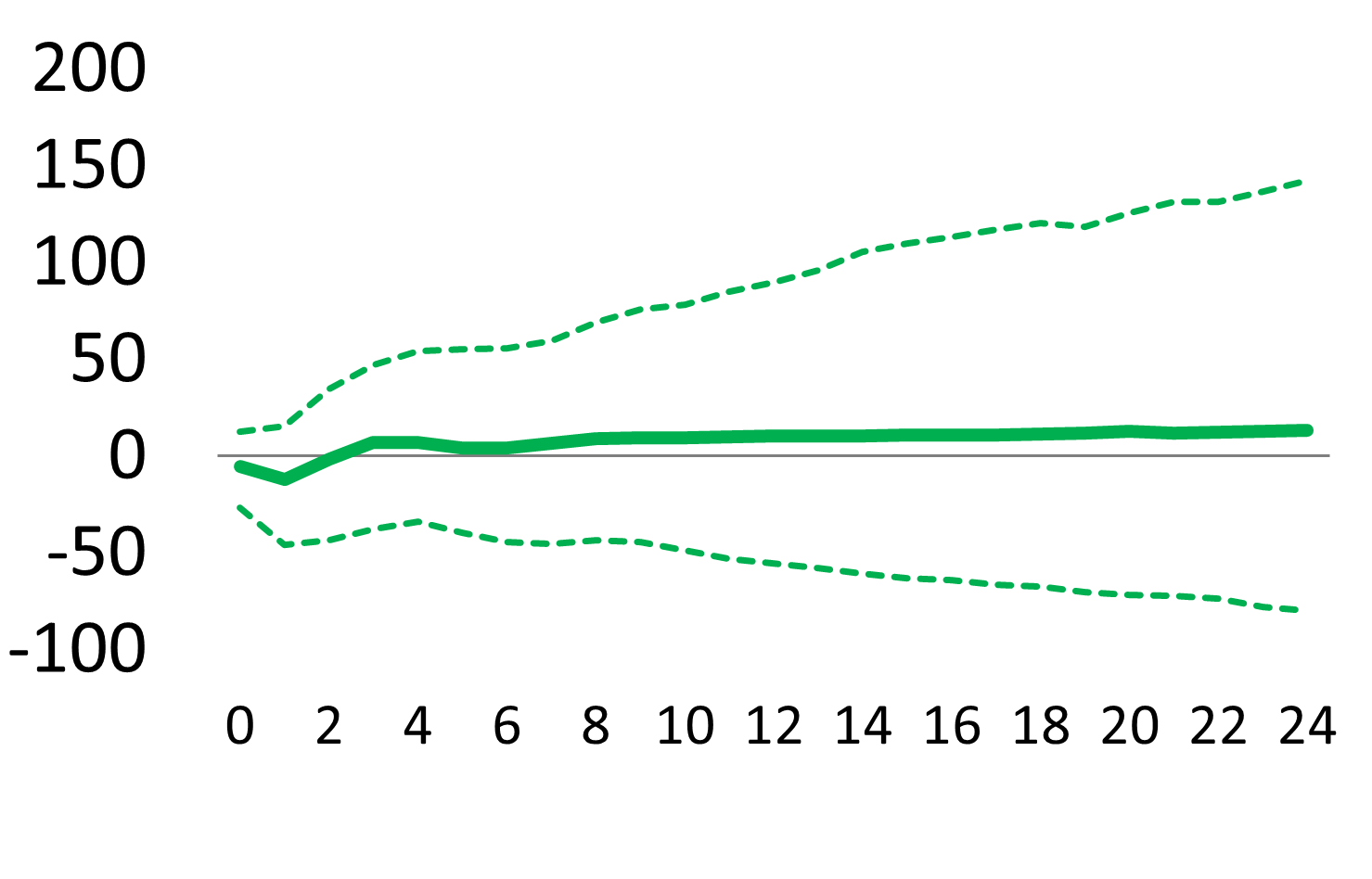}
			\includegraphics[scale=0.11]{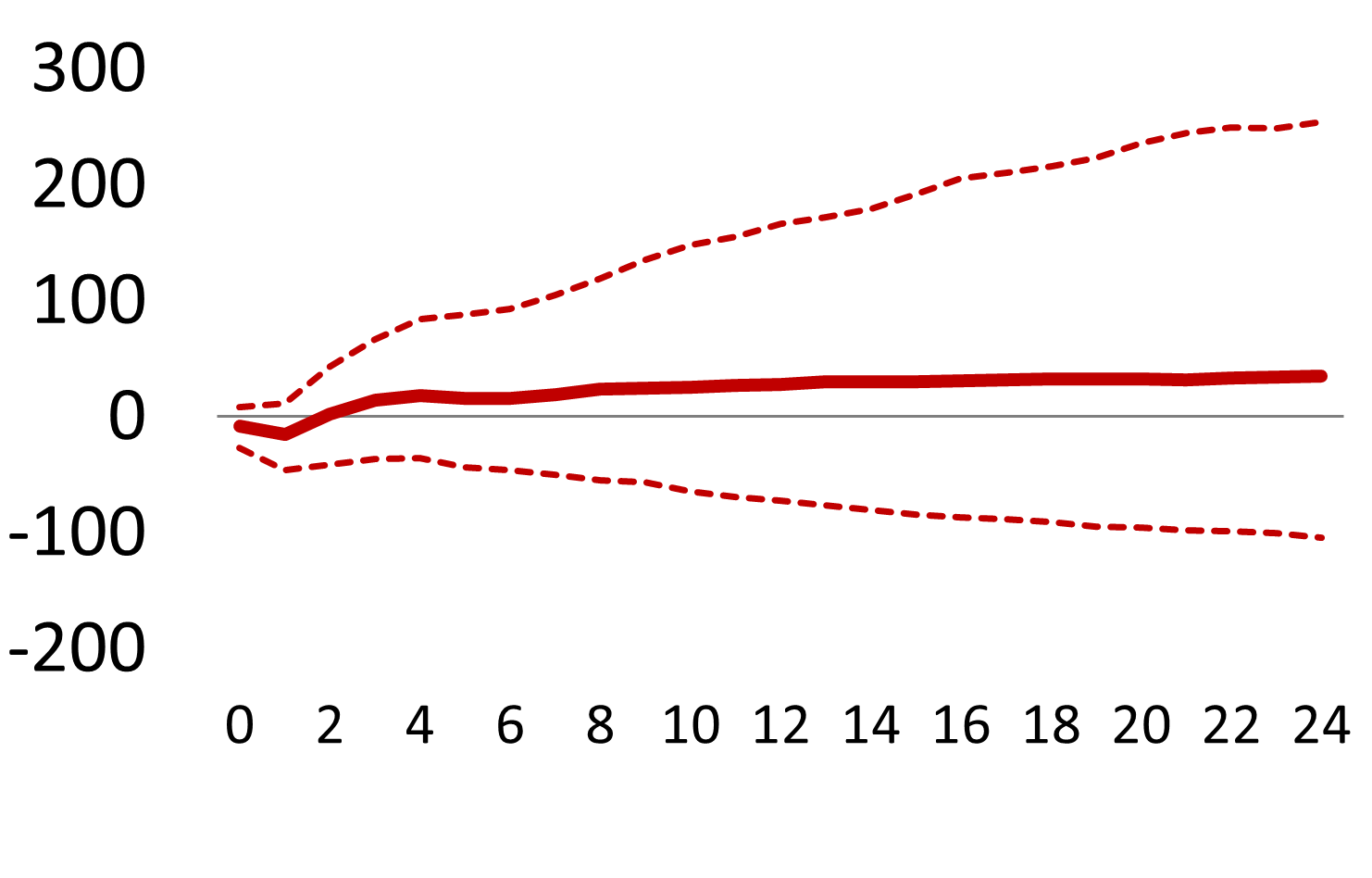}
		\end{minipage}%
		\subcaption*{\textit{Notes}: The figure shows the estimated generalized impulse responses to a positive 1 s.e. shock (left panel) and bootstrap 90\% confidence intervals and medians (right panels).}
		\label{fig:girf_gegov_cds}
	\end{figure}

\begin{figure}[H]
		\centering		
		\caption{GIRF of bank CDS spreads, shock to ITA government yield}
		\begin{minipage}{.5\textwidth}
			\hspace{-15pt}
			\includegraphics[scale=0.15]{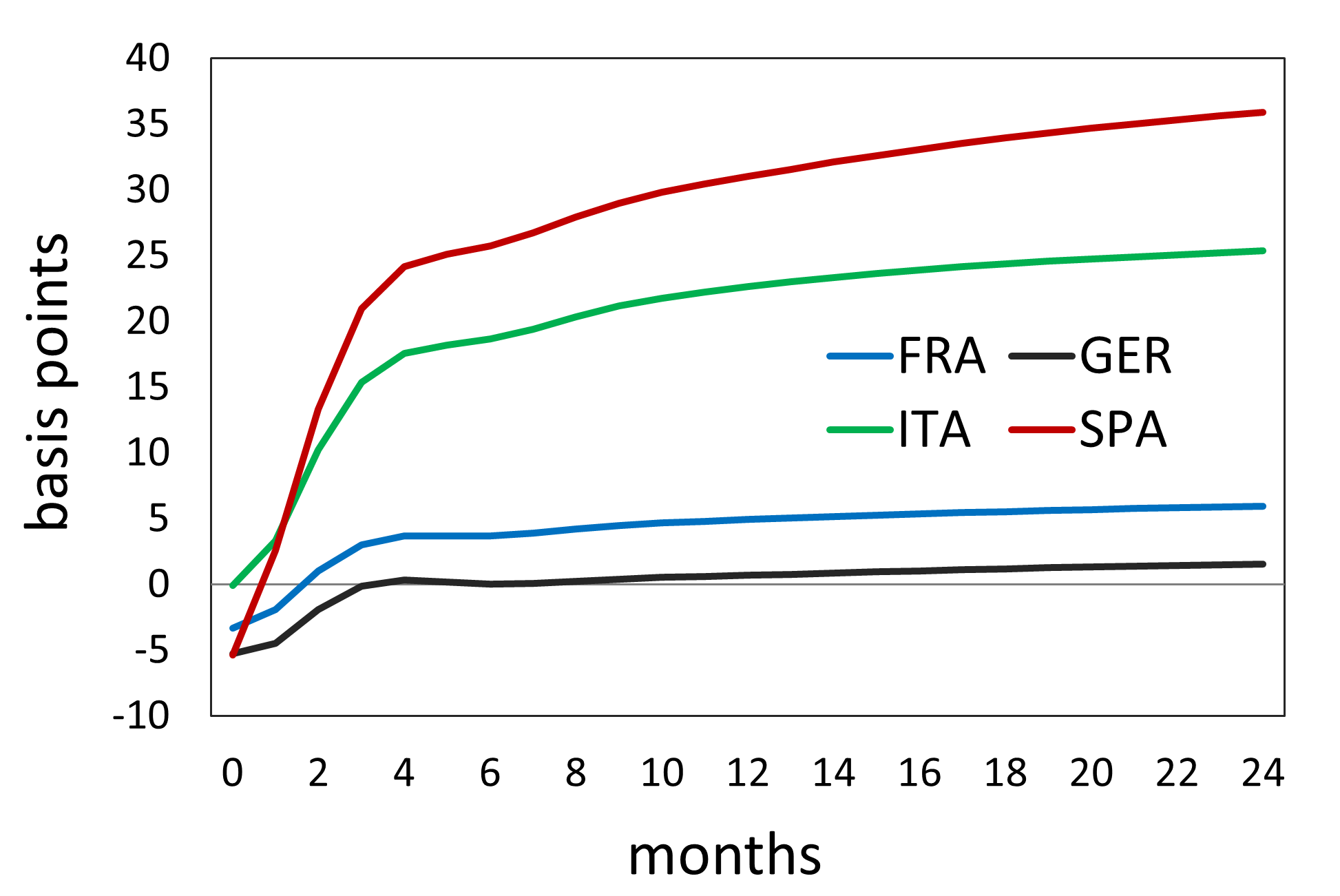}
		\end{minipage}%
		\begin{minipage}{.60\textwidth}
			\includegraphics[scale=0.11]{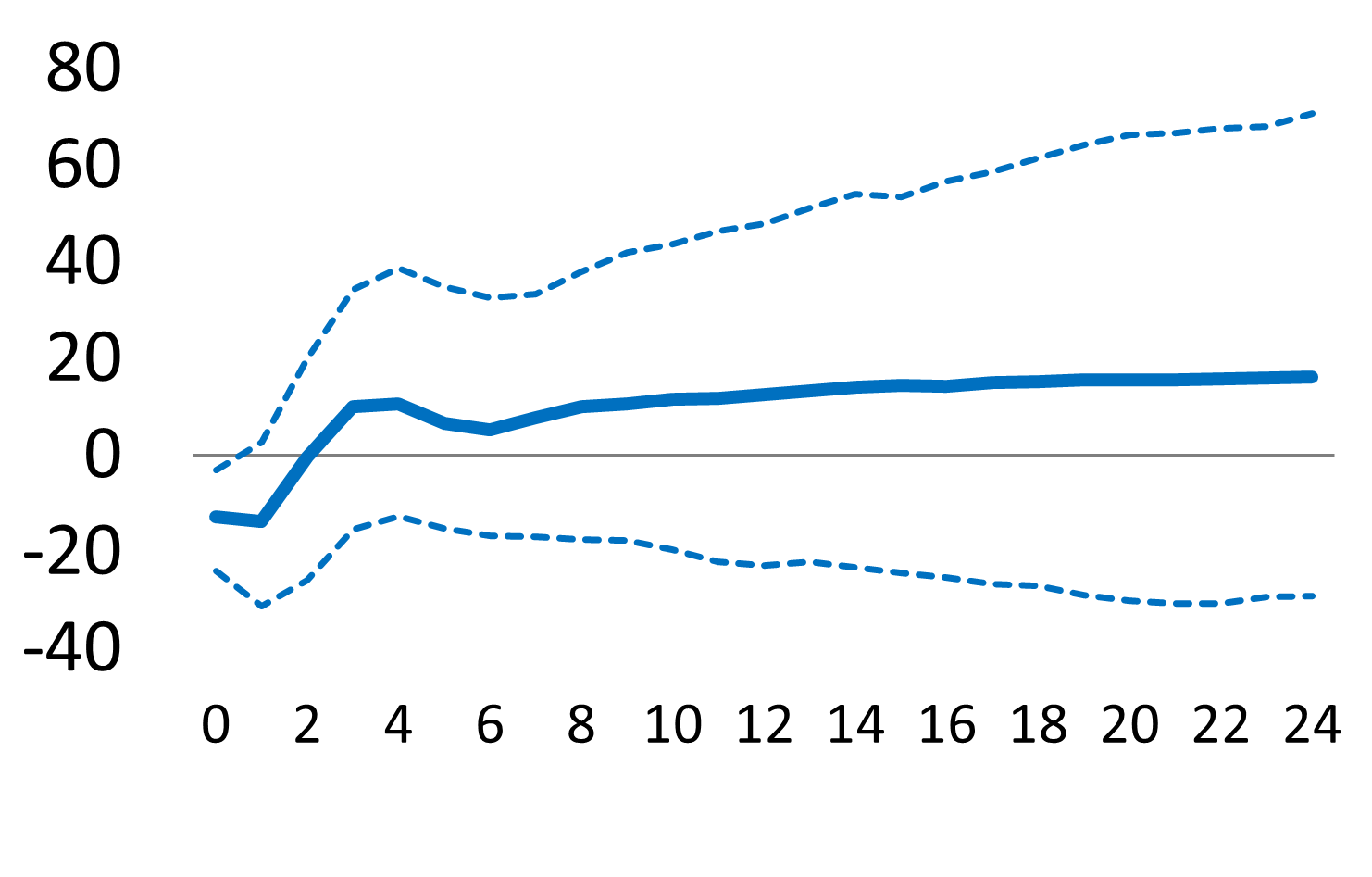}
			\includegraphics[scale=0.11]{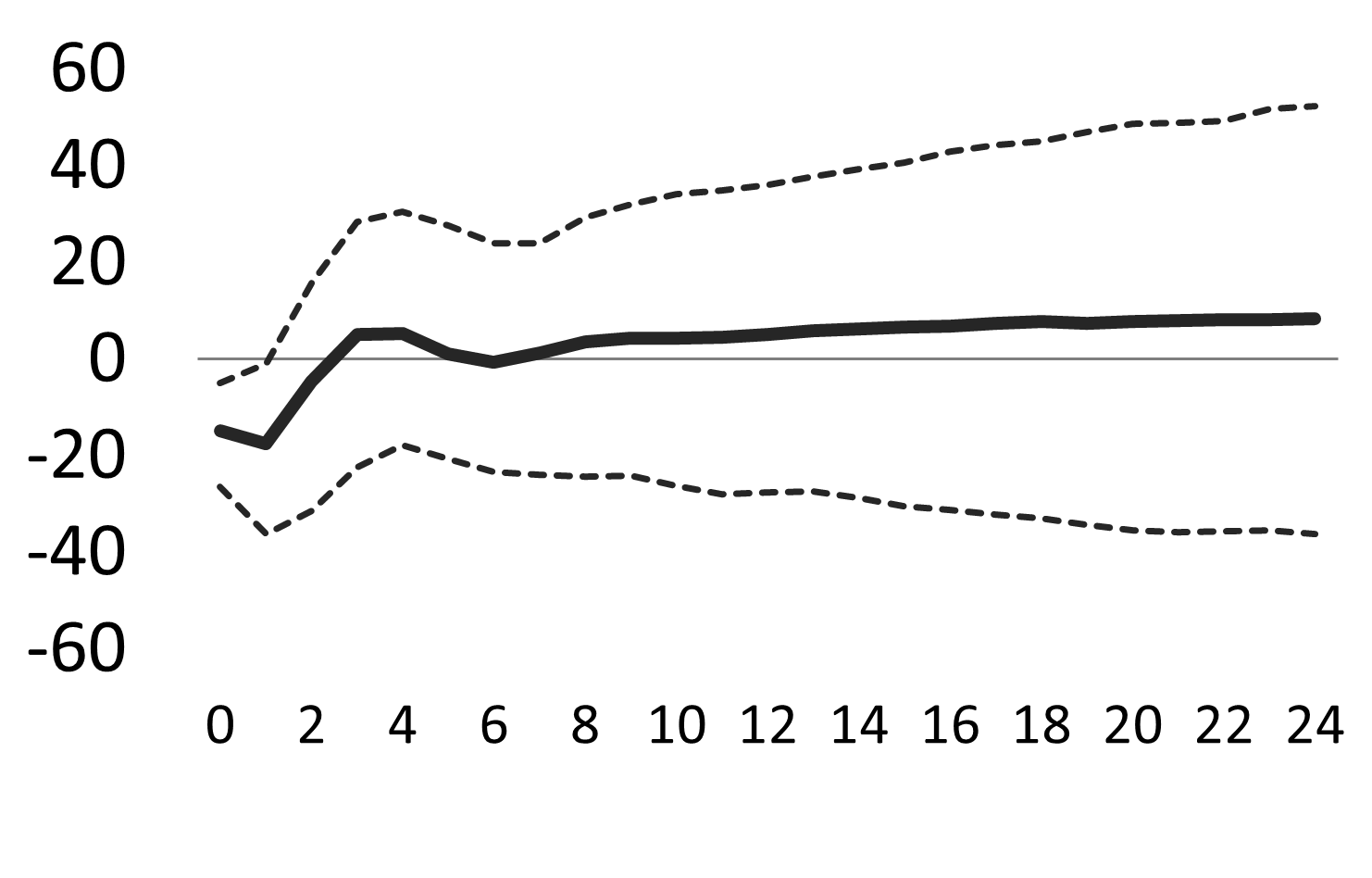}\\
			\includegraphics[scale=0.11]{girf_itgov_itcds_boot.png}
			\includegraphics[scale=0.11]{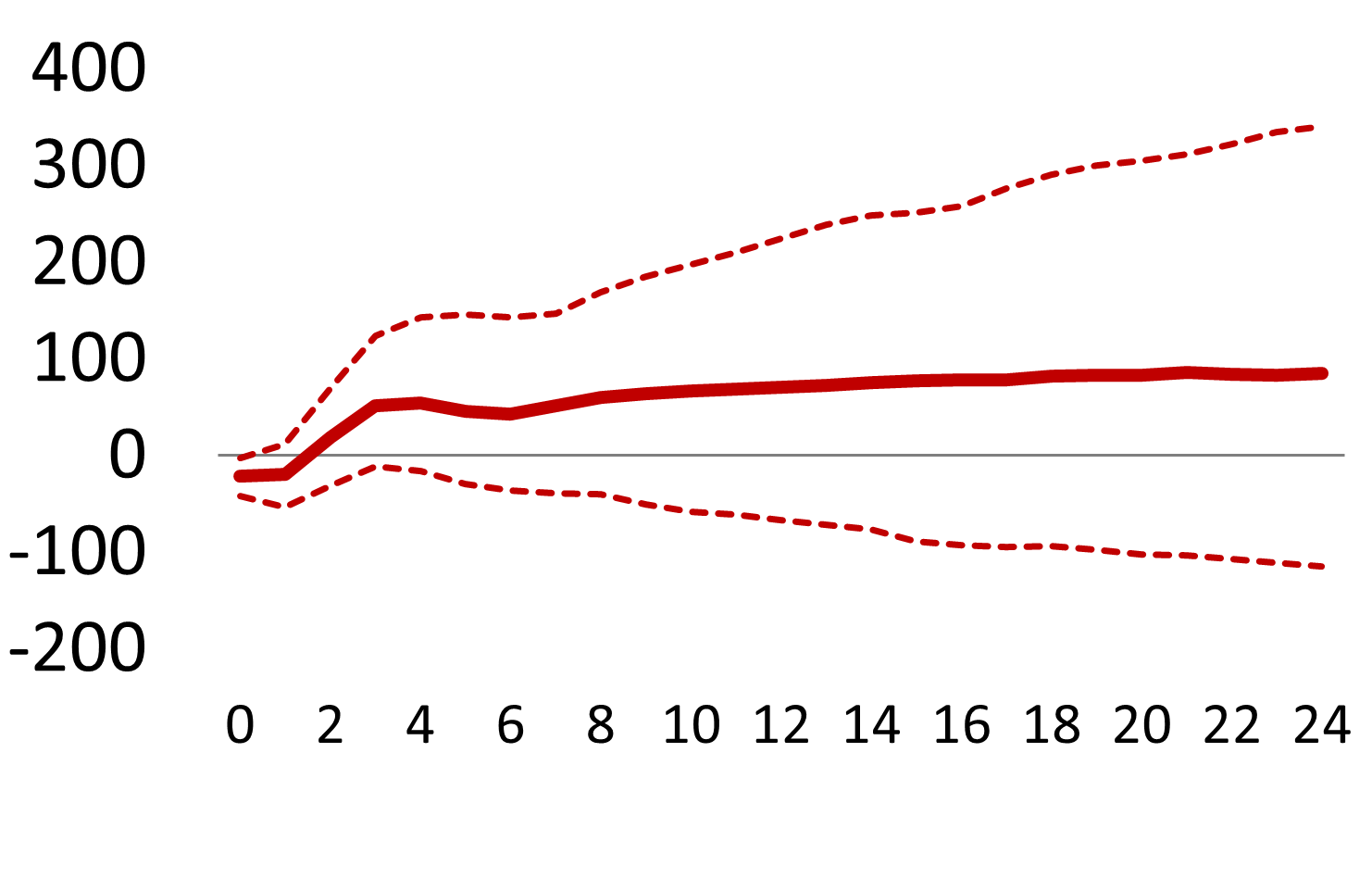}
		\end{minipage}%
		\subcaption*{\textit{Notes}: The figure shows the estimated generalized impulse responses to a positive 1 s.e. shock (left panel) and bootstrap 90\% confidence intervals and medians (right panels).}
		\label{fig:girf_itgov_cds}
	\end{figure}

\begin{figure}[H]
		\centering		
		\caption{GIRF of government yields, shock to ITA CDS spread}
		\begin{minipage}{.5\textwidth}
			\hspace{-15pt}
			\includegraphics[scale=0.15]{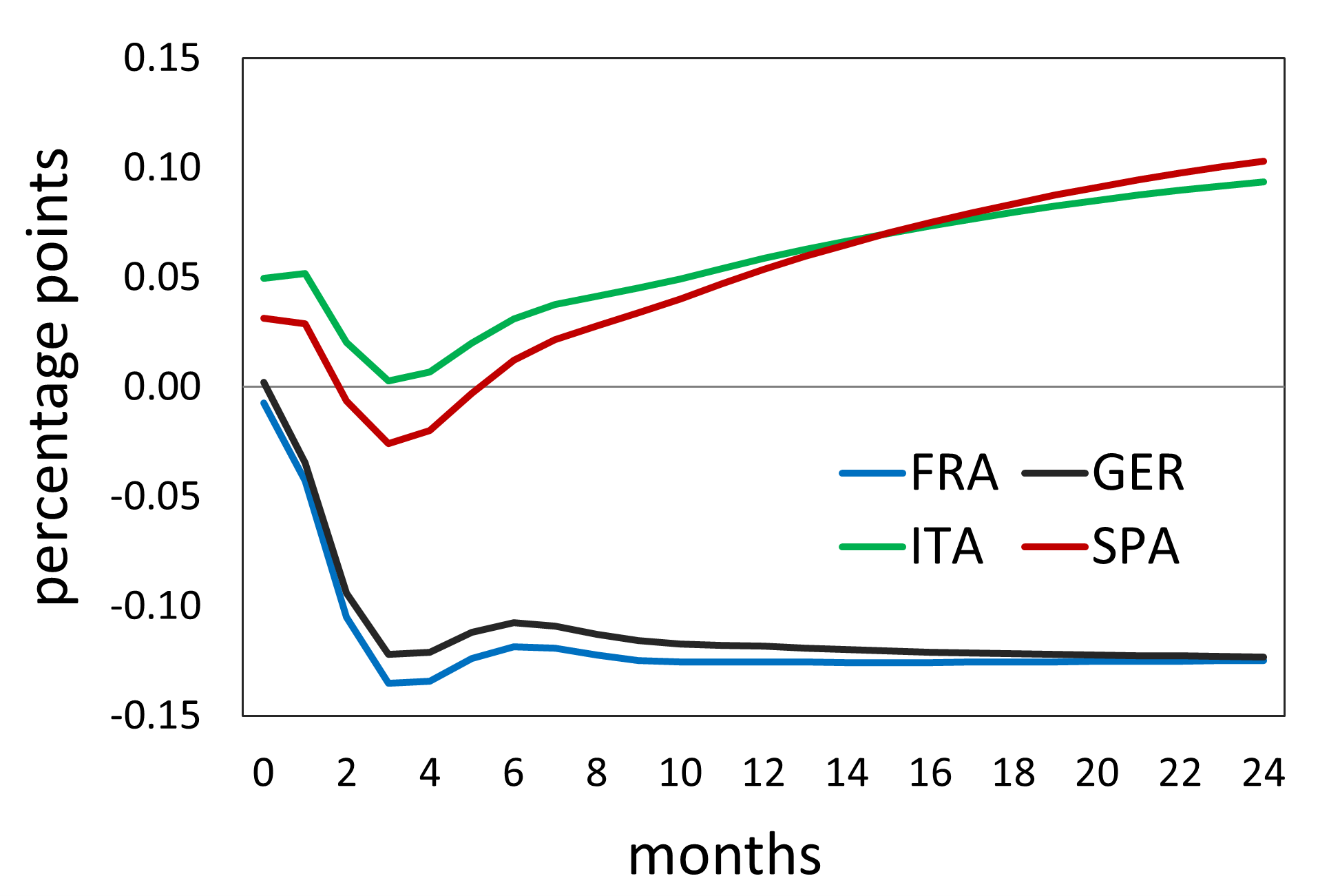}
		\end{minipage}%
		\begin{minipage}{.60\textwidth}
			\includegraphics[scale=0.11]{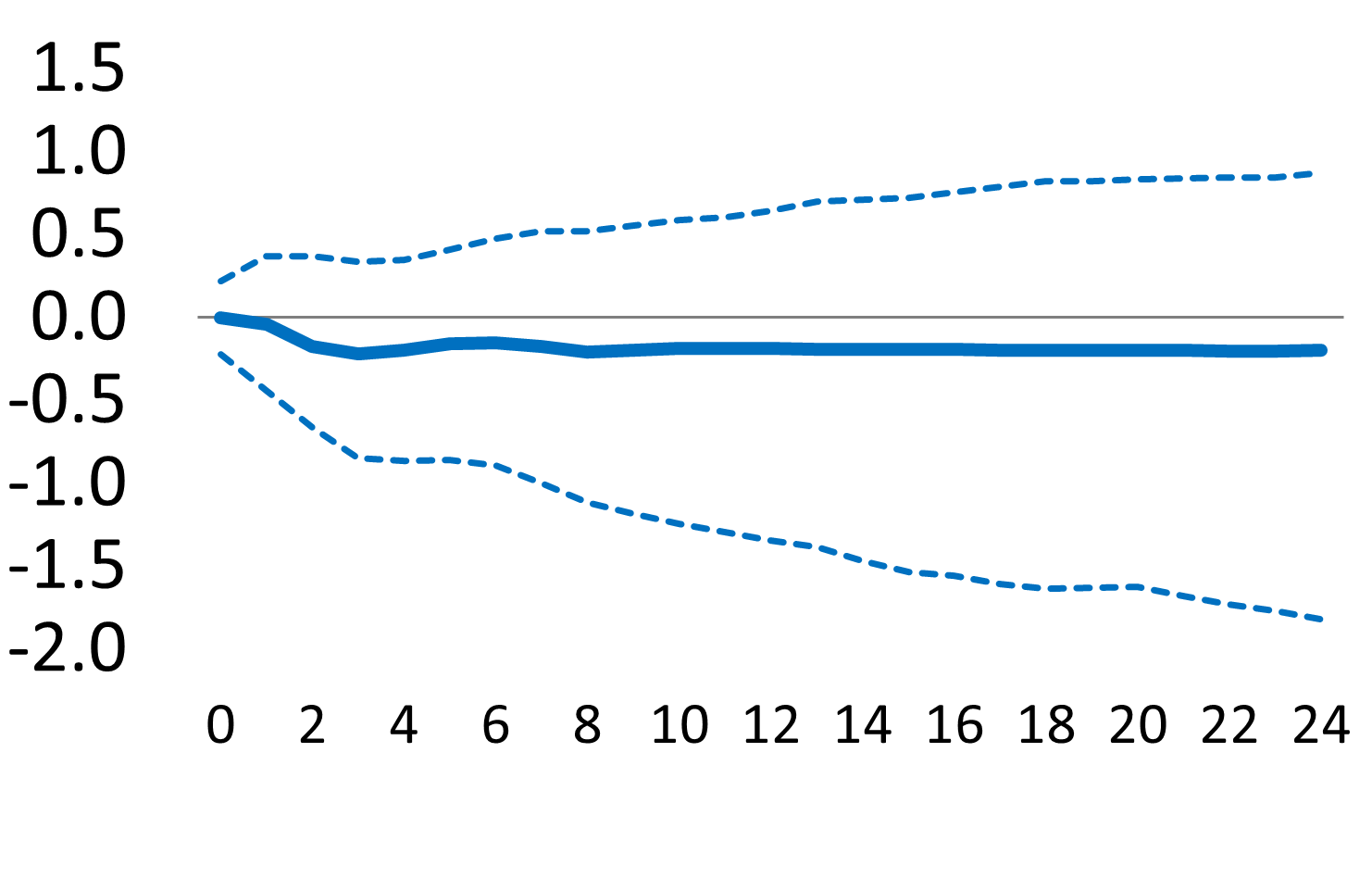}
			\includegraphics[scale=0.11]{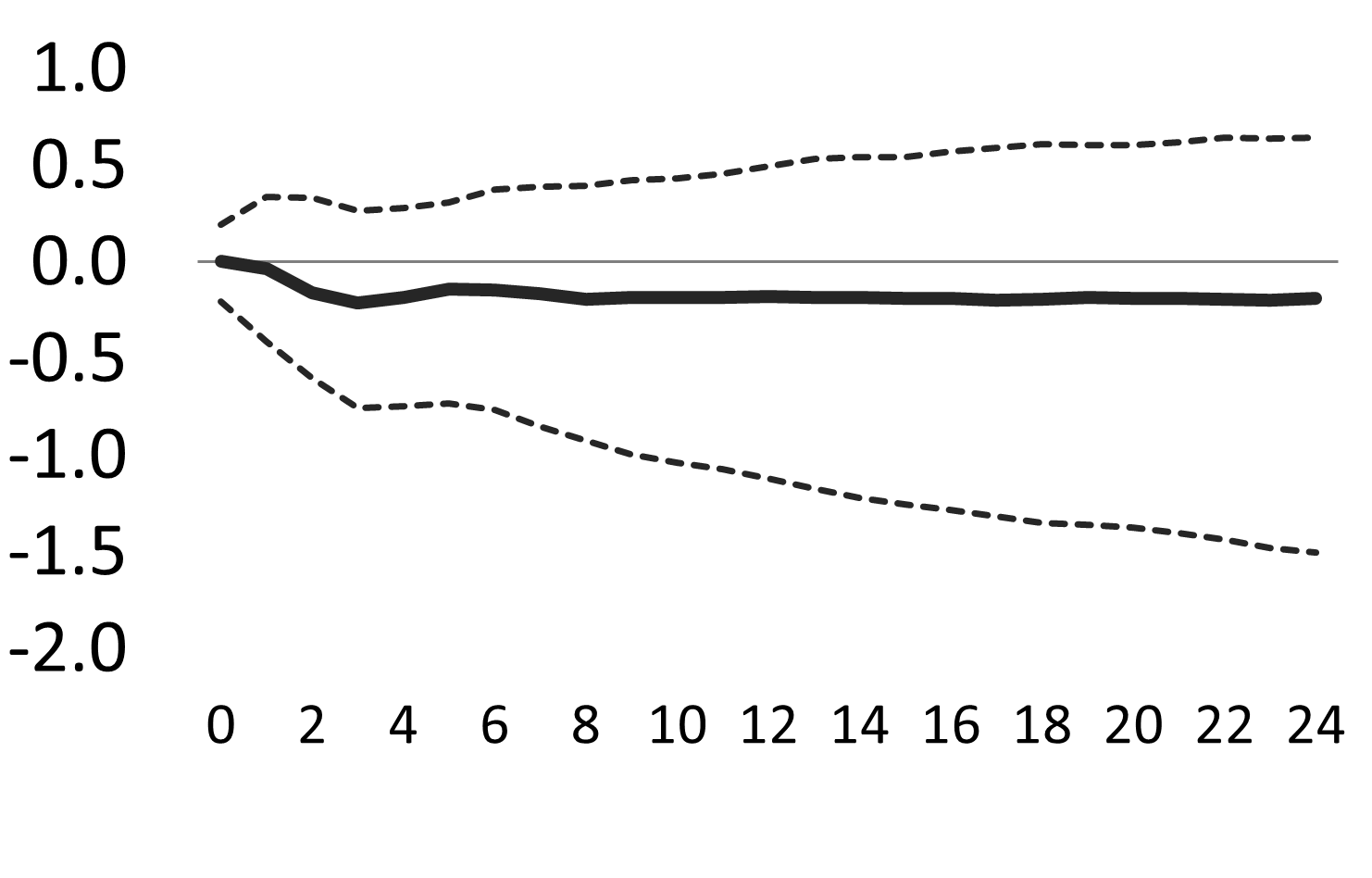}\\
			\includegraphics[scale=0.11]{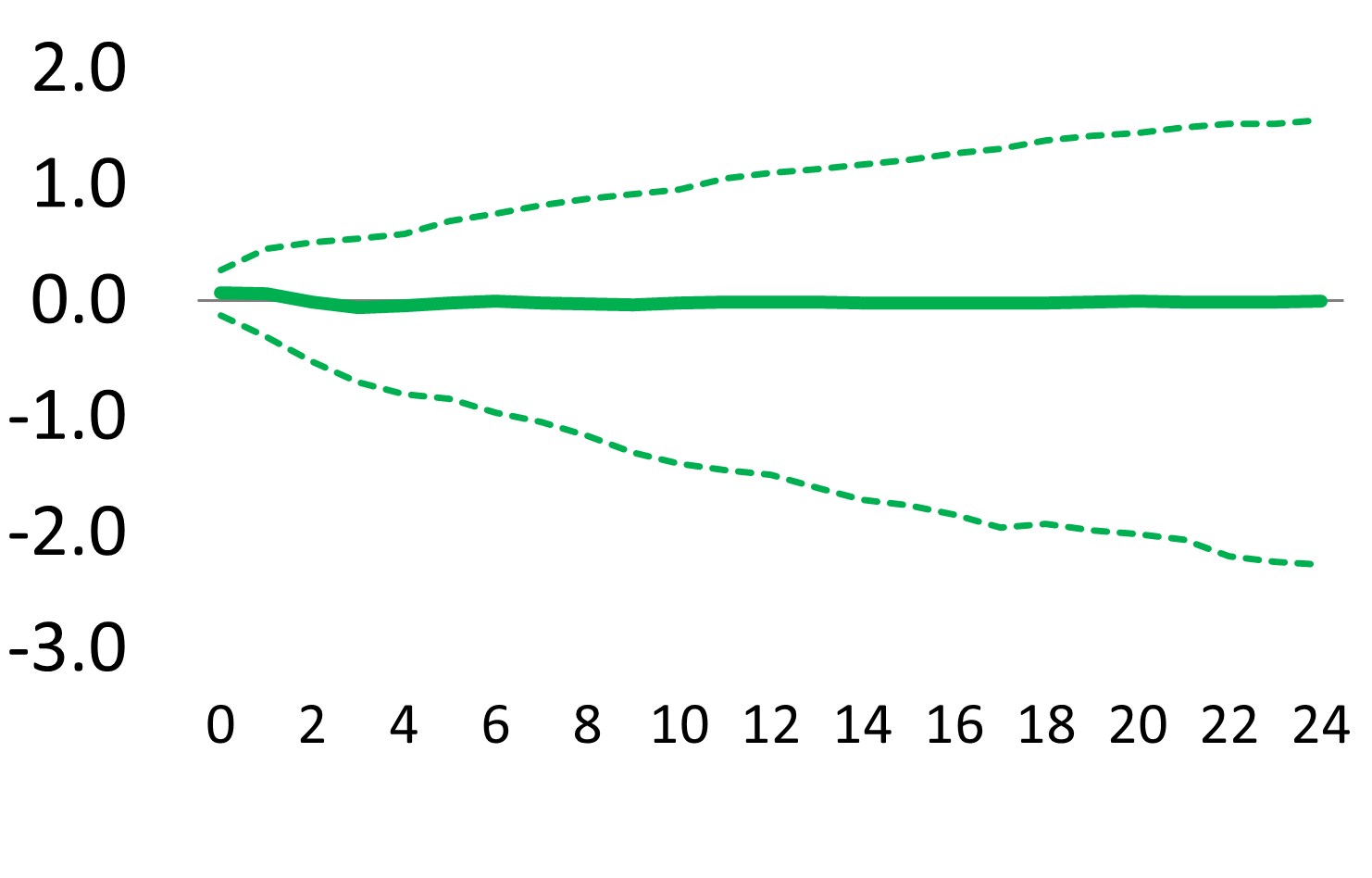}
			\includegraphics[scale=0.11]{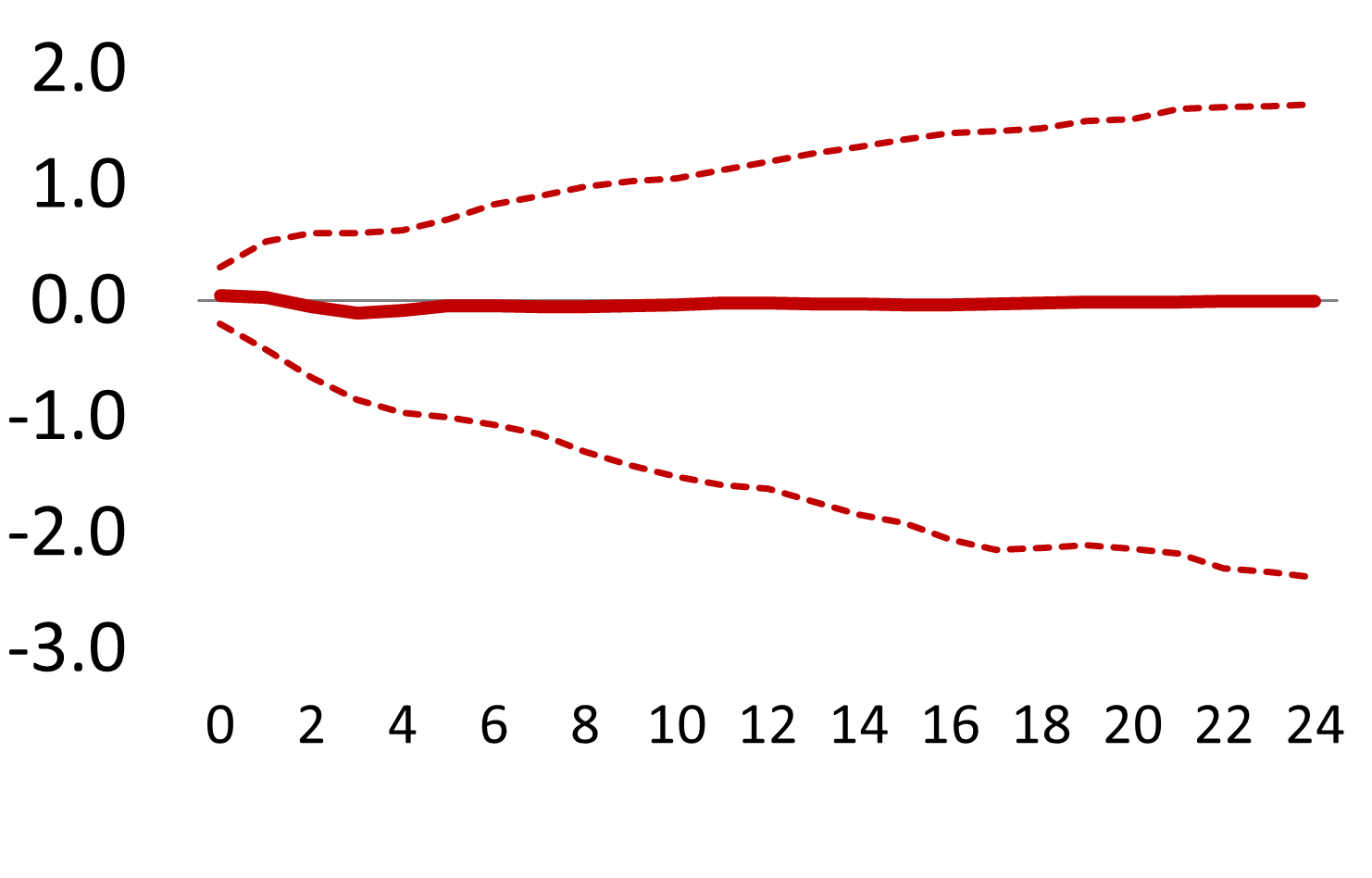}
		\end{minipage}%
		\subcaption*{\textit{Notes}: The figure shows the estimated generalized impulse responses to a positive 1 s.e. shock (left panel) and bootstrap 90\% confidence intervals and medians (right panels).}
		\label{fig:girf_itcds_gov}
	\end{figure}

	Figures \ref{fig:girf_gov_cds}-\ref{fig:girf_itcds_gov} summarize the dynamic relationships between government bond yields and bank CDS spreads in the four countries. ``Core-periphery" heterogeneity and flight to quality are particularly evident here.
	As shown in Figure \ref{fig:girf_gov_cds}, following increases in domestic government yields, bank CDS spreads increase in Spain (reaching +80 bps after 2 years) and Italy (+20 bps), while the estimated effects are almost negligible in Germany. 
	Increases in the German government yield have little or no impact on bank CDS spreads of all countries, with an estimated negative sign in the first months (Figure \ref{fig:girf_gegov_cds}), which seems consistent with the assumption that higher Bund rates tend to be associated with improving economic conditions. 
	A shock to the Italian sovereign yield results in higher Italian and Spanish CDS spreads (+25 bps and +35 bps, respectively, after 24 months), while French and German CDS spreads are almost unaffected (Figure \ref{fig:girf_itgov_cds}). 
	Following a shock to the Italian bank CDS spreads, government yields tend to decrease in Germany and France (-0.12\%, approximately, in point estimates) and to increase in Italy and Spain (+0.1\%, approximately), as shown in Figure \ref{fig:girf_itcds_gov}.

	\vspace{0pt}

	\subsection{Shocks to the exogenous ECB programmes}
	The effects of shocks to the exogenous variables (ECB policies) are assessed by simulating several ``shock scenarios". In each scenario, a single ECB programme is shocked.  For programmes involving actual asset purchases, i.e., LTRO, SMP and APP, the shock scenario consists in a 50\% reduction in the amounts of assets held by the ECB over the period 2015M1-2017M3, while for the OMT it is defined by setting the OMT dummy variable to zero (termination of the programme). 
	
\begin{figure}[H]
		\centering		
		\caption{Response of government yields, shock to LTRO}
		\begin{minipage}{.5\textwidth}
			\hspace{-15pt}
			\includegraphics[scale=0.15]{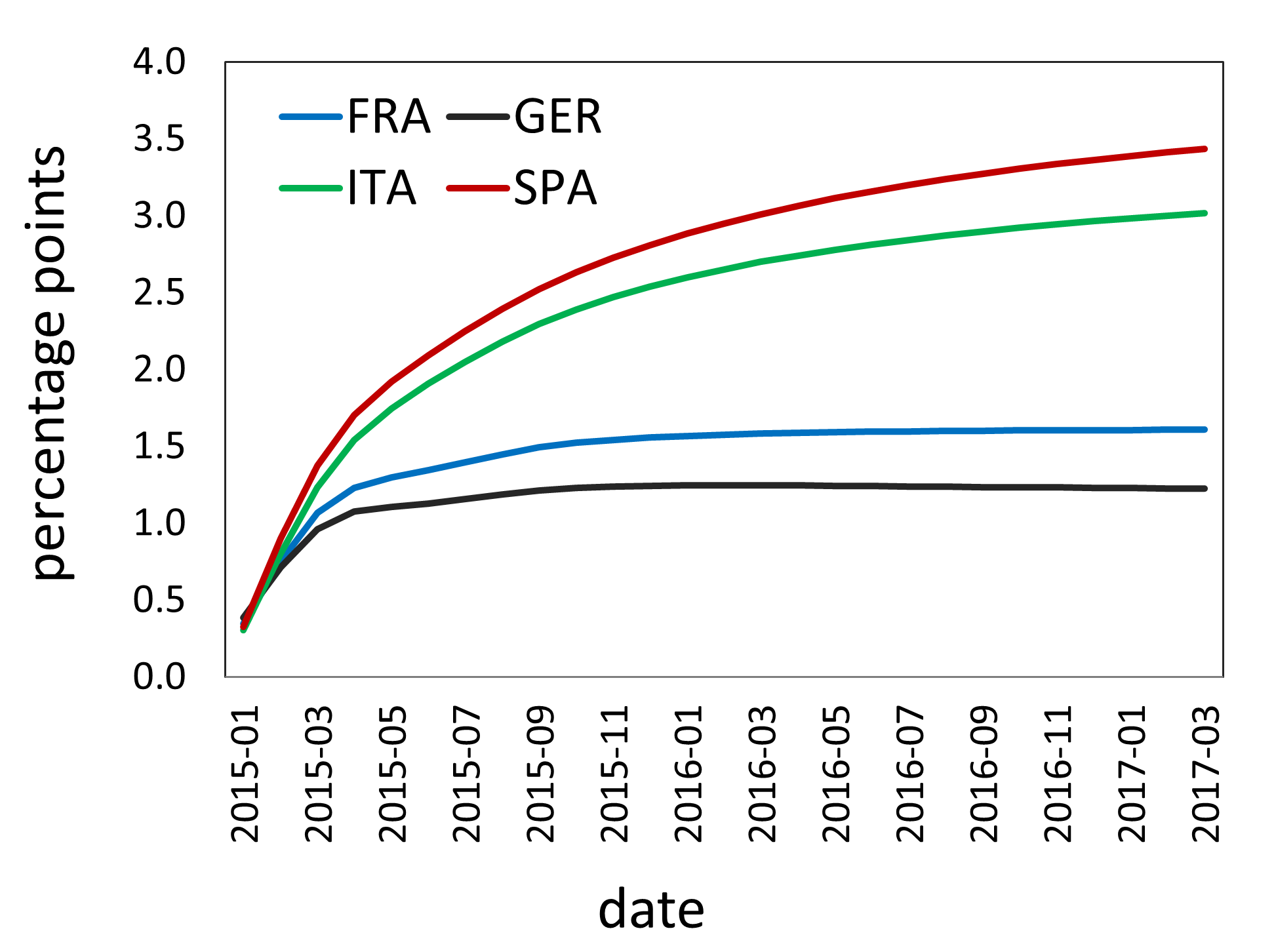}
		\end{minipage}%
		\begin{minipage}{.60\textwidth}
			\includegraphics[scale=0.11]{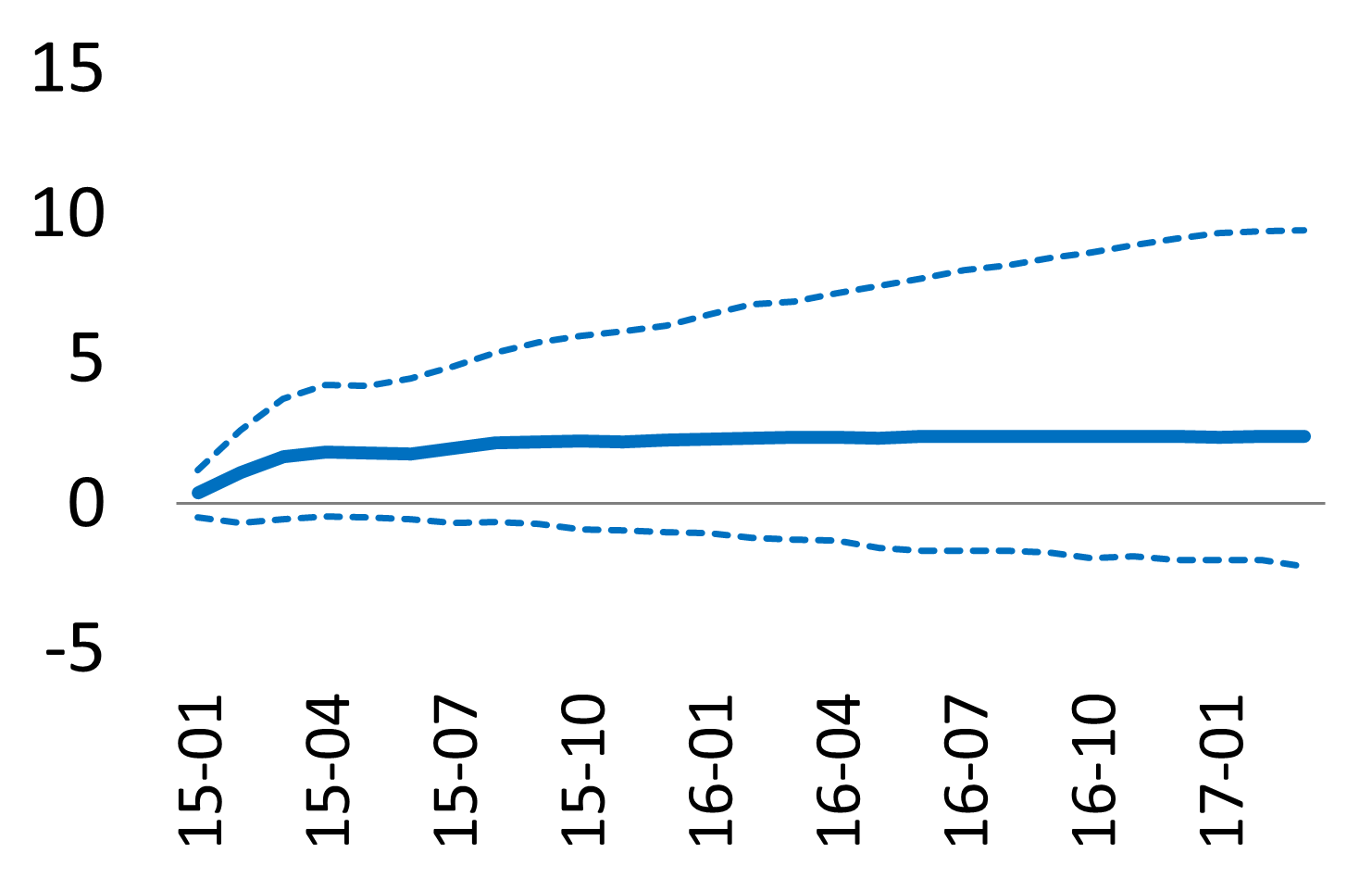}
			\includegraphics[scale=0.11]{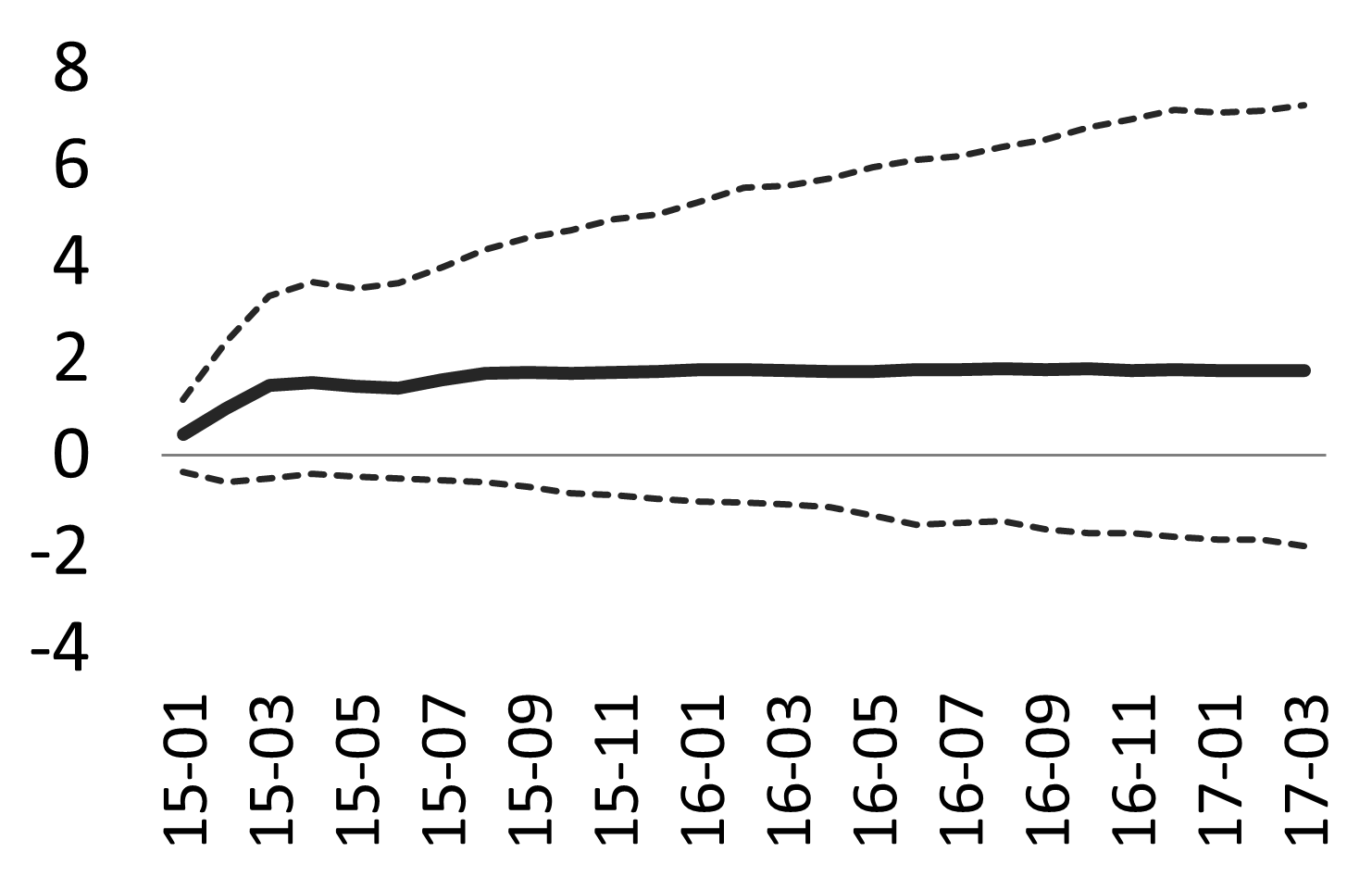}\\
			\includegraphics[scale=0.11]{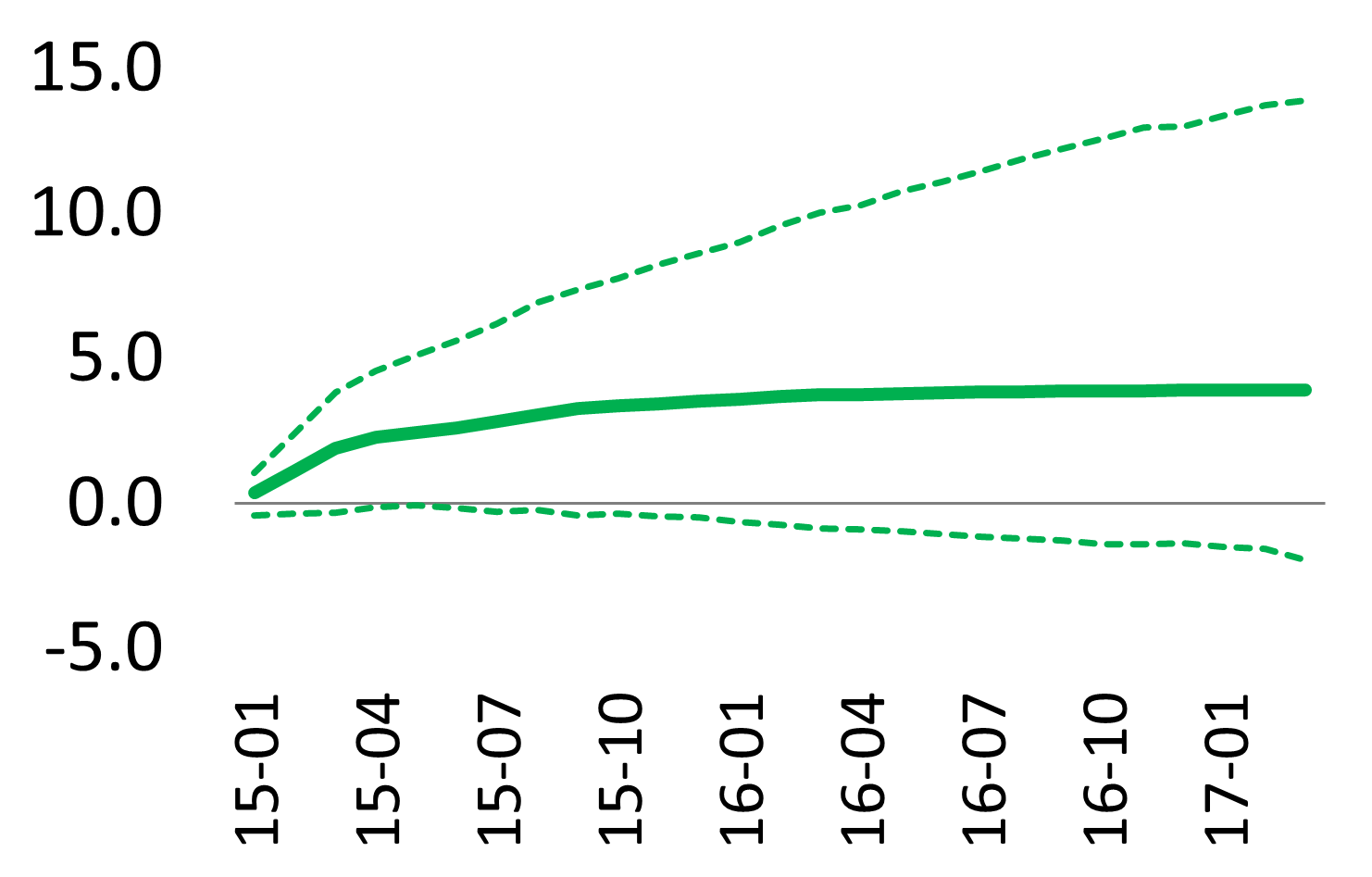}
			\includegraphics[scale=0.11]{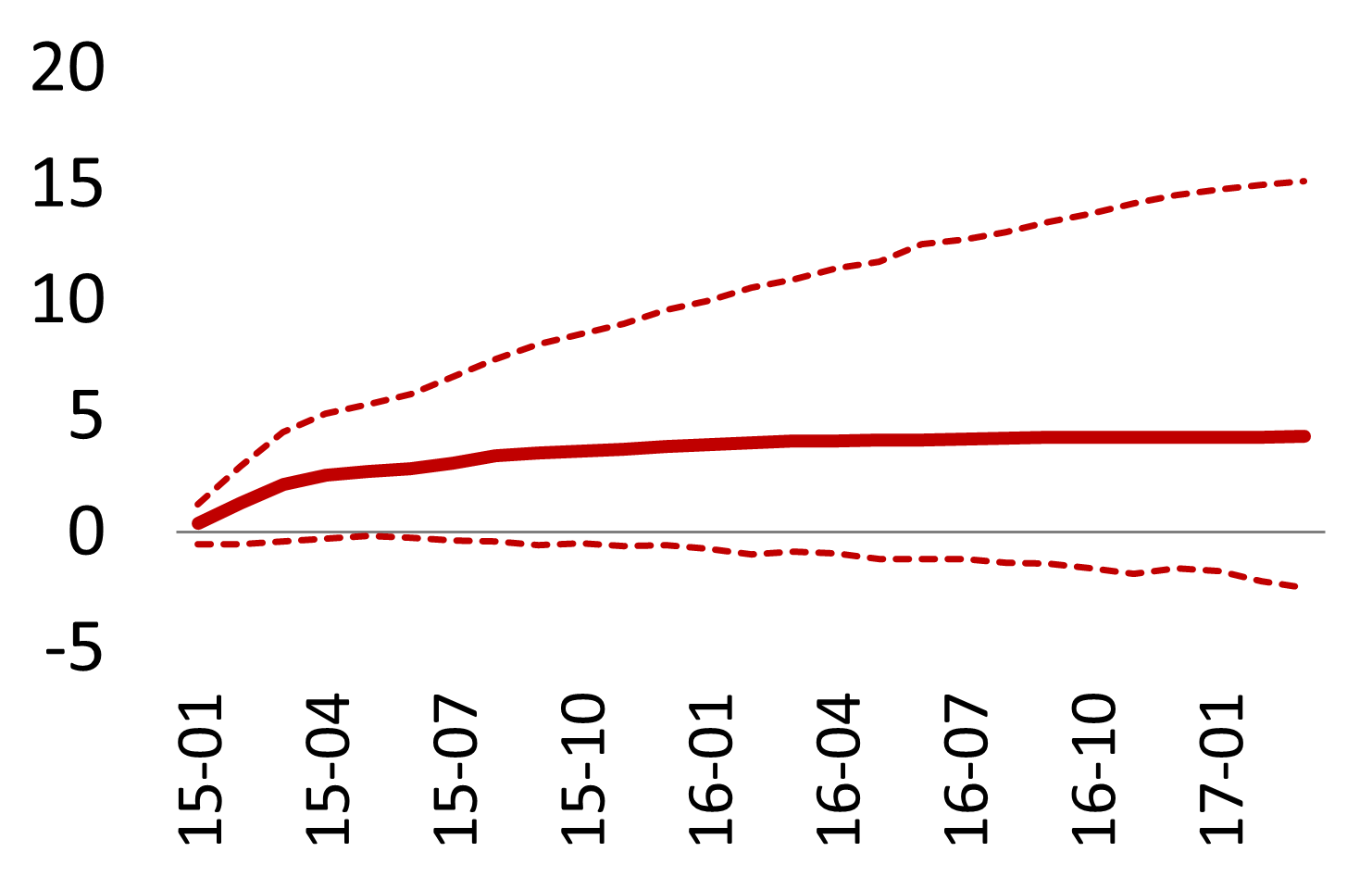}
		\end{minipage}%
		\subcaption*{\textit{Notes}: The figure shows the estimated responses to the shock scenario for the LTRO programme (left panel) and bootstrap 90\% confidence intervals and medians (right panels).}
		\label{fig:shock_ltro_gov}
	\end{figure}

\begin{figure}[H]
		\centering		
		\caption{Response of repo rates, shock to APP}
		\begin{minipage}{.5\textwidth}
			\hspace{-15pt}
			\includegraphics[scale=0.15]{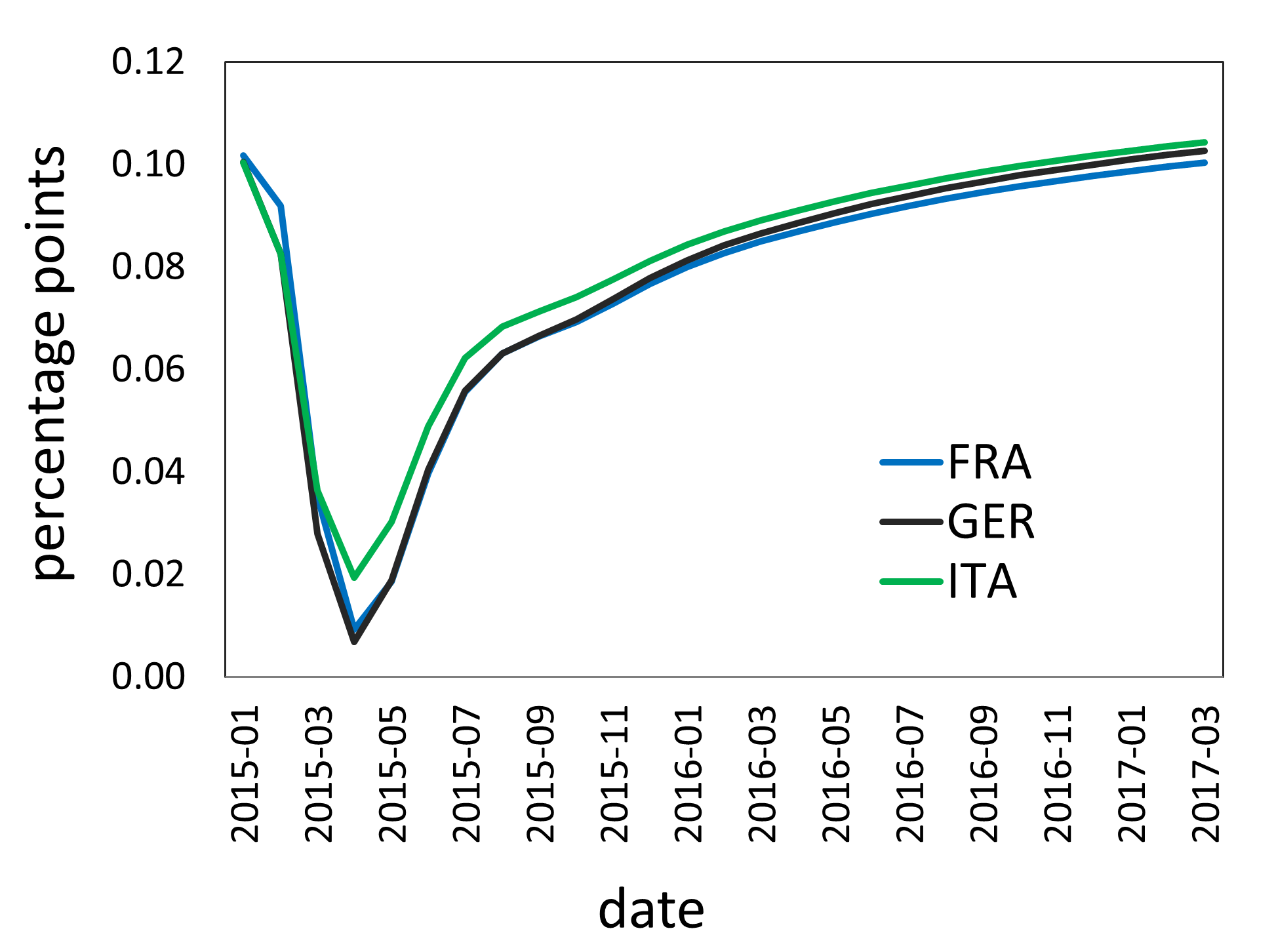}
		\end{minipage}%
		\begin{minipage}{.60\textwidth}
			\includegraphics[scale=0.11]{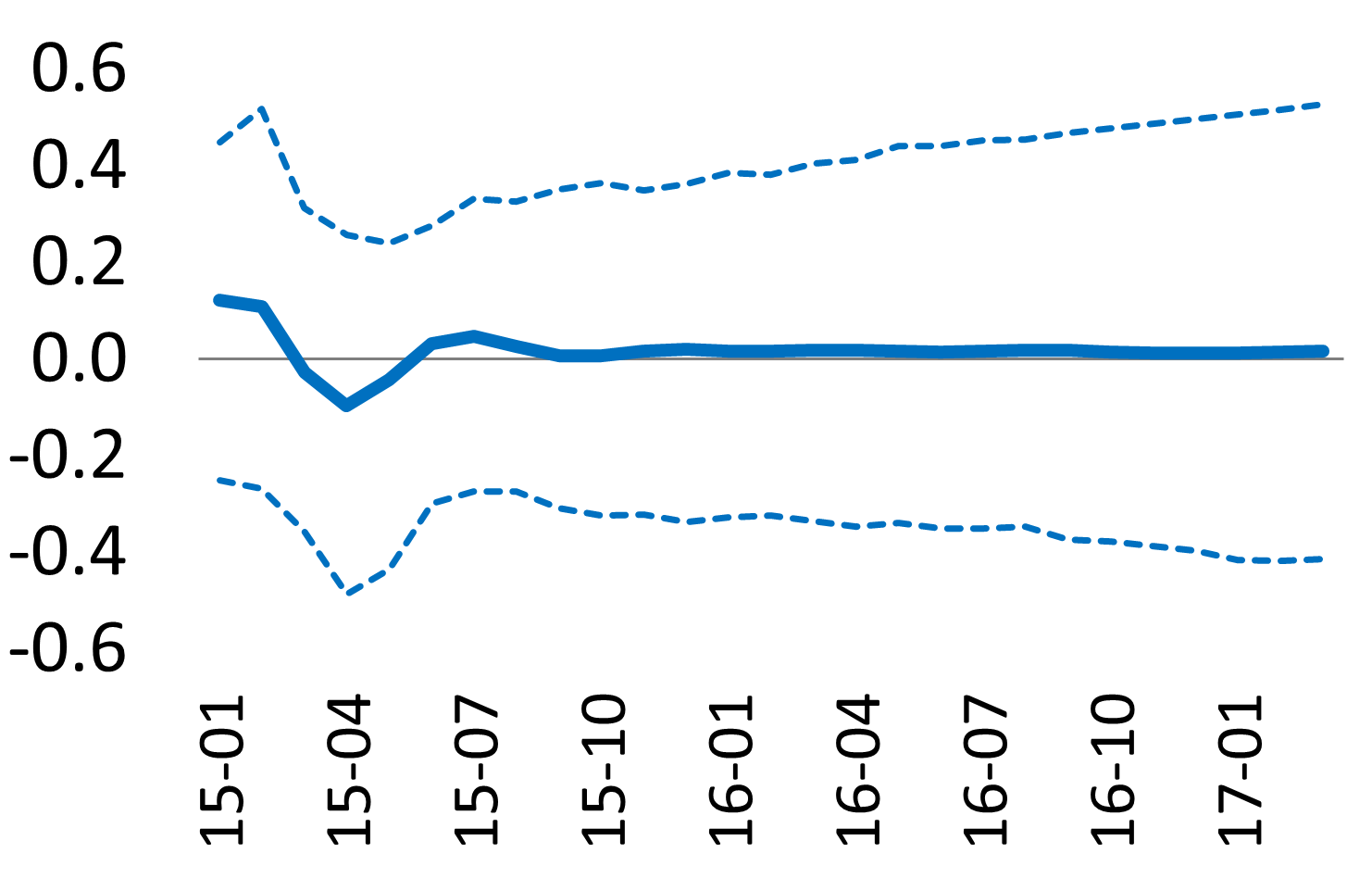}
			\includegraphics[scale=0.11]{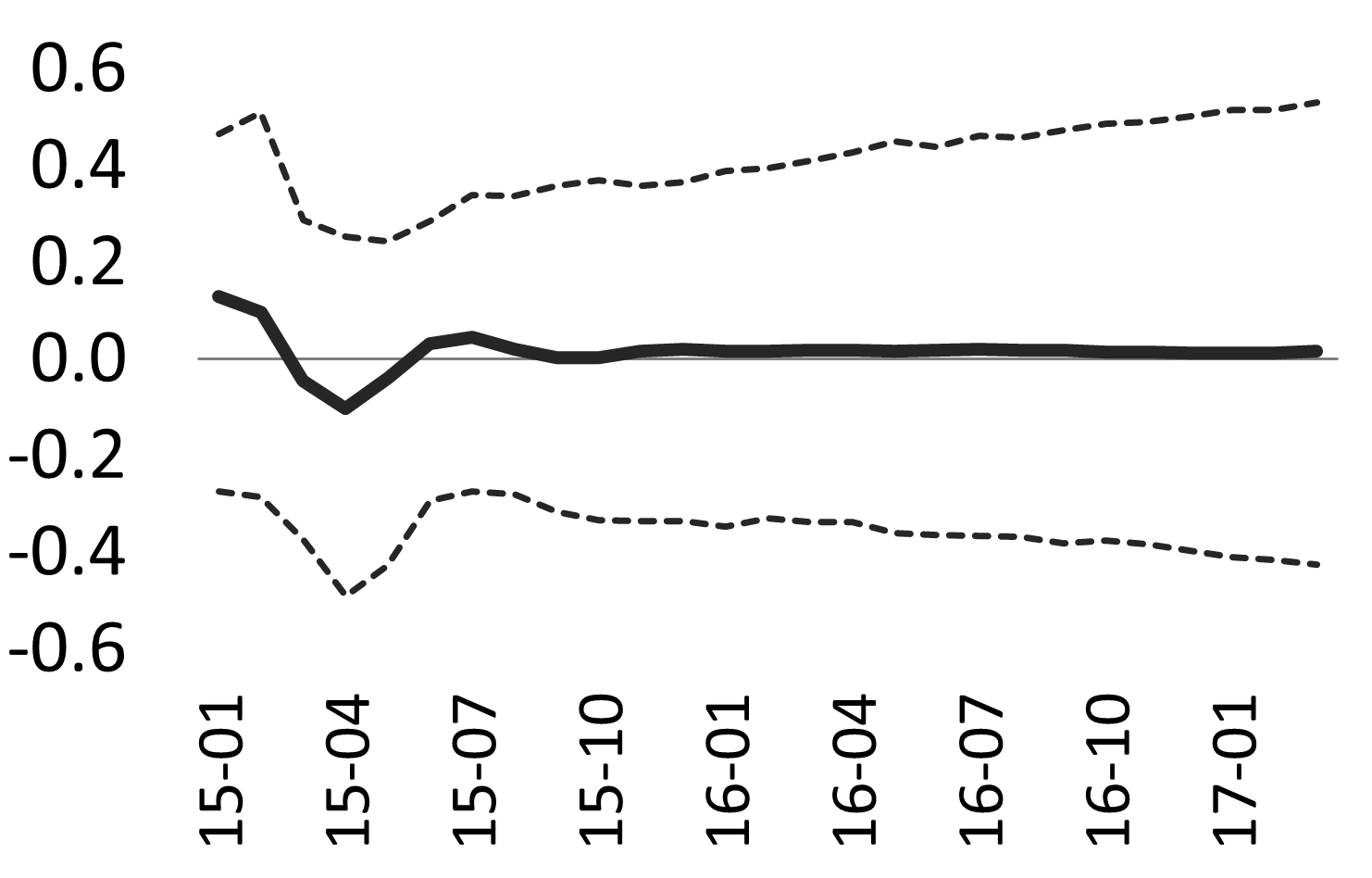}\\
			\includegraphics[scale=0.11]{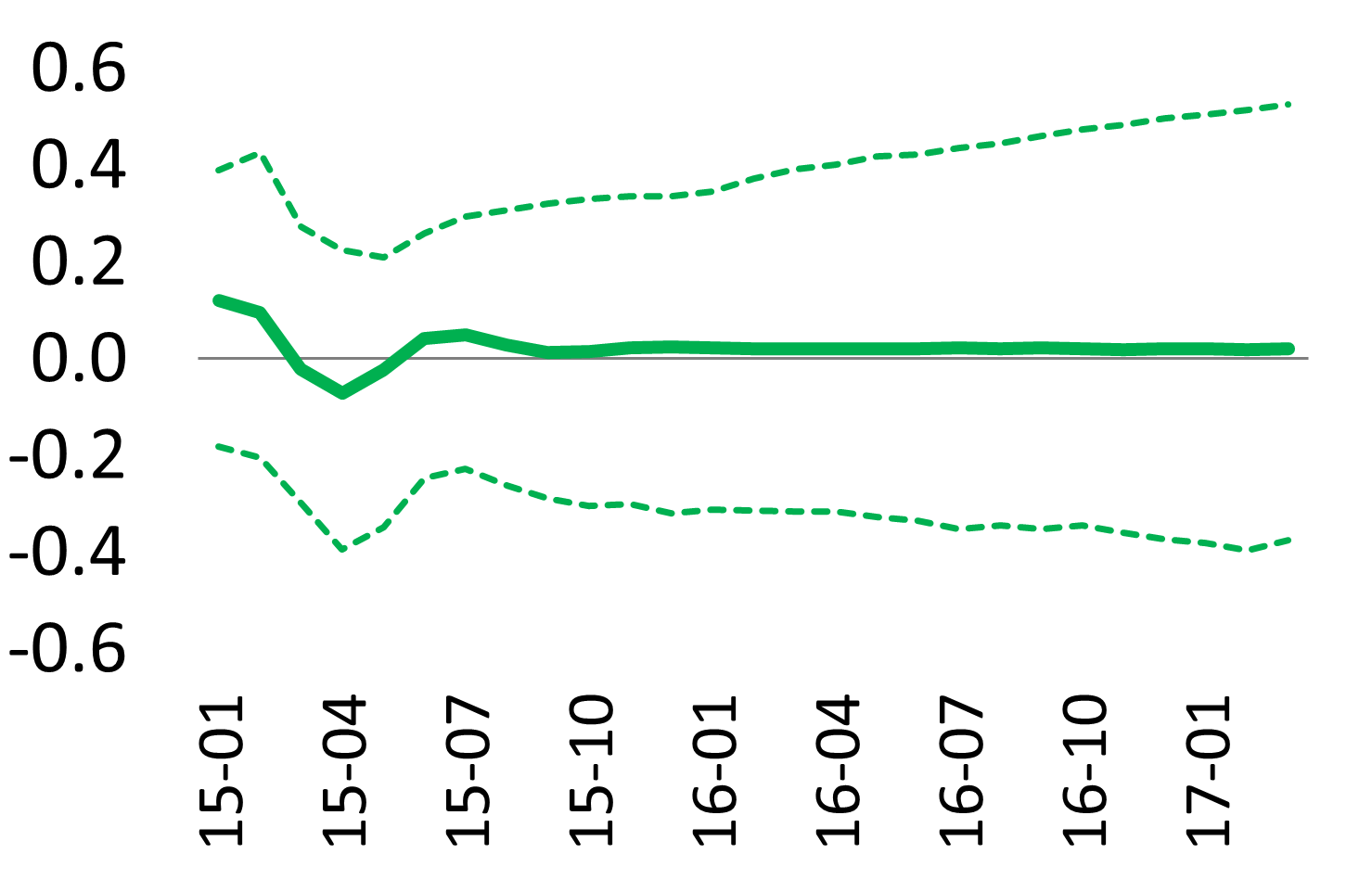}
		\end{minipage}%
		\subcaption*{\textit{Notes}: The figure shows the estimated responses to the shock scenario for the APP programme (left panel) and bootstrap 90\% confidence intervals and medians (right panels).}
		\label{fig:shock_app_rep}
	\end{figure}

	Figures \ref{fig:shock_ltro_gov}-\ref{fig:shock_app_rep} show the responses of the endogenous variables to these shocks. Downsizing of any programme (or termination, in the case of OMT) tends to result in higher government bond yields and bank CDS spreads, especially in Italy and Spain, as well as in lower repo trading volumes and higher repo interest rates in all countries.%
	
	A 50\% reduction in LTRO causes the Italian 10-year yield to soar by 2.5\% at the end of 2015 and by almost 3\% at the end of the simulation period. The Spanish yield climbs up by almost 3.5\% over the two-year simulation horizon. The German and French yields also rise, but to a lesser extent: the German yield rises by 1\%, the French yield by 1.5\%. Accordingly, Italian and Spanish sovereign spreads rise. This ``core-periphery" type of heterogeneity is also observed in the effects on bank CDS. The Spanish and Italian CDS indices rise by 450 basis points and 200 basis points, respectively, at the end of simulation period, while the German and French CDS increase by approximately 50 bps. While positive on impact (especially for Germany), all estimated responses of trading volumes become negative within 9 months, although with very large confidence intervals. In point estimates, all volumes drop by more than 15\% by March 2017. All repo rates increase significantly on impact and their final increases are around 1.1\%-1.2\%.
	
	Shocking the SMP delivers results which are qualitatively similar but quantitatively smaller.  
	In response to the shock, the Italian yield rises by up to 0.7\% in point estimates, the Spanish yield by around 0.8\%, the German yield by 0.3\% and the French yield by 0.4\%. The reponse of the bank CDS spreads eventually exceeds 100 bps for Spain and 40 bps for Italy, while French and German banks are almost unaffected.
	
	The simulated termination of the OMT delivers more dramatic increases in Italian and Spanish yields and CDS spreads. Yields are estimated to increase in both countries by around 6\% relative to the baseline scenario, but these estimates are subject to considerable uncertainty. CDS spreads shoot up (significantly) by 1500 bps for Spain and 900 bps for Italy. The German Bund is not affected, while the French yield increases by just 1\% in point estimate. Both ``core" countries experience 350-bps increases in CDS spreads, approximately. Repo trading volumes plummet (especially in Italy), while the responses of repo rates are negative on impact and their final increases are smaller than in the case of the LTRO shock. 
	
	Finally, the APP shock provides the smallest (and all statistically non-significant) effects. After reaching 30-40 bps on impact, the responses of government yields converge to 5-20 bps in point estimates. Italy and Spain still exhibit the largest responses. CDS responses are negative on impact but end up at around zero, except for Spain (+30 bps in 2017). All repo trading volumes rise on impact, then turn negative and eventually decline by no more than 3\%. Repo rates increase by just 10 bps at the end of the period.
	
	To conclude, the OMT shock scenario is associated with the largest estimated increases in Italian and Spanish government yields and CDS spreads, and with the largest declines in repo trading volumes. It should be stressed, however, that the nature of the OMT shock (complete termination of the programme) is different from that of the shocks to other ECB programmes (halving the amount of assets).

	\subsubsection{LTRO}

	\begin{figure}[H]
		\centering		
		\caption{Response of bank CDS spreads, shock to LTRO}
		\begin{minipage}{.5\textwidth}
			\hspace{-15pt}
			\includegraphics[scale=0.15]{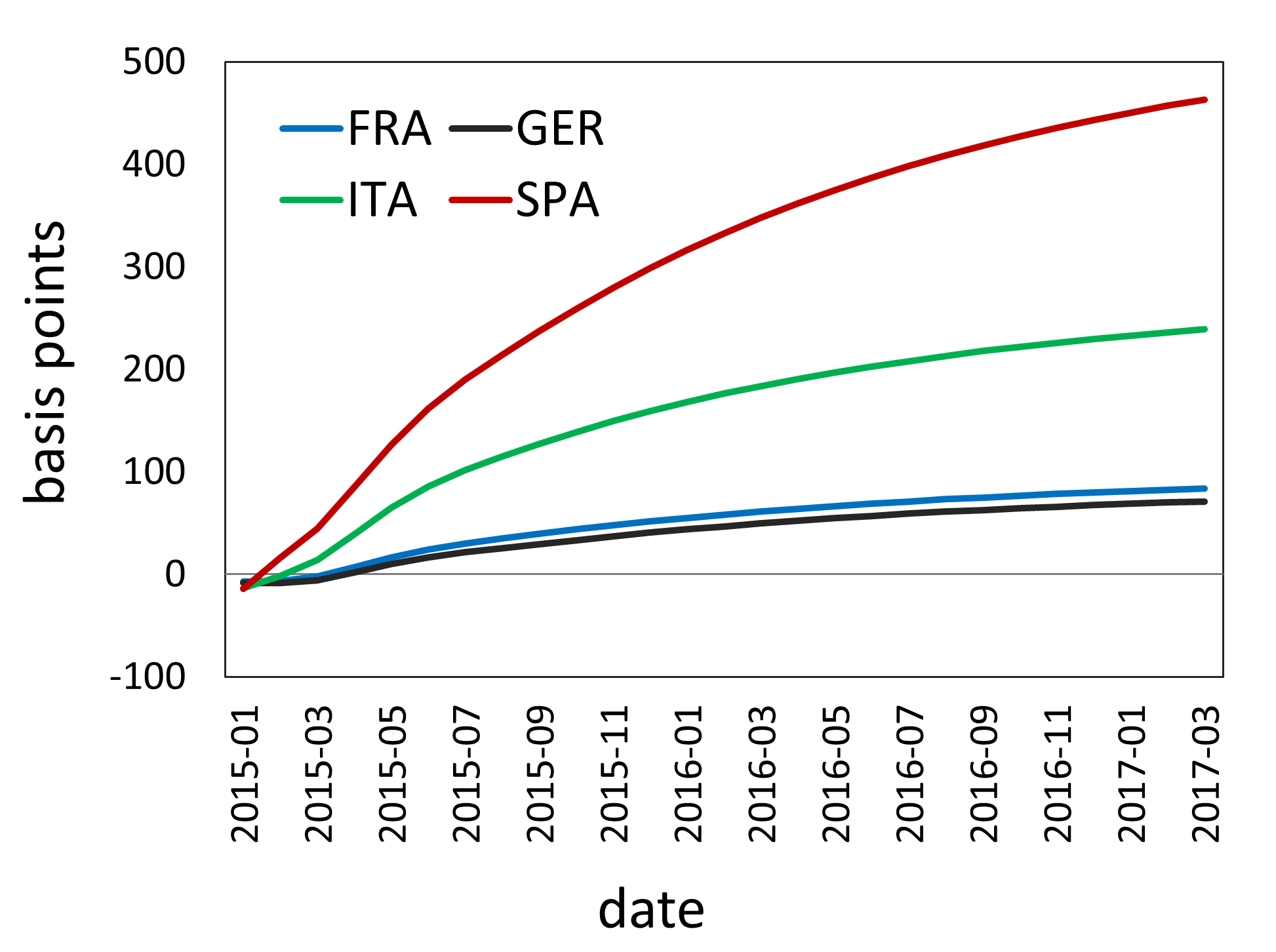}
		\end{minipage}%
		\begin{minipage}{.60\textwidth}
			\includegraphics[scale=0.11]{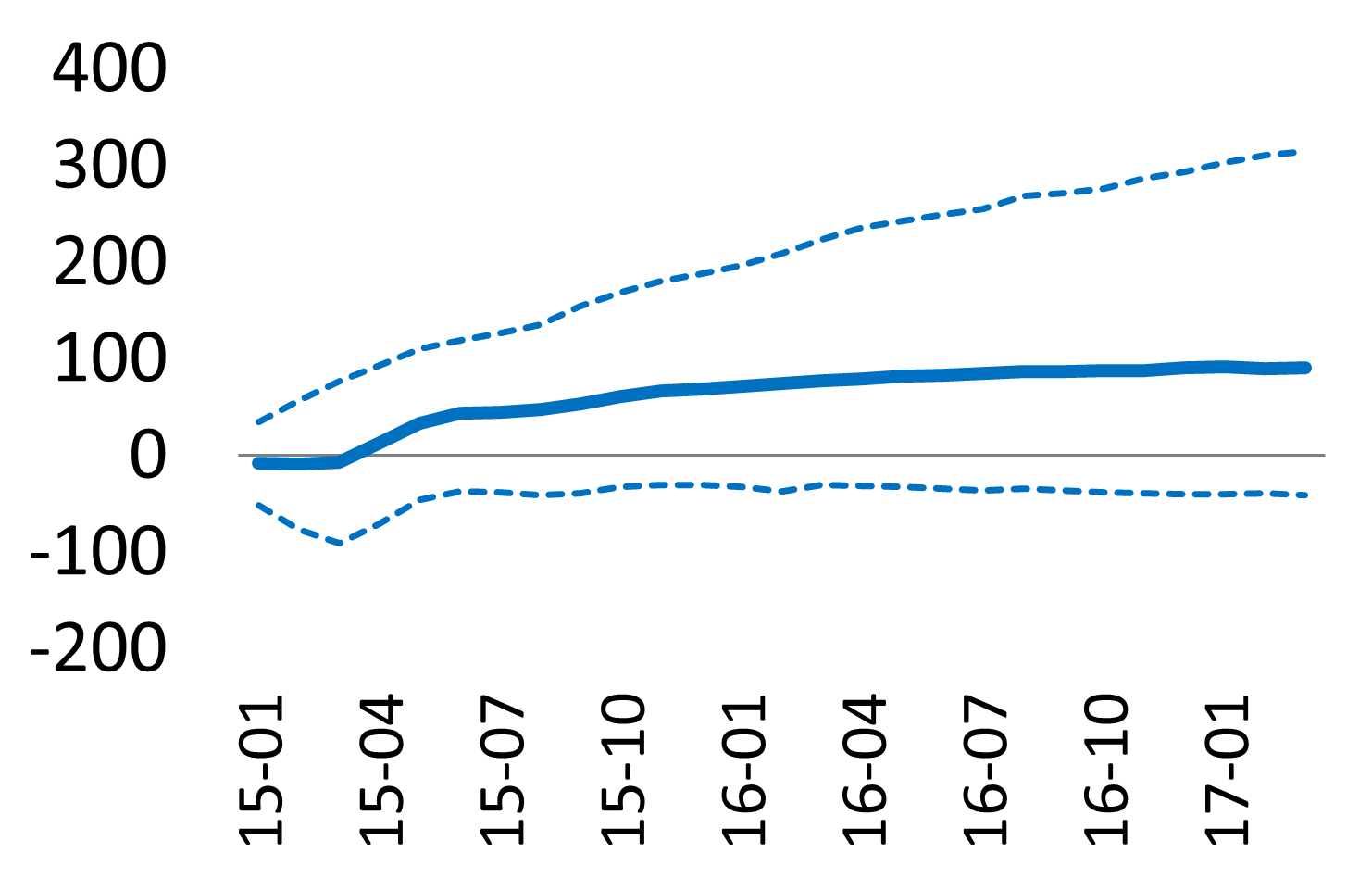}
			\includegraphics[scale=0.11]{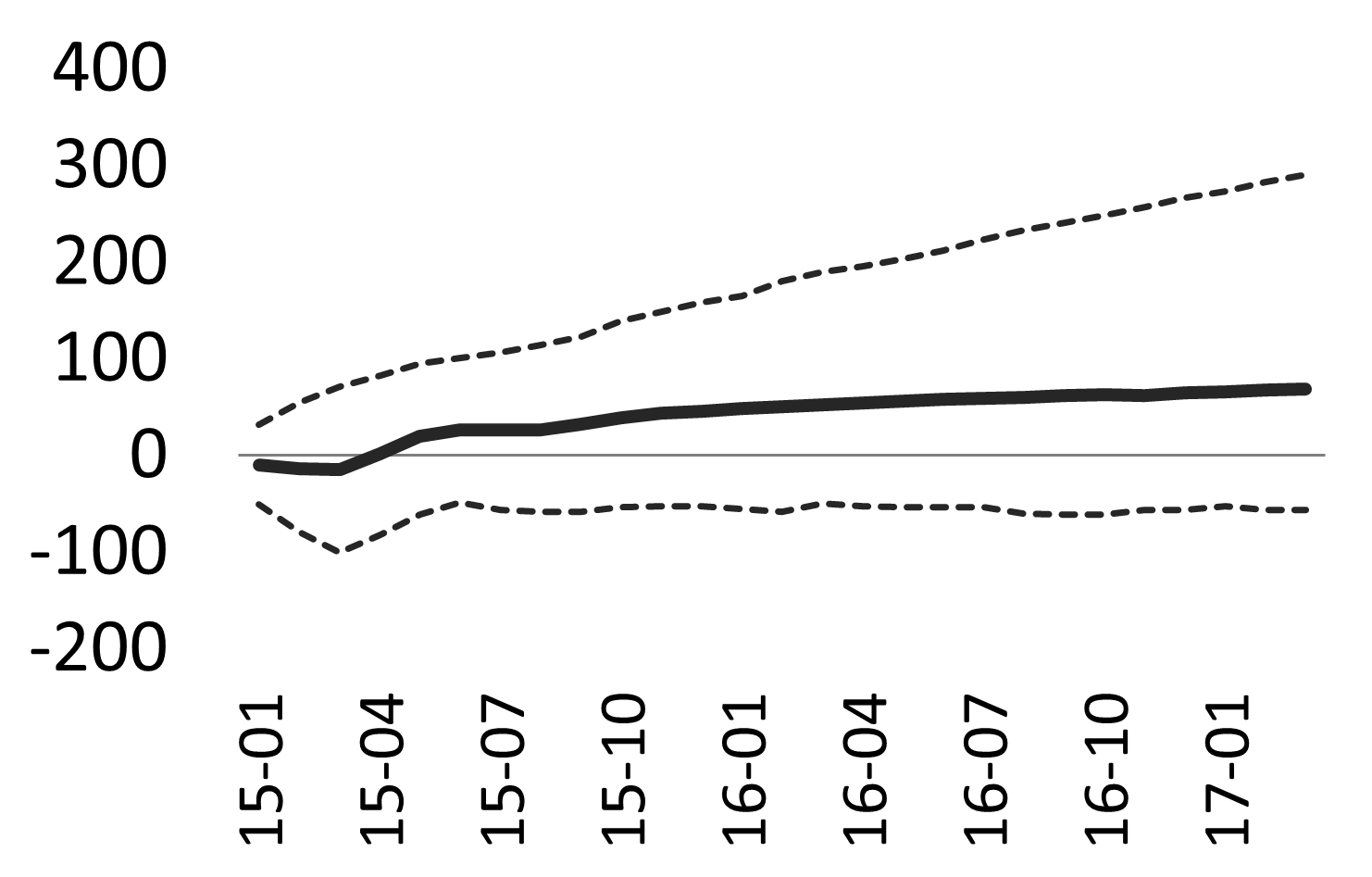}\\
			\includegraphics[scale=0.11]{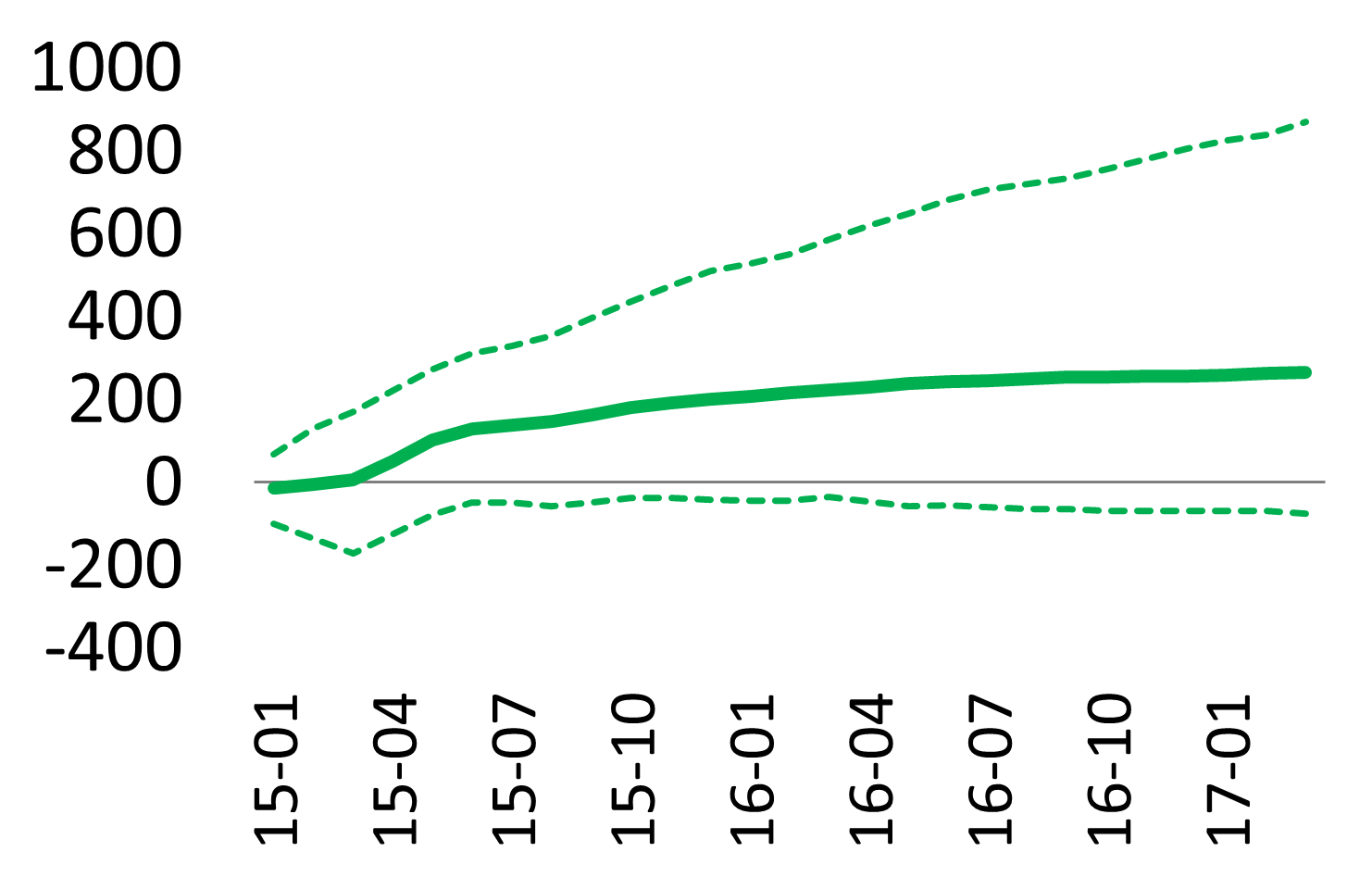}
			\includegraphics[scale=0.11]{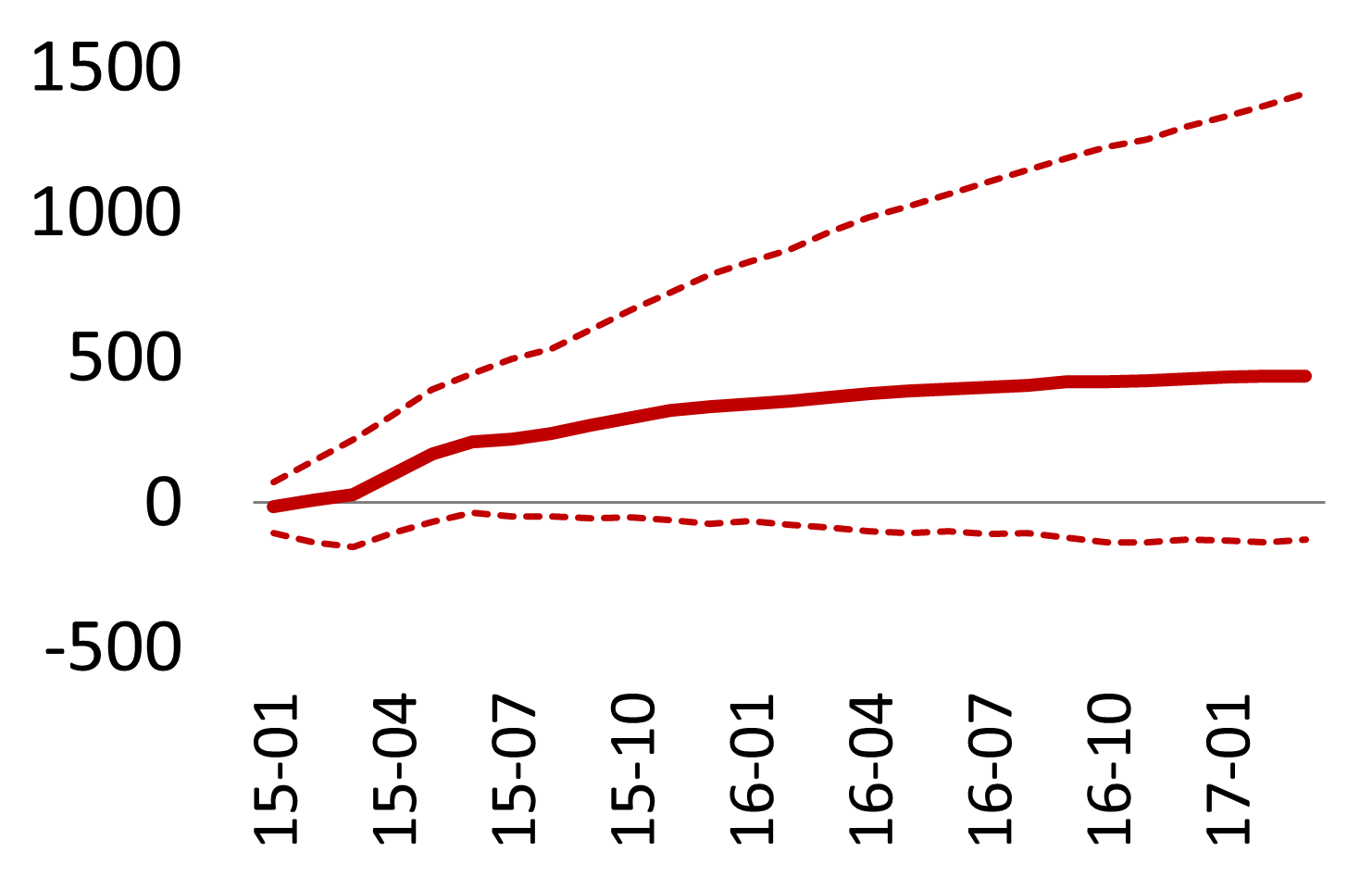}
		\end{minipage}%
		\subcaption*{\textit{Notes}: The figure shows the estimated responses to the shock scenario for the LTRO programme (left panel) and bootstrap 90\% confidence intervals and medians (right panels).}
	\end{figure}
	
	\begin{figure}[H]
		\centering		
		\caption{Response of repo trade volumes, shock to LTRO}
		\begin{minipage}{.5\textwidth}
			\hspace{-15pt}
			\includegraphics[scale=0.15]{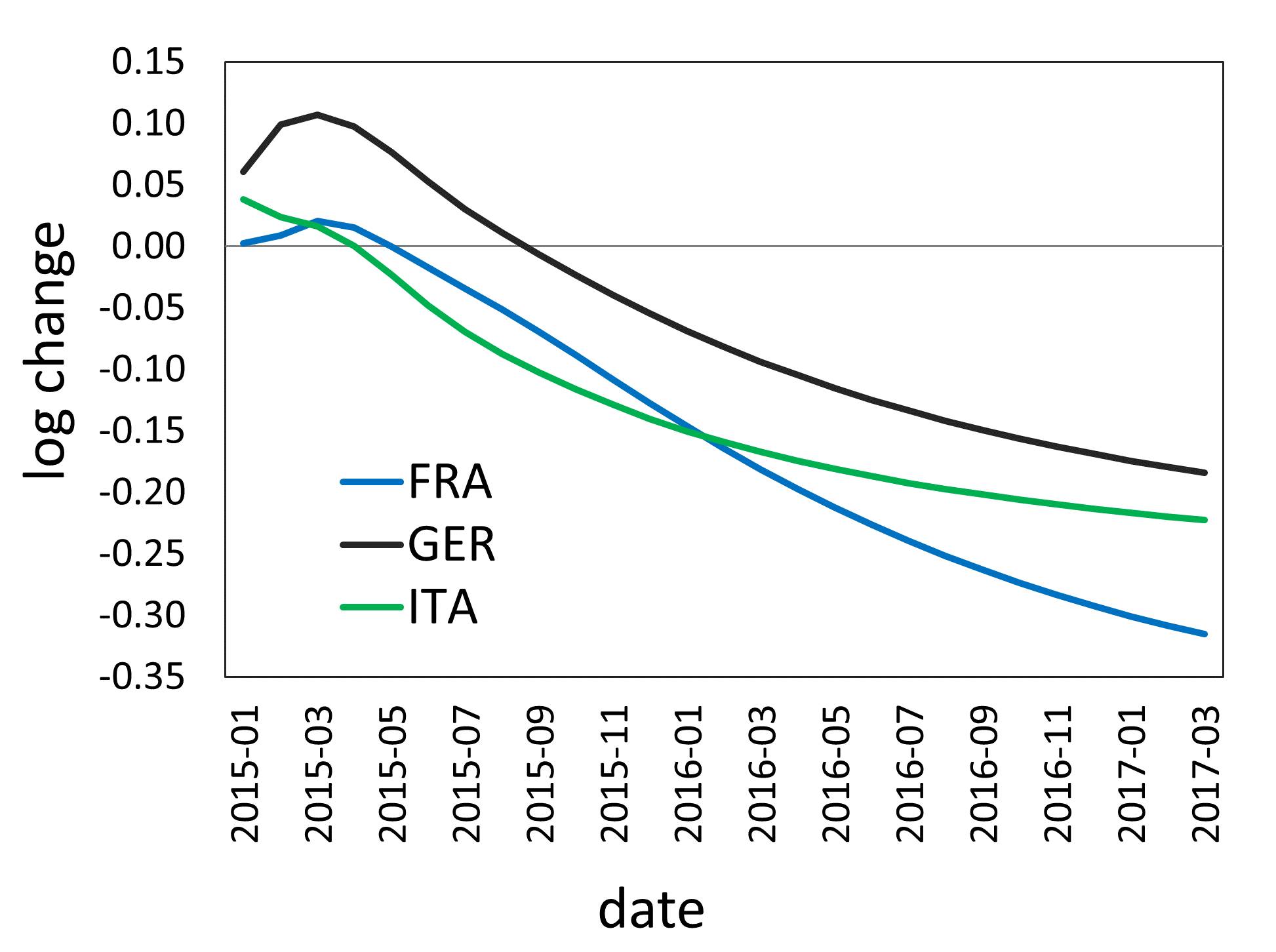}
		\end{minipage}%
		\begin{minipage}{.60\textwidth}
			\includegraphics[scale=0.11]{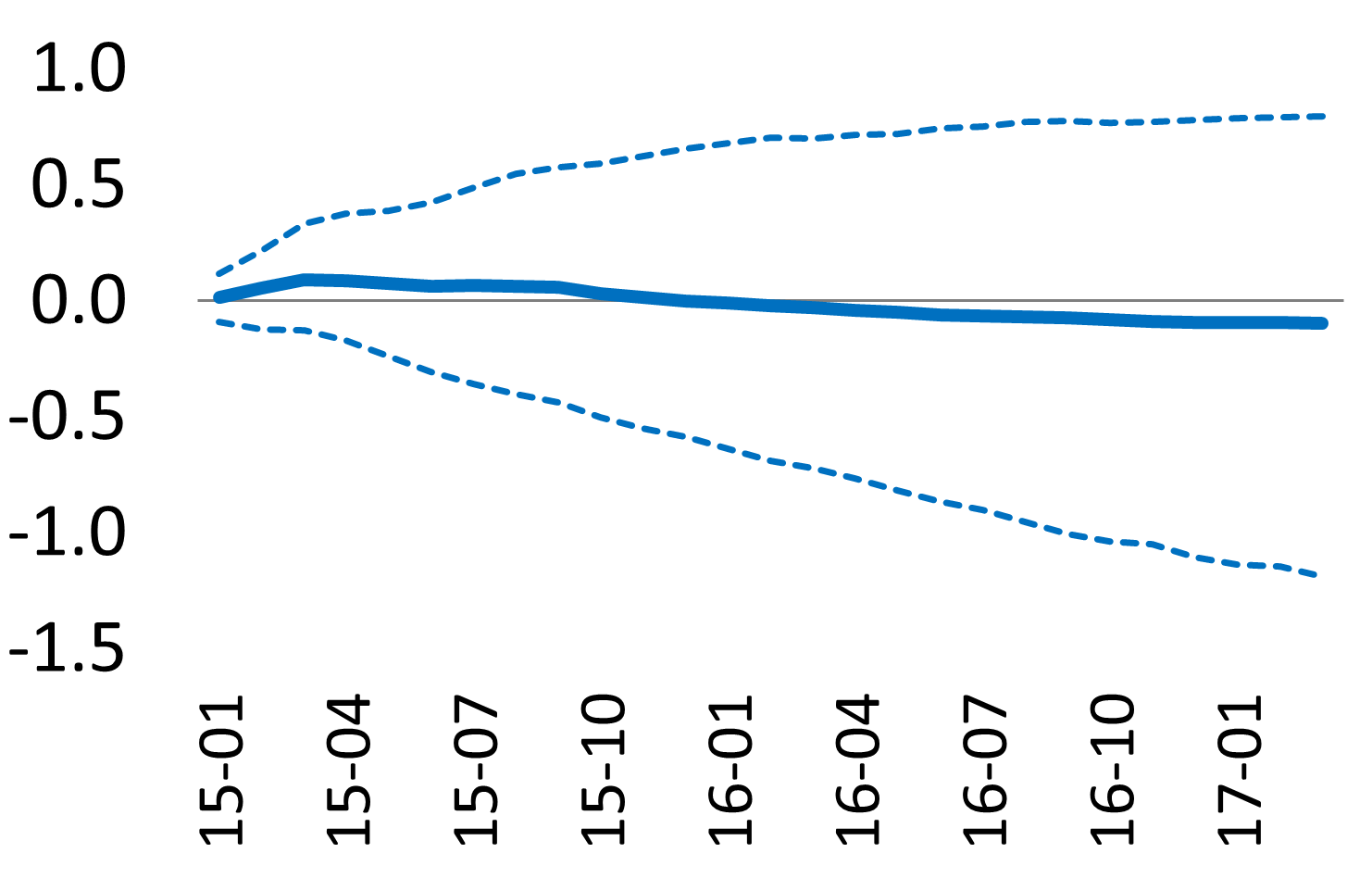}
			\includegraphics[scale=0.11]{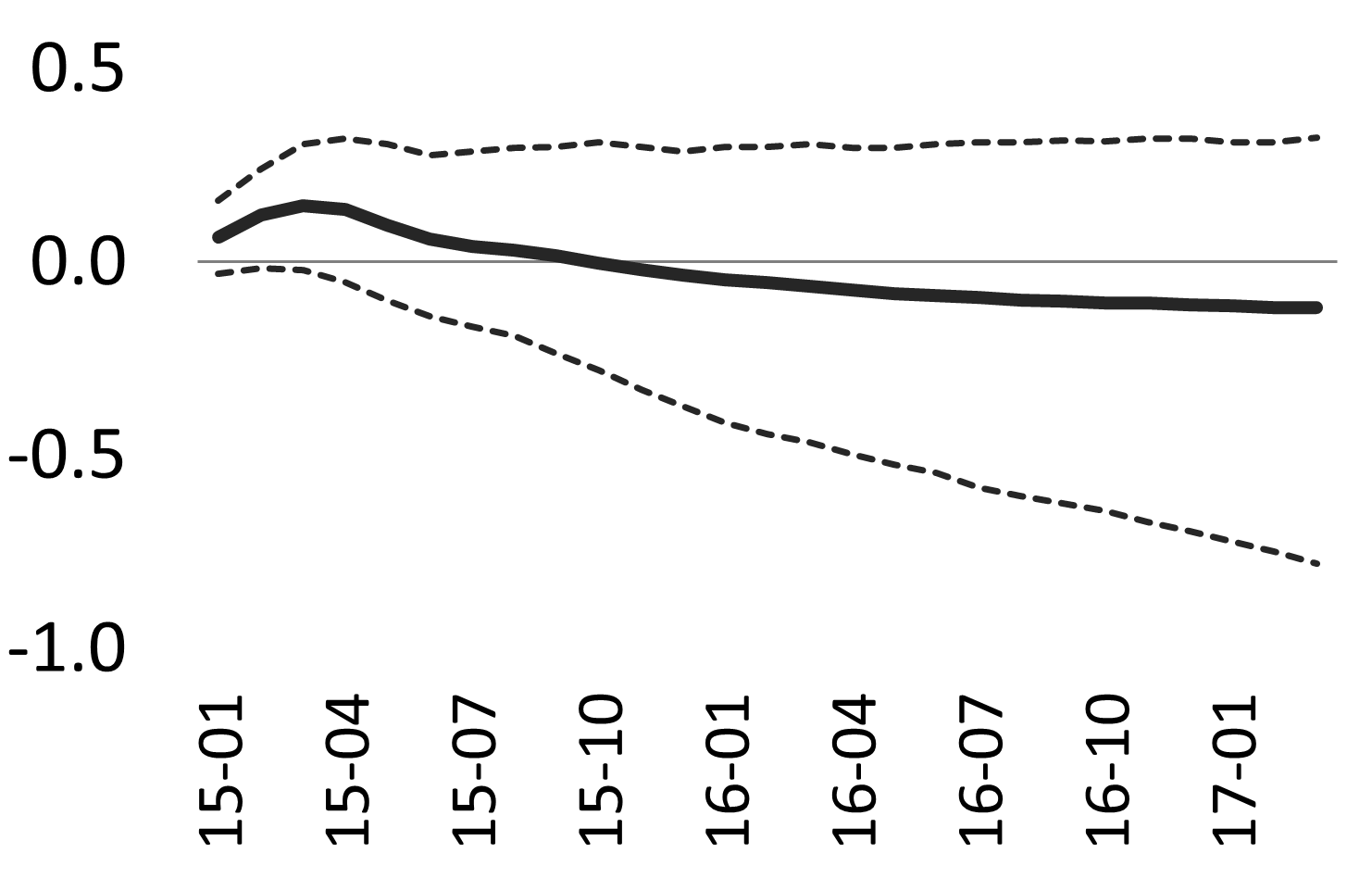}\\
			\includegraphics[scale=0.11]{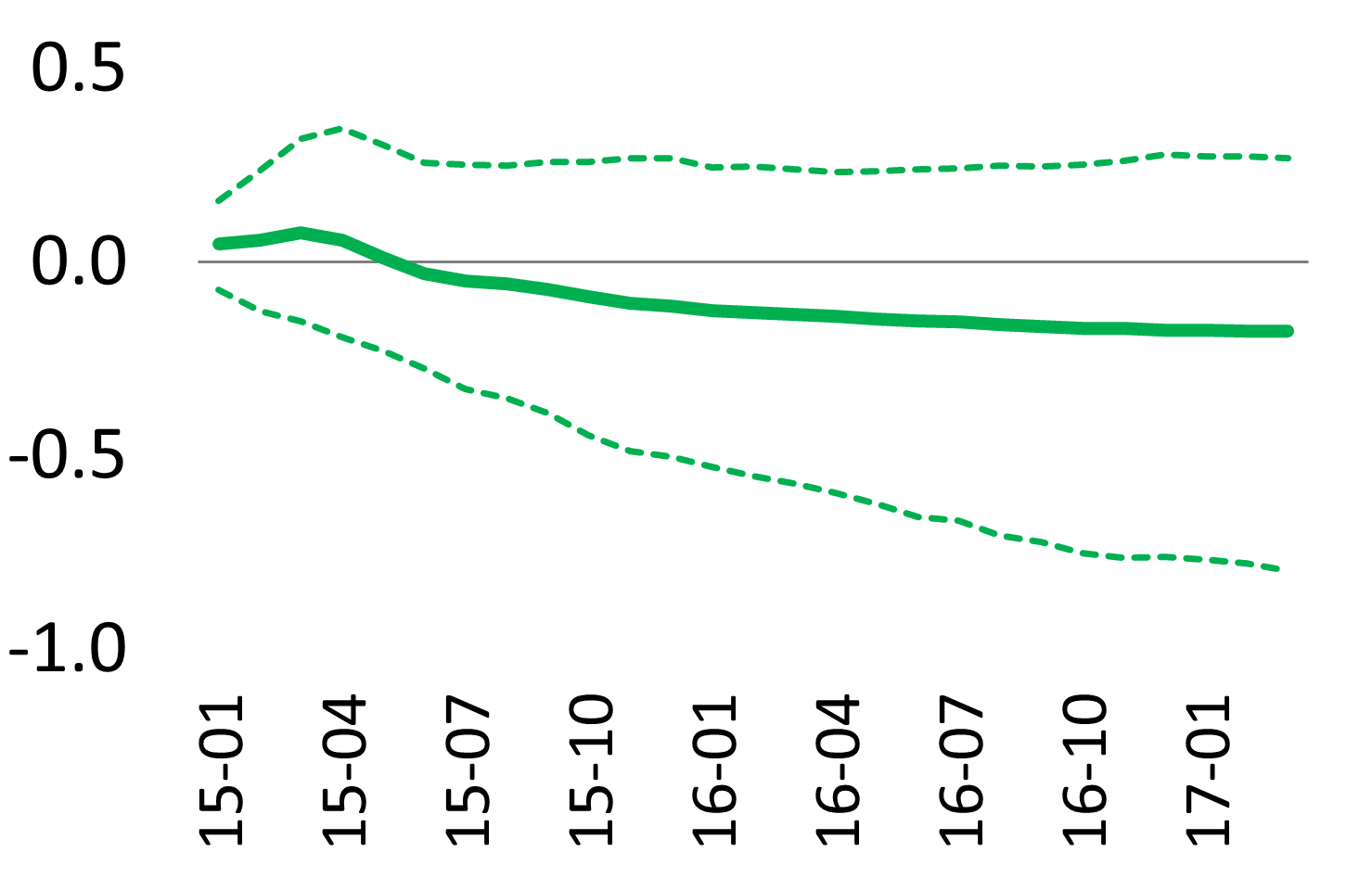}
		\end{minipage}%
		\subcaption*{\textit{Notes}: The figure shows the estimated responses to the shock scenario for the LTRO programme (left panel) and bootstrap 90\% confidence intervals and medians (right panels).}
	\end{figure}

	\begin{figure}[H]
		\centering		
		\caption{Response of repo rates, shock to LTRO}
		\begin{minipage}{.5\textwidth}
			\hspace{-15pt}
			\includegraphics[scale=0.15]{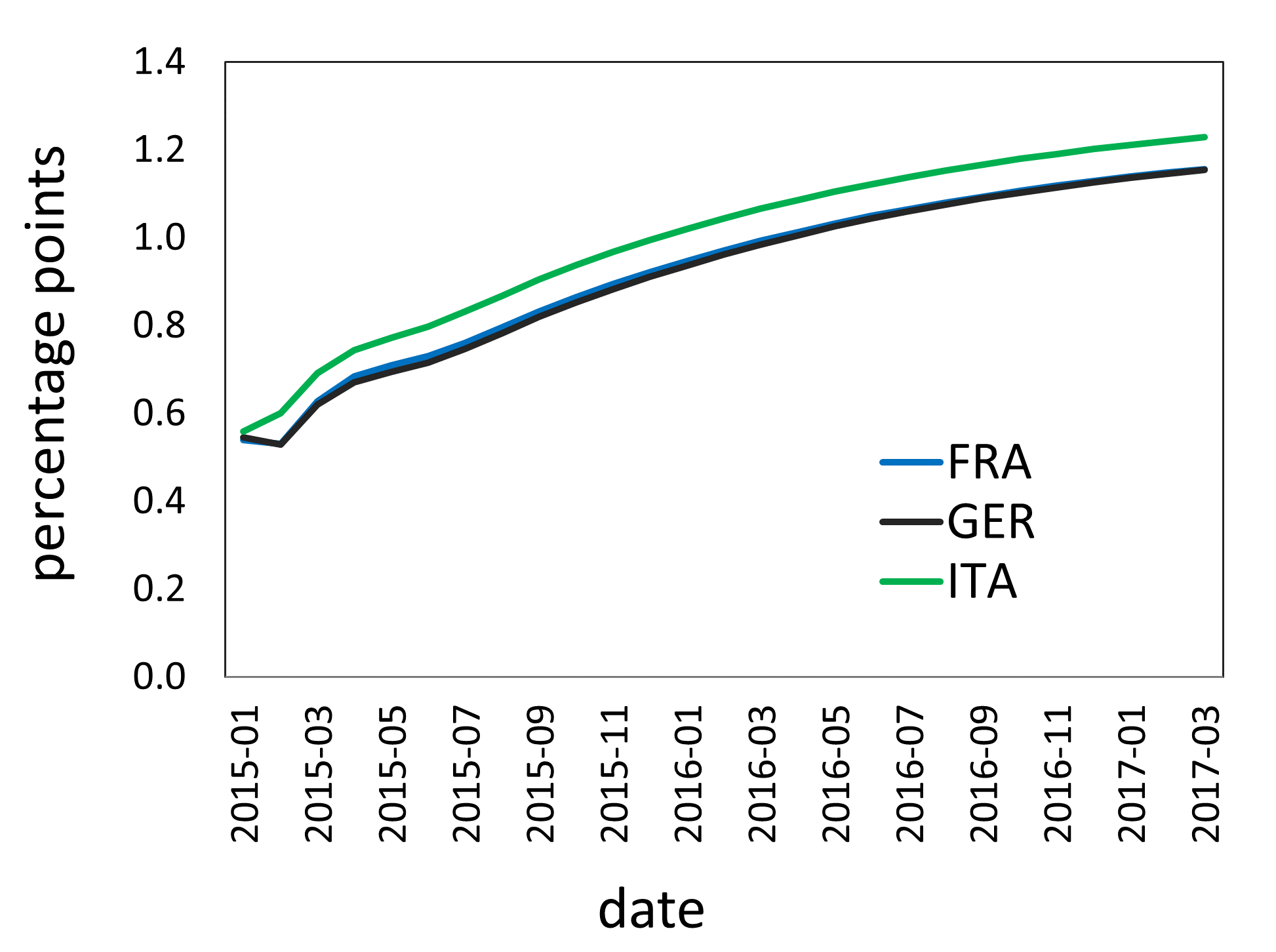}
		\end{minipage}%
		\begin{minipage}{.60\textwidth}
			\includegraphics[scale=0.11]{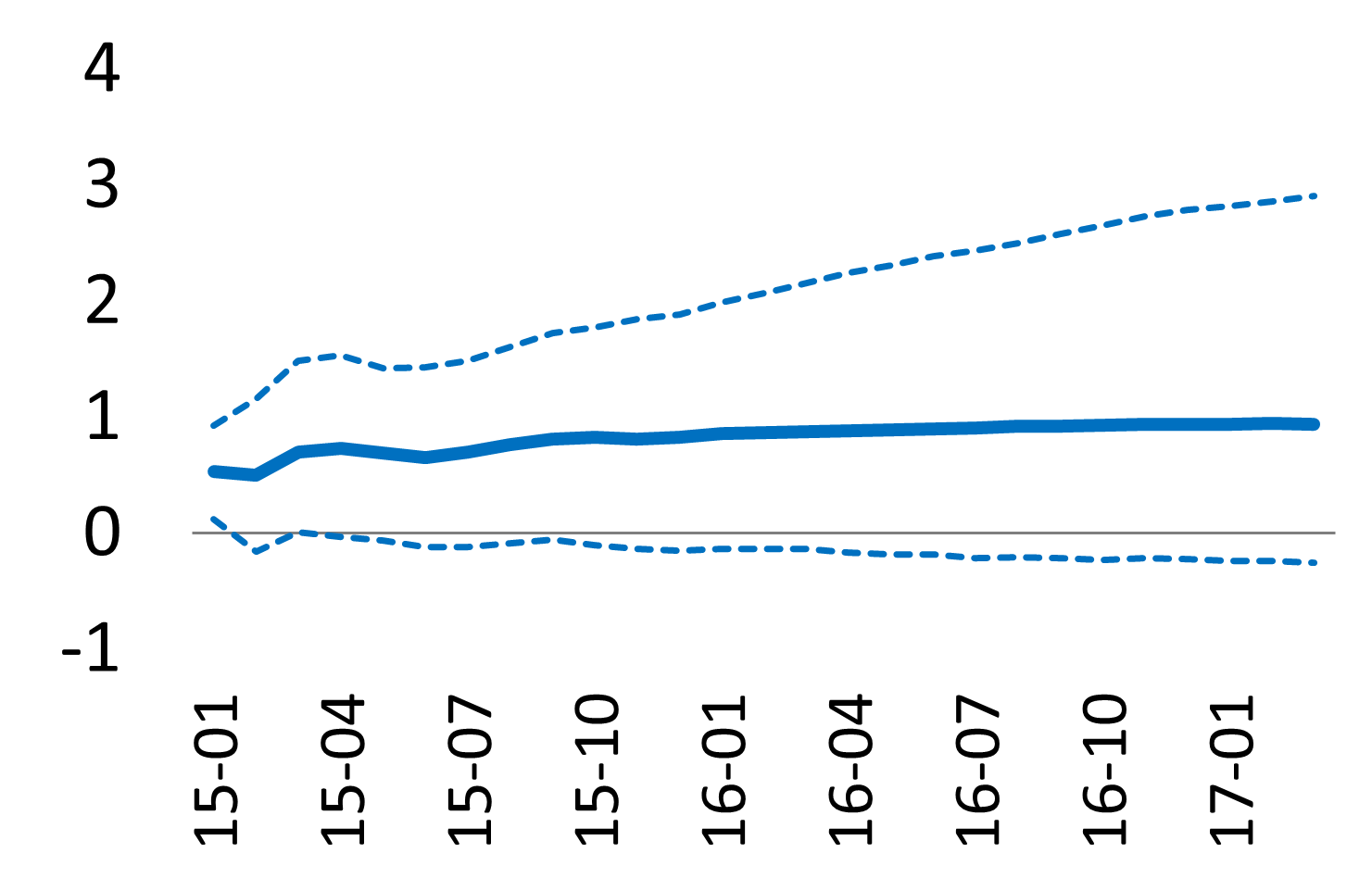}
			\includegraphics[scale=0.11]{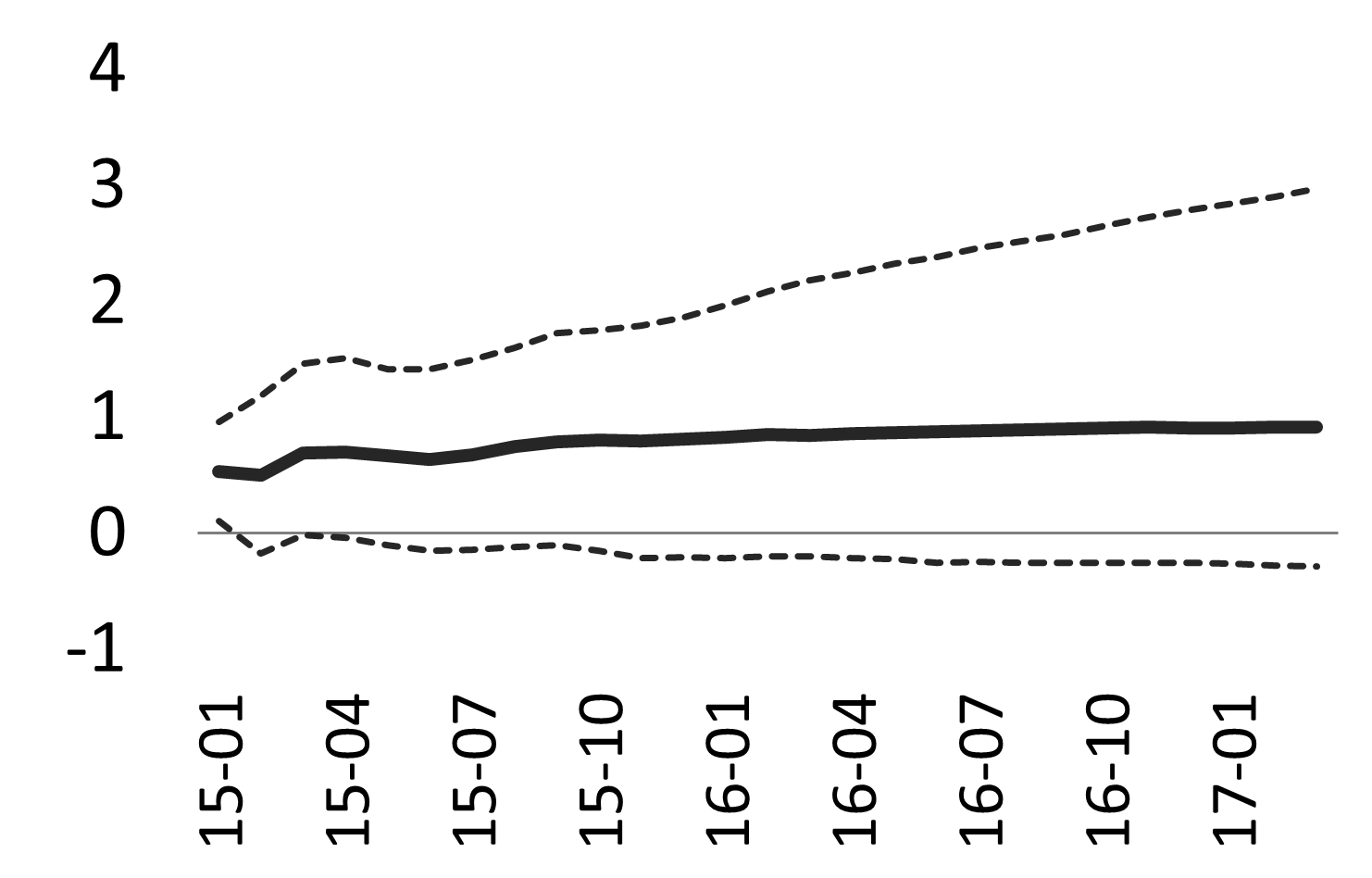}\\
			\includegraphics[scale=0.11]{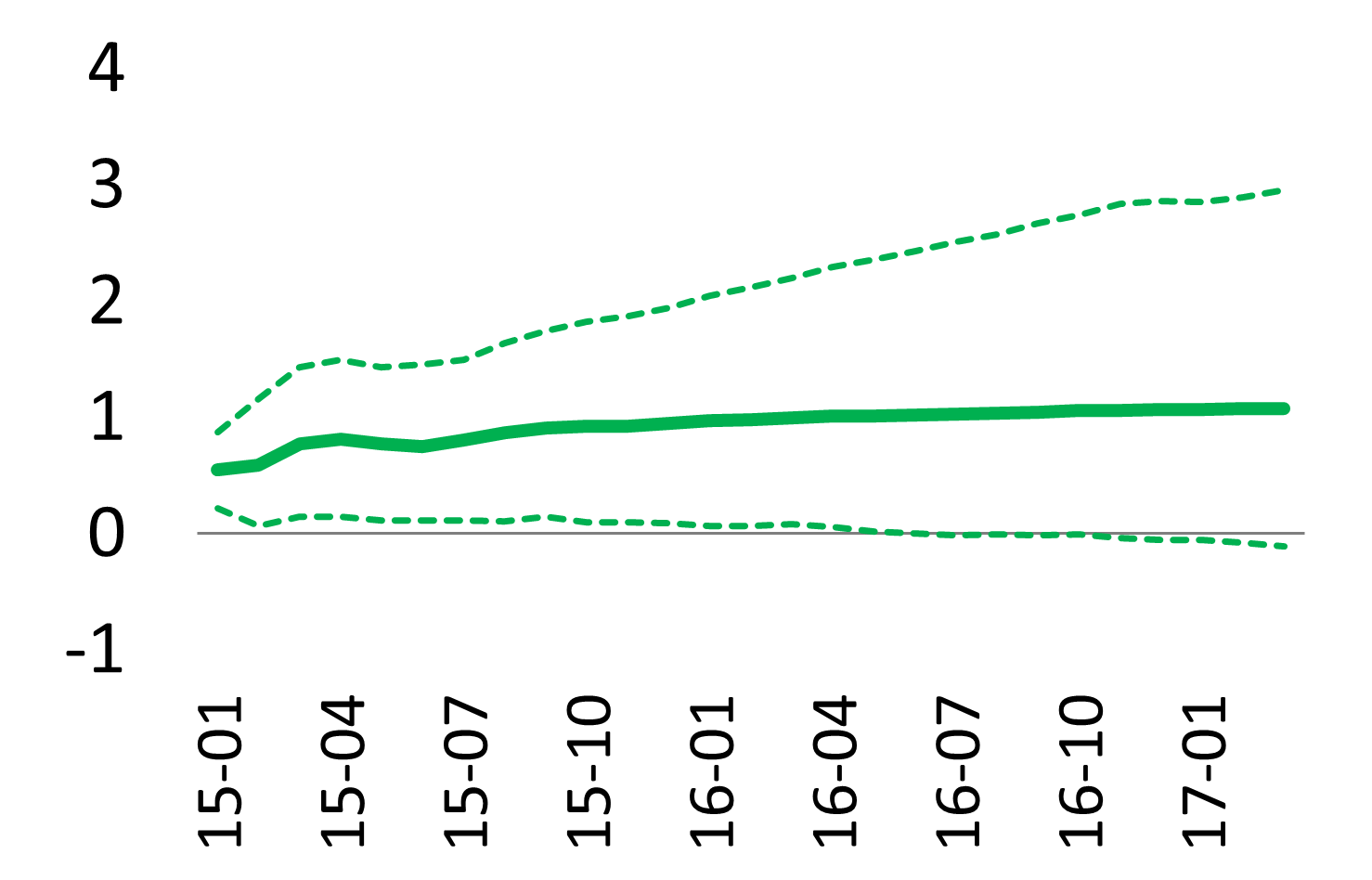}
		\end{minipage}%
		\subcaption*{\textit{Notes}: The figure shows the estimated responses to the shock scenario for the LTRO programme (left panel) and bootstrap 90\% confidence intervals and medians (right panels).}
	\end{figure}

	\subsubsection{SMP}

	\begin{figure}[H]
		\centering		
		\caption{Response of government yields, shock to SMP}
		\begin{minipage}{.5\textwidth}
			\hspace{-15pt}
			\includegraphics[scale=0.15]{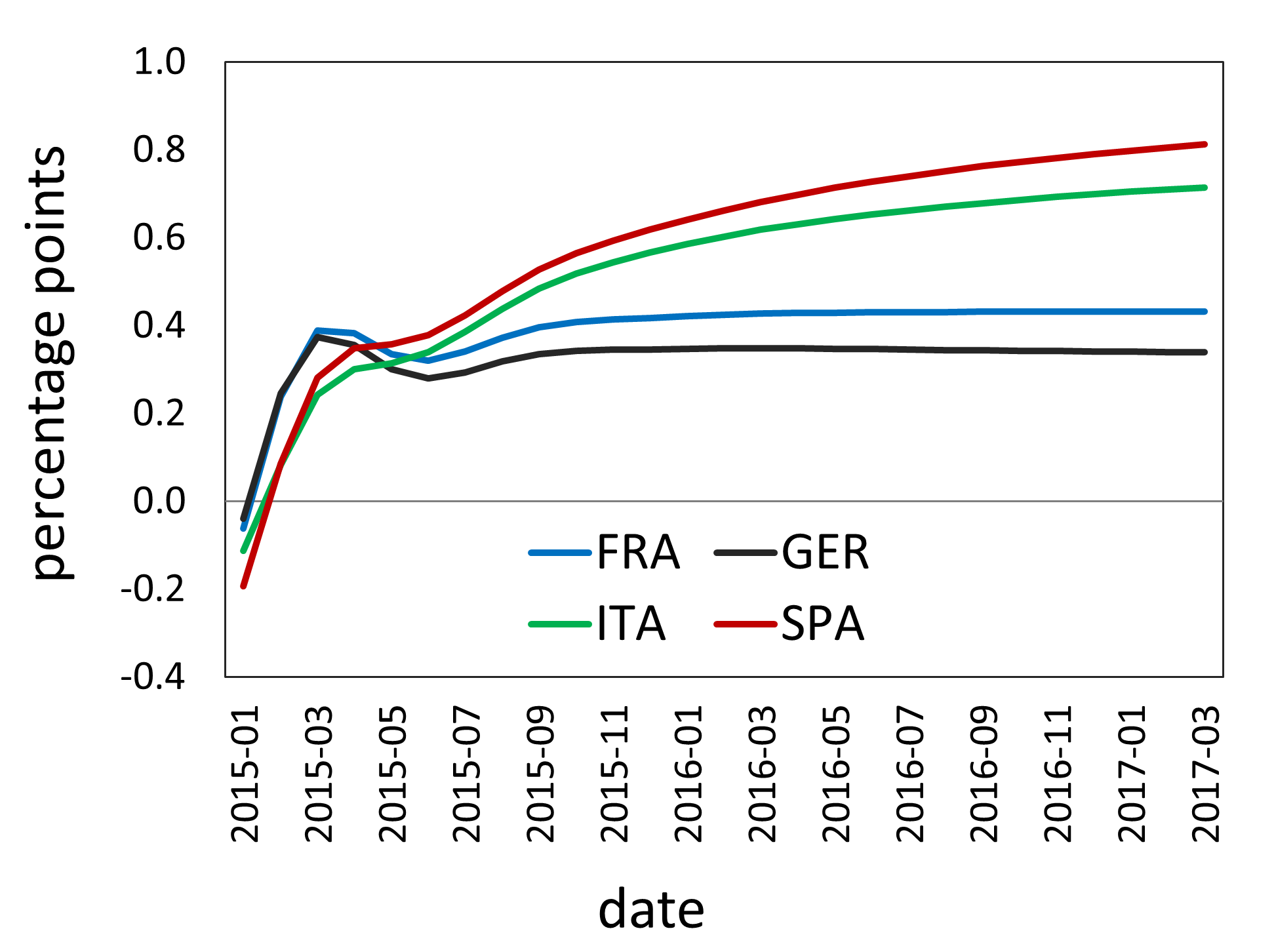}
		\end{minipage}%
		\begin{minipage}{.60\textwidth}
			\includegraphics[scale=0.11]{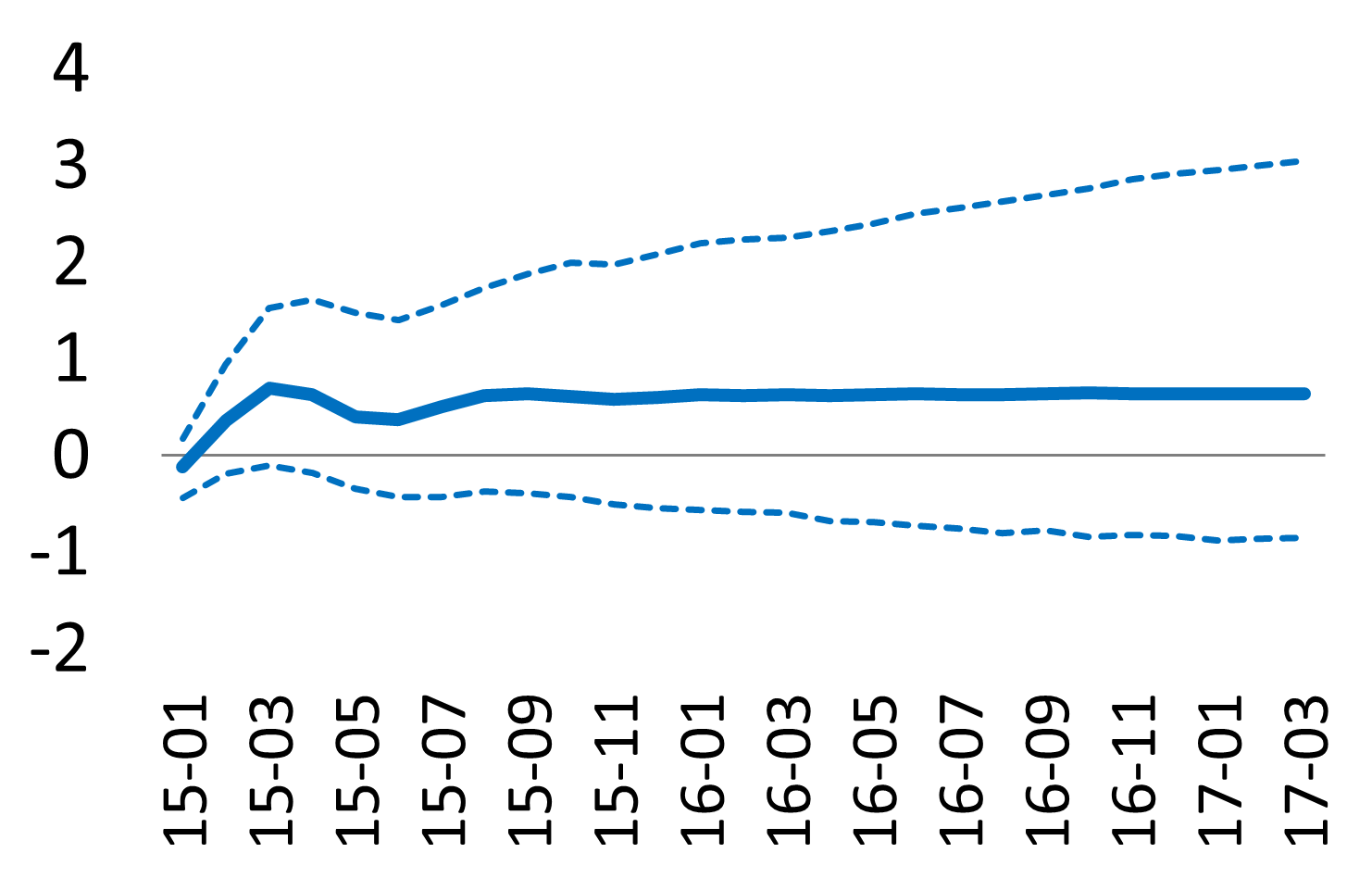}
			\includegraphics[scale=0.11]{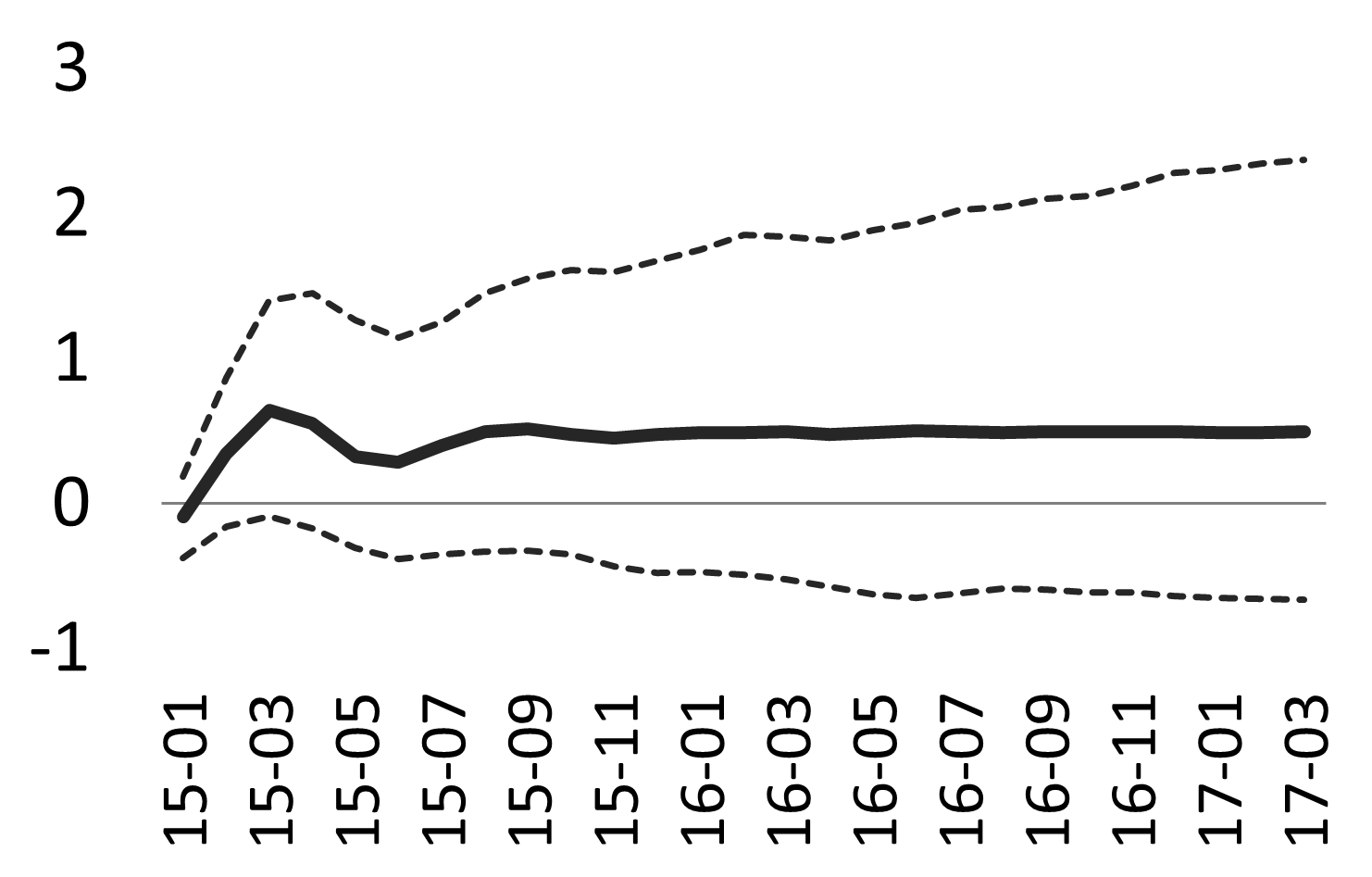}\\
			\includegraphics[scale=0.11]{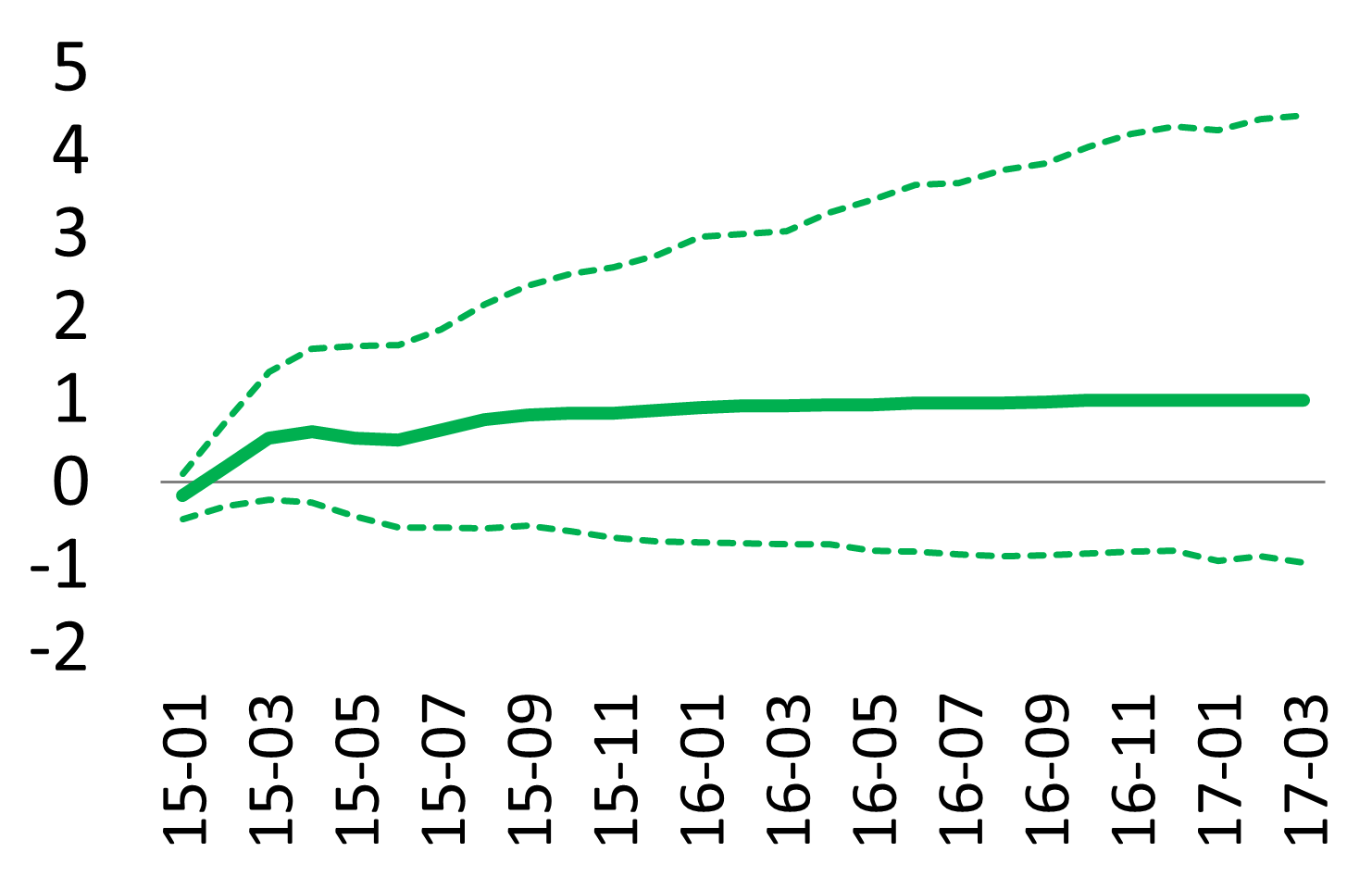}
			\includegraphics[scale=0.11]{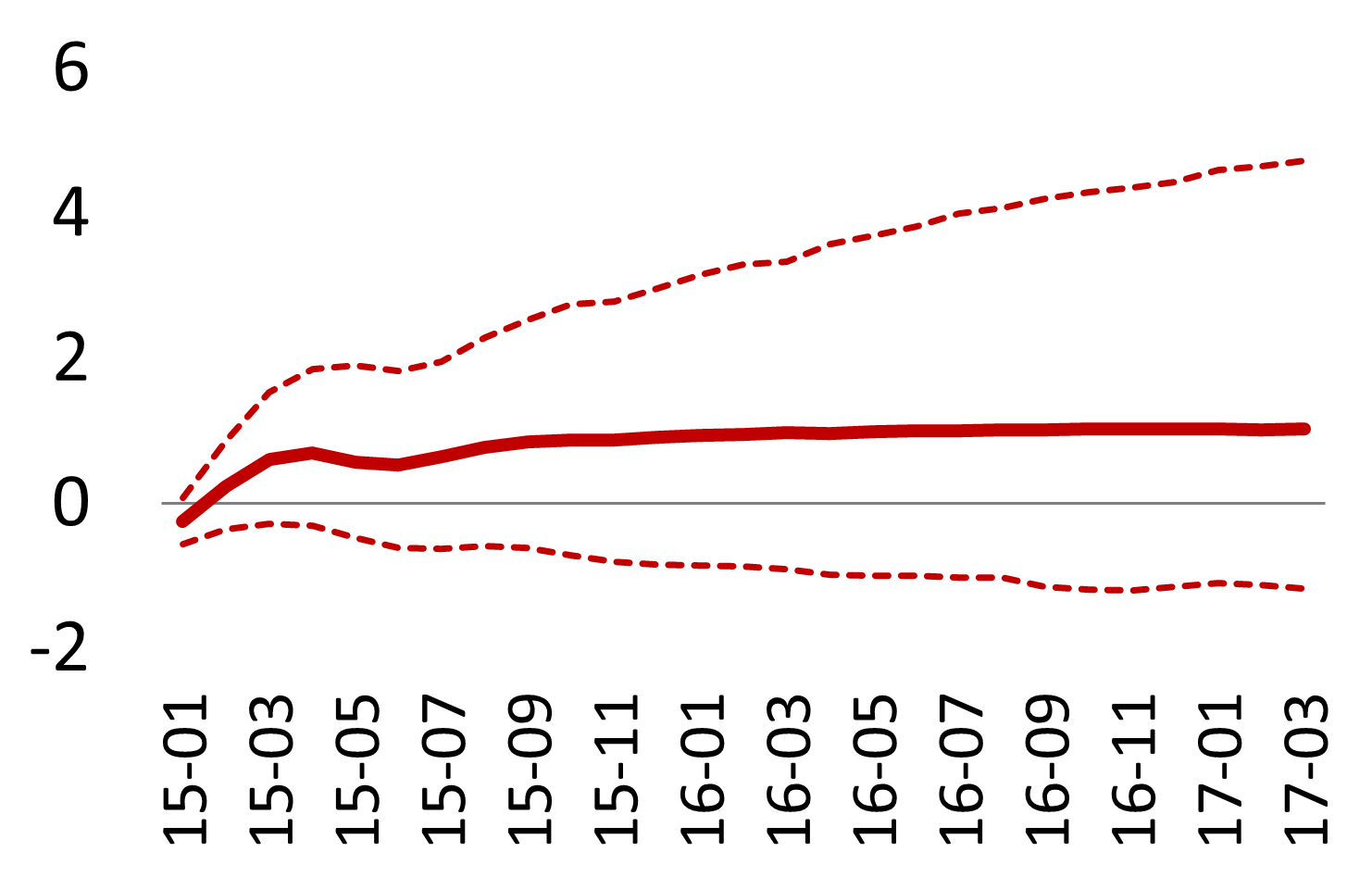}
		\end{minipage}%
		\subcaption*{\textit{Notes}: The figure shows the estimated responses to the shock scenario for the SMP programme (left panel) and bootstrap 90\% confidence intervals and medians (right panels).}
	\end{figure}
	
	\begin{figure}[H]
		\centering		
		\caption{Response of bank CDS spreads, shock to SMP}
		\begin{minipage}{.5\textwidth}
			\hspace{-15pt}
			\includegraphics[scale=0.15]{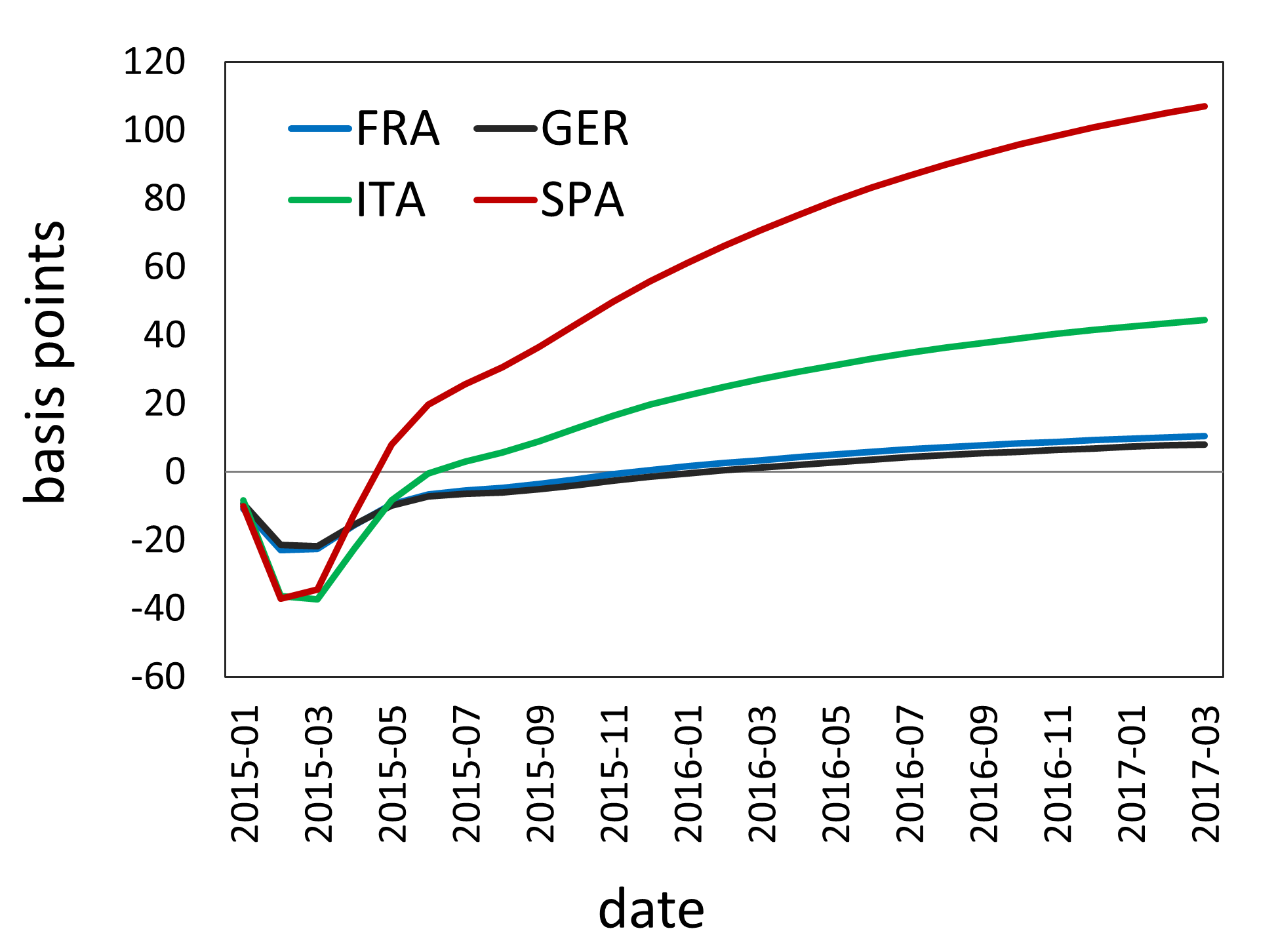}
		\end{minipage}%
		\begin{minipage}{.60\textwidth}
			\includegraphics[scale=0.11]{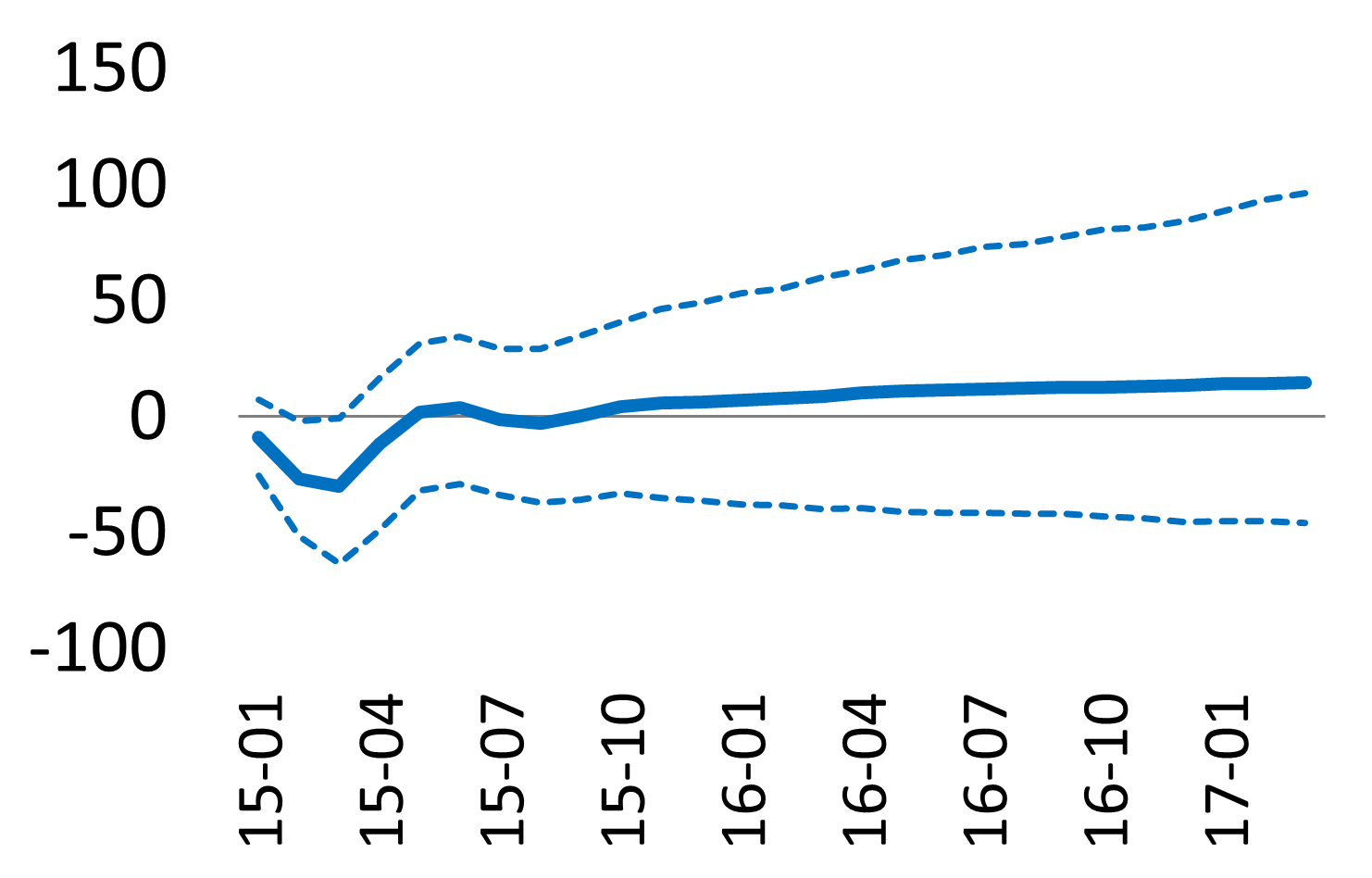}
			\includegraphics[scale=0.11]{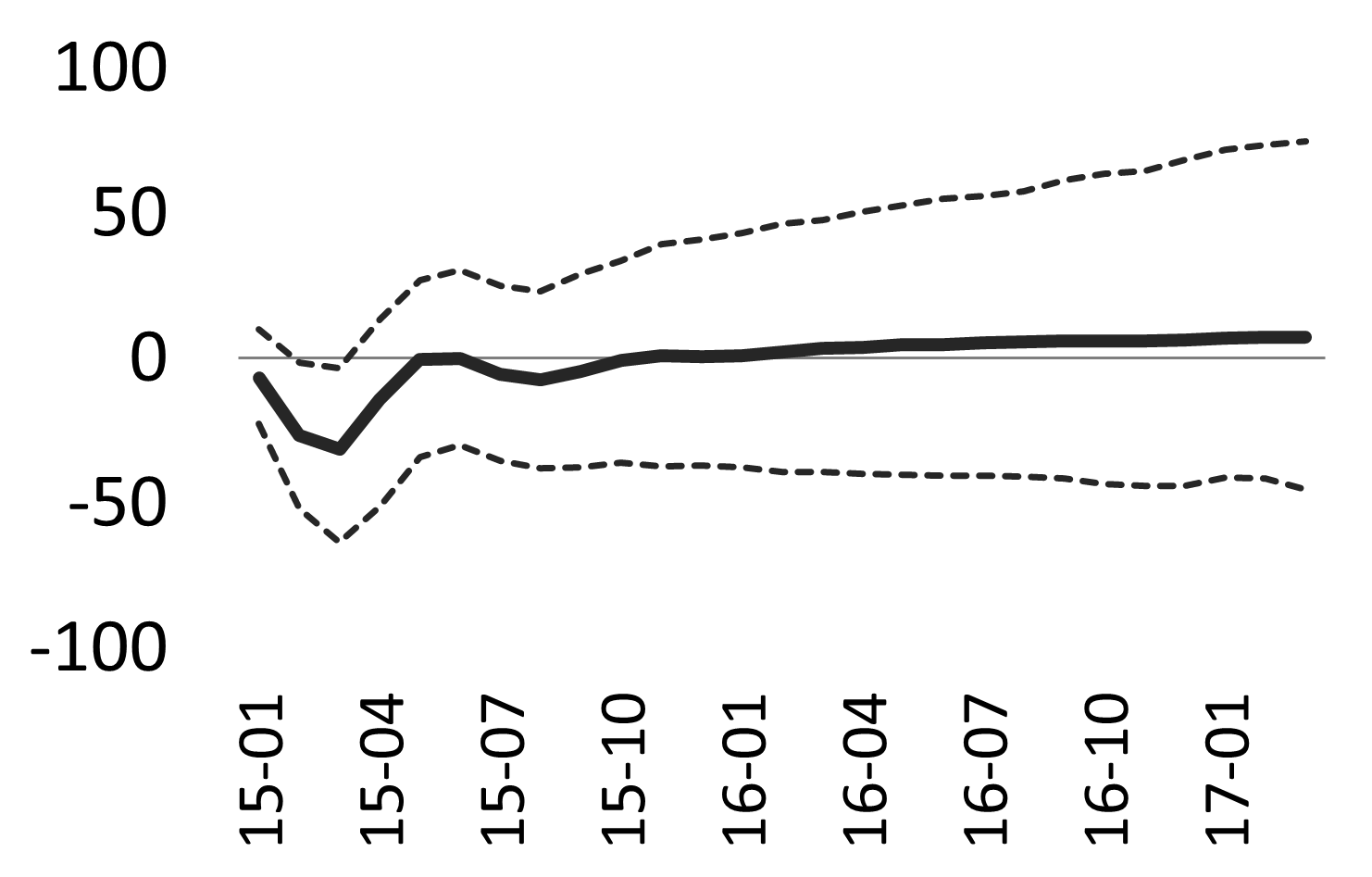}\\
			\includegraphics[scale=0.11]{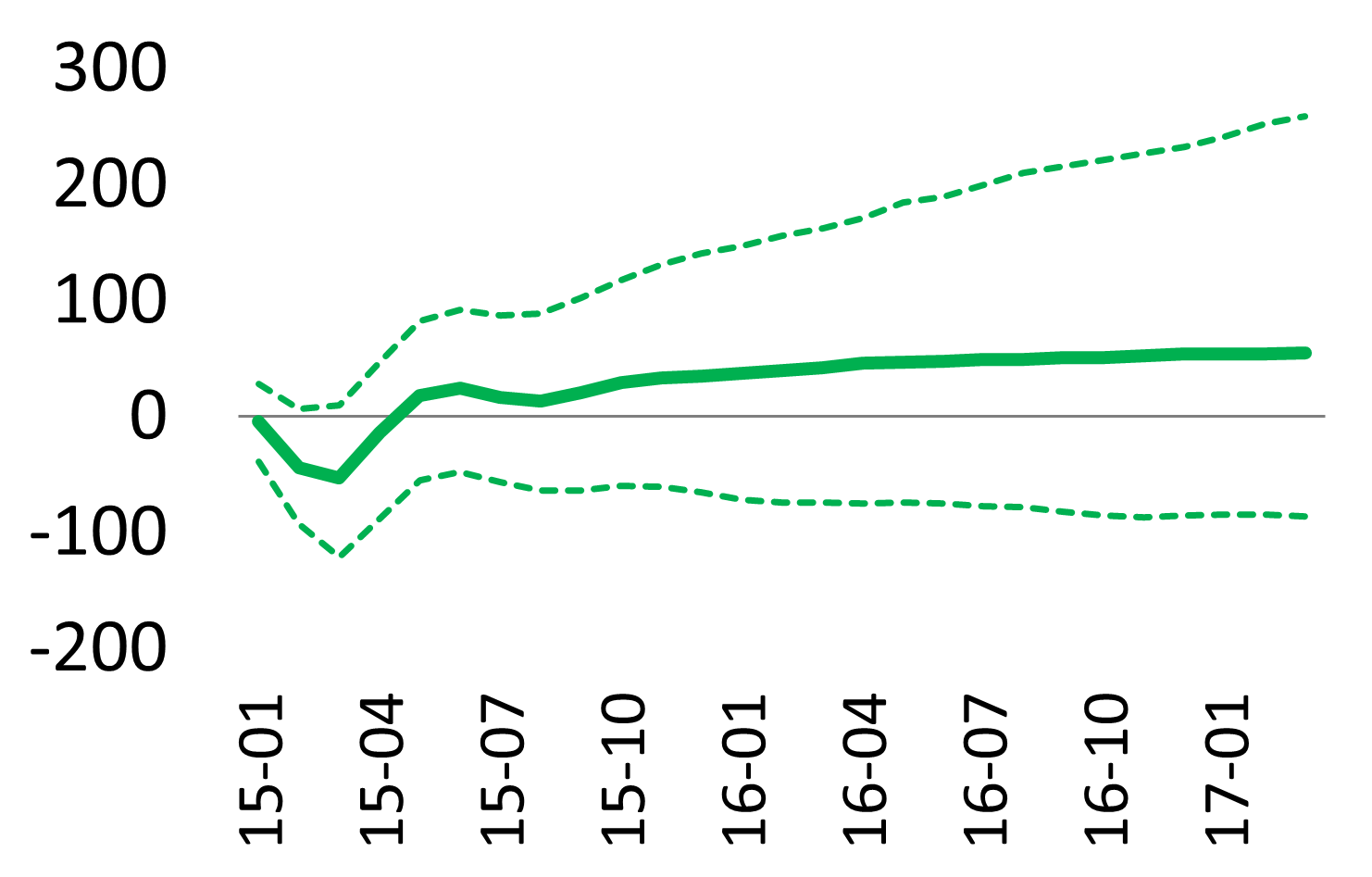}
			\includegraphics[scale=0.11]{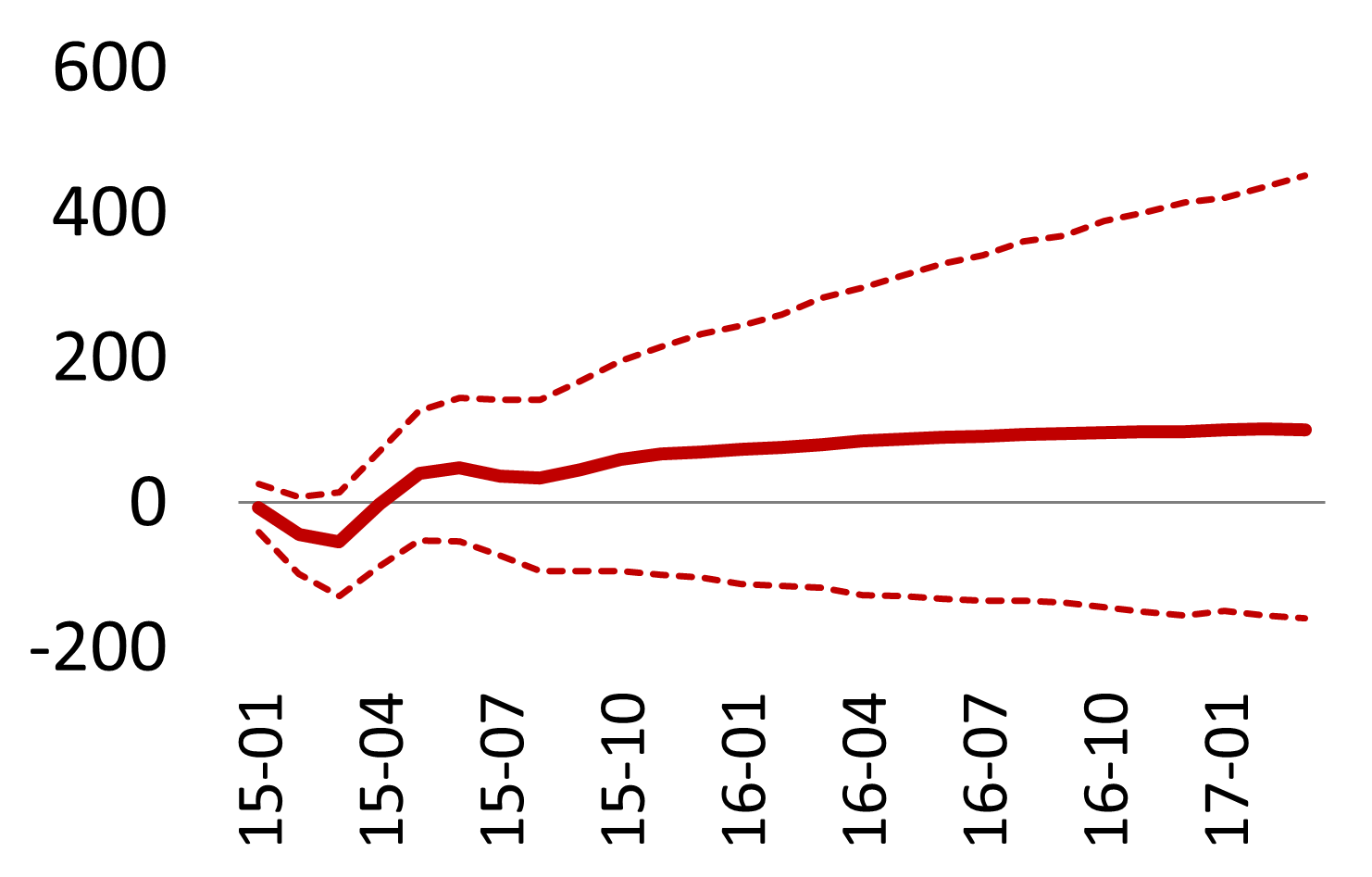}
		\end{minipage}%
		\subcaption*{\textit{Notes}: The figure shows the estimated responses to the shock scenario for the SMP programme (left panel) and bootstrap 90\% confidence intervals and medians (right panels).}
	\end{figure}
	
	\begin{figure}[H]
		\centering		
		\caption{Response of repo trade volumes, shock to SMP}
		\begin{minipage}{.5\textwidth}
			\hspace{-15pt}
			\includegraphics[scale=0.15]{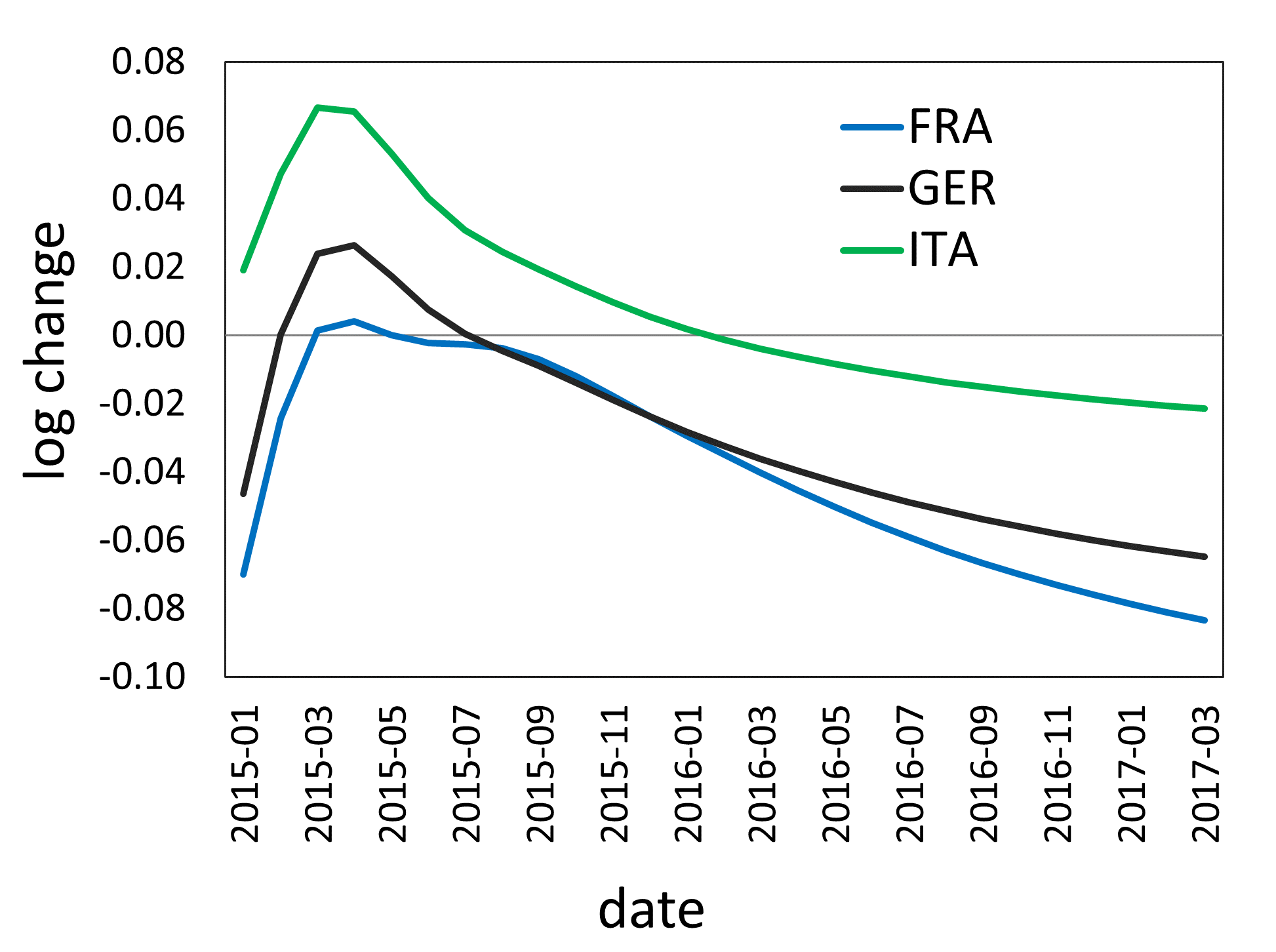}
		\end{minipage}%
		\begin{minipage}{.60\textwidth}
			\includegraphics[scale=0.11]{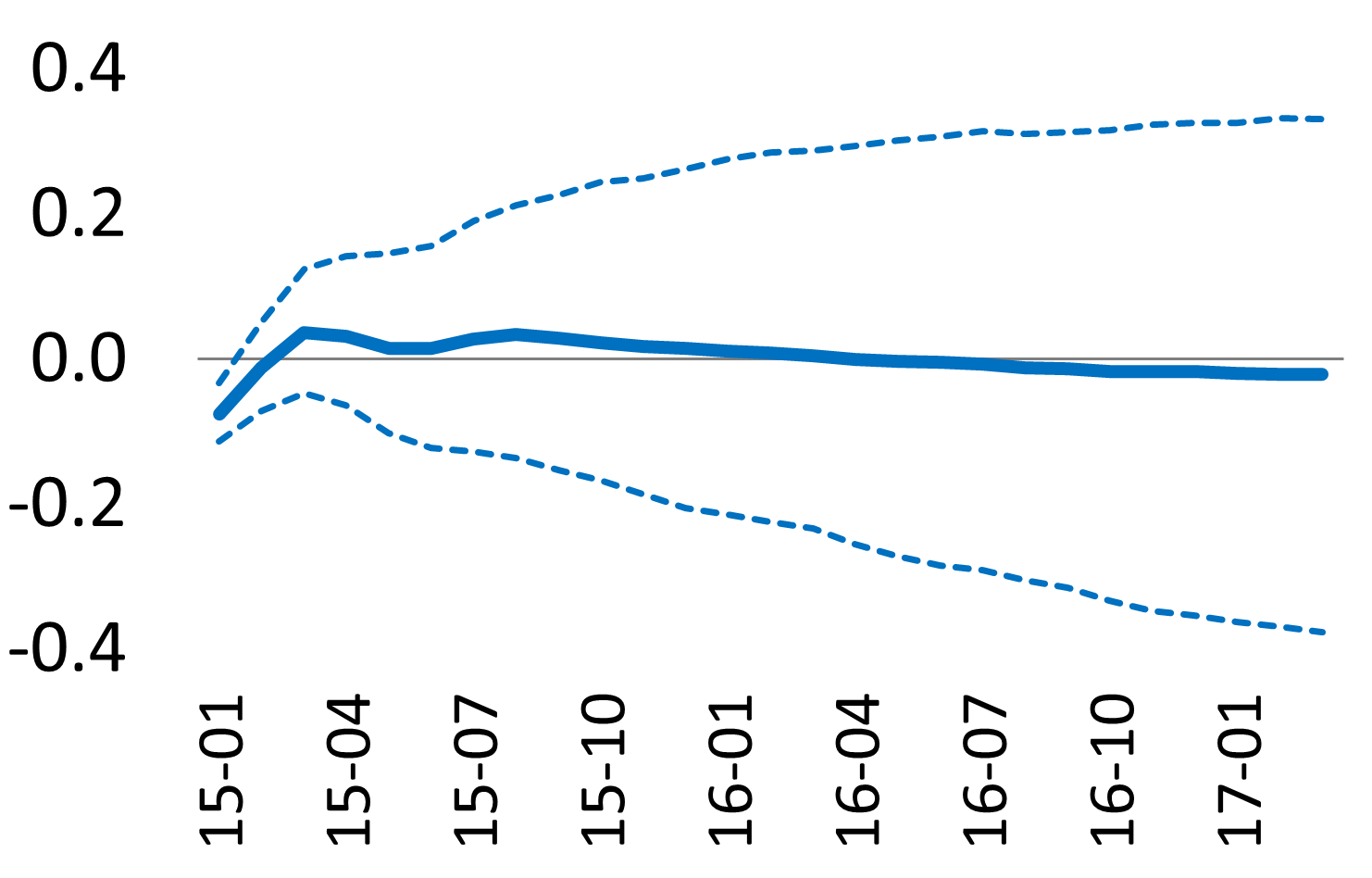}
			\includegraphics[scale=0.11]{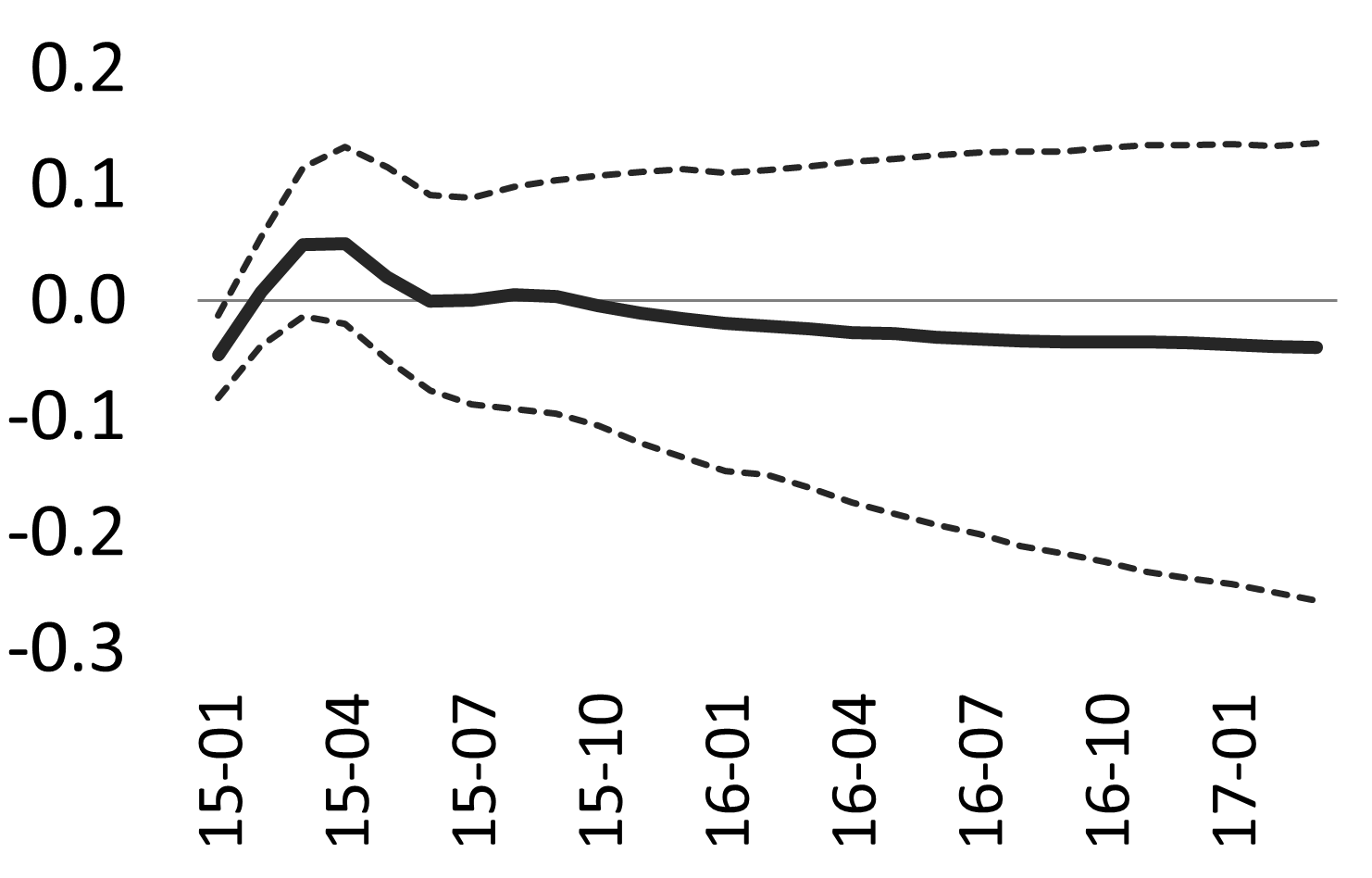}\\
			\includegraphics[scale=0.11]{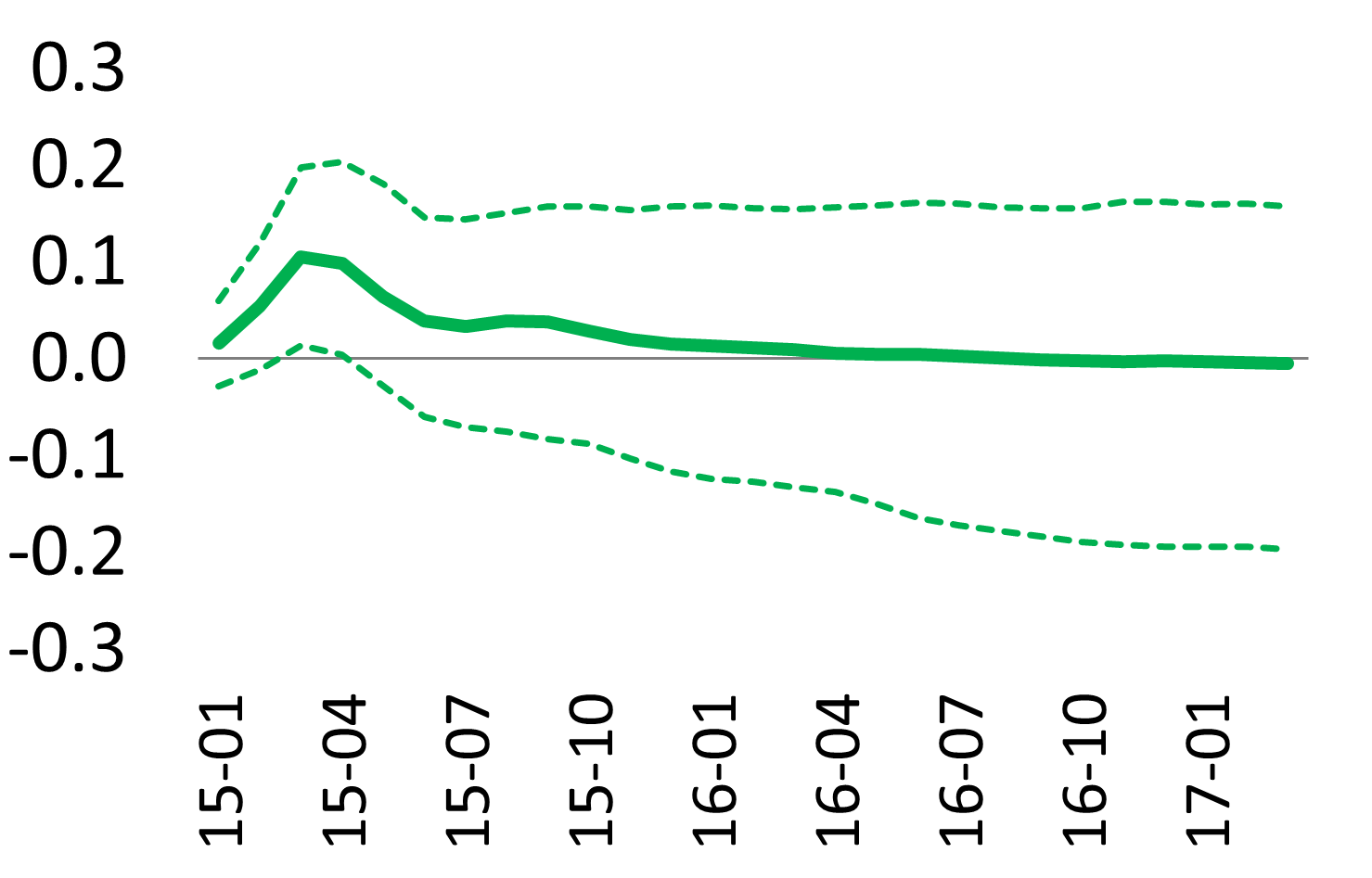}
		\end{minipage}%
		\subcaption*{\textit{Notes}: The figure shows the estimated responses to the shock scenario for the SMP programme (left panel) and bootstrap 90\% confidence intervals and medians (right panels).}
	\end{figure}

	\begin{figure}[H]
		\centering		
		\caption{Response of repo rates, shock to SMP}
		\begin{minipage}{.5\textwidth}
			\hspace{-15pt}
			\includegraphics[scale=0.15]{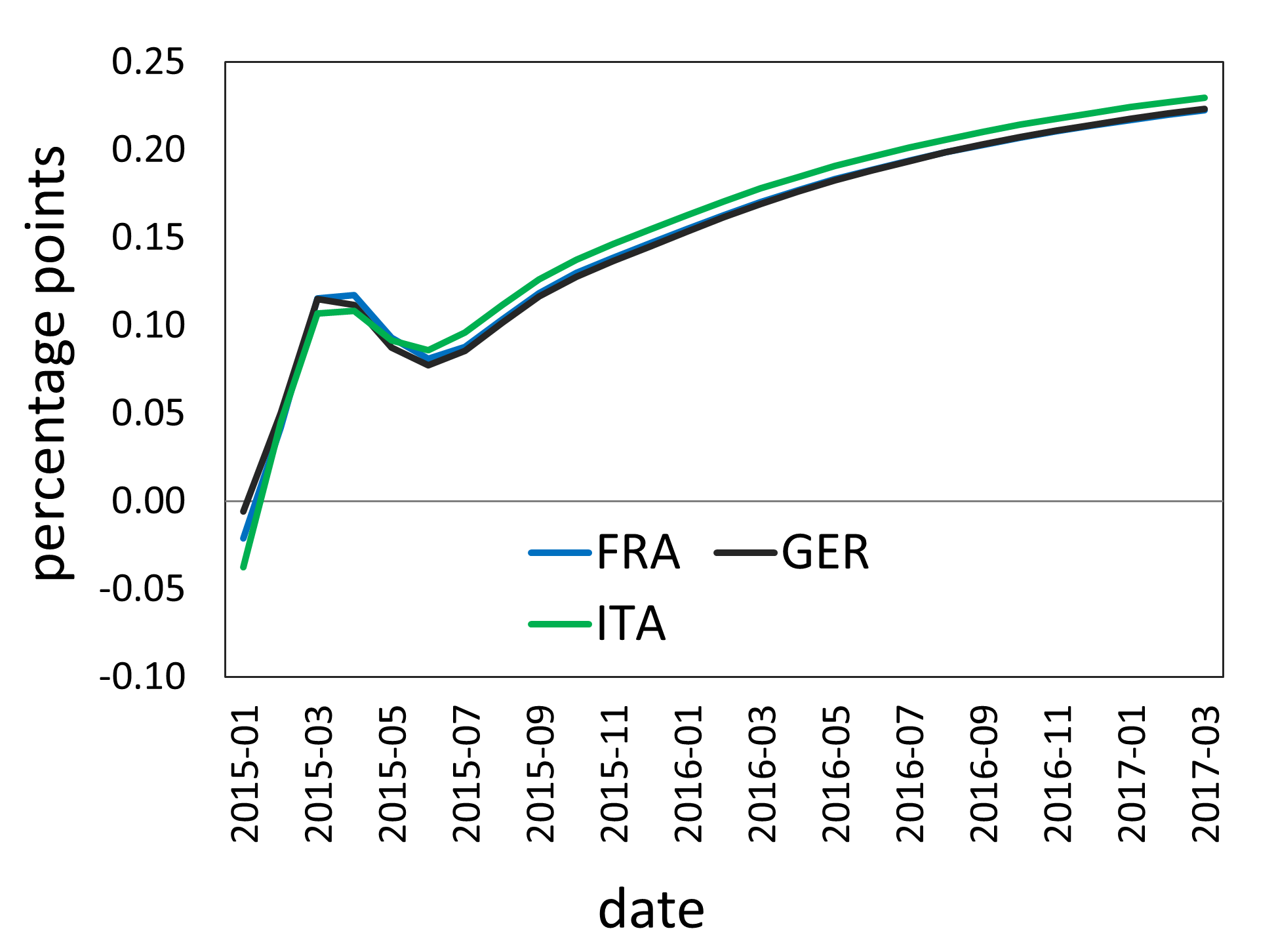}
		\end{minipage}%
		\begin{minipage}{.60\textwidth}
			\includegraphics[scale=0.11]{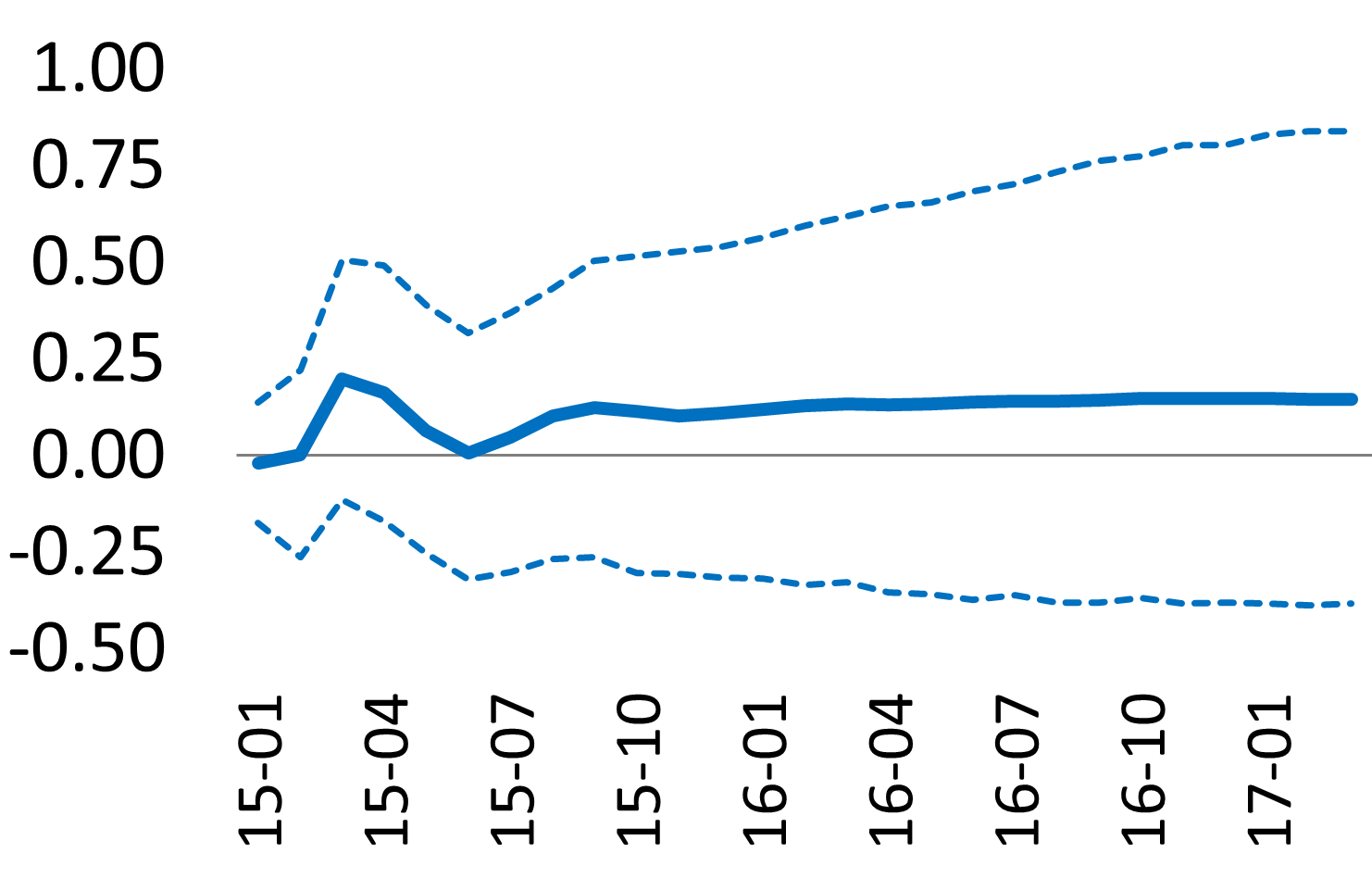}
			\includegraphics[scale=0.11]{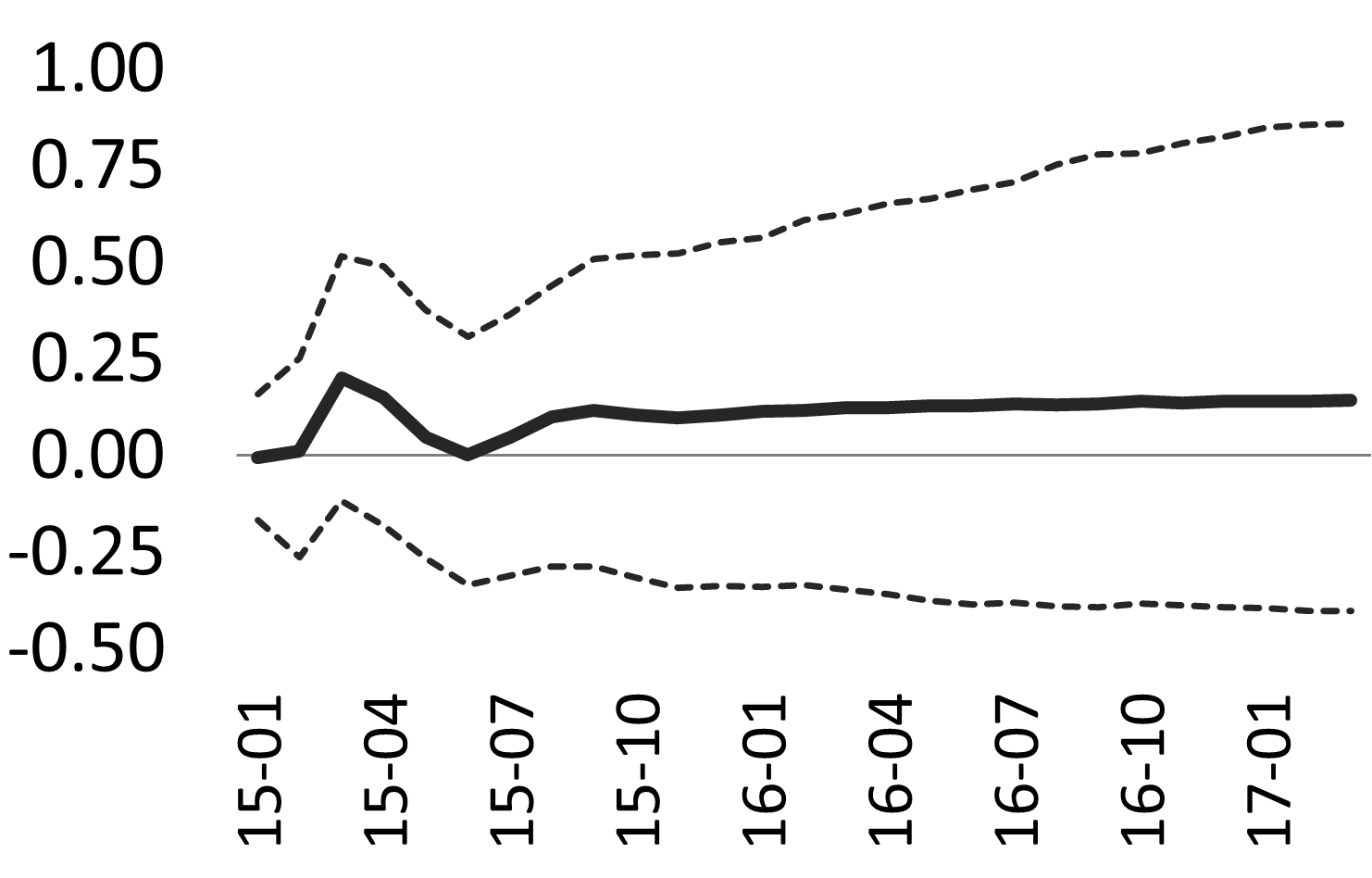}\\
			\includegraphics[scale=0.11]{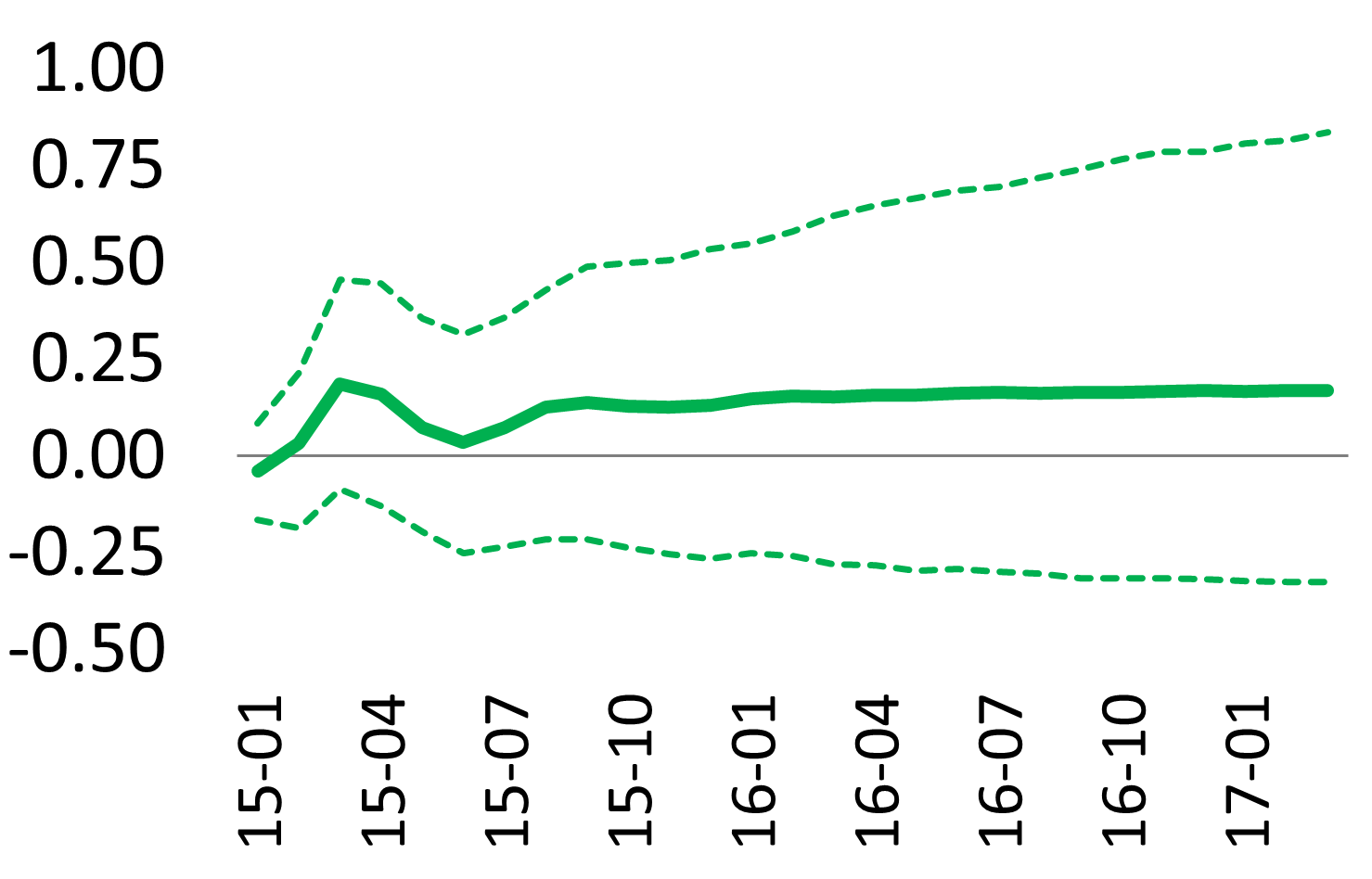}
		\end{minipage}%
		\subcaption*{\textit{Notes}: The figure shows the estimated responses to the shock scenario for the SMP programme (left panel) and bootstrap 90\% confidence intervals and medians (right panels).}
	\end{figure}

	\subsubsection{OMT}

	\begin{figure}[H]
		\centering		
		\caption{Response of government yields, shock to OMT}
		\begin{minipage}{.5\textwidth}
			\hspace{-15pt}
			\includegraphics[scale=0.15]{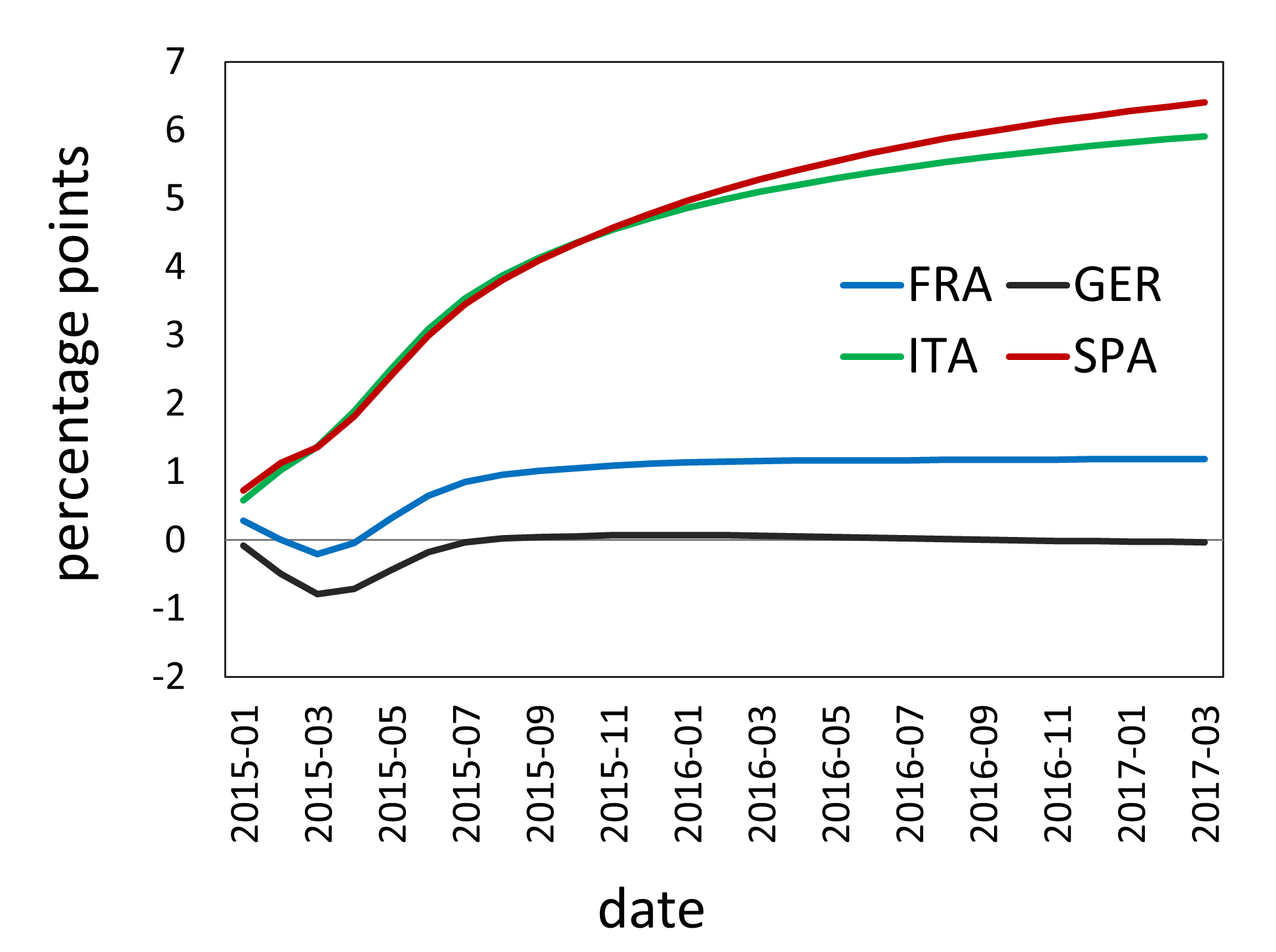}
		\end{minipage}%
		\begin{minipage}{.60\textwidth}
			\includegraphics[scale=0.11]{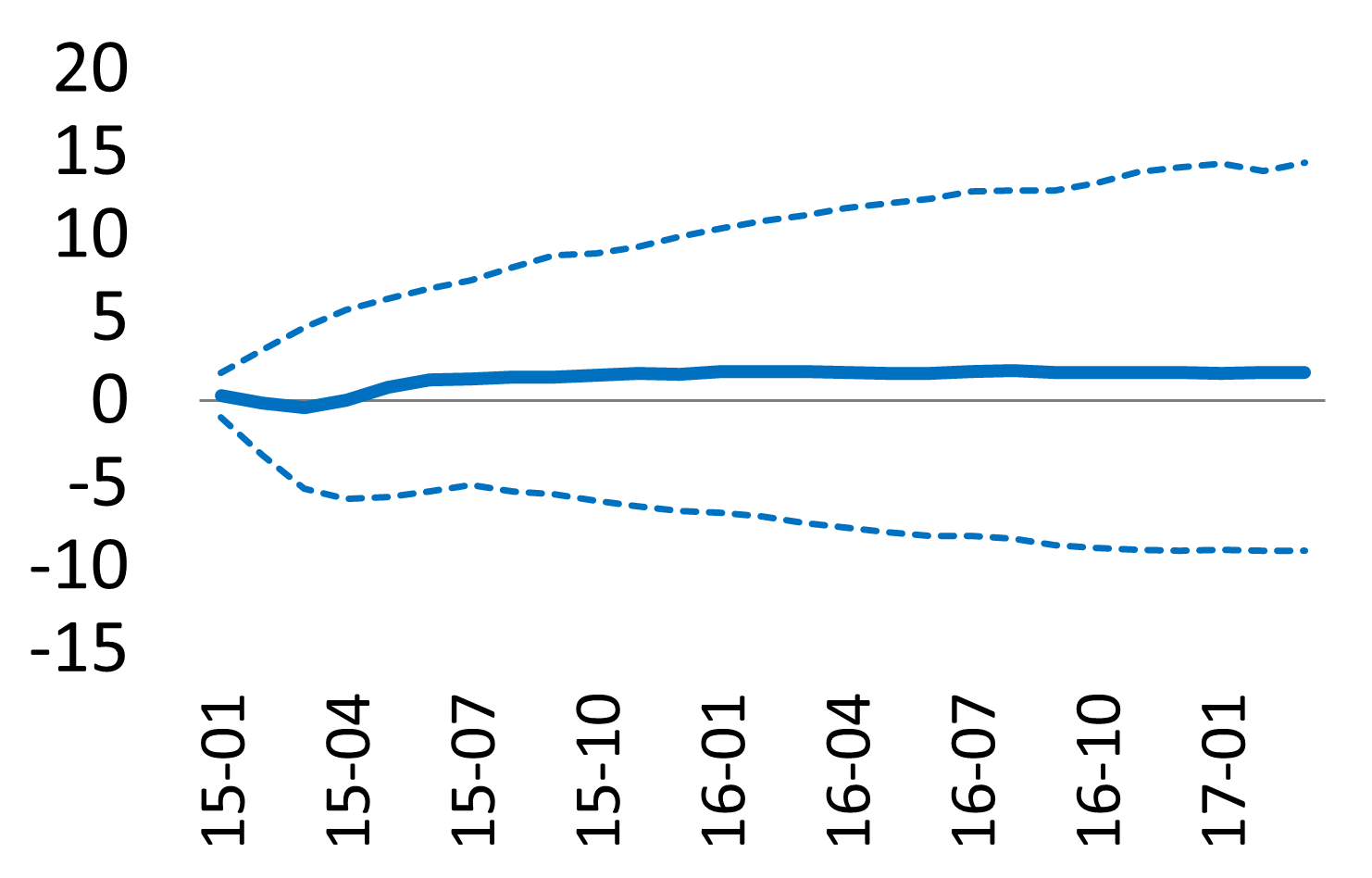}
			\includegraphics[scale=0.11]{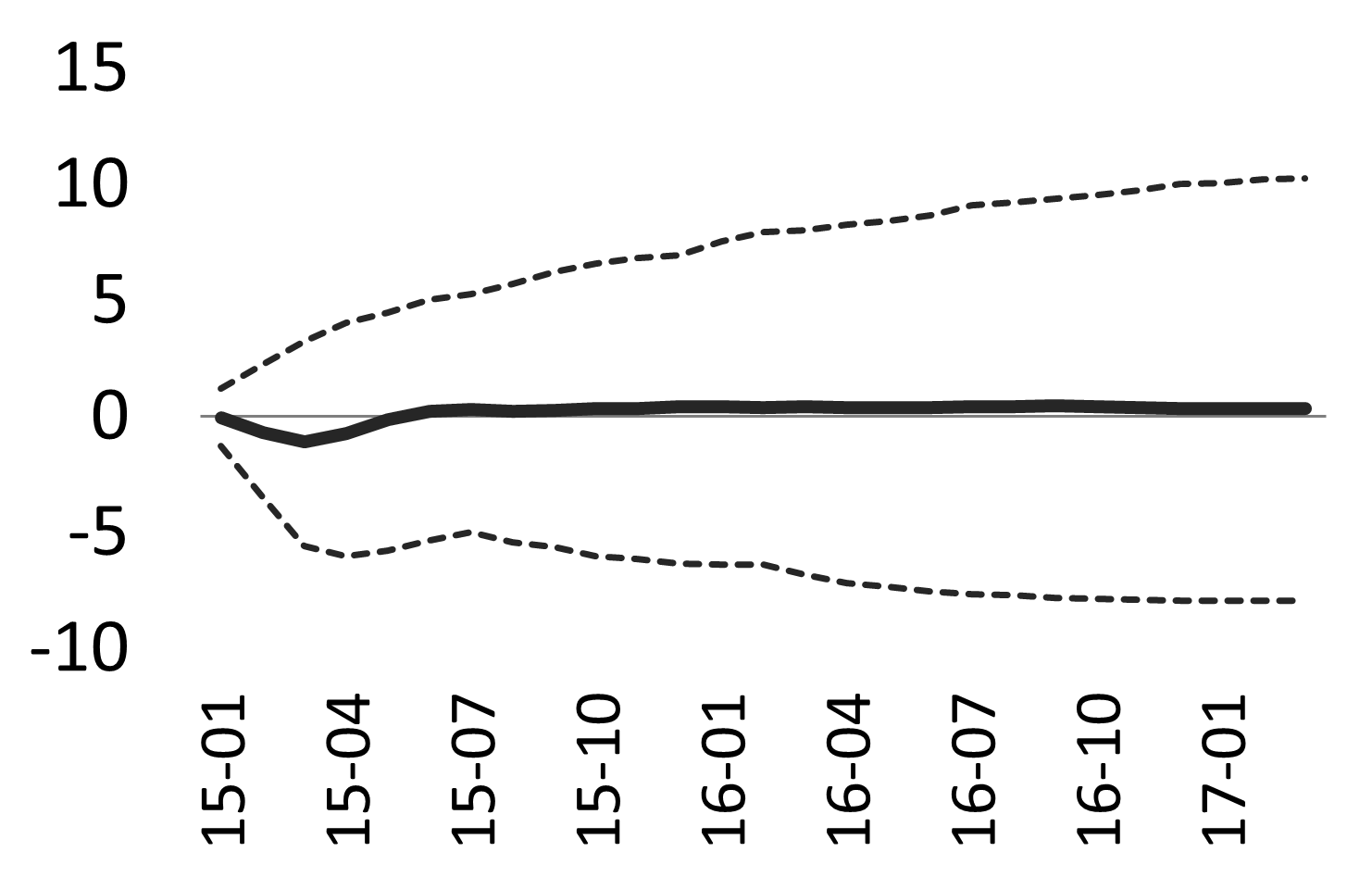}\\
			\includegraphics[scale=0.11]{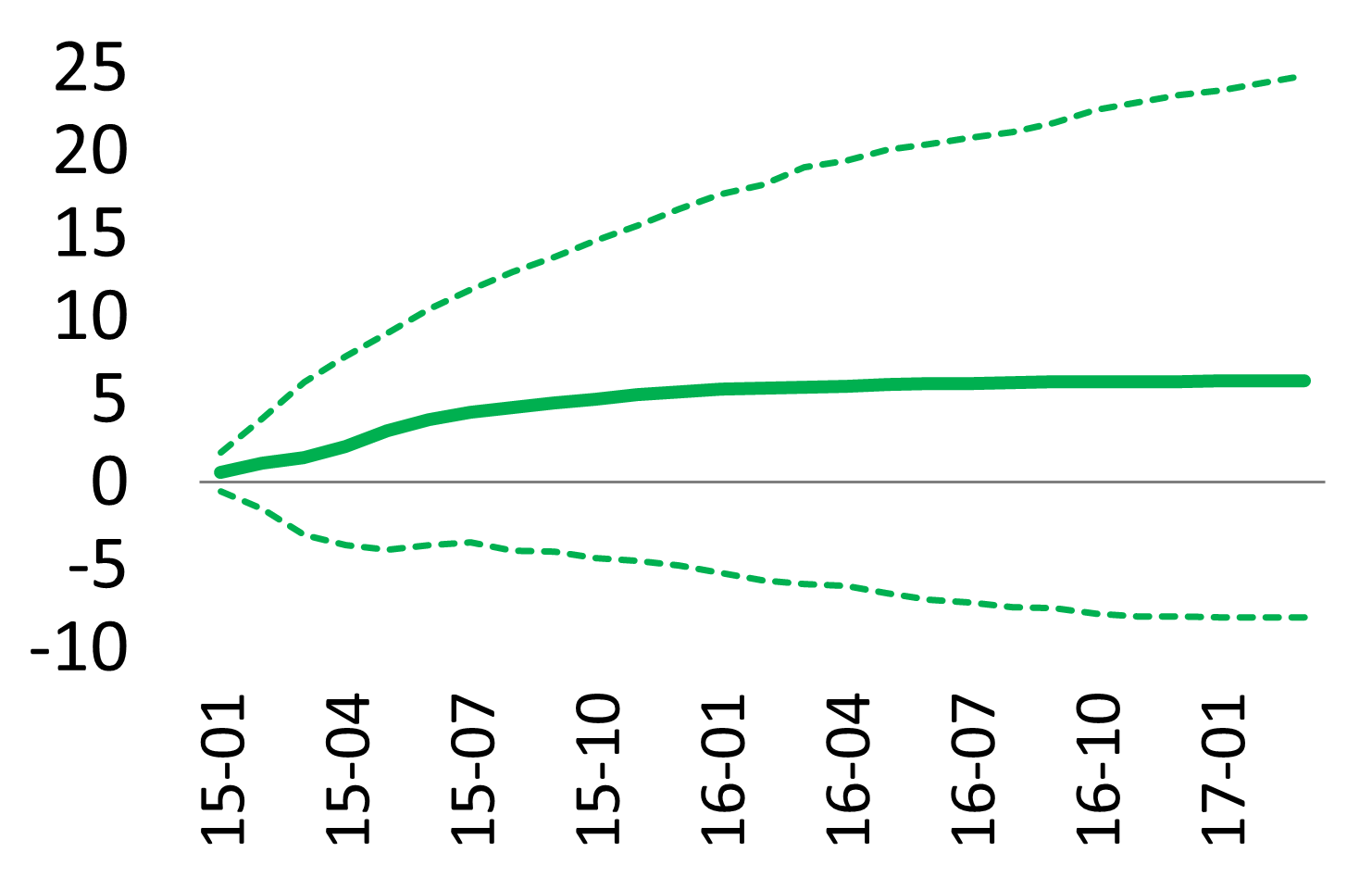}
			\includegraphics[scale=0.11]{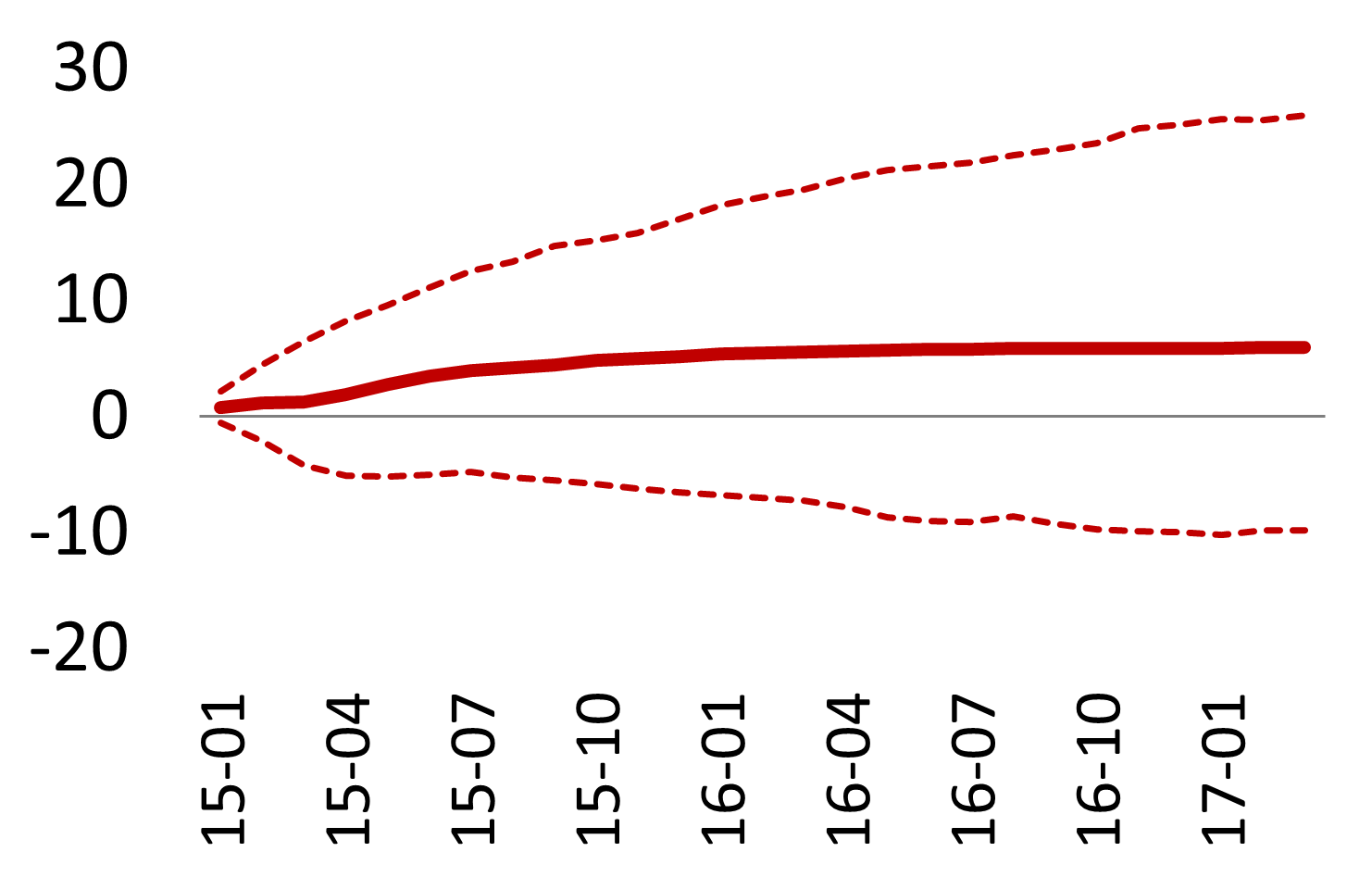}
		\end{minipage}%
		\subcaption*{\textit{Notes}: The figure shows the estimated responses to the shock scenario for the OMT programme (left panel) and bootstrap 90\% confidence intervals and medians (right panels).}
	\end{figure}
	
	\begin{figure}[H]
		\centering		
		\caption{Response of bank CDS spreads, shock to OMT}
		\begin{minipage}{.5\textwidth}
			\hspace{-15pt}
			\includegraphics[scale=0.15]{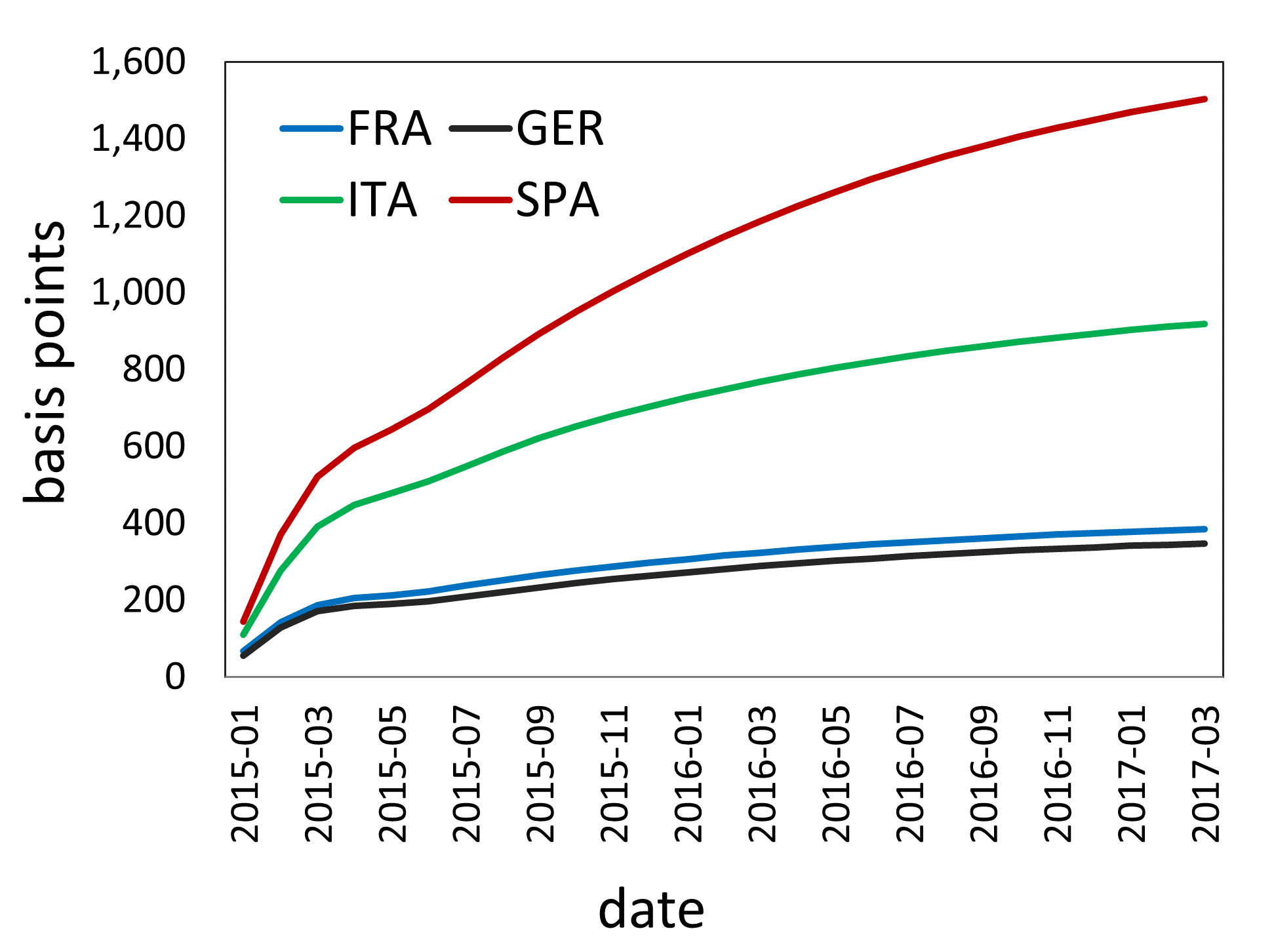}
		\end{minipage}%
		\begin{minipage}{.60\textwidth}
			\includegraphics[scale=0.11]{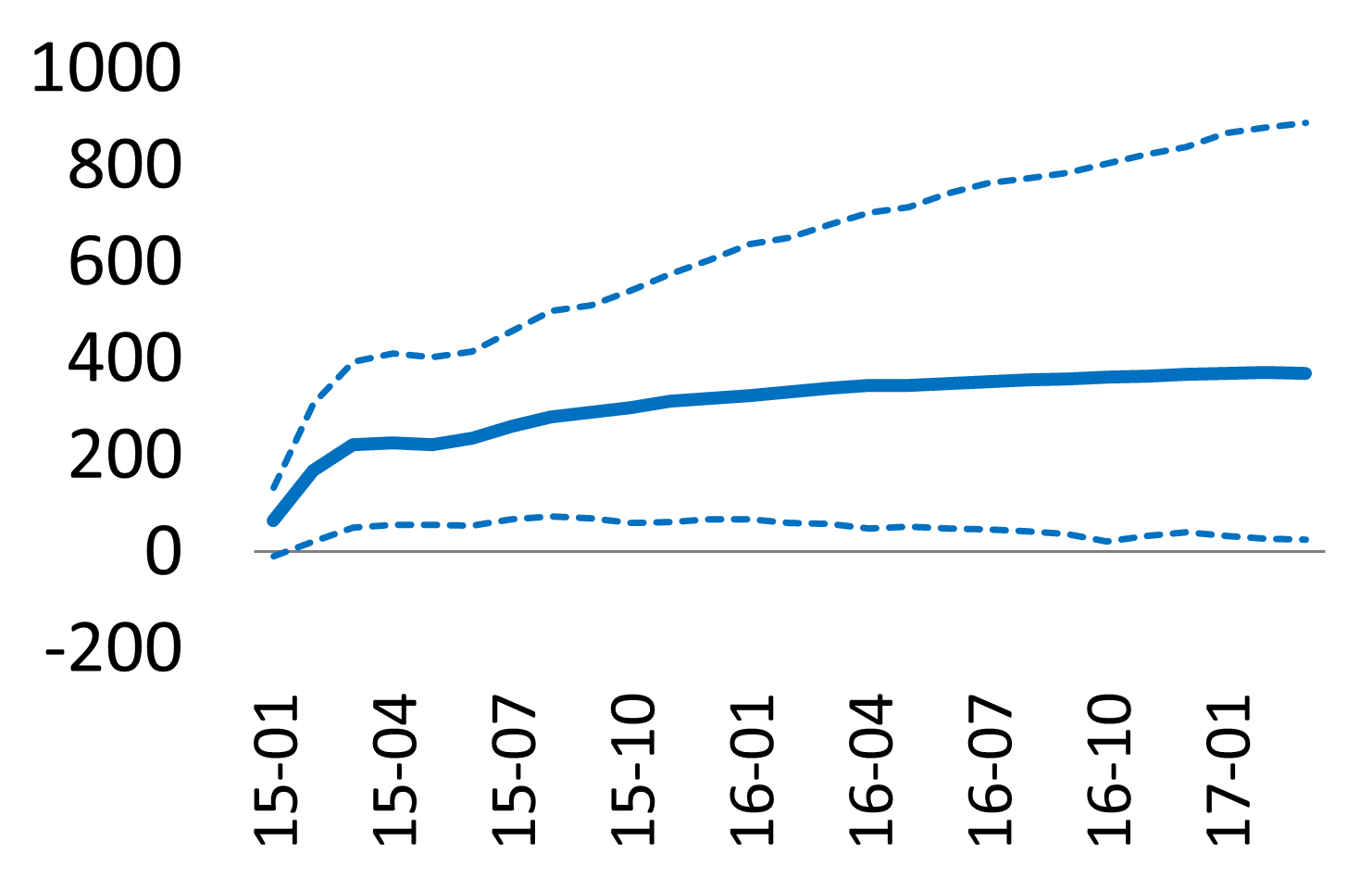}
			\includegraphics[scale=0.11]{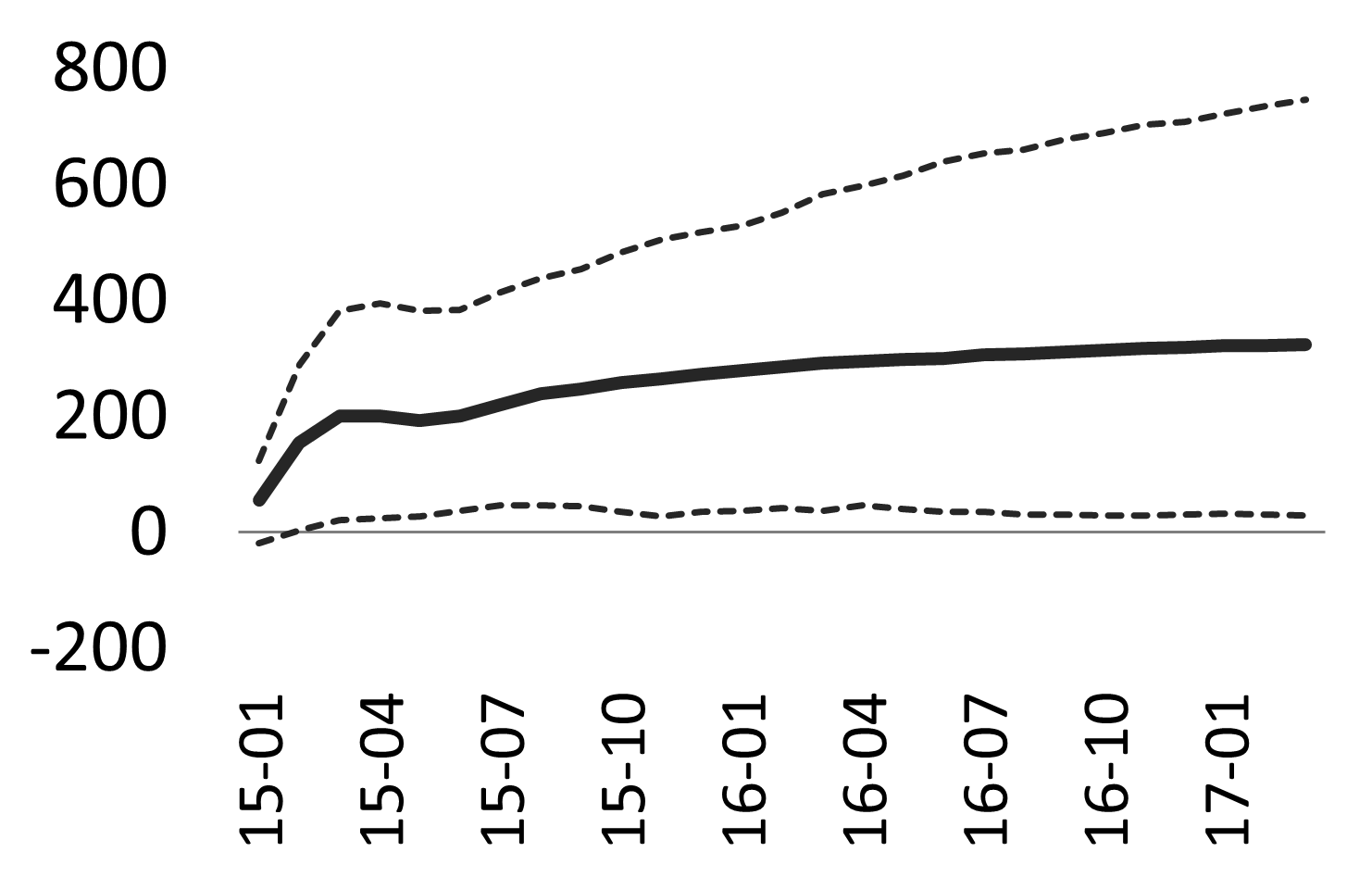}\\
			\includegraphics[scale=0.11]{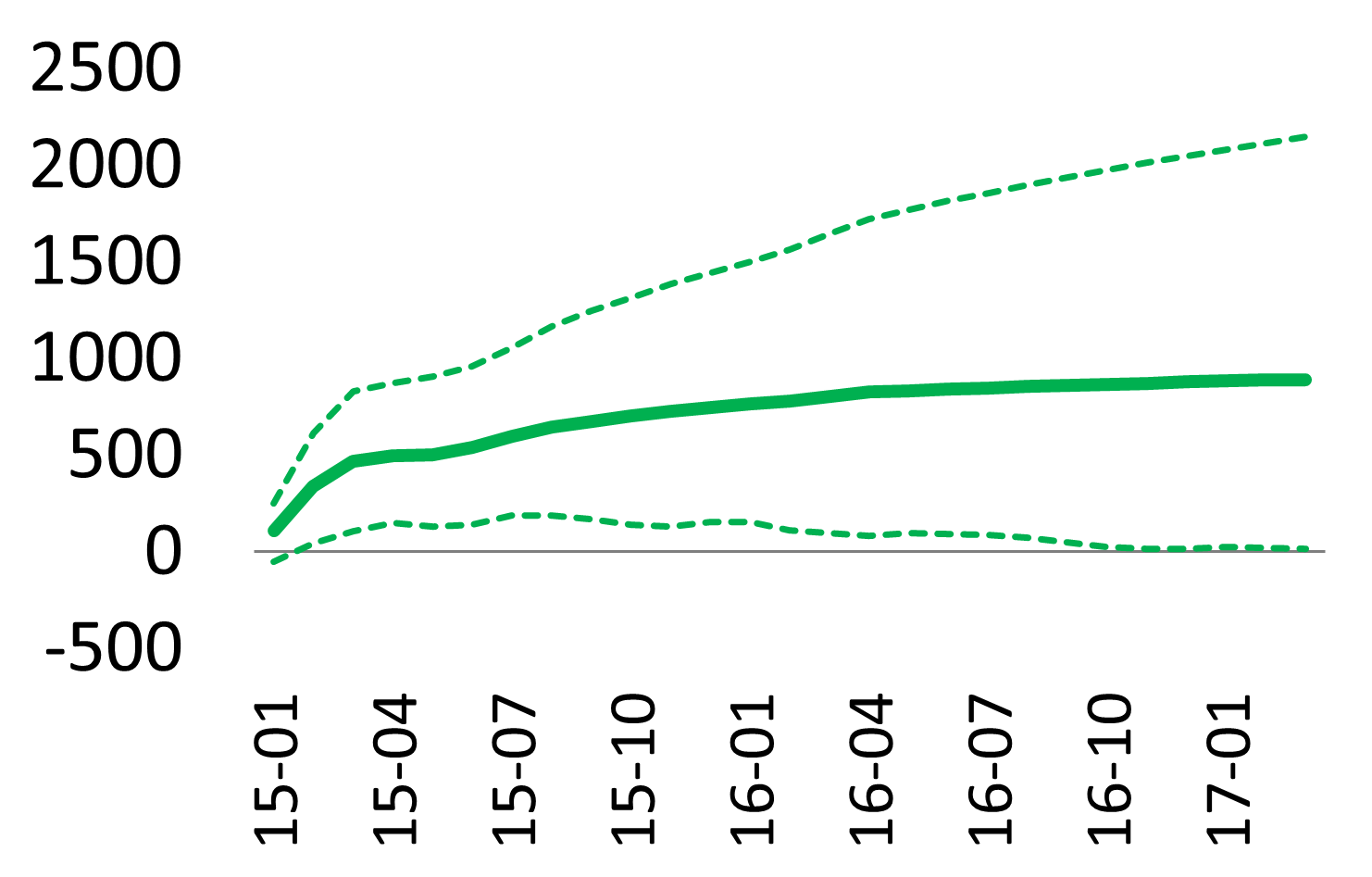}
			\includegraphics[scale=0.11]{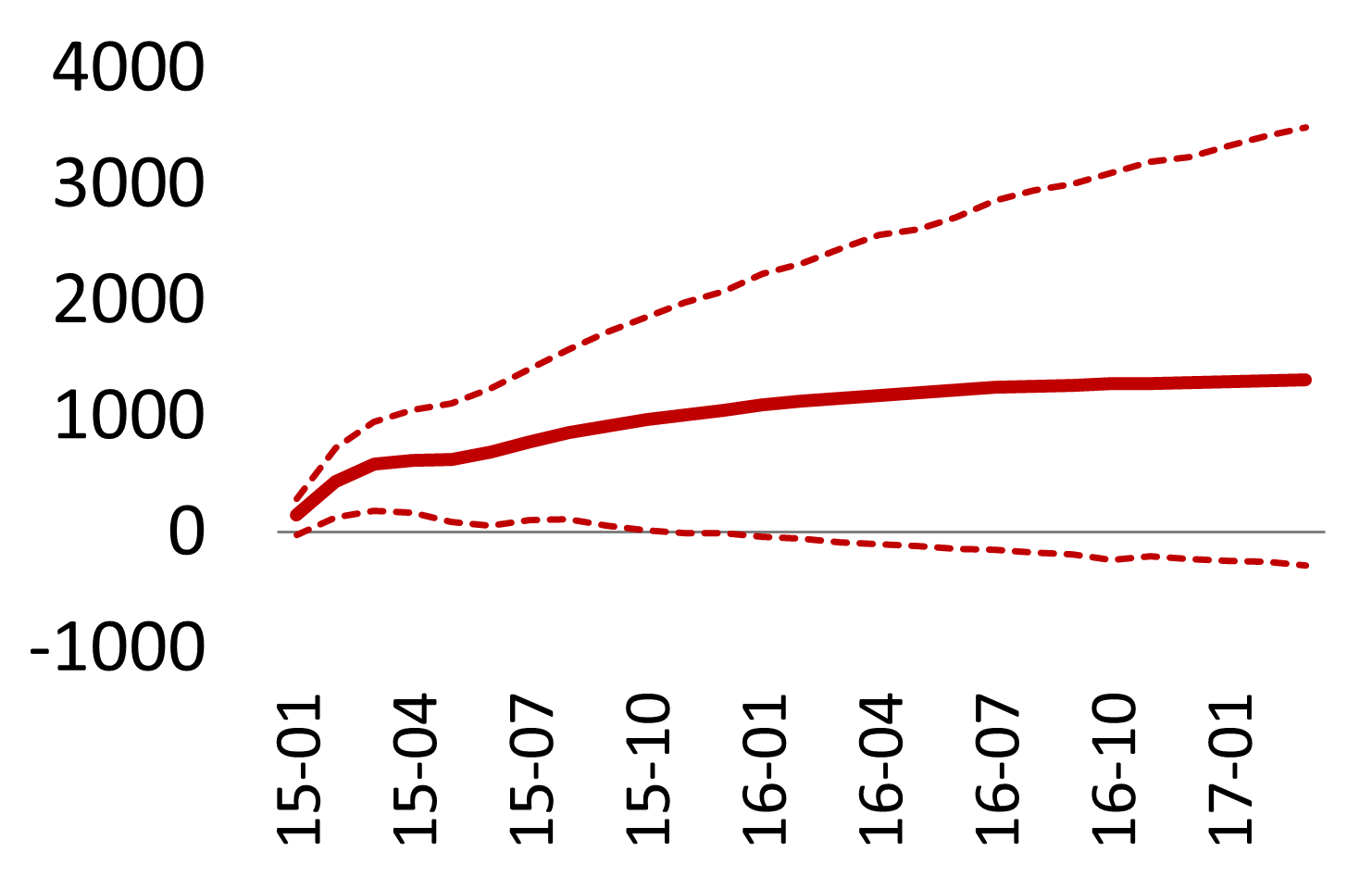}
		\end{minipage}%
		\subcaption*{\textit{Notes}: The figure shows the estimated responses to the shock scenario for the OMT programme (left panel) and bootstrap 90\% confidence intervals and medians (right panels).}
	\end{figure}
	
	\begin{figure}[H]
		\centering		
		\caption{Response of repo trade volumes, shock to OMT}
		\begin{minipage}{.5\textwidth}
			\hspace{-15pt}
			\includegraphics[scale=0.15]{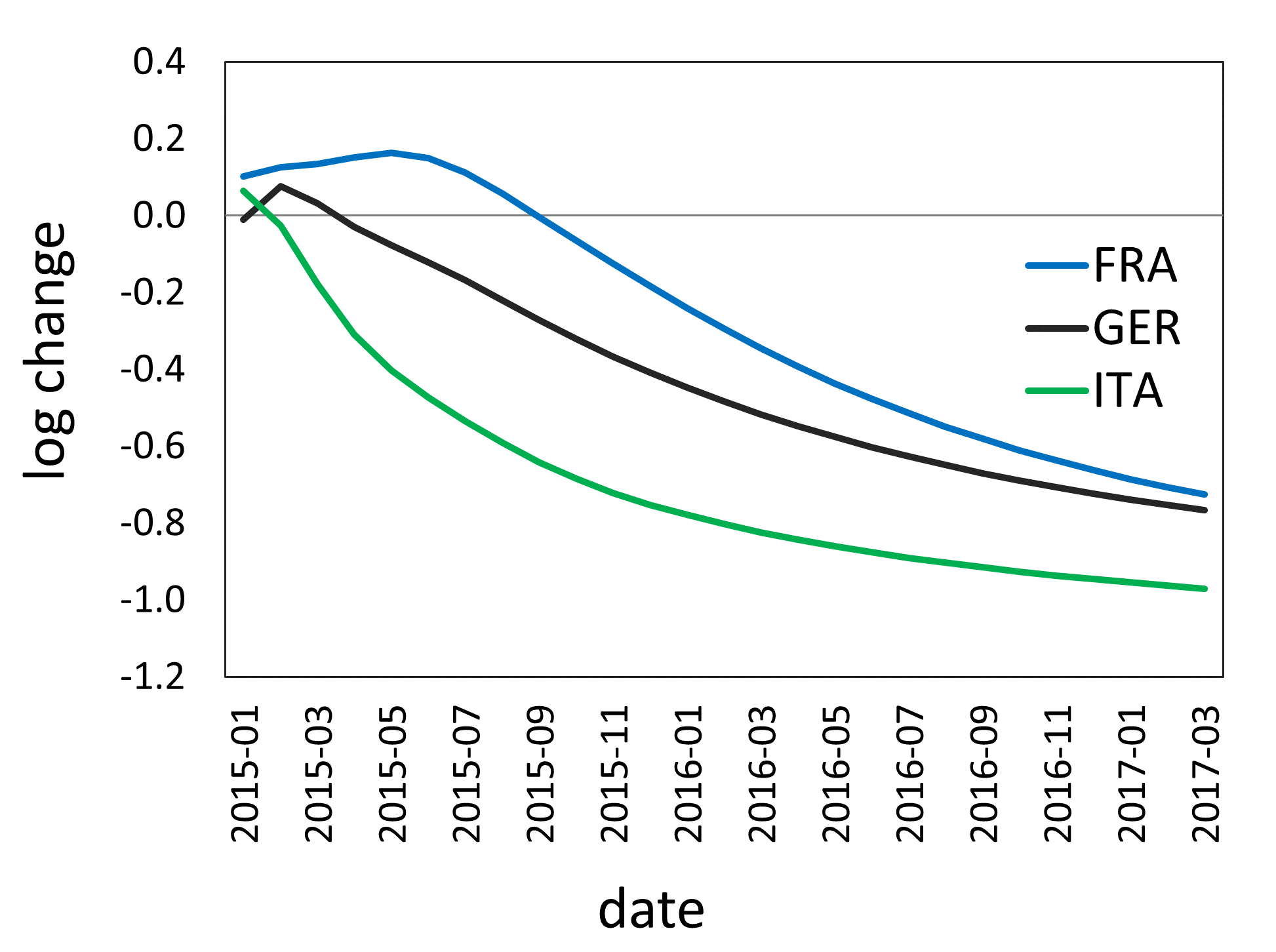}
		\end{minipage}%
		\begin{minipage}{.60\textwidth}
			\includegraphics[scale=0.11]{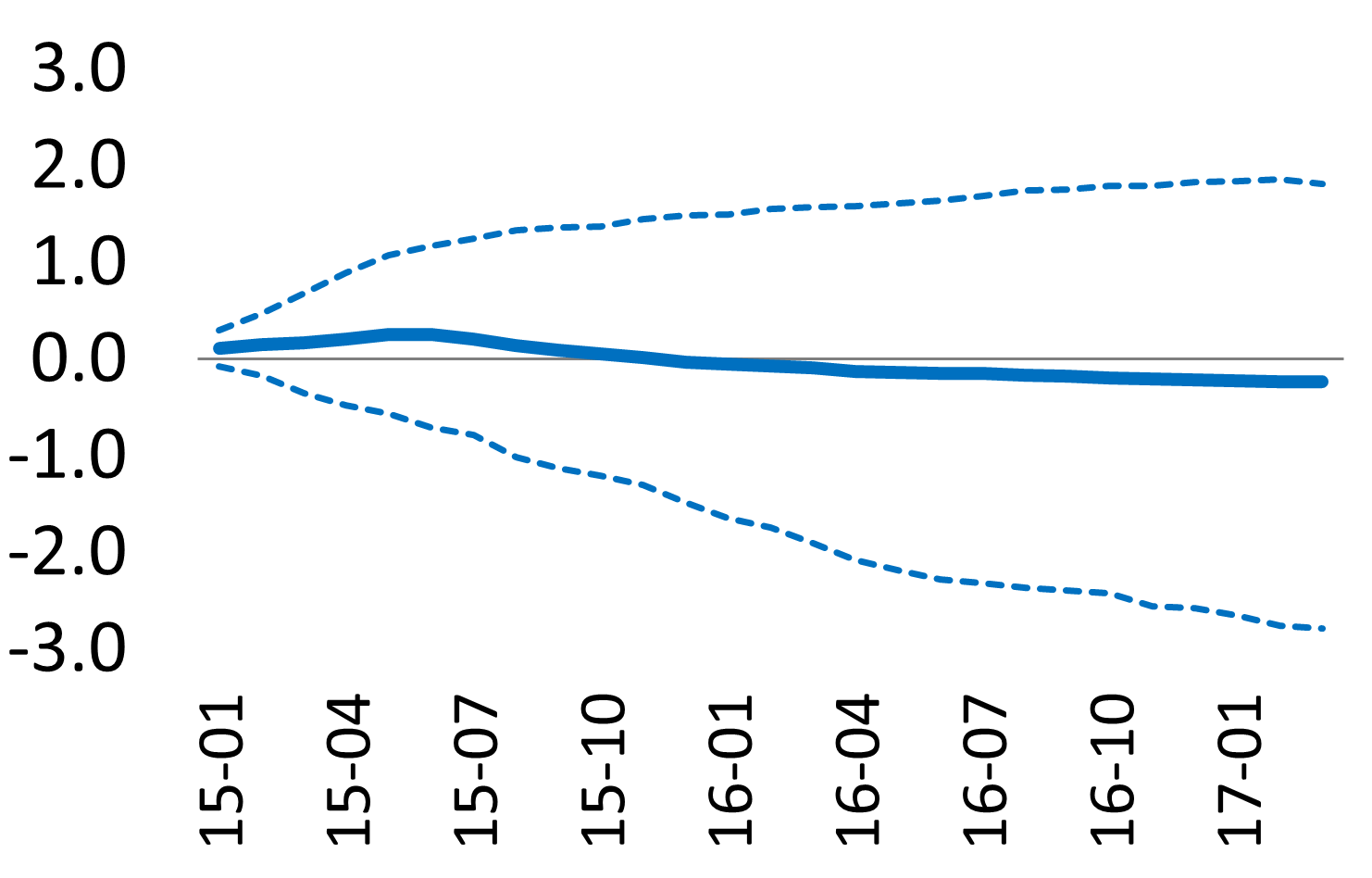}
			\includegraphics[scale=0.11]{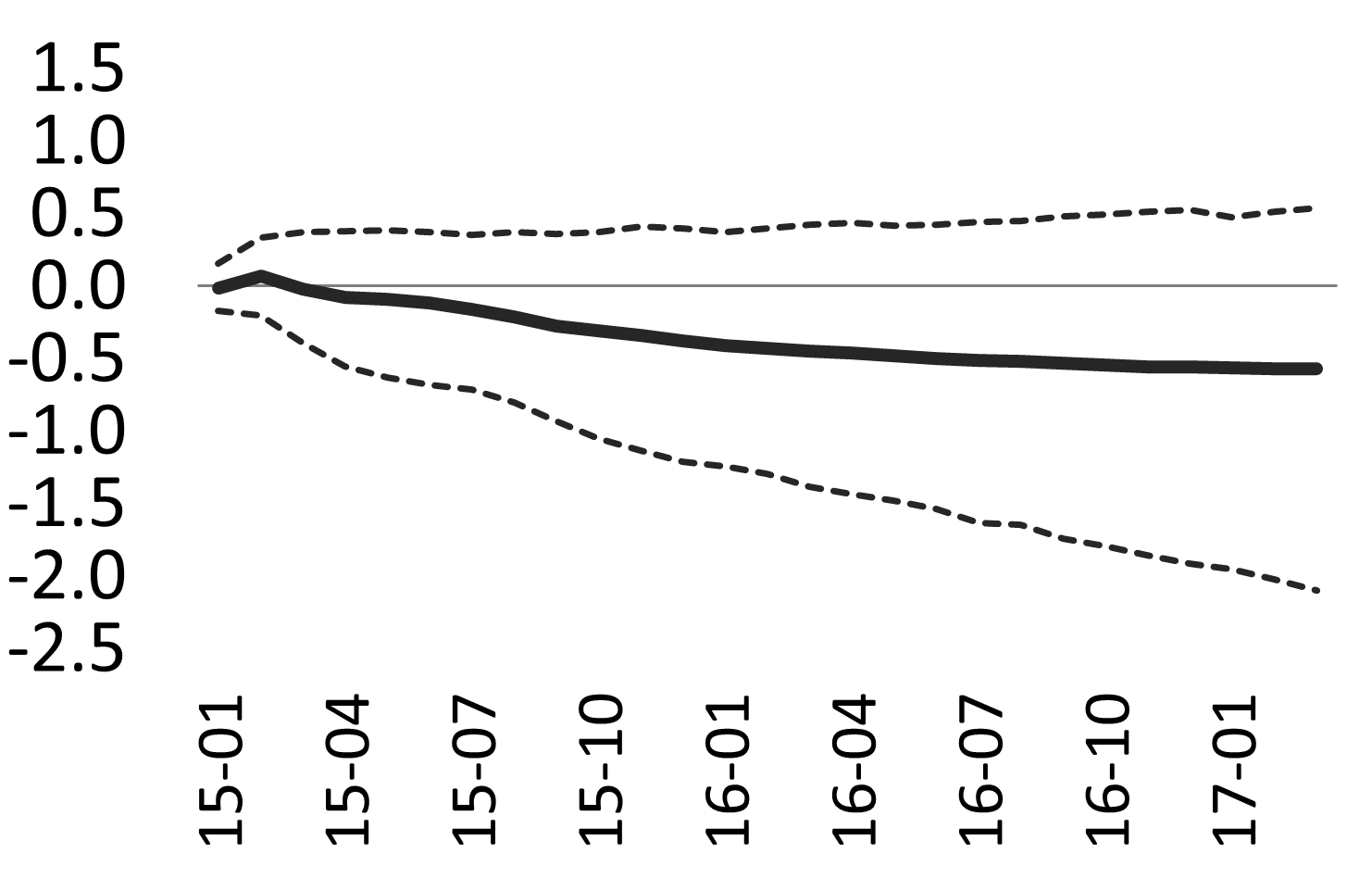}\\
			\includegraphics[scale=0.11]{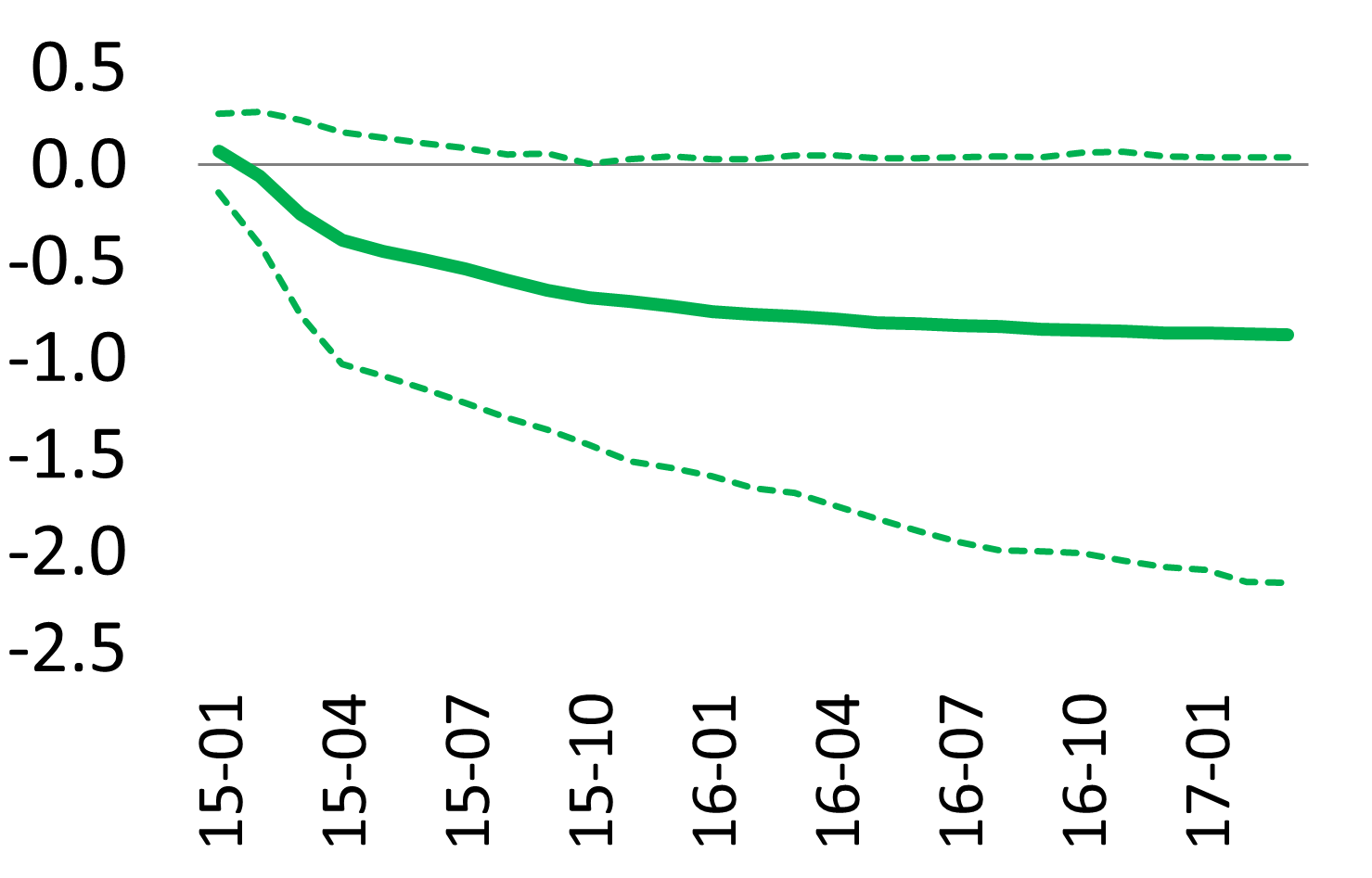}
		\end{minipage}%
		\subcaption*{\textit{Notes}: The figure shows the estimated responses to the shock scenario for the OMT programme (left panel) and bootstrap 90\% confidence intervals and medians (right panels).}
	\end{figure}

	\begin{figure}[H]
		\centering		
		\caption{Response of repo rates, shock to OMT}
		\begin{minipage}{.5\textwidth}
			\hspace{-15pt}
			\includegraphics[scale=0.15]{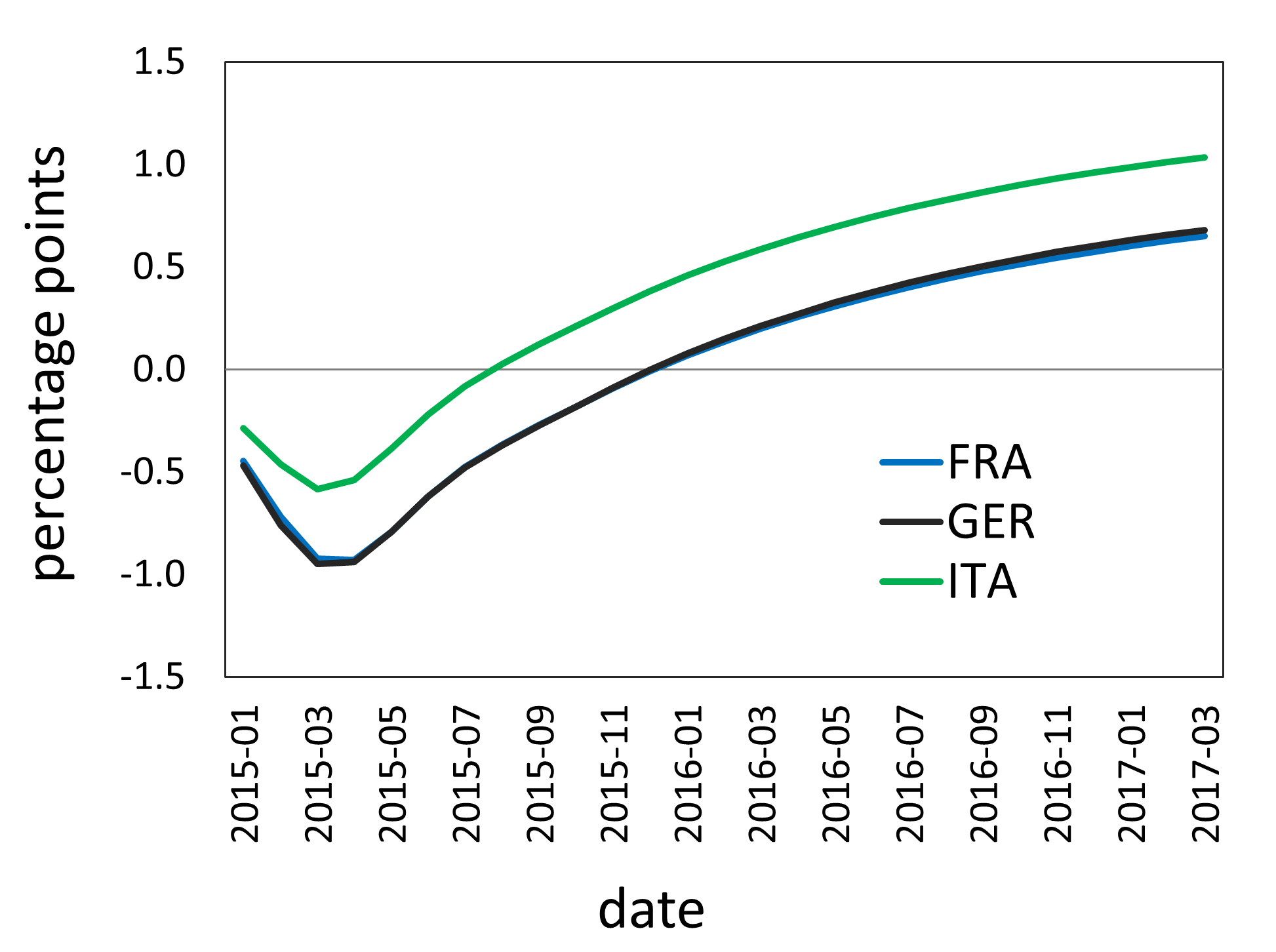}
		\end{minipage}%
		\begin{minipage}{.60\textwidth}
			\includegraphics[scale=0.11]{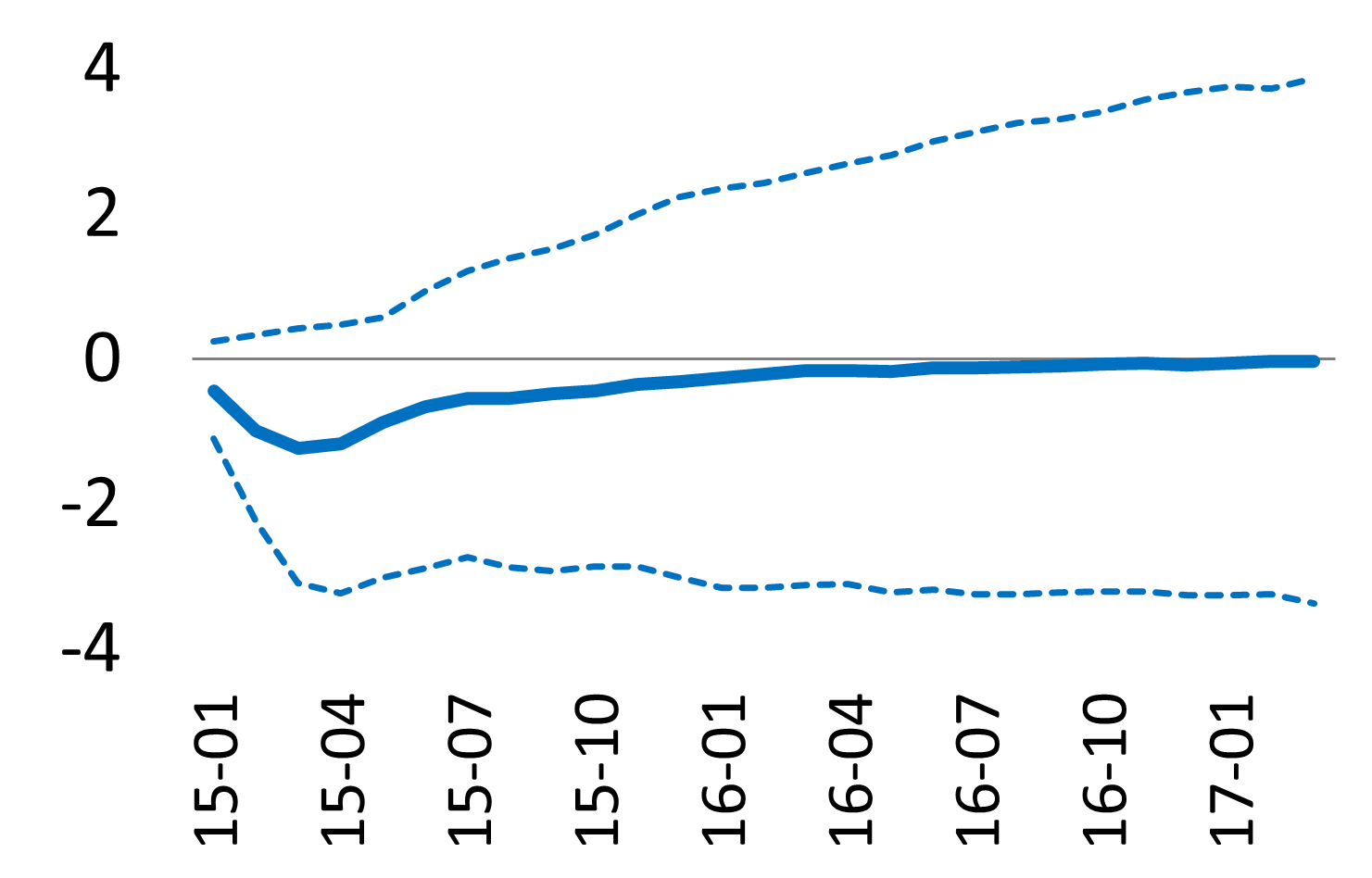}
			\includegraphics[scale=0.11]{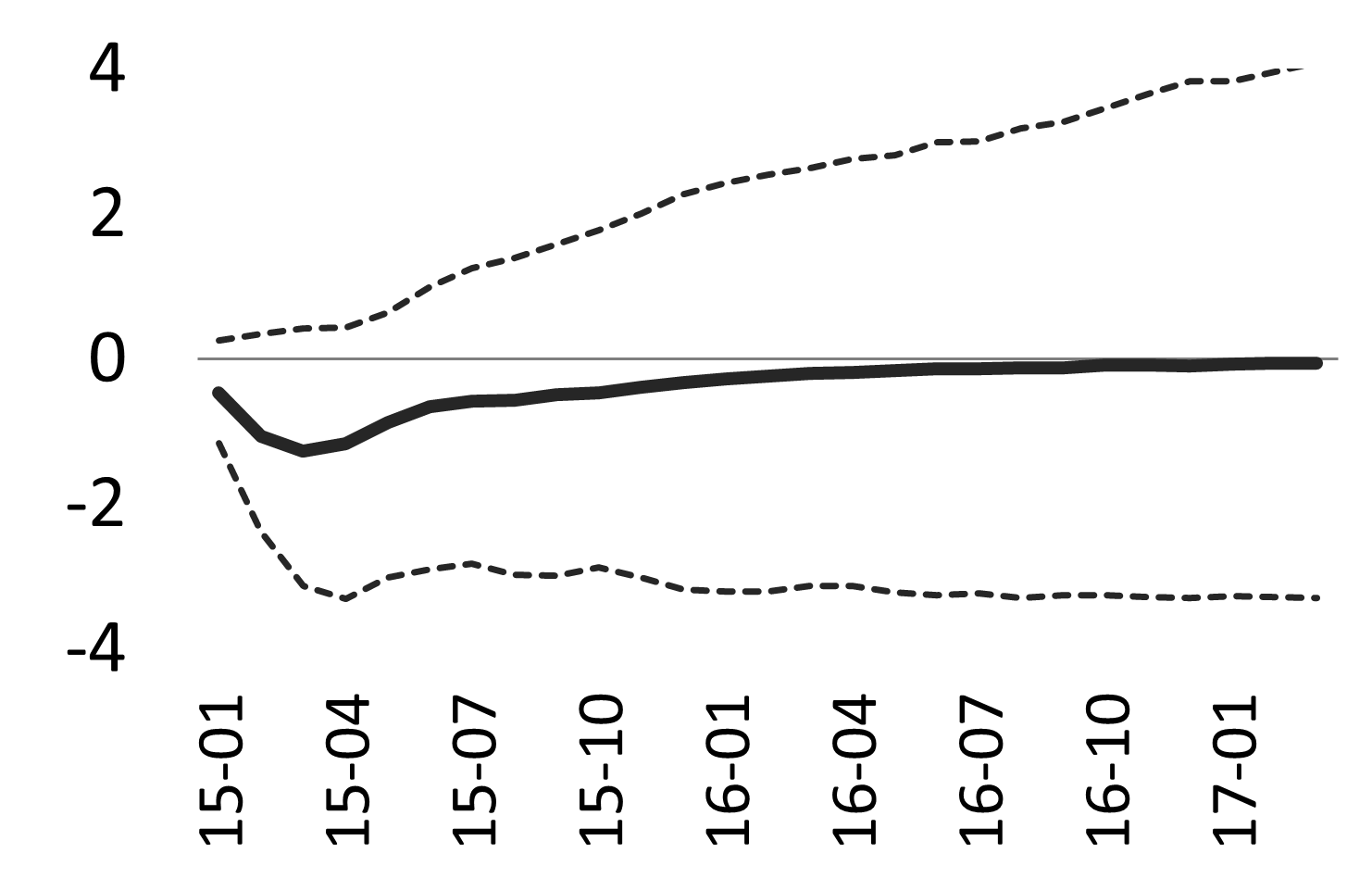}\\
			\includegraphics[scale=0.11]{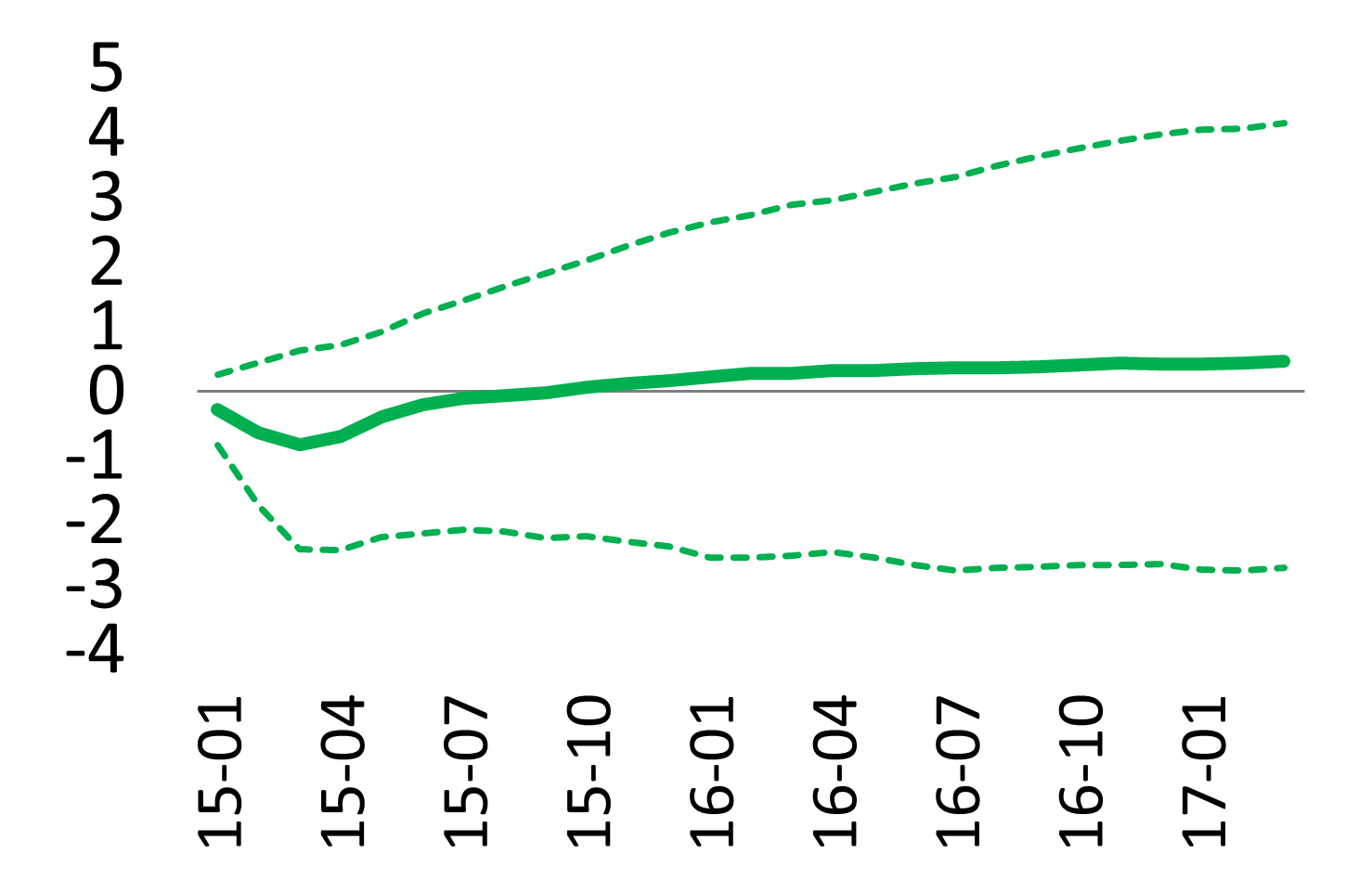}
		\end{minipage}%
		\subcaption*{\textit{Notes}: The figure shows the estimated responses to the shock scenario for the OMT programme (left panel) and bootstrap 90\% confidence intervals and medians (right panels).}
	\end{figure}

	\subsubsection{APP}

	\begin{figure}[H]
		\centering		
		\caption{Response of government yields, shock to APP}
		\begin{minipage}{.5\textwidth}
			\hspace{-15pt}
			\includegraphics[scale=0.15]{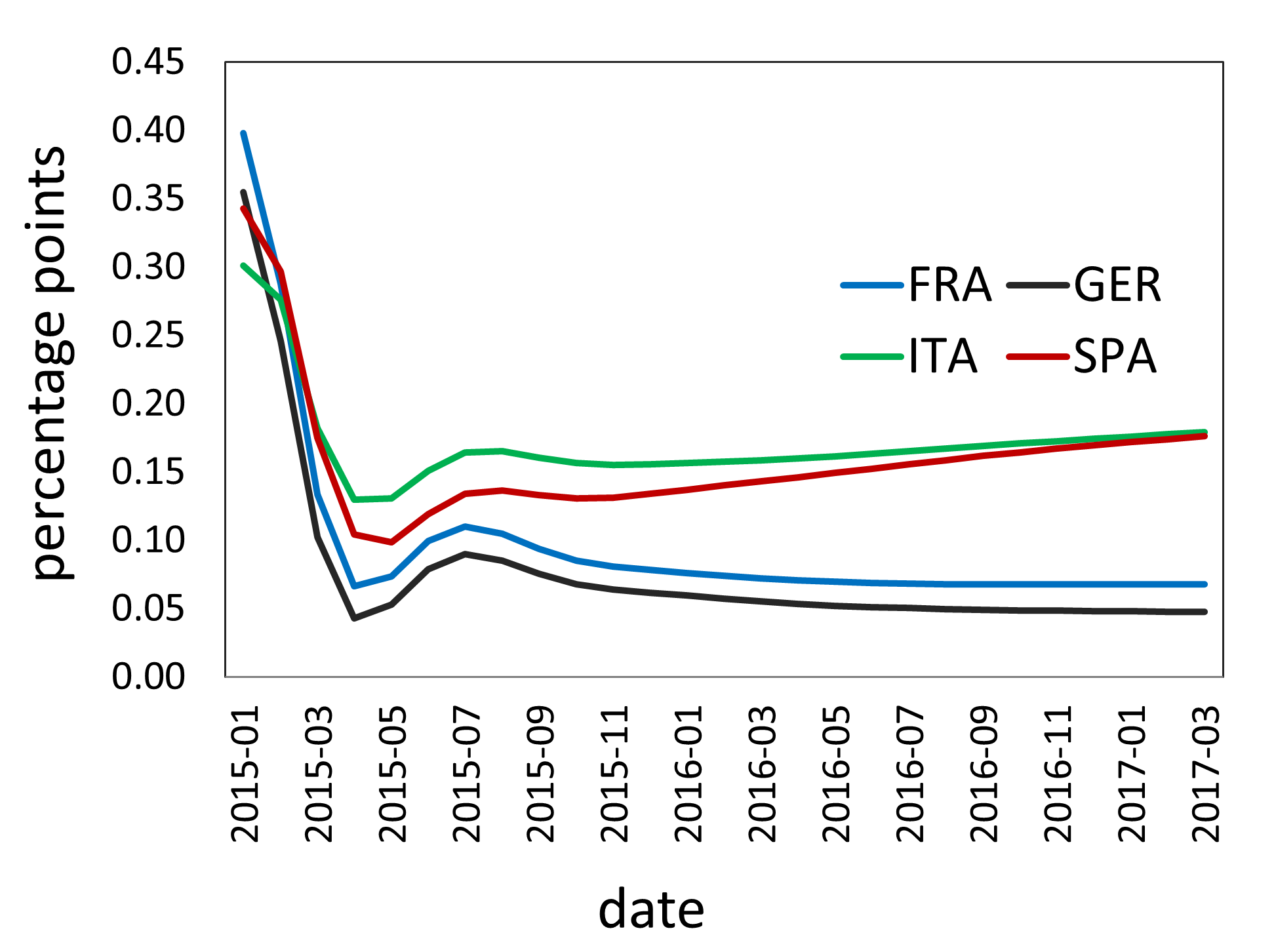}
		\end{minipage}%
		\begin{minipage}{.60\textwidth}
			\includegraphics[scale=0.11]{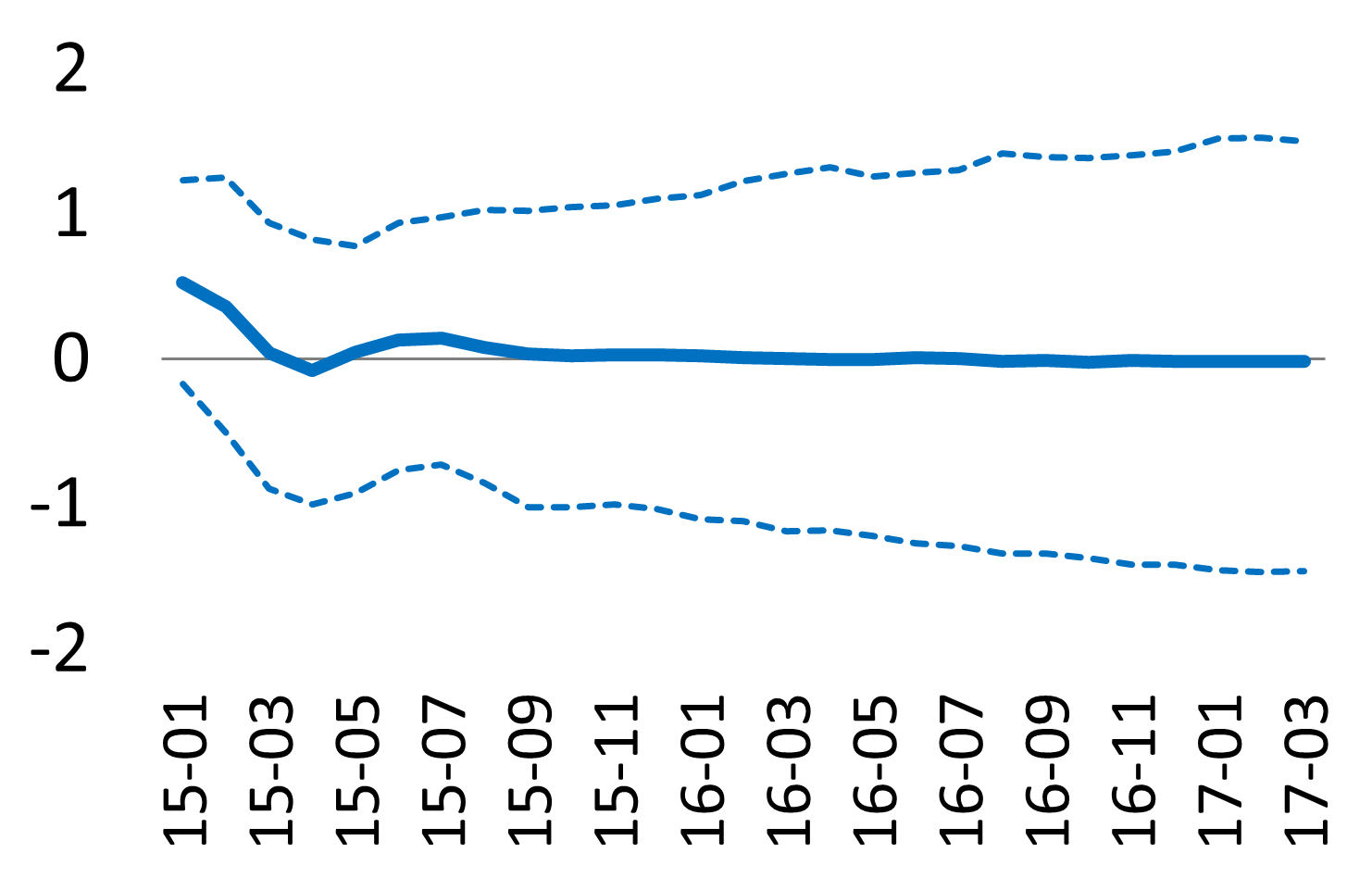}
			\includegraphics[scale=0.11]{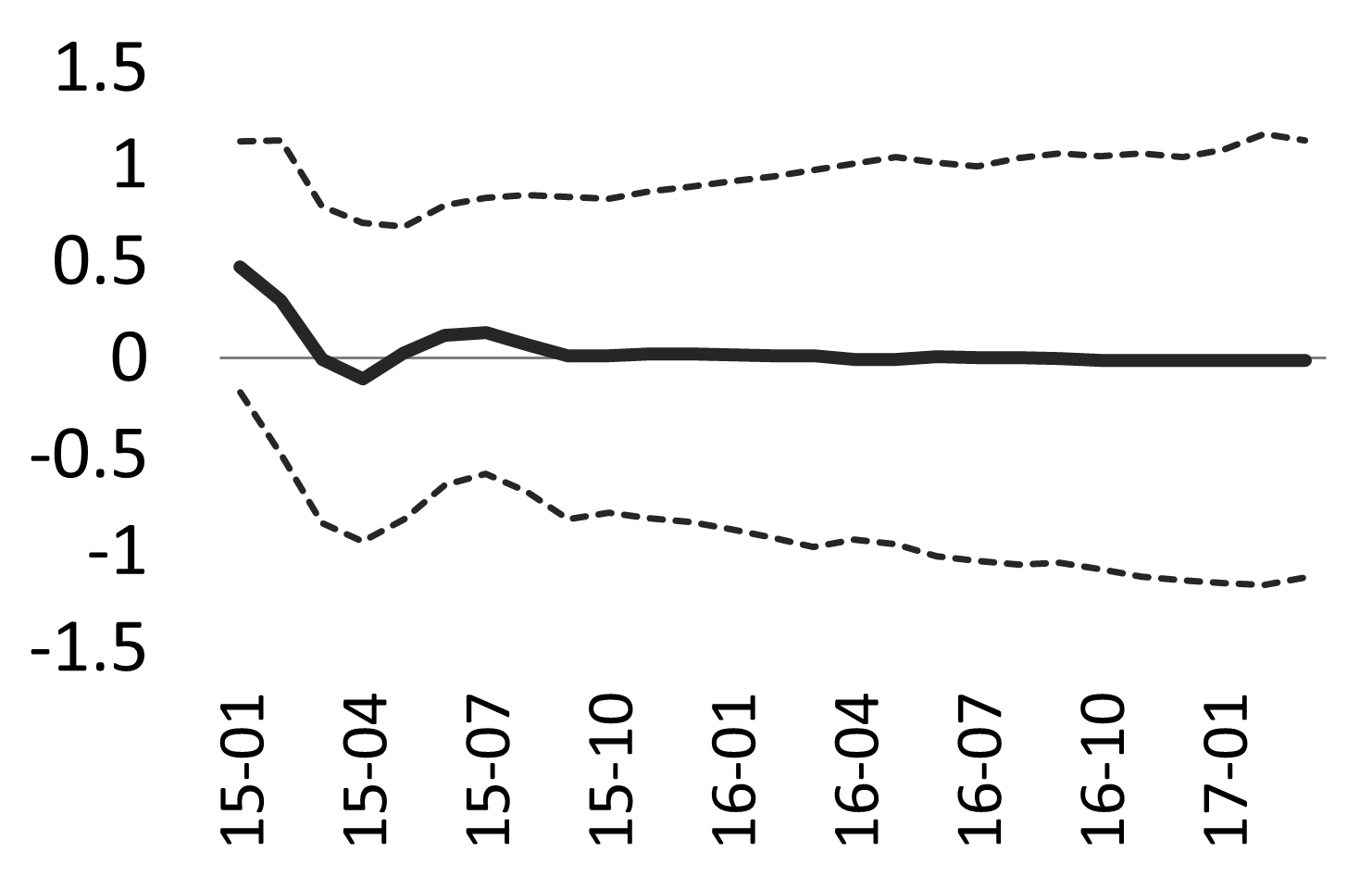}\\
			\includegraphics[scale=0.11]{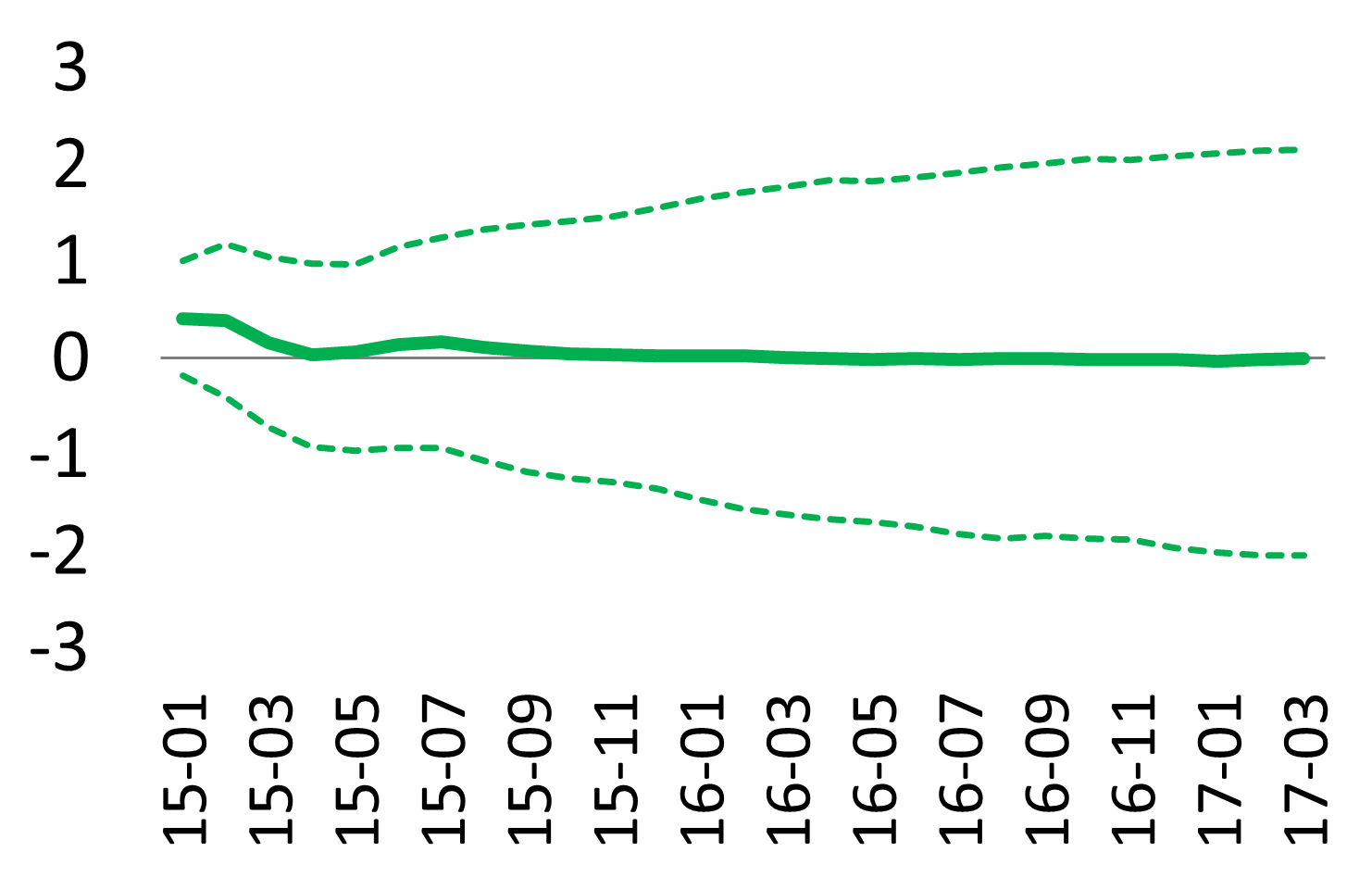}
			\includegraphics[scale=0.11]{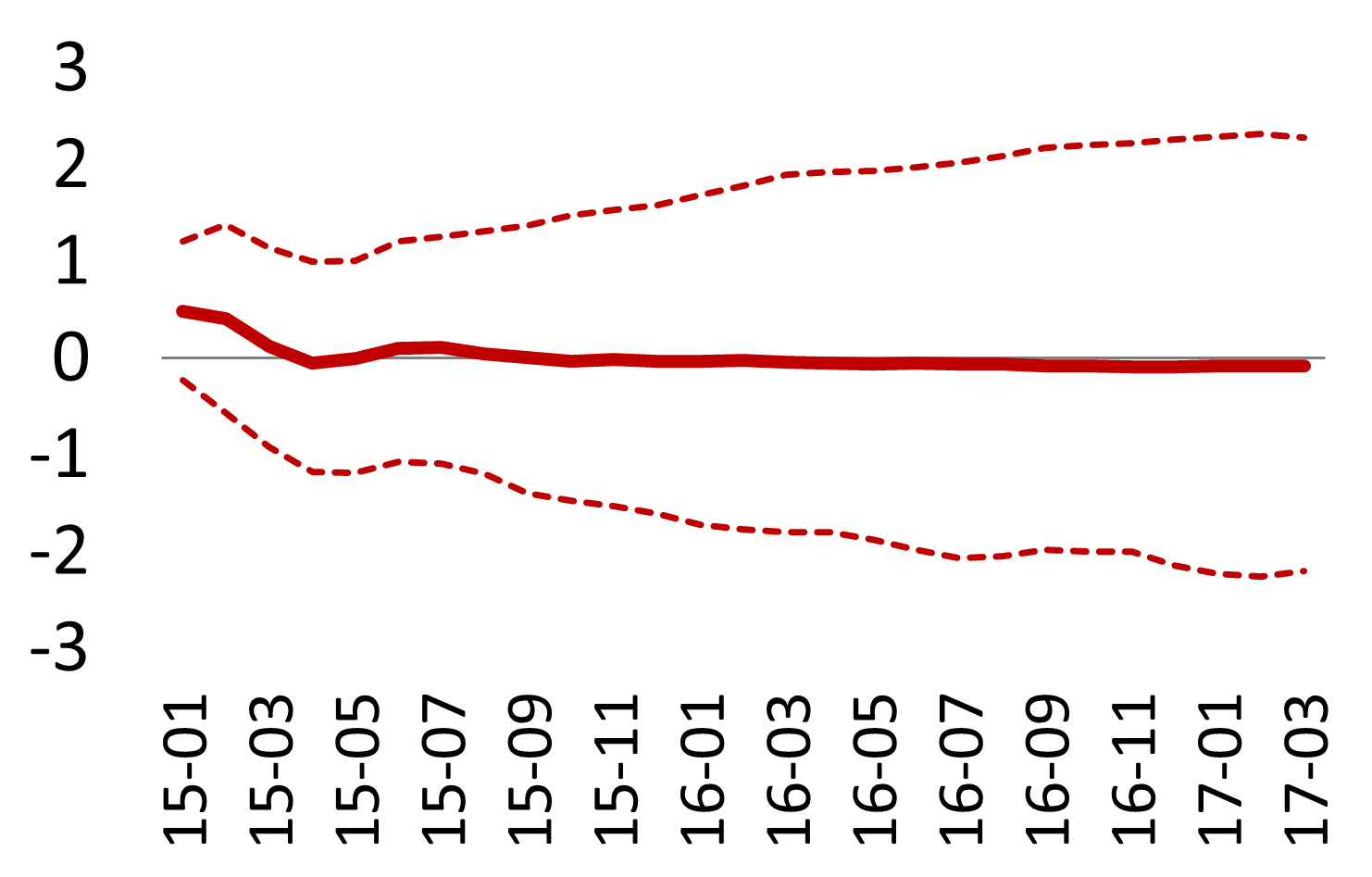}
		\end{minipage}%
		\subcaption*{\textit{Notes}: The figure shows the estimated responses to the shock scenario for the APP programme (left panel) and bootstrap 90\% confidence intervals and medians (right panels).}
	\end{figure}
	
	\begin{figure}[H]
		\centering		
		\caption{Response of bank CDS spreads, shock to APP}
		\begin{minipage}{.5\textwidth}
			\hspace{-15pt}
			\includegraphics[scale=0.15]{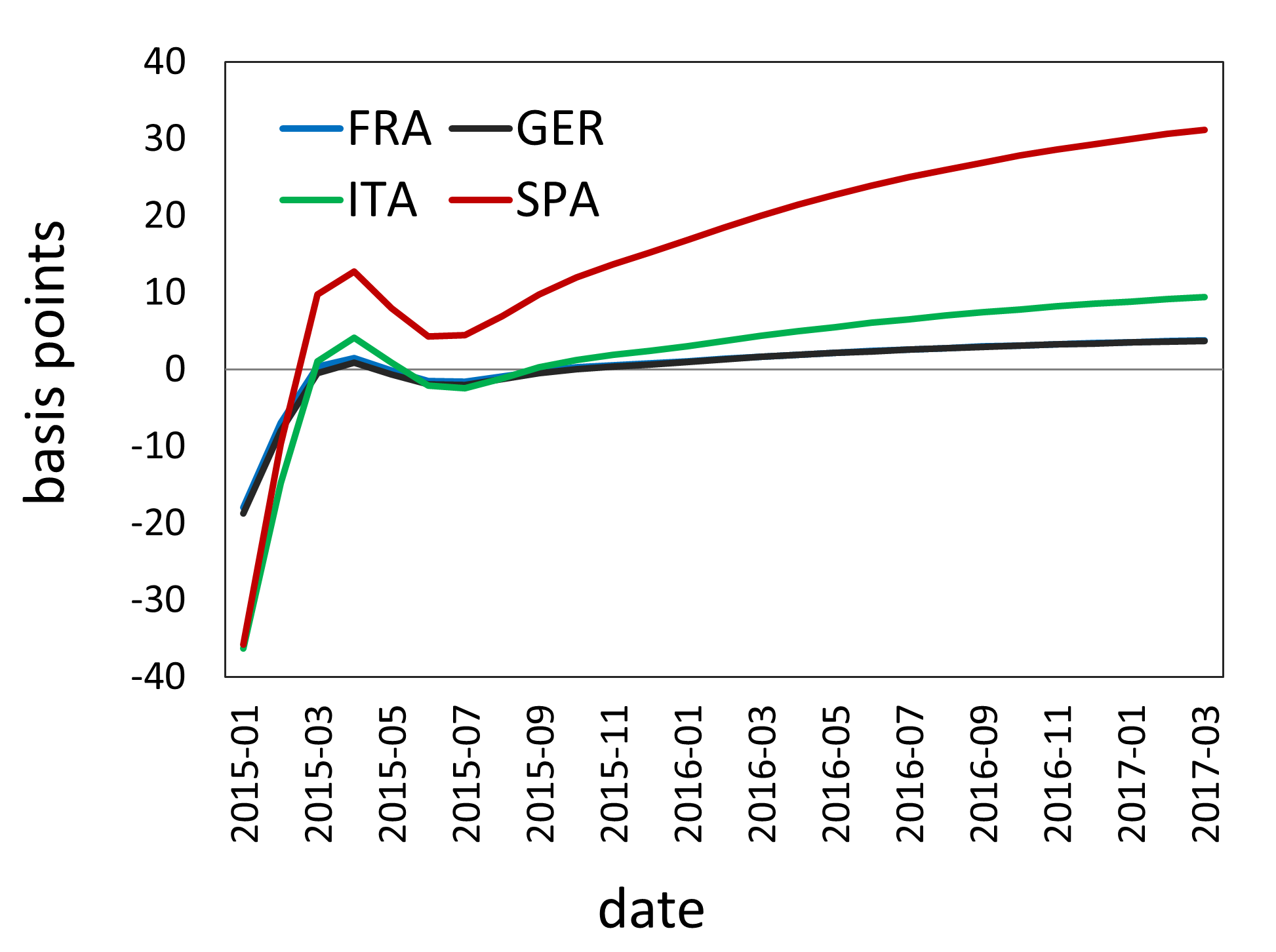}
		\end{minipage}%
		\begin{minipage}{.60\textwidth}
			\includegraphics[scale=0.11]{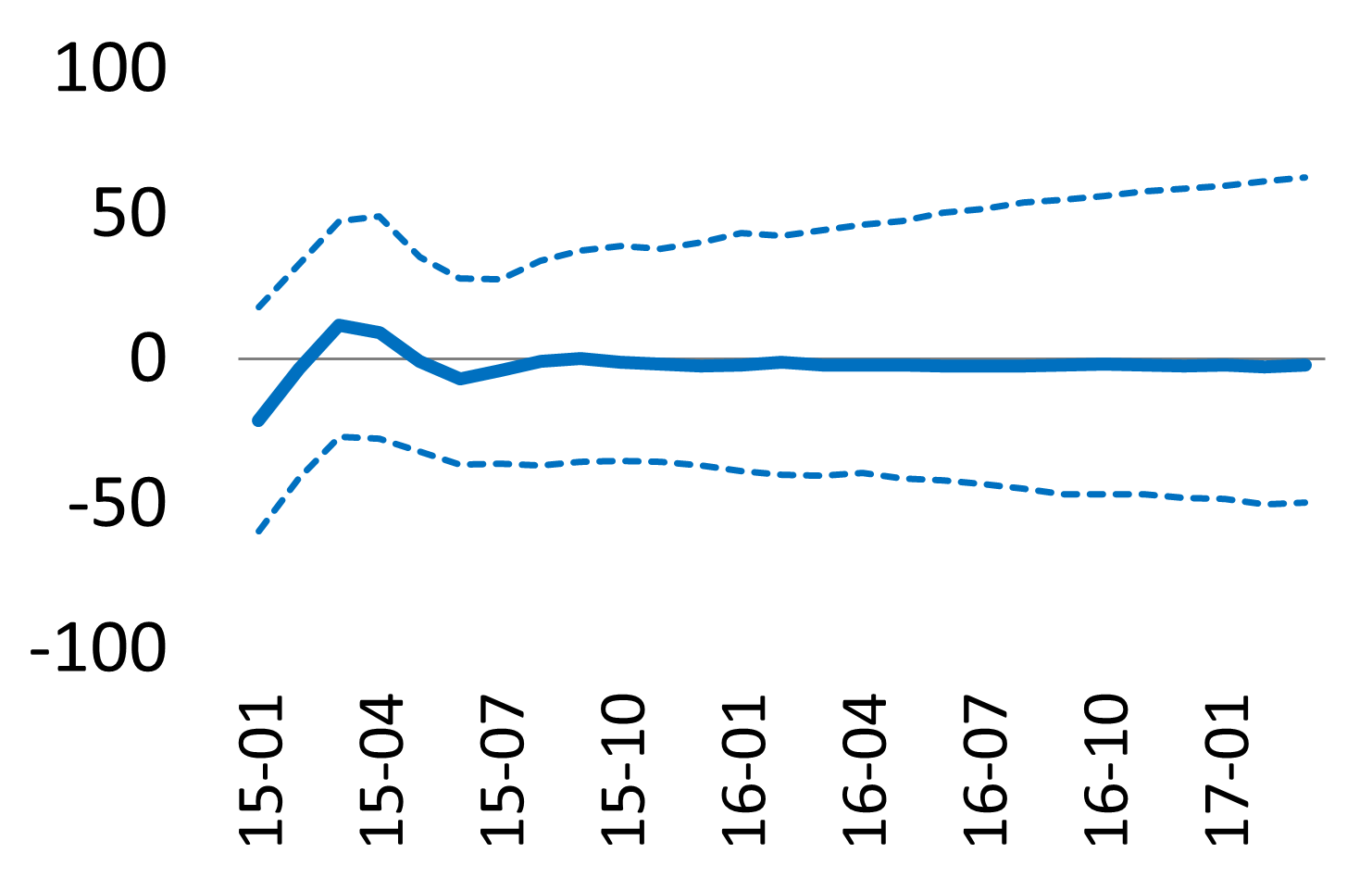}
			\includegraphics[scale=0.11]{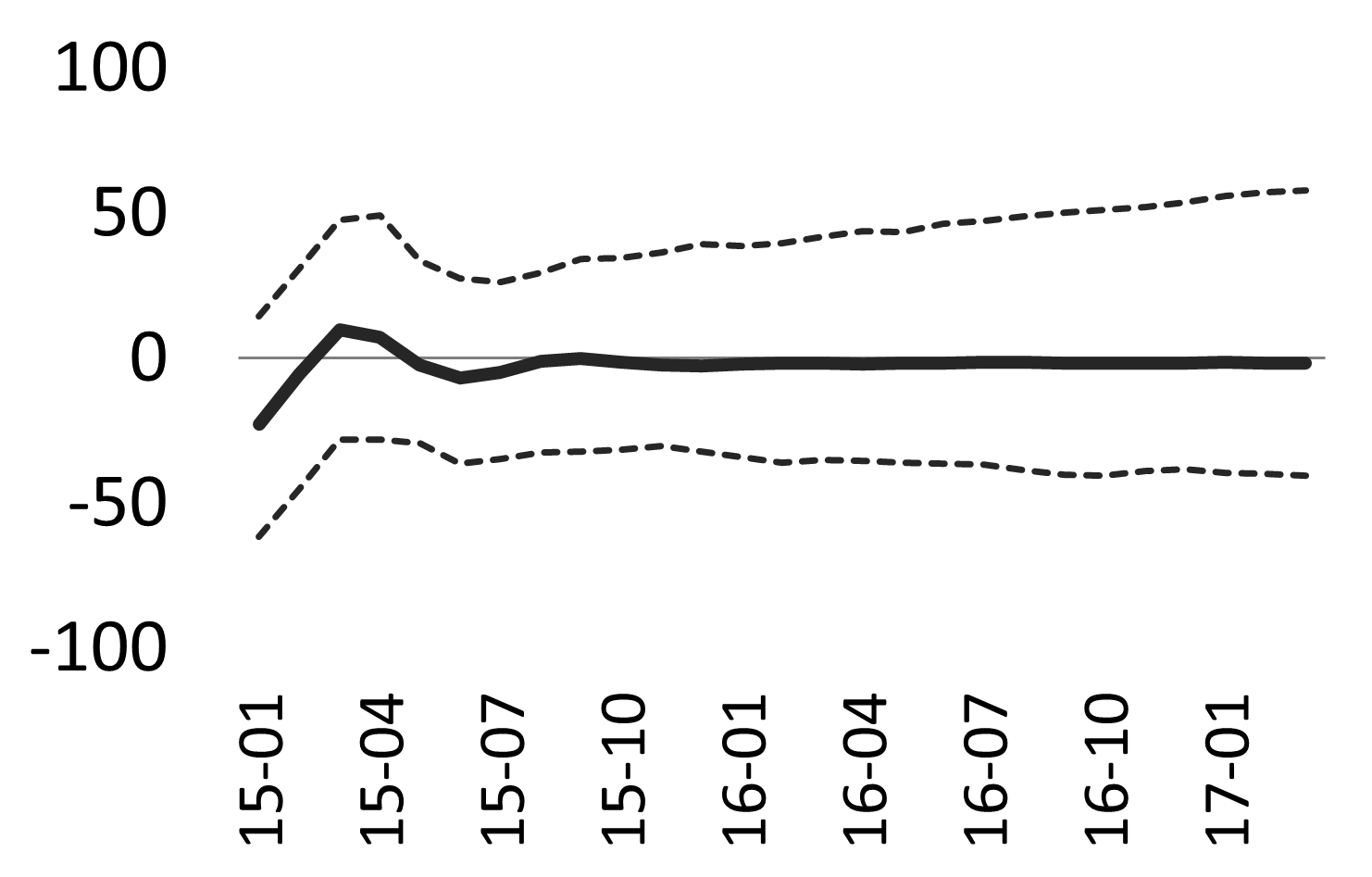}\\
			\includegraphics[scale=0.11]{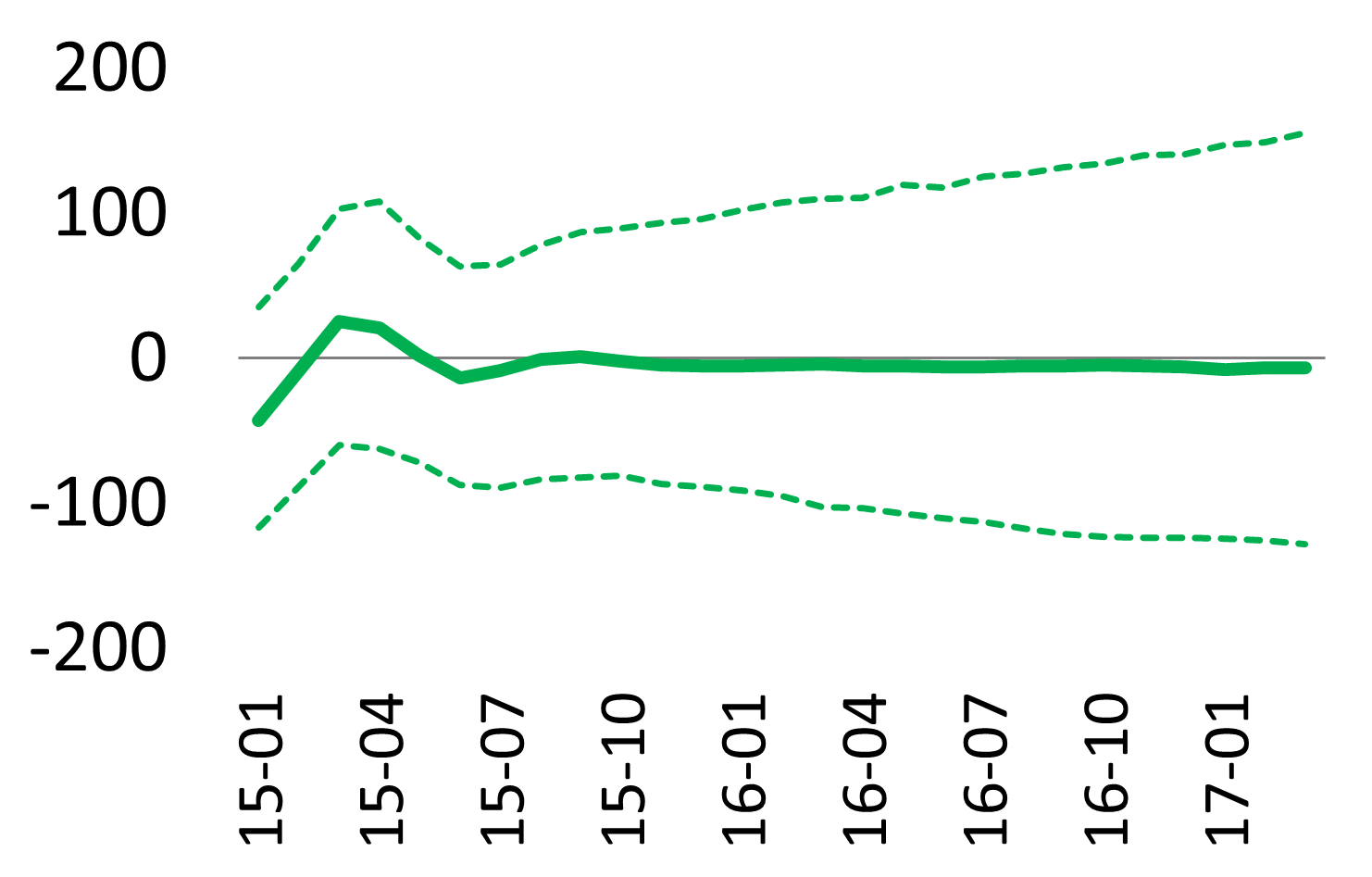}
			\includegraphics[scale=0.11]{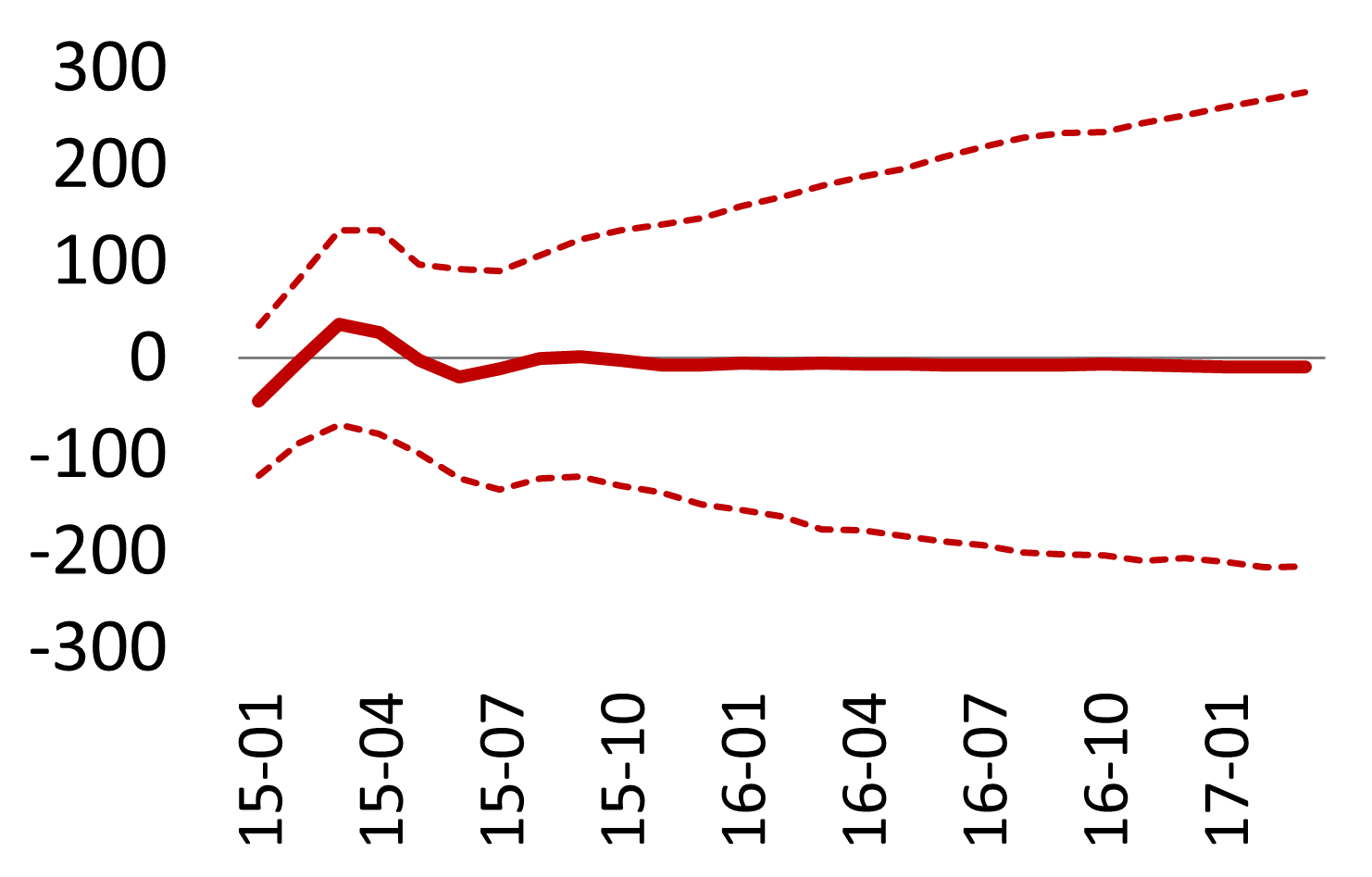}
		\end{minipage}%
		\subcaption*{\textit{Notes}: The figure shows the estimated responses to the shock scenario for the APP programme (left panel) and bootstrap 90\% confidence intervals and medians (right panels).}
	\end{figure}
	
	\begin{figure}[H]
		\centering		
		\caption{Response of repo trade volumes, shock to APP}
		\begin{minipage}{.5\textwidth}
			\hspace{-15pt}
			\includegraphics[scale=0.15]{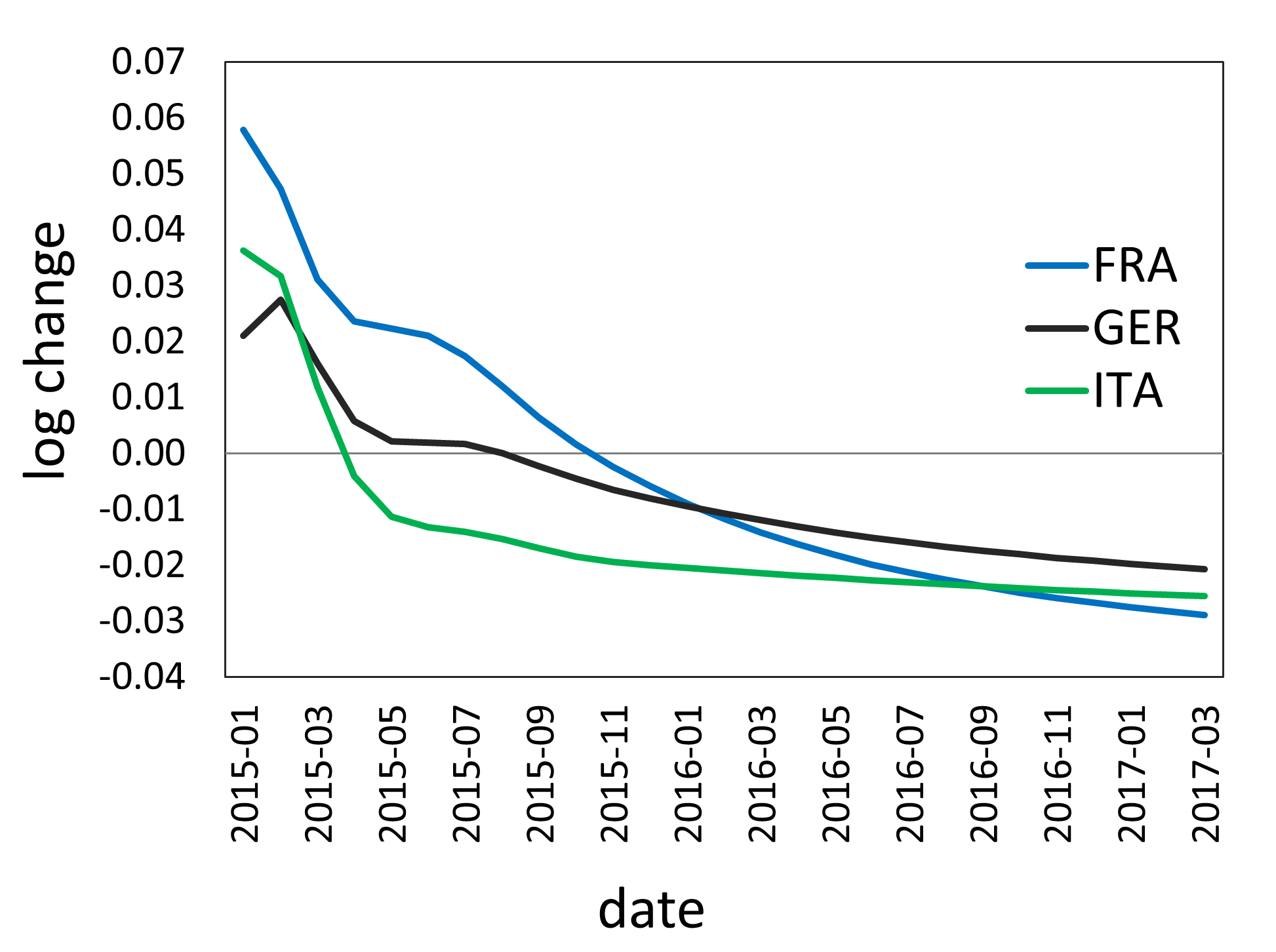}
		\end{minipage}%
		\begin{minipage}{.60\textwidth}
			\includegraphics[scale=0.11]{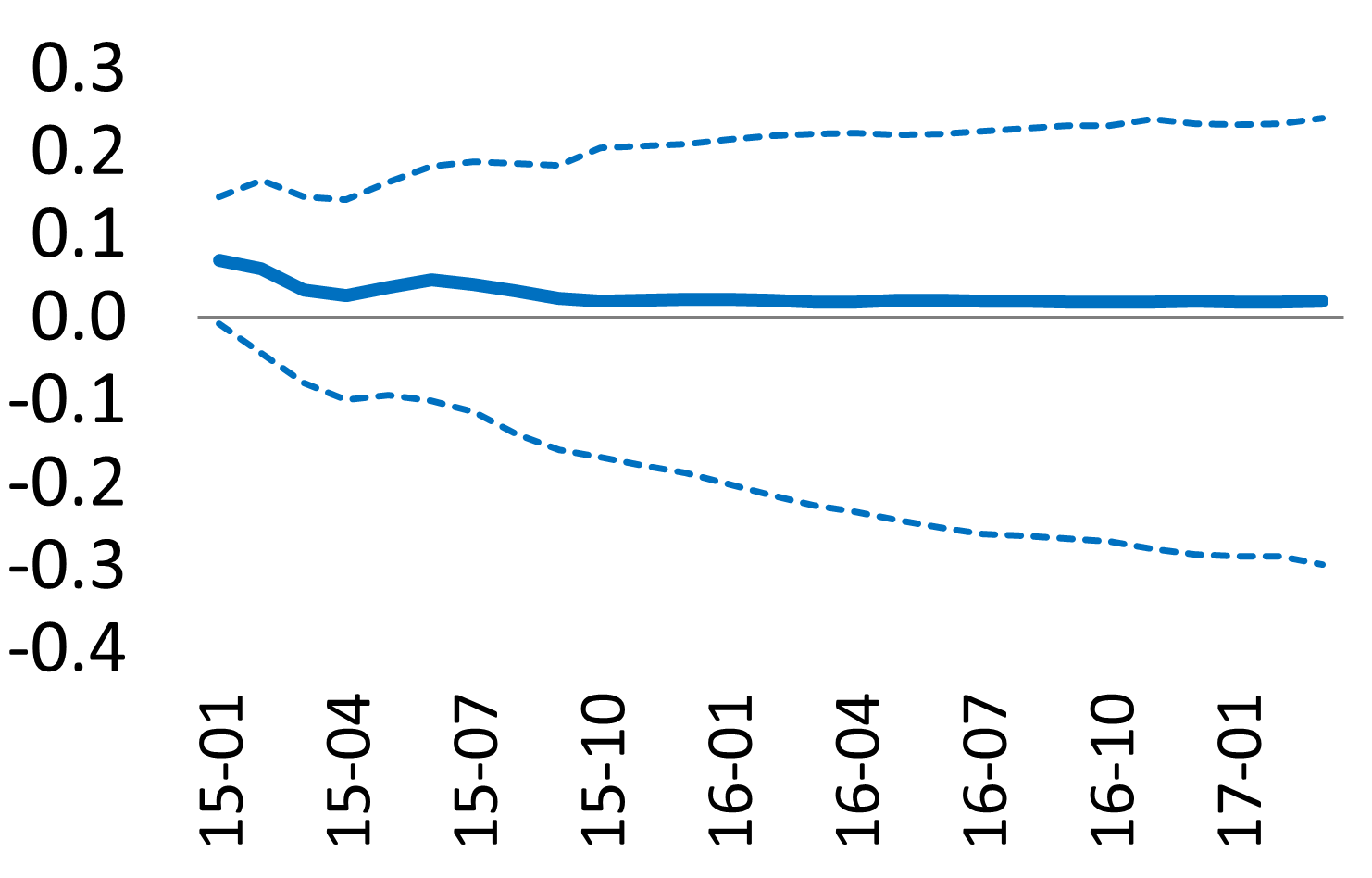}
			\includegraphics[scale=0.11]{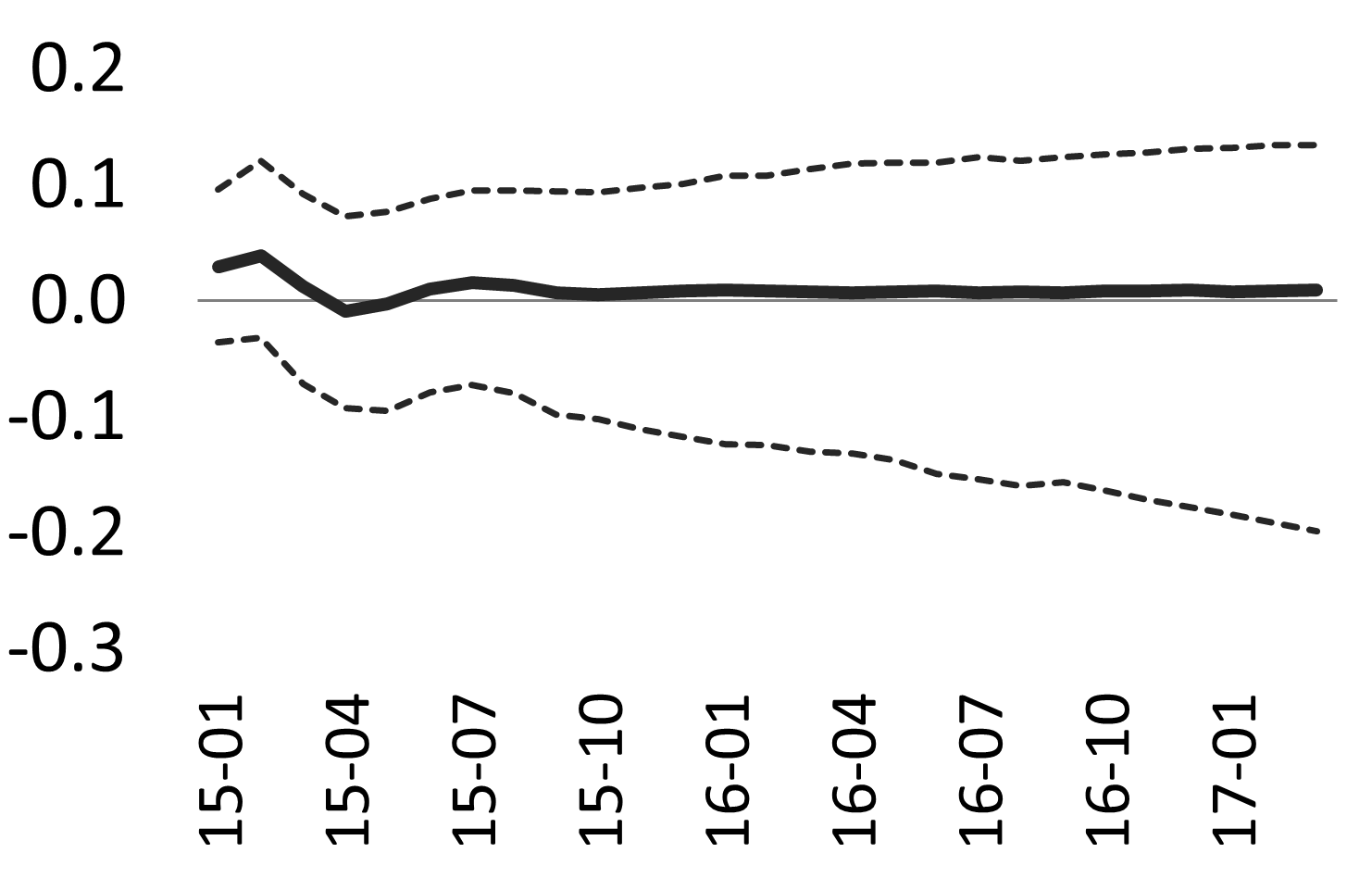}\\
			\includegraphics[scale=0.11]{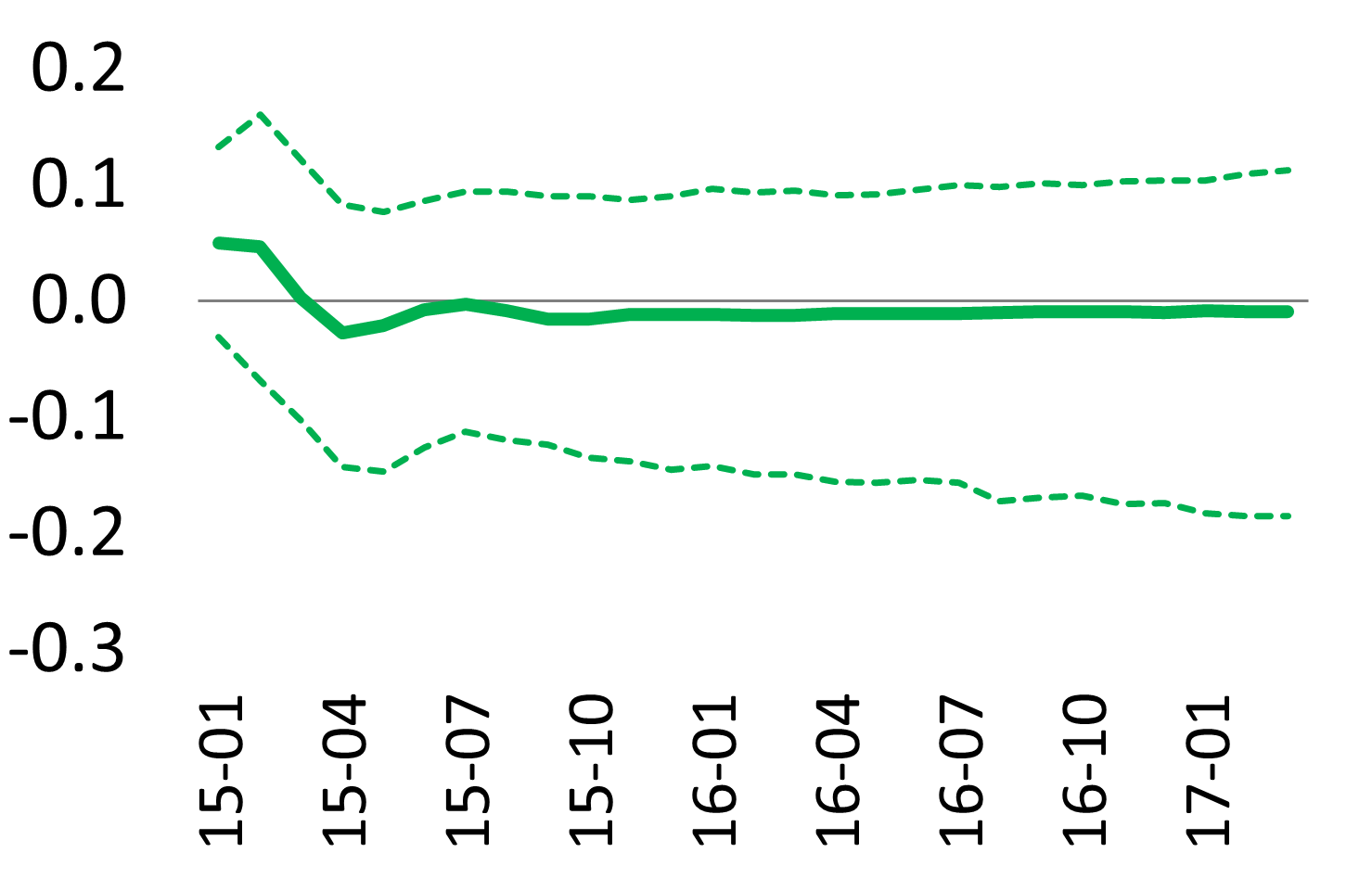}
		\end{minipage}%
		\subcaption*{\textit{Notes}: The figure shows the estimated responses to the shock scenario for the APP programme (left panel) and bootstrap 90\% confidence intervals and medians (right panels).}
	\end{figure}

	\section{Conclusions}
	
	In this paper, we have developed a financial GVAR model to study the empirical relationships between repo funding conditions, sovereign bond markets and banks' credit risk in major Euro area economies, as well as the effects of the ECB's asset purchase programmes. 
	Although affected by considerable parameter uncertainty, impulse response analysis suggests that the ECB's unconventional monetary policy prevented drops in repo funding liquidity in the Euro area as well as sizeable increases in Italian and Spanish government yield spreads and bank CDS spreads.
	Overall, the results show economically meaningful but only marginally significant differences in the dynamics of liquidity and credit risk between Germany and France, on one side, and Italy and Spain, on the other.

	\bibliography{Financial_GVAR}

\end{document}